\newcommand\DLB{{\cal D}_{\text{LB}}}
\newcommand\DNN{{\cal D}_{\text{NN}}}
\newcommand\Figref[1]{Figure~\ref{#1}}
\newcommand\LQCD{\Lambda_{\text{QCD}}}
\def\Mlb{{M_{\ell b}}}
\def\Mlbmin{{M_{{\ell b}_{\min}}}}
\def\Mlbminsq{{M^2_{{\ell b}_{\min}}}}
\def\Mlbsq{{M^2_{\ell b}}}
\newcommand\Secsref[1]{Sections~\ref{#1}}
\newcommand\Tabref[1]{Table~\ref{#1}}
\def\Vtb{|V_{tb}|}
\def\Vtd{|V_{td}|}
\def\Vts{|V_{ts}|}
\newcommand\aplan{{\cal{A}}}        %
\newcommand\bbar{\overline{b}}               %
\newcommand\bbbar{b\overline{b}}             %
\newcommand\brocket[1]{\left\langle #1 \right\rangle}
\newcommand\calD{{\cal D}}
\newcommand\ccbar{c\overline{c}}             %
\newcommand\chisq{\chi^2}         %
\newcommand\degree{^\circ}                      %
\newcommand\dzero{D\O}                         %
\newcommand\eqref[1]{Eq.~(\ref{#1})}
\newcommand\et{E_T}               %
\newcommand\etal{{\sl et al.}}                 %
\newcommand\etjet{\et^{\text{jet}}}
\newcommand\figref[1]{Fig.~\ref{#1}}
\newcommand\figsref[1]{Figs.~\ref{#1}}
\newcommand\gev{\unit{GeV}}
\newcommand\gevc{\unit{GeV} / c}
\newcommand\gevcc{\unit{GeV} / c^2}
\newcommand\htran{H_T}            %
\newcommand\ifb{\unit{fb}^{-1}}
\newcommand\ipb{\unit{pb}^{-1}}                     %
\newcommand\jet{\text{jet}}
\newcommand\jets{\text{jets}}
\newcommand\kev{\unit{keV}}
\newcommand\lumint{\int\! {\cal L}\, dt}       %
\newcommand\lumunits{\text{cm}^{-2}\,\text{s}^{-1}}        %
\newcommand\mb{m_b}                          %
\def\mch{m_{H^+}}
\newcommand\met{\mbox{${\hbox{$E$\kern-0.6em\lower-.1ex\hbox{/}}}_T$}}%
\newcommand\metcal{\mbox{${\hbox{$E$\kern-0.6em\lower-.1ex\hbox{/}}}_T^{\text{cal}}$}}%
\newcommand\mev{\unit{MeV}}
\newcommand\mfit{m_{\text{fit}}}
\newcommand\mt{m_t}                          %
\newcommand\pb{\unit{pb}}
\newcommand\pbar{\overline{p}}               %
\newcommand\ppbar{p\overline{p}}             %
\newcommand\progname[1]{{\sc\lowercase{#1}}}
\newcommand\pt{p_T}               %
\newcommand\qbar{\overline{q}}               %
\newcommand\qqbar{q\overline{q}}             %
\def\ra{\rightarrow}
\newcommand\secref[1]{Sec.~\ref{#1}}
\newcommand\sigtop{\sigma_{t \overline{t}}}%
\def\statsyst#1#2#3{#1\pm#2~\text{(stat)}\pm#3~\text{(syst)}}
\newcommand\tabref[1]{Table~\ref{#1}}
\newcommand\tabsref[1]{Tables~\ref{#1}}
\newcommand\tbar{\overline{t}}               %
\newcommand\tev{\unit{TeV}}
\newcommand\ttbar{t\overline{t}}             %
\newcommand\ugev{\text{GeV}}
\newcommand\ugevc{\text{GeV} / c}
\newcommand\ugevcc{\text{GeV} / c^2}
\newcommand\umevcc{\text{MeV} / c^2}
\def\unit#1{~\text{#1}}
\def\usefigure#1{\vcenter{\hbox{\epsfig{file=#1}}}}
\newcommand\vecmet{\vec{\met}} %
\newcommand\vs{{\sl vs.}}                      %


\documentstyle[11pt,preprint,tighten,aps,prb,floats,epsfig,mymcite,wsci,verbatim]{revtex}

\newbox\figbox

\begin{document}

\preprint{Fermilab-Pub-98/236}

\title{\Large \bf Top Quark Physics At The Tevatron}
\date{\today}

\author{Pushpalatha C. Bhat}  \address{Fermi      National
Accelerator  Laboratory\thanks{Operated  by the Universities  Research
Association,   under contract with   the  U.S.\ Department of  Energy.}
\\~P.O. Box  500,  Batavia, IL~60510, USA}

\author{Harrison~B.~Prosper} \address{Department  of Physics, Florida  State
University\\Tallahassee, FL~32306, USA} 

\author{Scott S. Snyder}  \address{Brookhaven      National
Laboratory\thanks{Operated  by Brookhaven Science Associates,
under contract with   the  U.S.\ Department of  Energy.}
\\~P.O. Box  5000,  Upton, NY~11973, USA}

\maketitle

\begin{abstract}

The discovery of the  top quark in 1995,
by the CDF and D\O\ collaborations at the Fermilab Tevatron,
marked the dawn of a new era in particle physics.
Since then, enormous efforts have been made to study the properties
of this remarkable particle,
especially its mass and production cross section.
In this article,
we review the status of top quark physics as studied by the two
collaborations using the $\ppbar$ collider data at $\sqrt{s}=1.8\tev$.
The combined measurement of the top quark mass,  $m_t=173.8\pm 5.0\gevcc$,
makes it known to a fractional precision better than any other quark mass.
The production cross sections are measured as
$\sigma_{\ttbar}=7.6_{-1.5}^{+1.8}\ \text{pb}$  by
CDF and $\sigma_{\ttbar}=5.5\pm~1.8\ \text{pb}$ 
by \dzero. Further investigations of $\ttbar$ decays and
future prospects are briefly discussed.  
   
\end{abstract}

\pagebreak
\tableofcontents

\pagebreak
\section{Introduction}
\label{introduction}

The discovery\cite{cdfdiscovery,cdftopprd94,d0discovery} 
of the top~quark in 1995 was a major triumph of the Standard
Model of particle
physics.\mcite{weinberg,*salam,*glashow,*gross1,*gross2,*politzer,*lepewg1}
It was the culmination of
nearly two decades of intense research at accelerators
around the world.
The direct measurement of a large mass for the top~quark, by far 
the heaviest fundamental particle 
known, has caused much excitement. That the mass is close to
the electroweak scale suggests the tantalizing possibility that the
top~quark may play a role in the breaking of electroweak symmetry and
therefore in the origin of fermion masses. The top~quark mass is one
of the most important parameters of the Standard Model.

Since the discovery, the CDF and \dzero\
collaborations have collected more data and performed detailed studies.
They have refined particle identification
techniques and adopted innovative analysis methods, resulting in 
precise
measurements of the top~quark mass and the $\ttbar$ production 
cross section.  There exist several excellent
reviews\cite{wimpenny96,*franklin97} of the work that led to the
discovery and the work done shortly thereafter. Although
we shall touch upon some of the highlights of that exciting
time, the focus
of this review is 
the current status of top~quark physics resulting from the recent
measurements from the Tevatron experiments.

In the remainder of this section, we give
a sketch of the 
Standard Model, followed by discussions 
of the arguments and evidence for the existence
of the top quark that predate the discovery, the
indirect measurements of the 
top quark mass from electroweak data,
and 
the significance of the heavy top quark in taking us beyond the Standard Model.
In \secref{production}, we outline the top quark production  
mechanisms in $\ppbar$ collisions, the decay signatures of $\ttbar$ pairs
and the Monte Carlo modeling of $\ttbar$ events.  The ingredients
involved in making and detecting the top quarks, such as
the Tevatron collider complex, the detectors, particle identification 
techniques, and the 
characteristics of the signal and background are discussed in 
\secref{detection}.  The saga of the early searches for the 
top quark and of its
discovery at the Tevatron is summarized in \secref{discovery}. 
\Secsref{topxs} and~\ref{topmass} describe, respectively, 
the measurements of the  $\ttbar$ production cross section and 
the top quark mass by the CDF and \dzero\ collaborations.  Several other 
studies that have been made using the present $\ttbar$ event samples are 
summarized in \secref{other}.
Finally, in \secref{prospects},
we discuss the prospects for the next collider run, scheduled to begin
in the year~2000.

\subsection{Synopsis of the Standard Model}

The Standard Model (SM), the prevailing theory of matter
and forces, has been in place for over two decades. The particles of matter
are spin-1/2 quarks ($q$) and leptons ($\ell$),
which seem to be elementary, at least
down to $10^{-18}$ meters.  There are six ``flavors'' of quarks, and
likewise of leptons, grouped in pairs into three generations. They
interact via the exchange of spin-1 gauge bosons: eight massless gluons,
the massless photon, and the massive $W^{\pm}$ and $Z^0$ bosons. The
top~quark was the important missing piece in the fermion sector of the
Standard Model. The building blocks of the Standard Model are shown in
\tabref{tab:sm}.

\begin{table}
\centering
\outertabfalse %
\caption{Particles of the Standard Model.\protect\cite{pdg96}}
\vskip 10pt
\label{tab:sm}
\begin{tabular}{rc|c|c|c|c|}
\cline{3-6}
             && symbol      & name              & mass ($\umevcc$) & charge ($e$)
\cr
\cline{3-6}
Quarks       && $u$         & up                & $\approx 5$     & $ 2/3$ \cr
(spin $=1/2$)&& $d$         & down              & $\approx 10$    & $-1/3$ \cr
             && $c$         & charm             & $\approx 1500$  & $ 2/3$ \cr
             && $s$         & strange           & $\approx 200$   & $-1/3$ \cr
             && $t$         & top               & $173.8\gevcc$   & $ 2/3$ \cr
             && $b$         & bottom            & $\approx 4500$  & $-1/3$ \cr
\cline{3-6} 
Leptons      && $\nu_e$     & electron neutrino & $< 10\unit{eV}$ &  $0$   \cr
(spin=$1/2$) && $e$         & electron          & 0.511           & $-1$   \cr
             && $\nu_{\mu}$ & muon neutrino     & $< 0.17$        &  $0$   \cr
             && $\mu$       & muon              & 105.7           & $-1$   \cr
             && $\nu_{\tau}$& tau neutrino      & $< 24$          &  $0$   \cr
             && $\tau$      & tau               & 1777            & $-1$   \cr
\cline{3-6}
Gauge bosons && $\gamma$    & photon            & 0               &  $0$   \cr
(spin $=1$)  && $W$         & $W$               & $80.3\gevcc$    &  $1$   \cr
             && $Z$         & $Z$               & $91.2\gevcc$    &  $0$   \cr
             && $g$         & gluon             & 0               &  $0$   \cr
\cline{3-6}
Higgs        && $H$         & Higgs             & ?               &  ?     \cr
\cline{3-6}
\end{tabular}
\end{table}

A vital part of the Standard Model that awaits experimental
evidence is the ``Higgs mechanism.''
The Standard Model is based on the gauge group 
$SU(3)_C \times SU(2)_L \times U(1)_Y$ and accommodates 
electroweak and flavor symmetry breaking
 by introducing a weak-isospin doublet of 
fundamental scalar fields ${\Phi = {\phi^+ \choose  \phi^0}} $
with the potential function
\begin{equation}
 V(\Phi^\dagger\Phi) = \mu^2(\Phi^\dagger\Phi)+
 |\lambda| (\Phi^{\dagger} \Phi)^2,
\label{eq:higgspotential}
\end{equation}
where $\lambda$ is the 
self coupling of the scalar field.
With $\mu^2$ chosen to be negative, the electroweak
symmetry is spontaneously broken (that is, the vacuum state fails
to display the symmetry of the theory) when the 
scalar field is expanded about its (non-zero) vacuum
expectation value
$v=\sqrt{{-\mu^2}/{\lambda}}=(G_F\sqrt{2})^{-\frac{1}{2}}=246\gev$
(referred to as the  electroweak scale).
The spontaneous breaking of electroweak symmetry endows
the $W ^ \pm $ and $Z ^ 0 $ bosons with masses, 
$M_W^2=\pi\alpha/G_F\sqrt{2}\sin^2\theta_W$ and
$M_Z^2=M_W^2/\cos^2\theta_W$
 and also gives rise to a 
spin-0 (scalar) particle called the Higgs~boson. 
(Here, $\alpha$ is the fine structure constant, 
$G_F$ is the Fermi (weak) coupling constant, and 
$\theta_W$ is the weak angle.)
Each
quark and lepton $f$ has its own Yukawa coupling to the Higgs~boson
$G_f$ and thus acquires a mass $m_f=G_f v/\sqrt{2}$.

The Standard Model has been extremely successful, so far!
\mcite{ellis97,*altarelli98}
It has withstood scores of very stringent experimental tests in a
variety of high-energy interactions.  No significant discrepancies
between experimental data and the Standard Model have yet been found.  But 
several critical issues remain
unresolved.  The inclusion of the Higgs mechanism is artificial: there
is no explanation for the form of the Higgs potential, and therefore 
for neither electroweak
symmetry breaking nor the breaking of flavor symmetry.
Hence, physics beyond the Standard Model seems
inevitable,\cite{wilczek98}
and it is entirely plausible that the top~quark might
be our window to that new physics.

\subsection{Why Must the Top Quark Exist?}

\begin{figure}
\centering
\epsfig{file=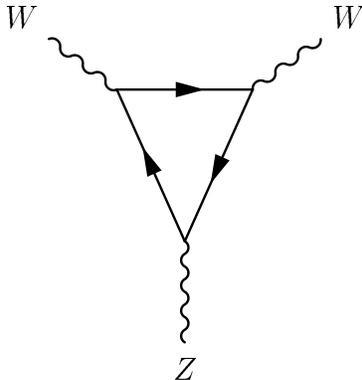}
\caption{An example of a triangle diagram that gives
rise to a chiral anomaly in the Standard Model.}
\label{fig:triangle}
\end{figure}

Long before the top~quark was observed, there were compelling
arguments for its existence. The renormalizability of the Standard Model
requires the cancellation of triangle anomalies --- a problem that arises
from the interaction of three gauge bosons via a closed loop of fermions
as shown in \figref{fig:triangle}. It turns out that the fermion contributions
within each generation cancel if the electric charges of all left-handed
fermions sum to zero:
\begin{equation}
\sum Q_L = -1 + 3 \times \left[\left(\frac{2}{3}\right) +
                              \left(-\frac{1}{3}\right)\right] = 0.
\end{equation} 
The factor 3 is the number of color charges for each quark flavor. For this to
work for the third generation, the top~quark with $Q = 2/3$ must exist.

There is ample indirect experimental evidence for the
existence of the top~quark.
The experimental limits on
flavor changing neutral current (FCNC) decays of the
$b$-quark\mcite{kane82,*bean87}
such as $b\rightarrow s\ell^+\ell^-$ 
and the absence of large tree level (lowest order)
$B_d ^ 0 {\bar B _d ^ 0} $ mixing at the $\Upsilon (4S) $
resonance~\mcite{roy90,albrecht87,*albrecht94,*bartelt93}
rule out the hypothesis of an isosinglet $b$-quark.  In
other words, the $b$-quark must be a member of a left-handed weak isospin
doublet.

The most compelling experimental evidence comes from the wealth of data
accumulated at $e^+e^- $ colliders in recent years,
particularly the detailed studies of the $Zb\bbar$ vertex
near the $Z$ resonance.
These studies have yielded
a measurement of the isospin
of the $b$-quark.  The $Z$~boson is coupled to the $b$-quarks (as well as
to other quarks) through vector and axial vector charges ($v_b$ and $a_b$)
with  strength\cite{quigg}
\begin{equation}  
  \usefigure{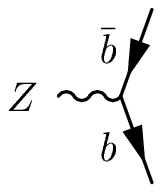}  = \sqrt{\frac {G_F M_Z^2}{2\sqrt{2}}}
\gamma_{\mu} (v_b - a_b\gamma_5),
\end{equation}
where $v_b$ and $a_b$ are given by
\begin{equation}
v_b = 2[I_3^L(b) + I_3^R(b)]  - 4e_b \sin^2\theta_W, \quad\hbox{and}
\end{equation}
\begin{equation}
a_b = 2[  I_3^L(b) + I_3^R(b)   ].
\end{equation}
Here,  $I_3^L(b)$ and $ I_3^R(b)$
are the third  components of the isospin for the left-handed 
and right-handed $b$-quark fields.
The electric  charge of the $b$-quark, $e_b = -1/3 $,  has been 
well established from the $\Upsilon$ leptonic width 
as measured by the DORIS $e^+e^-$ 
experiments.\mcite{bcharge1,*bcharge2,*bcharge3}
The Born approximation in  the limit of 
a zero mass $b$-quark gives for the partial $Z$ boson decay rate
\begin{equation}
\Gamma_{b\bar b} \equiv \Gamma(Z\rightarrow b\bar b)
\approx
\frac {G_F M_Z^3}{8 \sqrt{2}\pi} (v_b^2 + a_b^2).
\end{equation}
The partial width $\Gamma_ {b\bar b}$ is expected to be thirteen times 
smaller if $I_3^L(b)$=0.0.  The LEP measurement
of the ratio of this partial width to the full hadronic decay
width, $R_b = \Gamma_b /\Gamma_{\text{had}} = 0.2170 \pm 0.0009$,
is in excellent agreement with  the SM  
expectations (including the effects of the top~quark) of  
0.2158,\cite{lepewg} ruling out 
$I_3^L(b)$=0.0.  In addition, the forward-backward asymmetry in 
$e^+e^- \rightarrow \bbbar$ at the $Z$ resonance,
\begin{equation}
A_{FB} = {3 a_e v_e a_b v_b \over (v_e^2 + a_e^2) (v_b^2 + a_b^2)},
\end{equation}
is sensitive to the relative size of the vector and axial vector
couplings of the 
$Z\bbbar$ vertex. The sign ambiguity for the two contributions
can be resolved by the $A_{FB}$ measurements from low energy experiments that 
are sensitive  to the interference between neutral current 
and electromagnetic amplitudes.  So, from the measurements of
$\Gamma_{b\bar b}$ and $A_{FB}$ at LEP, SLC, and the low  energy experiments
(PEP, 
PETRA and TRISTAN), 
one obtains\cite{schaile}
\begin{eqnarray}
I_3^L(b) & = & -0.490_{-0.012}^{+0.015},  \\
I_3^R(b) & = & -0.028 \pm 0.056, \nonumber
\end{eqnarray}
for the third component of the isospin of the $b$-quark.
This implies that the $b$-quark must have a weak isospin partner,
i.e., the top~quark, with $I_3^L(t) = +1/2$.

\subsection{Indirect Constraints on the Top Quark Mass}

An upper bound on the top quark mass can be obtained by requiring that
partial wave unitarity be respected at tree level in the reactions
$\ttbar \rightarrow W^+W^-$, $ZZ$, $HZ$, and $HH$.  
This  leads to a condition on the
top~quark mass, which sets the scale $m_t(G_F \sqrt 2)^{-\frac{1}{2}}$ of
$H\ttbar$ Yukawa couplings and a constraint $m_t\leq 500\gevcc$.
\mcite{chanowitz78,*chanowitz79}

\begin{figure}
\centering
\epsfig{file=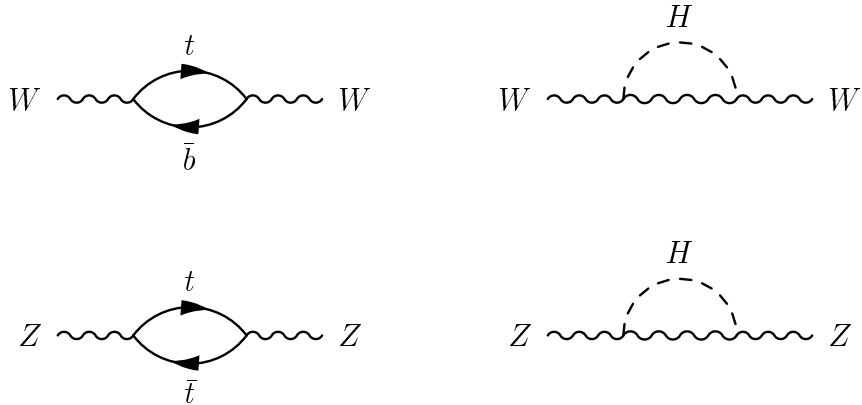}
\caption{Self-coupling loops contributing higher order quantum corrections.}
\label{fig:loop}
\end{figure}

Since virtual top~quarks are involved in higher order electroweak
processes, tighter constraints on the top~quark mass can be obtained
from precision electroweak measurements. The higher-order (radiative)
corrections to many electroweak variables depend on the masses of the
top~quark and Higgs boson via loop diagrams such as those shown in
\figref {fig:loop}.

At one loop, for example, the $\rho$ parameter,
\begin{equation}
\rho = \frac {M_W^2}{M_Z^2(1 -  {\rm sin}^2\theta_W)}
\equiv 1+\Delta r,
\end{equation}
 which relates the $W$ and
$Z$ boson masses and the weak angle, gets a radiative correction
\begin{equation}
\Delta r = \frac{3G_F}{8\pi^2 \sqrt{2} }m_t^2 + 
 \frac{\sqrt{2}G_F}{16\pi^2 }M_W^2
\left[\frac{11}{3} \ln \left(\frac{M_H^2}{M_W^2}\right) + \ldots\right]+\ldots,
\end{equation}
which is quadratic in the top~quark mass.  Note, however, 
that the dependence on the mass of the Higgs~boson $M_H$ is only logarithmic.
Therefore, the top~quark mass, especially if large, is the dominant
parameter in corrections to electroweak processes.  This relation was
used to set early constraints on the mass of the top~quark;\cite{ehlq84}
for example, ignoring the Higgs contribution, $\Delta r < 0.02 (0.01)$
implies that $m_t < 250 (180) \gevcc$.
Additional constraints
can be derived from the large body of precision electroweak data.
Taking as inputs
the extremely high precision measurement of the $Z$~boson mass,
$\sin^2\theta_W$ from the
$Z$~boson's decay rate, the forward-backward asymmetry and left-right
polarization measurements in $Z$~boson decay, and the
$\sin^2\theta_W$ measurement from the $\nu N$ scattering,
a fit\cite{lepewg}
to the Standard Model predictions with $m_t$ and $M_H$ as
free parameters yields $m_t = 157^{+10}_{-9}\gevcc$ and
$M_H=41^{+64}_{-21}\gevcc$.
When the $M_W$ measurements from the Tevatron and LEP are included, 
the resulting mass\cite{martinez98} of the top~quark is
$m_t = 181.3^{+6.1}_{-6.2}~^{+15.7}_{-17.3} \gevcc$, where
the first uncertainty is, as before, the statistical error from the fit and
the second uncertainty reflects the assumed variation of $M_H$ in
the range 70--$1000\gevcc$.
These precision data, along with the direct measurements of the
masses of the top~quark
and the $W$~boson, provide an indirect measurement
of the mass of the Higgs boson.

\subsection{Significance of a Heavy Top Quark}

The top~quark is the heaviest elementary particle yet discovered.
Its mass, of the same order as the electroweak scale ($\sim v/\sqrt{2}$),
is about twice that of the $W$ and $Z$~bosons
and about 40 times larger than its isospin partner, the $b$-quark.
Because of its large Yukawa coupling to the Higgs boson ($G_t \sim 1$),
and hence to the mechanism of electroweak symmetry breaking,
the top~quark may have unique dynamics.
Its mass has already set severe
constraints on extensions to the Standard Model,
including any new theories of strong interactions,
leading to the development of top condensation,
topcolor, and related ideas.

But the most intriguing observation of all is that in
supersymmetric models with grand unification, a large top quark
mass will automatically break electroweak symmetry in the
required manner.\mcite{simmons,*ibanez,*gaume}
At the grand unification scale $M_{\text{GUT}}$, well above
the weak scale $M_W$, all the supersymmetric scalars (the
squarks and sleptons, denoted $\tilde{f}$) will have the same mass:
\begin{equation}
  M_{\tilde{f}}^2 (M_{\text{GUT}}) = m_0^2 .
\end{equation}
The two Higgs scalars will also have the same mass:
\begin{equation}
  M_{h,H}^2 (M_{\text{GUT}}) = m_0^2  + \mu^2 .
\end{equation}
As one moves to smaller energy scales, these masses evolve
according to the renormalization group equations.  For the mass
of the Higgs scalar at a scale $Q$, one finds:
\begin{equation}
  M_h^2 (M_{\text{GUT}}) - M_h^2(Q) \propto m_t^2
                                    \ln {\left(M_{\text{GUT}}\over Q\right)}.
\end{equation}
For a sufficiently large top quark mass ($\mt \sim 175\gevcc$), 
it is therefore possible
that $M_h^2(M_W) < 0$ at the weak scale, which is required to break
electroweak symmetry.  The squark masses evolve in a similar
manner, but with a smaller proportionality constant.  Thus,
one can avoid breaking the $SU(3)$ color symmetry.

\section {Top Quark Production and Decays}
\label{production}

\begin{figure}
\centering
\epsfig{file=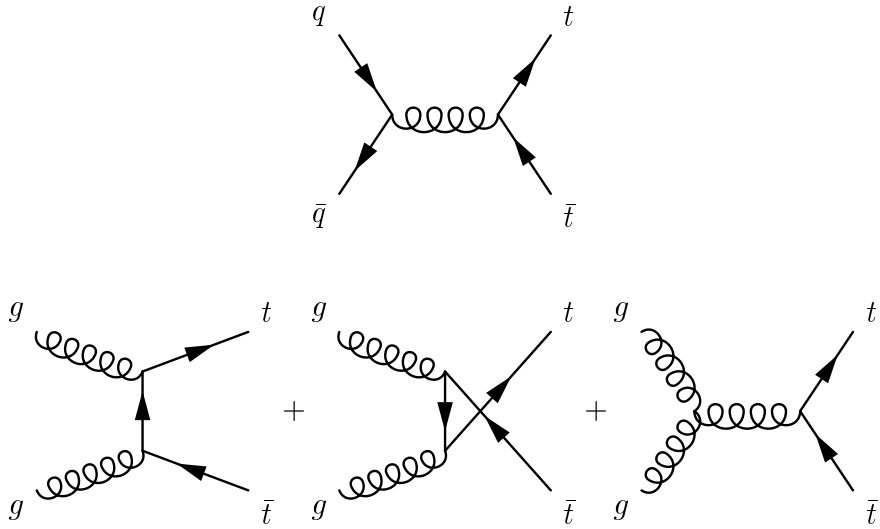}
\caption{Lowest order processes for QCD~$\ttbar$ production.}
\label{fig:topprod}
\end{figure}

\begin{figure}
\centering
\epsfig{file=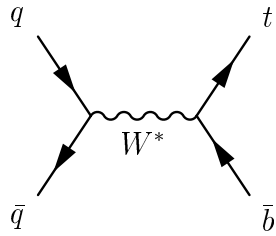}
\caption{Single top quark production via the Drell-Yan process.}
\label{fig:topprod-drellyan}
\end{figure}

\begin{figure}
\centering
\epsfig{file=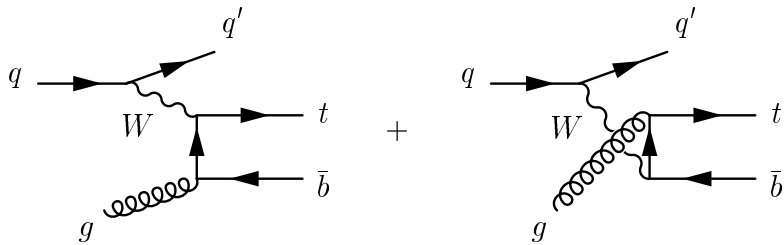}
\caption{Single top quark production via the $W$-gluon fusion process.}
\label{fig:topprod-wgfusion}
\end{figure}

The dominant production mechanism for top quarks at a hadron collider
is pair production (that is, $\ttbar$).  
They can also be produced singly, but with a
rate calculated to be
about half that for pair production.
The main characteristics of these processes are as follows:
\begin{enumerate}
\item \textbf{Pair production of top~quarks} through the
  quantum chromodynamic (QCD) processes $\qqbar \ra \ttbar$
  and $gg \ra \ttbar$.  (See \figref{fig:topprod}.)  At the Tevatron,
  the relative contributions of these two processes are about 90\% and
  10\%, respectively.
  There are also contributions with
  intermediate photons or $Z$~bosons, but they are much smaller and
  can safely be ignored.
\item \textbf{Drell-Yan production of single top~quarks} through
  $\ppbar \ra W^*+X \ra t\bbar+X$.
  (See \figref{fig:topprod-drellyan}.)  This process would have been
  dominant at $\sqrt{s} = 630\gev$
  if $m_t$ were less than ($M_W - m_b$), which is why earlier
  top~quark searches at CERN\mcite{albajar90,*akesson90}
  were based on detecting this mode (see \secref{discovery}).  
  However,
  it still contributes significantly to the inclusive top quark production 
  cross section for $m_t\sim 175\gevcc$.
\item \textbf{Single top quark production via $W$-gluon fusion}.  
  (See \figref{fig:topprod-wgfusion}.)   Photon-gluon and
  $Z$-gluon processes are also allowed, but, again, the rates are very small.
 
\end{enumerate}

In the rest of this section we will discuss QCD pair production of
top~quarks.  The single top~quark production modes will be discussed
in \secref{singletop}.

\subsection{QCD Pair Production of Top Quarks}

Several reviews\cite{cacciari97,frixione97} cover
recent developments in the calculation of heavy-quark production cross
sections.
Here, we will summarize the main issues involved in
the calculation of the $\ttbar$ production cross section in perturbative QCD.

The total production cross section for $\ppbar \ra \ttbar$, where the
proton and antiproton each have momentum $P$ ($=\sqrt{s}/2$), can be
factorized in the standard way:\cite{collins86}
\begin{equation}
\sigma(\ppbar \rightarrow \ttbar) = 
\sum_{a,b} \int dx_a dx_b f_a^p(x_a, \mu ^2)
f_b^{\pbar}(x_b, \mu ^2) \hat {\sigma}(ab\rightarrow \ttbar ; \hat {s},
\mu ^2, m_t),
\end{equation}
where the summation indices $a$ and $b$ run over light quarks and
gluons.  This formula expresses the total cross section in terms of
the parton-parton processes $ab \ra \ttbar$, where $a$ and $b$ are
partons contained in the initial proton and antiproton carrying
momentum fractions of $x_a$ and $x_b$, respectively.  The parton
distribution
functions $f_a^p$ and $f_b^{\pbar}$ are the probability densities of
finding a parton with a given momentum fraction in a proton or
antiproton, and $\hat\sigma$ is the subprocess cross section at a
parton-parton center-of-mass energy of
$\hat {s} = 4 x_a x_b P^2 = x_a x_b s$.  The renormalization and
factorization scales, here chosen to
be the same value $\mu$, are arbitrary parameters.  The first
is introduced by the renormalization procedure, and the second
by the splitting of the total cross section
into perturbative ($\hat\sigma$) and nonperturbative ($f^p$, $f^{\pbar}$)
parts.  The dependence of
observables on $\mu$ is an artifact
of truncating the perturbation expansion at finite order;  if the
calculations could be carried out to all orders, the dependence on $\mu$ would
vanish.  For $\ttbar$ production, one usually takes $\mu \sim m_t$.
Theorists typically estimate the error introduced by truncating the series
by varying $\mu$ within some (arbitrary) range, such as
$m_t/2 < \mu < 2 m_t$.
However, it should be recognized that neither the choice of $\mu$,
nor the range over which it is allowed to vary, have physical significance.

The first calculations of the parton-parton cross section
$\hat\sigma$ were
to $O(\alpha_s^2)$, that is, to leading order~(LO).
\mcite{georgi78,*jones78,*gluck78,*babcock78,*hagiwara79,*combridge79}
At the Tevatron,
the $\qqbar\ra\ttbar$ process dominates, contributing
$90\%$ of the cross section, while the $gg$ process contributes the
other $10\%$.  (The difference between the strengths of these two
subprocesses arises mainly from differences in the parton
distribution functions for quarks and gluons, rather than from
differences in the parton-parton cross sections.)
Subsequently, several
groups calculated the complete $O(\alpha_s^3)$ next-to-leading
order~(NLO) cross section.
\mcite{nason88,*nason89,*beenakker89,*beenakker91,*mangano92,*altarelli88,*ellis91}
The NLO cross section is about $30\%$ higher than the LO~cross
section; the estimated uncertainty (from the sensitivity to
variations in $\mu$ and changes in the parton distribution functions)
is typically 10--$20\%$.

In a regime where perturbation theory is
valid, the NLO contribution
should be small compared to the LO terms.  However, for top~quark
production at the Tevatron, the NLO contribution is
worryingly large: for the $gg$ process it
is about $70\%$ of the size of the LO terms.\cite{laenen94}  (The
situation is better for the $q\qbar$ process, where the
NLO contribution is about $20\%$ that of LO.)  The
large difference between the LO and NLO calculations 
is mainly due to processes
involving the emission of soft initial state gluons.  Fortunately,
it is possible, through a technique called \emph{resummation}, to
calculate the sums of the dominant logarithms from soft gluon emission
to all orders in perturbation theory.
This was first carried out by Laenen,
Smith, and van Neerven
(LSvN).\mcite{laenen94,*laenen92,*kidonakis95}  A difficulty
arises, however, in that the resummed gluon series is divergent
due to nonperturbative effects as $\alpha_s$ becomes
large.  LSvN solved
this problem by introducing a new scale $\mu_0 \gg \LQCD$ which is
used as a cutoff to remove this divergence.  They predict an increase
in the cross section by $10\%$ over the NLO prediction; the uncertainty,
however, is relatively large ($>10\%$), due to the dependence on $\mu_0$.

\begin{figure}
  \begin{center}
    \epsfig{file=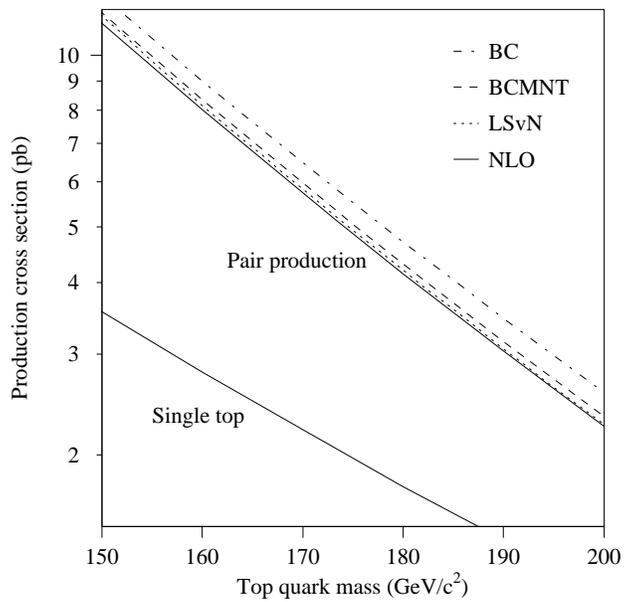}
    \caption{Results of top quark production cross section
      calculations for $\sqrt{s} = 1800\gev$.
      The upper group of curves are for $\ttbar$
      production,\protect\cite{nason88,laenen94,berger98,bonciani98} and the
      lower curve is for single top quark
      production.\protect\cite{heinson97b}  Note that the LSvN result uses an
      older set of structure functions, and is thus systematically low
      compared to the other results.\protect\cite{berger96}}
    \label{fig:xsec1800}
  \end{center}
\end{figure}

More recently, two other groups have performed this calculation using
methods which avoid the need for an arbitrary cutoff.
Berger and
Contopanagos (BC),\mcite{berger95,berger96,*berger97,berger98}
use the technique of
principal value resummation.\mcite{contopanagos93,*contopanagos94}
They also find an increase
of about $10\%$ over the NLO prediction, with an estimated uncertainty of
about $5\%$.  Bonciani, Catani, Mangano, Nason, and
Trentadue (BCMNT)\mcite{catani96,bonciani98} use a slightly different
scheme to avoid
the divergence and treat the subleading log terms differently.
They find a much smaller enhancement of the NLO cross section, on the
order of $1\%$, with estimated uncertainties also of about $5\%$.  A full
discussion of the differences between these calculations is beyond the
scope of this review, but the subject has been discussed
in detail in the literature.\cite{frixione97,berger98,bonciani98}
Note, however, that
if one is comparing these calculations to the present experimental
results, the discrepancy between them
is of no practical importance, as the
difference between them is substantially smaller than the uncertainties
on the experimental measurements ($\approx 30\%$).  The results of
various calculations for $m_t = 175\gevcc$ are compared in
\tabref{tab:xsec-comparison}.  A plot of the various cross sections is
given in \figref{fig:xsec1800}.

\begin{table}
  \caption{Results of several different $\ppbar\ra\ttbar$
    calculations, for $m_t = 175\gevcc$ and $\sqrt{s} = 1.8\tev$.
    Note that the LSvN result uses an older set of structure functions,
    which makes it systematically low compared to the other
    results.\protect\cite{berger96}}
  \begin{center}
    \begin{tabular}{lccc}
      Calculation & Type & Structure Function   & $\sigma_{\ttbar}$         \\
      \hline
      (1) Exact NLO\cite{nason88,bonciani98}&NLO only& MRSR2\cite{martin96}
          & $4.87^{+0.30}_{-0.56}\pb$ \\
      (2) LSvN\cite{laenen94}   & Resummed & MRSD$'$\cite{martin93}
          & $4.94^{+0.71}_{-0.45}\pb$ \\
      (3) BC\cite{berger98}     & Resummed & CTEQ3\cite{cteq95}
          & $5.52^{+0.07}_{-0.42}\pb$ \\
      (4) BCMNT\cite{bonciani98}& Resummed & MRSR2\cite{martin96}
          & $5.06^{+0.13}_{-0.36}\pb$ \\
    \end{tabular}
  \end{center}
  \label{tab:xsec-comparison}
\end{table}

It should be appreciated that these cross sections are extremely
small --- about ten orders of magnitude smaller than 
the total inelastic $\ppbar$
cross section.  Of the five~trillion or so collisions  which have
occurred at each of the CDF and \dzero\ interaction regions 
during Run~1, one
expects about~600 $\ttbar$ pairs to have been produced.
It should also be noted that all these calculations
assume a minimal form of the Standard Model.  Certain extensions
to the SM, such as models with two Higgs~doublets or supersymmetry,
predict $\ttbar$ production cross sections which are different from
that of the SM by a few percent.
\mcite{hollik97,*lampe97,*moretti98}

\subsection{Top Quark Hadronization and Decay}
\label{topdecay}

Within the Standard Model, the dominant decay of a top~quark is via
$t\ra Wb$, with a branching ratio of nearly $100\%$.  The decays
$t\ra Ws$ and $t\ra Wd$ are also allowed, but are suppressed by
factors of $10^{-3}$--$10^{-4}$ by the Cabibbo-Kobayashi-Maskawa (CKM)
mixings.\cite{pdg96}  Other decays, such as flavor-changing neutral
current (FCNC) decays, are predicted to be many orders of magnitude smaller.
We will discuss these further in \secref{other}.

Owing to the large mass of the top~quark, its lifetime is extremely
short
($\Gamma \approx 1.5\gev$, corresponding to
$\tau \approx 5\times 10^{-25}\unit{s}$);
so short, in fact, that at Tevatron energies it decays before it has
a chance to hadronize.\mcite{orr91,*bigi86}  This implies
that a decaying top~quark can be treated as a free particle.  Note,
however, that it isn't just the large mass of the top~quark which gives it
its short lifetime, but also the fact that it has a CKM-allowed 
decay into a $b$-quark.  A fourth-generation down-type quark
$b'$ of mass comparable to the top~quark might still have a long
lifetime if all its decay modes were suppressed by the CKM mixings.

A $\ttbar$ final state contains two $t\ra Wb$ decays.
The two $b$-quarks will
form jets, while each $W$~boson will decay into either a
lepton-neutrino or a quark-antiquark pair.  To a good
approximation, each possible decay of the $W$~boson is equally
probable; however, one must remember to count each quark flavor three
times, since quarks come in three colors.  Therefore, the probability for
a $W$~boson to decay into each of the three lepton flavors is about $1/9$,
while the probability for it to decay into the two available
quark final states is about $2/3$.

\begin{table}
\centering
\outertabfalse %
\def\wdec#1{$W \rightarrow #1$}
\def\wldec#1{\wdec{#1\nu_{#1}}}
\caption{Decay modes for a $\ttbar$ pair and branching fractions.}
\vskip 10pt
\begin{tabular}{|lr|c|c|c|c|}
\cline{3-6}
\multicolumn{2}{c|}{}& \wldec{e}  & \wldec{\mu}& \wldec{\tau} & \wdec{q\qbar}\\
\multicolumn{2}{c|}{}& (1/9)      & (1/9)      & (1/9)        & (2/3)        \\
\hline
\wdec{q\qbar} & (2/3)& $12/81$    & $12/81$    & $12/81$      & $36/81$      \\
              &      &($e+\jets$)&($\mu+\jets$)&($\tau+\jets$)&(all jets)    \\
\hline
\wldec{\tau}  & (1/9)& $2/81$     & $2/81$     & $1/81$ &\multicolumn{1}{c}{}\\
              &      & ($e\tau$)  & ($\mu\tau$)& ($\tau\tau$)
                                                        &\multicolumn{1}{c}{}\\
\cline{1-5}
\wldec{\mu}   & (1/9)& $2/81$     & $1/81$     & \multicolumn{2}{c}{}        \\
              &      & ($e\mu$)   & ($\mu\mu$) & \multicolumn{2}{c}{}        \\
\cline{1-4}
\wldec{e}     & (1/9)& $1/81$     & \multicolumn{3}{c}{}                     \\
              &      & ($ee$)     & \multicolumn{3}{c}{}                     \\
\cline{1-3}
\end{tabular}
\label{tab:topclass}
\end{table}

Since there are two top~quarks in each event, and since the $W$~bosons
decay independently of each other, the events can be classified
according to
how the $W$~bosons decay (see \tabref{tab:topclass}).
\begin{itemize}
\item Events in which both $W$~bosons decay leptonically are called
  \emph{dilepton} events.  Since tau~leptons are difficult to identify,
  the particular dilepton channels which have been
  most studied are the $ee$, $\mu\mu$, and $e\mu$ channels.  These
  final states have the signature of two high-$\pt$ leptons, a large
  imbalance in the total transverse momentum (``missing-$\et$,''
  or $\met$), and two $b$-jets.
  These events are expected to have small backgrounds (especially the
  $e\mu$ channel).  However, as can be seen from \tabref{tab:topclass}, they
  also have small branching fractions, with all three of these
  channels comprising only about $4/81 \approx 4.9\%$ of $\ttbar$
  decays.  (There is also a small contribution from the $\tau$~lepton
  channels.  For example,
  $\ttbar \ra b\bbar\,e\nu\,\tau\nu \ra b\bbar\,e\nu\,e\nu\nu$ can contribute
  to the $ee$ channel.)
  Dilepton events also have the drawback of containing two unobserved
  neutrinos in the final state, which prevents complete
  reconstruction of the event kinematics.

  Several recent analyses have also considered
  the $e\tau$ and $\mu\tau$ channels.
  These will be discussed in \secref{cdfxs}.

\item Events in which one $W$~boson decays leptonically and the other
  decays into quarks are called \emph{lepton+jets} events.  Those
  which have been studied are the $e + \jets$ and
  $\mu + \jets$ channels.  They are characterized by a final state
  containing one high-$\pt$ lepton, large $\met$, and four jets, two
  of which are $b$-jets.  Compared to the dilepton channels, the
  lepton+jets channels have a much larger cross section --- the
  branching ratio for each is about
  $4/27 \approx 15\%$.  (Again, there is a small additional
  contribution from $\tau$~lepton channels.)  The final state
  contains only one neutrino, so there is sufficient information to
  completely reconstruct the event (once a particular set of assignments
  of jets to the final state partons
  is assumed).
  The disadvantage of these channels, however, is a large
  background from inclusive $W$~boson production with associated jets,
  plus a smaller background from QCD jet production.

\item Events in which both $W$~bosons decay into quarks are called
  \emph{all-jets} events.  This final state consists of six jets,
  of which two are $b$-jets,
  no high-$\pt$ leptons, and small $\met$.  This channel boasts the
  largest branching ratio ($\approx 44\%$ of the total).
  Unfortunately, that is more than countered by a huge
  background from QCD multijet processes.  Nevertheless, it is still
  possible to isolate a $\ttbar$ signal in this channel.  The
  techniques for doing so will be discussed in \secref{topxs}.
\end{itemize}

\subsection{Modeling Top Quark Events}

Accurate modeling of the kinematics of $\ttbar$ production and decay
is essential for extracting reliable information from the data.
The most widely used  general-purpose model is that provided by the
\progname{HERWIG} Monte Carlo program.\cite{herwig}
\progname{HERWIG} models $t\tbar$ production starting with the
leading-order hard process, choosing the parton momenta according to
the weight given by the matrix element of the process.
Gluon emission from both the initial
and final states
is modeled using leading-log QCD evolution,\cite{altarelli77} keeping
track of the correlations induced by
color strings between partons.  Each top~quark is then
decayed to a $W$ boson and a $b$-quark, and partons remaining in the
final state are
hadronized into jets. Products of interactions among the beam remnants,
called the \emph{underlying event}, are also
included in the model.  Detector effects are then added using
a model of the detector response to the physical objects.

\begin{figure}
\centering
\epsfig{file=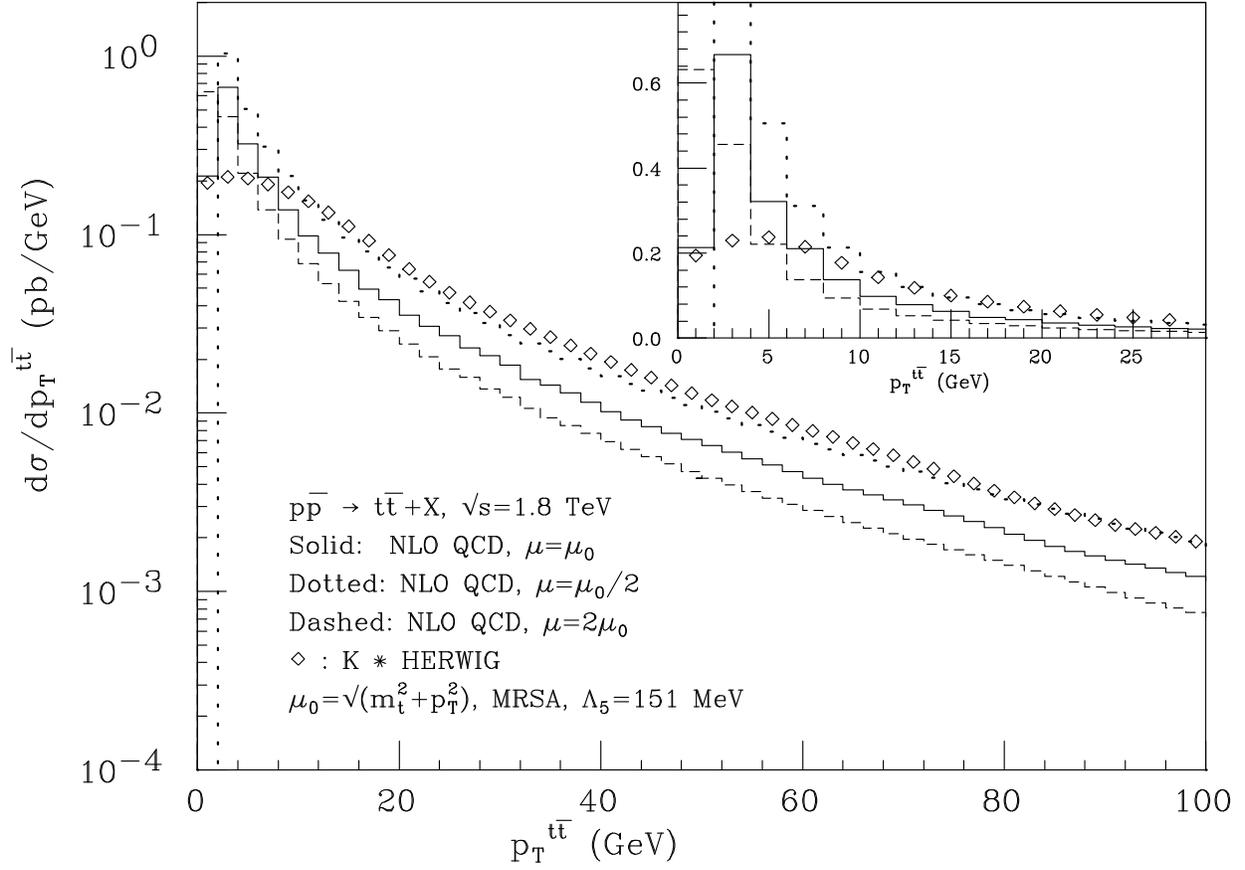,width=\hsize}
\caption{Transverse momentum of the $\ttbar$ pair: NLO calculation
  compared to \progname{herwig}.  The inset magnifies the low $\pt$
  portion of the plot on a linear scale.
  The \progname{herwig} prediction has
  been scaled by a constant factor of $K = 1.34$.
  From Ref.\protect\onlinecite{frixione97}.}
\label{fig:ttbpt-frixione}
\end{figure}

There have been several studies comparing the predictions of
parton shower Monte Carlo programs such as \progname{herwig} to the more
explicit calculations.
Frixione \etal\cite{frixione95}
compare \progname{herwig} to the full NLO calculation.  They distinguish
two types of quantities: those which are delta functions at
leading order, such as $\pt(\ttbar)$ and $\Delta\phi(\ttbar)$,
and those which are nontrivial at leading order, such as
$\pt(t)$, $\eta(t)$, and $m_{\ttbar}$.  For the latter, nontrivial,
set of quantities they find good agreement between the Monte Carlo
predictions and the full NLO calculation.  For the other quantities,
however, they find significant disagreement in the low-$\pt$ region 
(see \figref{fig:ttbpt-frixione}).
They interpret this as a deficiency in
the NLO calculation due to the lack of resummation effects, and
conclude that \progname{herwig} is more reliable than the NLO
calculation in that region.  Orr \etal\cite{orr97} compare
an $O(\alpha_s^3)$ calculation of gluon emission in $\ttbar$ events
to \progname{herwig}; they find that \progname{herwig} seems to
generate too much final state radiation.  On the other hand,
Mrenna \etal~\cite{mrenna97}
compare a resummed calculation for $\ttbar$
production to a different Monte Carlo program,
\progname{pythia},\cite{pythia}
and find that the latter
generates too little radiation.  These discrepancies are probably not
large enough to be important for the current experimental results.
But they will have to be understood better for the next round of
experiments --- radiation effects are already one of the dominant
uncertainties for the present measurements of the mass of the
top quark, so the precision of the measurement may not improve
much until these effects are better understood.

\section{Detecting the Top Quark}
\label{detection}

In this section, we discuss some of the experimental aspects
of detecting the top quark.  We first summarize the apparatus used:
the Tevatron collider and the CDF and \dzero\ detectors.  We then
discuss the procedures used by the experiments to identify final
state objects, such as electrons, muons, and jets.  We conclude this
section with a summary of the distinctive characteristics of the
top quark signal and its principal backgrounds.

\subsection{The Accelerator}

The Fermilab $\ppbar$ collider\mcite{tevrev,*tev1rep}
in Batavia, Illinois is the world's
highest energy particle accelerator, with a center-of-mass energy of
$1800\gev$.
It is the only facility at present capable of producing
top quarks for direct study.

\begin{figure}
\centering
\epsfig{file=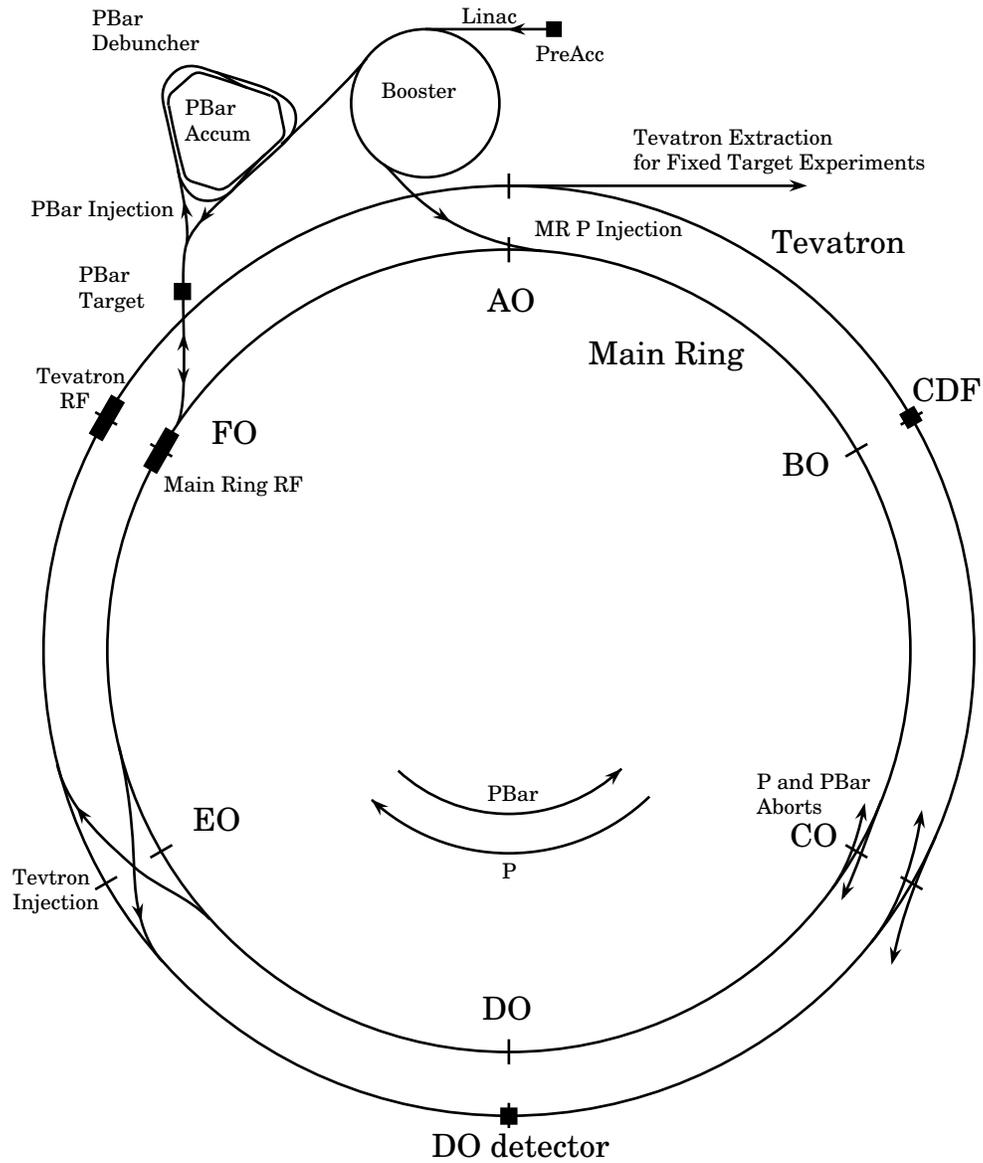}
\caption{Schematic of the Fermilab accelerator complex (not to scale).
         From Ref.\protect\onlinecite{joeythesis}.}
\label{fig:tevatron}
\end{figure}

A schematic of the accelerator complex is shown in
\figref{fig:tevatron}.  Negative hydrogen ions are first accelerated
to $750\kev$ by an electrostatic Cockroft-Walton accelerator and then
further boosted to $400\mev$ by a $150\unit{m}$ long linear
accelerator.  At the end of this accelerator, the electrons are stripped
from the ions and the resulting protons enter the booster.  This
$75\unit{m}$ radius synchrotron 
accelerates the protons to $8\gev$.
From there they are injected into the Main Ring.

The Main Ring is a large ($1000\unit{m}$ radius) synchrotron, composed
of conventional electromagnets, which accelerates protons and antiprotons
up to $150\gev$ for injection into the Tevatron.  It also serves as a
source of $120\gev$ protons for producing antiprotons.\cite{apsource}
The antiprotons are collected from the production target using a
lithium lens, momentum-selected around $8\gev$, and then directed
first into the Debuncher and then into the Accumulator.  These are two
concentric storage rings with radii of about $80\unit{m}$.  There,
the antiprotons are stochastically cooled\cite{stcool} to
reduce their momentum spread.  When enough antiprotons have been
accumulated (stacked), they are extracted into the Main Ring,
accelerated, and injected into the Tevatron.

The Tevatron, a synchrotron made from superconducting magnets, is
situated just below the Main Ring.
In collider mode, the Tevatron is
filled with six bunches of protons and six bunches of antiprotons,
circulating in opposite directions.  The beams are accelerated to the
maximum energy of $900\gev$ each and brought into collision at the
CDF and \dzero\ experimental areas. The beams are typically
kept colliding for about 20 hours, after which the
machine is emptied and refilled with new batches of protons and
antiprotons.  The length of each bunch is about $50\unit{cm}$,
dictated by the accelerator RF system, giving a luminous region
at each interaction point which is roughly Gaussian with a longitudinal
width of
about $30\unit{cm}$.  This relatively long bunch length produces a
degradation of the transverse energies at the trigger level
because the information about the position of the interaction point is
not available at the early stages of triggering.  At CDF, it also
results in a substantial loss of acceptance of the vertex detector for
$b$-quarks (see \secref{svxtag}).

The Main Ring lies mostly in a plane, except at the CDF and \dzero\
experimental areas where it is bent into overpasses to allow room for
the detectors.  The separation between the Main Ring and the Tevatron
is $5.8\unit{m}$ at CDF and $2.3\unit{m}$ at \dzero.  At CDF, the overpass
clears the detector.  At \dzero, however, the Main Ring overpass goes
through the outer (coarse hadronic) part of the calorimeter.
This is unfortunate
because during normal collider data-taking, the Main Ring is
used for antiproton production and losses from the Main Ring may deposit
energy in the detectors, thereby increasing background.
\dzero\ rejects much of this background at the trigger level by rejecting
triggers that occur during injection into the Main Ring, when losses
are large.  Some triggers are also disabled whenever a Main Ring bunch
passes through the detector or when losses are registered in scintillation
counters around the Main Ring.  This results in a loss of about $15\%$
of the available livetime.  The problem of Main Ring contamination is
far less severe at CDF; even so, CDF observes occasional events
with extra energy from the Main Ring.  These events are rejected
offline.

The collider was commissioned with a short run in~1985, followed by
the first high luminosity run in 1988--1989.  Only the CDF detector took
data during that run, which had a peak luminosity of about
$2\times 10^{30}~\lumunits$ (a factor of two greater than the
design luminosity).  The second series of runs
took place over the period 1992--1996,
during which a peak luminosity of $2\times 10^{31}~\lumunits$ was
achieved.  This period was divided into three runs, designated Run~1a,
Run~1b, and Run~1c; the delivered integrated luminosities for these runs
were about $23\ipb$, $122\ipb$, and $17\ipb$, respectively.
The results covered in this review are from this running period.
\Tabref{tab:tevatron} reviews the major parameters of the collider.

\begin{table}
\caption{Parameters of the Fermilab Tevatron collider for Run 1.}
\begin{tabular}{ll}
Accelerator radius            & $1000\unit{m}$   \\
Maximum beam energy           & $900\gev$  \\
Injection energy              & $150\gev$  \\
Peak luminosity               & $\approx 2\times 10^{31}~\lumunits$ \\
Number of bunches             & 6 $p$, 6 $\pbar$\\
Intensity per bunch           & $\approx 10^{11} p$,
                                $\approx 5\times 10^{10} \pbar$ \\
Crossing angle                & $0\degree$                      \\
Bunch length                  & $50\unit{cm}$                   \\
Transverse beam radius        & $\approx 25\unit{$\mu$m}$       \\
Fractional energy spread      & $0.15\times 10^{-3}$            \\
RF frequency                  & $53\unit{MHz}$                  \\
$\pbar$ stacking rate         & $\approx 3.5\times 10^{10}/\text{hour}$ \\
Beam crossing frequency       & $290\unit{kHz}$                 \\
Period between crossings      & $3.5\unit{$\mu$s}$              \\
\end{tabular}
\label{tab:tevatron}
\end{table}

\subsection{The CDF and \dzero\ Detectors}

\begin{figure}
\hbox to \hsize{%
\begin{minipage}[b]{.46\linewidth}
\epsfig{figure=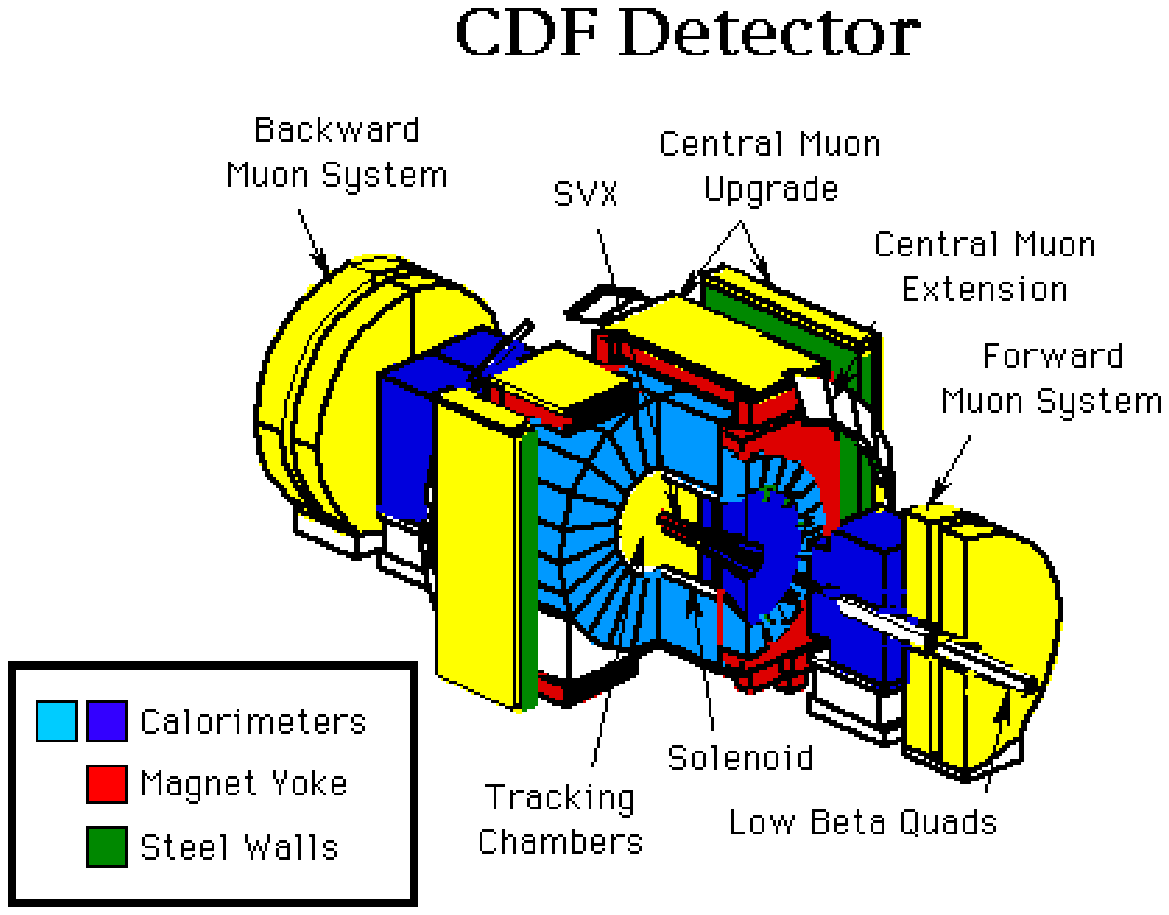,width=\hsize}
\end{minipage}
\hfill
\begin{minipage}[b]{.46\linewidth}
\epsfig{figure=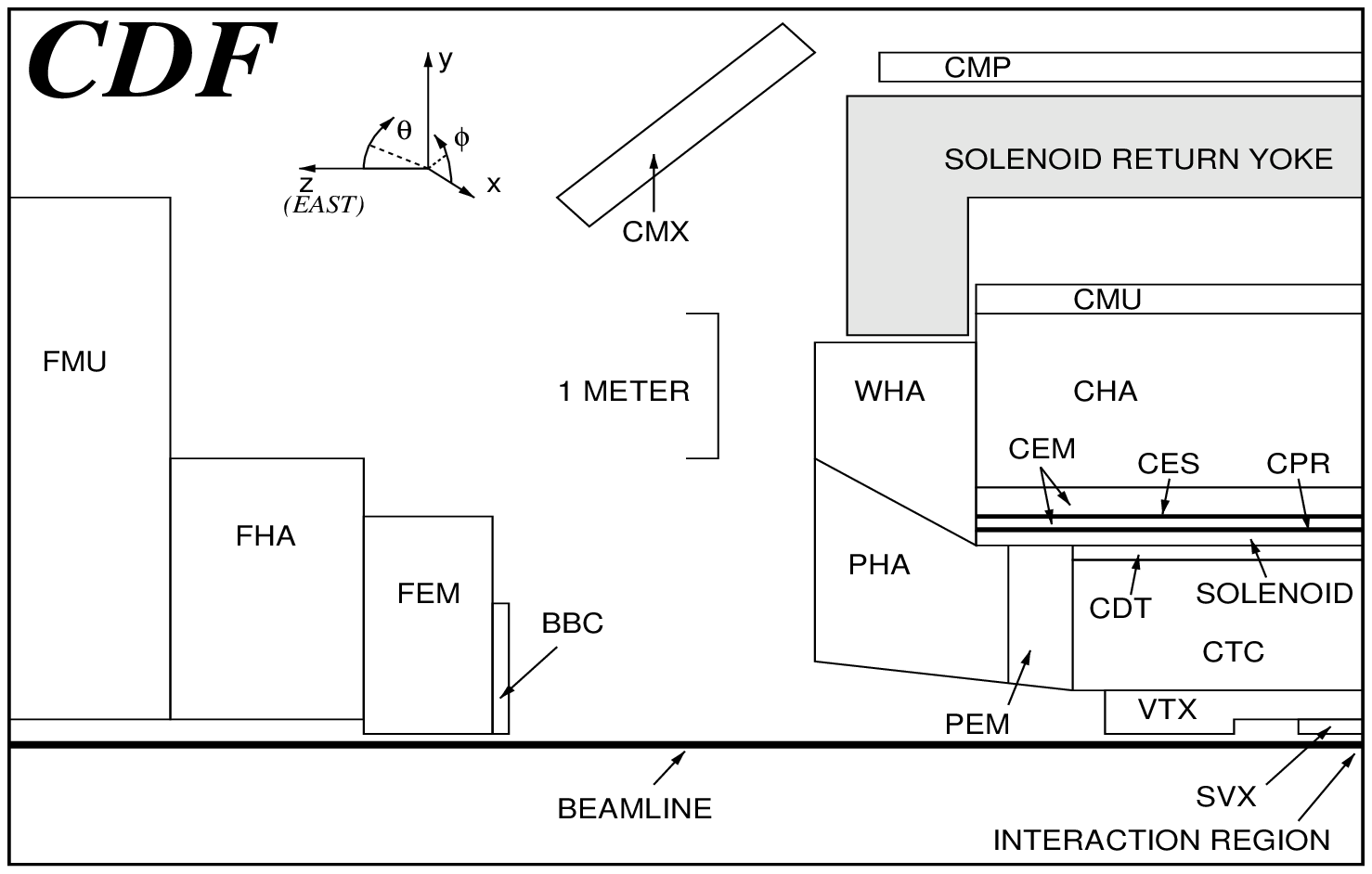,width=\hsize}
\end{minipage}}
\caption{The CDF detector, in isometric view (left) and cross sectional view
  of one quadrant (right).
  The detector is forward-backward symmetric about the interaction
  region, which is at the lower-right corner of the cross sectional view.
  SVX, VTX, CTC, and CDT are tracking detectors, CEM, CHA, WHA, PEM,
  PHA, FEM, and FHA are calorimeters, CMU, CMP, CMX, and FMU are muon
  chambers, BBC is a scintillation counter, and CPR and CES are
  multiwire proportional chambers.
  From Ref.\protect\onlinecite{cdftopprd94}.}
\label{fig:cdfdet}
\end{figure}

\begin{figure}
\hbox to \hsize{%
\begin{minipage}[b]{.46\linewidth}
\epsfig{figure=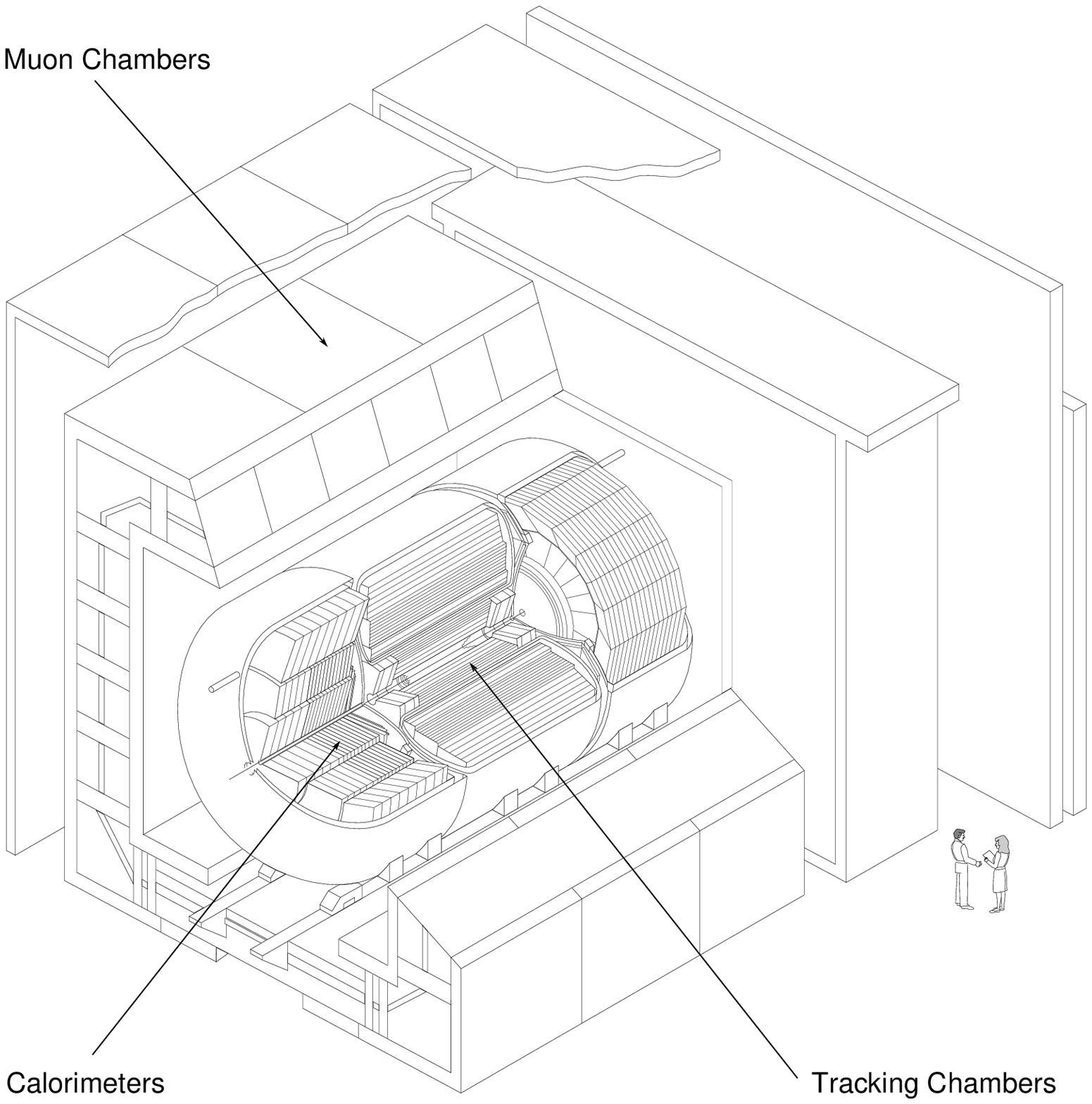,width=\hsize}
\end{minipage}
\hfill
\begin{minipage}[b]{.46\linewidth}
\epsfig{figure=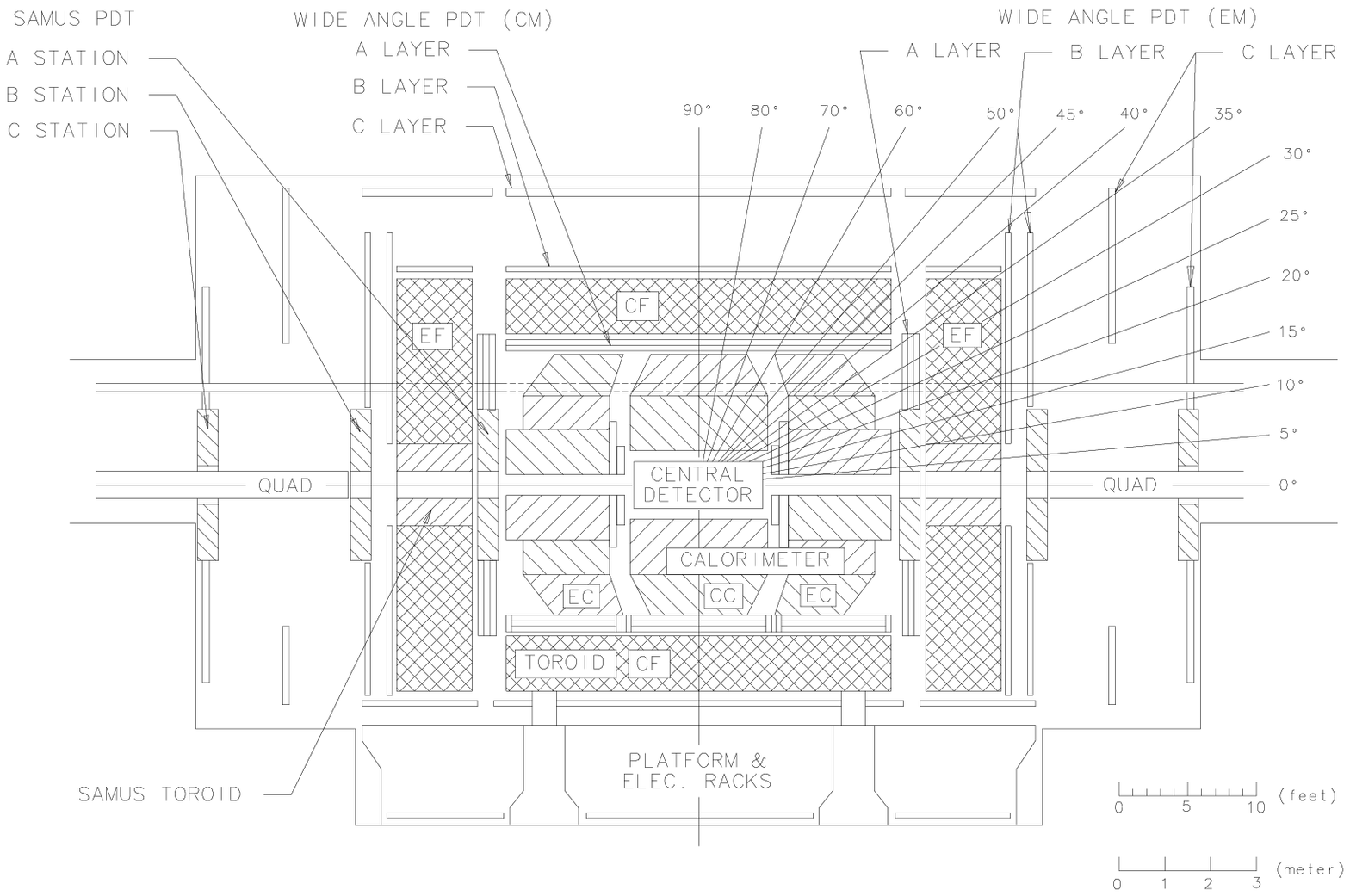,origin=c,angle=270,width=\hsize}
\end{minipage}}
\caption{The \dzero\ detector, in both isometric and cross sectional
  views.  CF, EF, and SAMUS are the three regions of the muon system
  (the small angle SAMUS system is not used in top analyses).
  From Ref.\protect\onlinecite{d0detector}.}
\label{fig:d0det}
\end{figure}

The two collider experiments at the Fermilab Tevatron,
CDF~\mcite{cdfdetector,*cdfsvx,cdfsvx2} and \dzero,\cite{d0detector} are
illustrated in \figsref{fig:cdfdet} and~\ref{fig:d0det}.
Both were designed to study high-$\pt$
interactions and feature large angular coverage and good
identification and measurement of electrons, muons, and jets.  The
layouts of both detectors are broadly similar.  Moving outwards from
the interaction point, one first encounters tracking detectors,
which measure the trajectories of charged particles,
then calorimeters, which measure the energies of jets and of
electromagnetic showers, and finally an outer set of tracking
chambers, which identify and measure muons that penetrate the
calorimeter.  There are tradeoffs between these various systems:
of the two detectors, CDF puts relatively more emphasis on
tracking, while \dzero\ emphasizes calorimetric
measurements.

We use a coordinate system centered on the
detector, with the $z$-axis along the beam direction, and the $x$- and
$y$-axes defining the transverse plane.  We also use the polar
angle $\theta$, the
azimuthal angle $\phi$, and the pseudorapidity
$\eta$, defined as $\eta\equiv - \ln (\tan{\theta/2})$.
It is also common to calculate the angle between directions of two objects 
in terms of the distance between them in the $(\eta,\phi)$ plane, as
$R = \sqrt{(\Delta\eta)^2 + (\Delta\phi)^2}$.

\subsubsection{CDF}
\label{sec:cdfdetector}

At the core of the CDF detector is a cylindrical tracking volume
contained within a large ($4.8\unit{m}$ long by $1.5\unit{m}$ radius) 
superconducting solenoid that generates a field of $1.4\unit{T}$.
Immediately surrounding the beam pipe is the four-layer silicon vertex
detector (SVX).  This $51\unit{cm}$ long device provides
precise track reconstruction in the plane normal to the beam, measuring
track impact parameters with a resolution of
$\sigma = (13 + 40\ (\ugevc) / \pt)\unit{$\mu$m}$.  This is
sufficiently precise to identify displaced vertices
from $b$-~and $c$-quark decays.  The luminous region at CDF has a
$1\sigma$ width of about $30\unit{cm}$, giving the SVX a geometrical
acceptance of about $67\%$.  The original SVX detector suffered
significant radiation damage during Run~1a, but was replaced before
the start of Run~1b with a similar detector equipped with
radiation-hard electronics.\cite{cdfsvx2}
Outside the SVX is a vertex drift chamber (VTX), which is used to
measure the $z$ position of the interaction vertex to a precision of
$1\unit{mm}$.  The VTX in turn is mounted inside of the central
tracking chamber (CTC), a large cylindrical drift chamber which
measures the curvature of tracks passing through the magnetic field.

Surrounding the tracking volume are the calorimeters.  There are three
distinct calorimeter regions: central ($|\eta| < 1.1$),
end-plug, and forward.
In each region, there is an electromagnetic calorimeter
backed by a hadronic calorimeter.
The electromagnetic calorimeters use lead as an absorber, while the hadronic
calorimeters use iron.  The active media are scintillator tiles in the
central region and gas proportional chambers in the end-plug and
forward regions.  The tower geometry is projective --- that is,
the towers point to the nominal interaction point --- with a cell size
in the central region of
$\Delta\eta\times\Delta\phi \approx 0.1\times0.25$.  A layer of
proportional wire chambers is located at shower maximum in the
central electromagnetic calorimeter, and
additional proportional chambers located between the solenoid and the
CEM sample the early development of electromagnetic showers in the
solenoid.  The energy resolution, $\sigma(E)/E$, 
of the calorimeters in the central region is
about $13.7\%/\sqrt{E} \oplus 2\%$ 
for electromagnetic showers
and about $50\%/\sqrt{E} \oplus 3\%$ for single hadrons.  The
thickness of the central calorimeters is about 18 radiation lengths for
the electromagnetic section and 4.5 interaction lengths for the
hadronic section.

Muons are identified using drift chambers surrounding the
calorimeter, the central muon system and central muon upgrade.
There is about $0.6\unit{m}$ of steel between these two sets
of chambers.  Together, they provide coverage out to about
$|\eta| < 0.6$.  Coverage is extended out to about $|\eta| < 1.1$ by
the chambers of the central muon extension.

\subsubsection{\dzero}

\dzero\ is the newer of the two detectors, commissioned during the
summer of~1992.
Its tracking volume is relatively compact ($1.35\unit{m}$
long by $0.78\unit{m}$ radius) and nonmagnetic.  Nested around
the beam pipe are the vertex drift chamber~(VTX), a
transition radiation detector~(TRD), and the central drift
chamber~(CDC).  The tracking volume is capped on the ends by two forward drift
chambers~(FDCs).  The trajectories of tracks of charged particles can be
measured with a resolution of $2.5\unit{mrad}$ in~$\phi$ and
$28\unit{mrad}$ in~$\theta$, and the $z$-coordinate of the interaction
vertex can be measured with a resolution of about $8\unit{mm}$.  The
central tracking system also measures the ionization of tracks in
order to distinguish between single charged particles and $e^+e^-$
pairs from photon conversions.

The calorimeter is divided into three parts: the central
calorimeter~(CC) and the two end calorimeters~(ECs).  They each
consist of an inner electromagnetic (EM) section, a fine hadronic (FH)
section, and a coarse hadronic (CH) section.  The absorber in the EM
and FH sections is depleted uranium; in the CH section, it is a
mixture of stainless steel and copper.  The active medium in all cases
is liquid argon.

The EM sections of the calorimeters are about 21 radiation lengths deep, and
are read out in
four longitudinal segments (layers).  The hadronic sections are 7--9
interaction lengths deep, with either four (CC) or five (EC)
layers.  The transverse segmentation is pseudoprojective (that is,
although each cell is nonprojective, they form towers which are), 
with a cell
size of $\Delta\eta\times\Delta\phi = 0.1\times0.1$.  In the third
layer of the EM calorimeter, near the shower maximum, the segmentation
is twice as fine in each direction, with a cell size of
$\Delta\eta\times\Delta\phi = 0.05\times0.05$.  The energy resolution
is about $15\%/\sqrt{E} \oplus 0.4\%$ for electromagnetic showers
and $50\%/\sqrt{E}$ for single hadrons.  The resolution is
substantially worse, however, in the transition regions between the CC
and the ECs, due to the presence of a 
large amount of uninstrumented material.  Some
of the energy that would otherwise be lost is collected
in extra argon gaps mounted on the
ends of the calorimeter modules (``massless gaps'') and in
scintillator tiles mounted between the CC and EC cryostats
(intercryostat detectors, or ICDs).

The \dzero\ muon system consists of a ``wide angle'' system,
covering $|\eta| < 1.7$, and a ``small angle'' (forward) system,
extending the coverage out to $|\eta| < 3.3$.  For studying top quark
decays, \dzero\ uses only the  wide angle system.  This system
consists of four planes
of proportional drift tubes in front of magnetized iron toroids, with
a magnetic field of $1.9\ \text{T}$, and two groups of three planes
of proportional drift tubes behind the toroids.  The magnetic
field lines and the wires in the drift tubes are oriented transversely
to the beam direction.  The muon momentum is measured from the muon's
deflection angle in the magnetic field of the toroid.
The total amount of material in the calorimeter and iron toroids
varies between 13 and 19 interaction lengths, making the background
from hadronic punchthrough negligible.  In addition, the compact
central tracking volume reduces backgrounds to prompt muons from
in-flight decays of $\pi$ and $K$~mesons.
During Run~1b, the forward muon chambers suffered
radiation damage that reduced their efficiency.  Midway through the
run, however, the damage was repaired.  As a result, the \dzero\ top quark
analyses do not use forward muons for the first half of Run~1b.

\subsection{Particle Identification}
\label{particle-id}

This section summarizes the algorithms used by the two experiments to
identify the various final-state objects in candidate $\ttbar$
events.  For more details, see
Refs.\onlinecite{cdftopprd94,d0topprd}, and \onlinecite{d0ljtopmassprd}.

\subsubsection{Quarks and Gluons}
\label{jetid}

As a quark or gluon leaves the site of a hard scattering it cannot
remain free, but instead \emph{hadronizes} (or \emph{fragments}) into a
collection or \emph{jet} of (colorless) hadronic particles.  This
collection tends to lie in a cone around the direction of motion
of the original parton, and will show up in a calorimeter as an
extended cluster of energy.
In order to compare measurements with theoretical
predictions it is necessary to have a precise definition 
of a jet: that is, one must specify how calorimetric
energy depositions (cells) are to be clustered into jets.
This algorithm is, in principle,
arbitrary.  However, at hadron colliders, it is conventional to
define jets by taking all calorimeter cells which lie within a cone of
fixed radius $R$ in the $(\eta,\phi)$ plane.  This choice is
convenient because jets are approximately circular in these variables;
further, the $R$-width of jets of a given $\et$ is independent of the
jet rapidity.  More importantly, this definition can be readily
implemented in phenomenological calculations, thereby facilitating the
comparison of theory with experimental data.\cite{snowmass}

In principle, not only is the jet algorithm arbitrary but also
the cone radius $R$.  In practice, the choice of cone radius 
involves several competing considerations.
Jets are extended objects, composed of a collection of particles
from hadronization of the progenitor parton.  The jet will
be further broadened as the particles undergo showering in
a calorimeter.  Consequently, if $R$ is too small, a substantial portion of
the energy from the progenitor parton will lie outside of the jet cone.  
This effect can be corrected for on average.  However,  the
smaller the cone radius, the larger the energy correction that
must be applied and, therefore, the
worse the energy resolution of the corrected jet.
On the other hand, if $R$ is made too large, one cannot
resolve the energy depositions arising from closely spaced partons; 
instead, the depositions get merged
together into a single jet.  This is of particular concern for $\ttbar$
events, which tend to have many jets in the final state.  The optimum
choice for $R$ for $\ttbar$ physics depends somewhat on the structure
of the calorimeter, but appears to be around $0.4$--$0.5$: CDF chooses
$R=0.4$, and \dzero\ uses $R=0.5$.

Although both experiments have a resolution for single hadrons that
scales as $\sigma(E)/E\sim 50\%/\sqrt{E}$,
the resolution achievable for
jets is typically $\sigma(E)/E\sim 100\% / \sqrt{E}$.
Most of the
particles comprising a jet are of relatively low energy, in which
region nonlinear effects in calorimeter response become
important.  Jet resolutions are also degraded by effects such as
gluon radiation,
differences in calorimeter response to hadrons and electrons,
energy falling out of the jet cone, and contamination from hadrons from
the underlying event.

The measurement of jet energies is subject to numerous systematic
effects, for which one must correct.  These include:
\begin{itemize}
\item The intrinsic response of the calorimeter to
  jets.
\item Calorimeter nonuniformities, and regions with uninstrumented
  material (such as cracks between modules).
\item Energy from the underlying event
  and, at \dzero, noise from the radioactive decay of the uranium
  absorber.
\item QCD radiation of gluons outside of the jet cone.
\item The
  spreading of particle showers outside of the
  jet cone in the calorimeter.
\end{itemize}

The procedures involved in performing these corrections are quite
complicated;\mcite{cdftopprd94,cdf4jet93,*cafix96,*cafix98}
we shall therefore only summarize the strategies used.
\begin{itemize}
\item Dijet events, in which the transverse energies of the two jets
  should balance, can be used to calibrate one region of the detector
  relative to another, better characterized, region.
\item Events with an electromagnetic cluster recoiling against a jet
  can be used to calibrate hadronic calorimeters relative to
  electromagnetic calorimeters.
\item The absolute scale of electromagnetic calorimeters can be
  determined by comparing electron energies to their momenta measured
  in the tracking system (at CDF) or by using the known masses of
  resonances such as the $Z$, the $\pi^0$, and the $J/\psi$
  (at \dzero).\cite{d0wmassprd1b}
\item Contributions from the underlying event and noise can be studied
  using Monte Carlo simulations, and by comparing data taken under
  differing trigger conditions and luminosities.
\item The broadening of showers in calorimeters can be studied
  using test beams.\cite{tb90l1a,*tb90l1b}
  Monte Carlo simulations are used to model the
  distribution of particles produced during hadronization of partons.
\end{itemize}

CDF and \dzero\ apply jet corrections at different points in their
analyses.  This should be kept in mind when comparing selections involving
jet energies.  CDF does not use the jet corrections for 
measuring cross sections (except in the all-jets channel),
but does apply them for the
mass measurement, after the event sample has been selected.
\dzero, on the other hand, applies most
corrections before making any analysis selections.  The corrections
include effects of jets spreading in the calorimeter, but not of
particles originating from gluons radiated outside of the jet cone.  \dzero\
applies an additional correction in its mass analysis to include
this effect.

It is important to realize that
there is not necessarily any one-to-one
correspondence between quarks and gluons in the final state of the
hard scattering and the detected jets.  A jet may have insufficient
energy to be selected as a jet (a typical requirement is that the
jet energy be at least $15\gev$), or two partons may be
sufficiently close together that their energies are merged together during
jet reconstruction.  Conversely, if a parton radiates a gluon with
a large relative transverse momentum, then that gluon may be reconstructed as a
separate jet. Moreover, nonclassical effects, such as partonic interference,
are always present and place a fundamental limit on the validity
of identifying a given jet with a specific progenitor parton.

\subsubsection{Electrons}
\label{electronid}

Electron identification is based on finding
isolated clusters of energy in the electromagnetic sections of
the calorimeter, along with a
matching track in the central detector from a charged particle.  Additional
requirements are then made to further suppress background from
QCD~jets.  The exact requirements vary between experiments and
among different analyses.  However, typical requirements are:
\begin{itemize}
\item The fraction $f_E$ of the cluster energy in the electromagnetic
  sections of the calorimeter should be $\gtrsim 90\%$.
\item The shape of the cluster should be consistent with expectations
  from test beams or Monte Carlo simulations.
\item The track momentum $p$ should be consistent with the cluster
  energy $E$, $E/p \sim 1$ (CDF only).
\item The distance between the
  cluster and the extrapolated track should be small.
\item A track ionization consistent with a singly charged particle
  (\dzero\ only).  Photon conversions into $e^+e^-$ pairs
  typically deposit twice the charge expected from one minimum
  ionizing particle.
\item Transition radiation information consistent with an electron
  (used in \dzero\ dilepton analyses).
\item Isolation,  based either on calorimetry or
  (at CDF only) on tracking.
One typically requires that the energy or momentum in
  the region around the electron candidate be small.
\end{itemize}
CDF accepts electrons in the central calorimeter with
$|\eta| < 1.0$.  For the dilepton channels, one electron may also be
in the plug calorimeter, thereby providing extra coverage in the range
$1.20 < |\eta| < 1.35$.  \dzero\ accepts electrons out to
$|\eta| < 2.5$ for the dilepton channels and out to $|\eta| < 2.0$ for the
lepton+jets channels (although the efficiency is poor in the
transition region between the central and end cryostats,
$1.1 < |\eta| < 1.5$).

\subsubsection{Muons}
\label{muonid}

At CDF, muons are identified by
requiring a match between a CTC track and a track segment in the muon
chambers.  This provides coverage out to $|\eta| < 1.0$.
For dilepton channels, CDF also identifies muons in
sections of the detector where there is no coverage from muon
chambers but good central tracking.
This extends coverage out to $|\eta| < 1.2$,
and fills in coverage of azimuthal holes for $|\eta| < 1.0$.
Additional selections are made on the following variables:
\begin{itemize}
\item The energy deposited in the calorimeters along the muon track.
\item The distance of closest approach of the muon track to the beam
  line.
\item The distance in $z$ between the interaction vertex and the
  muon track.
\item The distance between the extrapolated CTC track and the track
  segment in the muon chambers.
\item Isolation, defined in a similar manner as for
  electrons except that calorimeter-based isolation is used
  when there is no matching track in the muon chambers.
\end{itemize}

\dzero\ identifies muons by tracks in the muon
chambers.  A matching central
detector track is not required, but if one is present, it
is taken into account in the momentum measurement.  The
following additional criteria are used for \dzero's muon selection:
\begin{itemize}
\item The line integral over the magnetic field 
  $> 2.0\unit{$\text{T}\cdot\text{m}$}$.  This rejects muons
  which pass through the thin portion of the toroid around
  $|\eta| \approx 0.9$.  Such muons bend less so their momenta
  are poorly measured.  They also have
  a larger contamination from hadronic punchthrough.
\item The energy deposited along a muon track in the calorimeter
  must be at least that expected for a muon ($\sim 2$--$3\gev$).
\item The impact parameter between the muon track and the interaction
  vertex must be small.
\item Muons are defined as either isolated or nonisolated,
  depending on whether the distance in the $\eta - \phi$ plane
  between the muon and the closest jet is greater than or less
  than $0.5$, respectively.
\end{itemize}
For the first portion of Run~1b, when the forward muon chambers were
affected by radiation damage, the muon selection was restricted to the
central region, with $|\eta| < 1.0$.  For other run periods, muon
acceptance extended out to $|\eta| < 1.7$.

\subsubsection{$\tau$ Leptons}
\label{tauid}

Tau leptons are difficult to identify.  A $\tau$ will decay into
an electron or a muon about $36\%$ of the time, and the only observable
difference between the decays $W \ra \ell \nu$ and
$W \ra \tau\nu \ra \ell\nu\nu\nu$ is that the lepton spectrum is somewhat
softer in the latter case.  If the $\tau$ decays hadronically, it can
be detected as a narrow jet with either one or three associated
tracks.  The branching ratios for these ``one-prong'' and
``three-prong'' hadronic decays are about $50\%$ and $14\%$,
respectively.\cite{pdg96}  The challenge is to reject the large
background from quark and gluon jets.  CDF manages this with two
complementary techniques, one ``track-based'' and the other
``calorimeter-based.''\cite{cdftoptotau97}

The track-based technique is sensitive only to one-prong $\tau$
decays.  It starts by finding an isolated, high-$\pt$ ($>15\gevc$),
central ($|\eta| < 1.0$) track.  The isolation requirement is based on
the sum of the $\pt$ values of all tracks in a $\Delta R$ cone around
the high-$\pt$ track.  This discriminates between $\tau$ leptons and jets.
The energy in the calorimeter around the track
is required to be consistent with the
track momentum.  In addition, candidates
consistent with being electrons (large fractional energy in
the electromagnetic calorimeter) or muons (energy deposition
consistent with a minimum ionizing particle) are rejected.

The calorimeter-based technique starts from a calorimeter cluster with
$|\eta| < 1.2$ with
either one or three isolated charged tracks pointing at it.  (The
tracks must have $\pt > 1\gevc$ and lie within a $10\degree$ cone
around the centroid of the cluster.)  About $73\%$ of one-prong and $41\%$ of
three-prong decays are expected to contain $\pi^0$s,
which can be identified from their $\pi^0 \ra \gamma\gamma$ decays.
The candidate is rejected if there are more than
two $\pi^0$s.  The $\tau$ lepton $\pt$ is defined as the sum 
of the transverse momenta of all candidate tracks plus the
transverse momenta of all $\pi^0$s; it must 
satisfy $\pt > 15\gevc$.  Also, the total invariant mass constructed
from the tracks and the $\pi^0$s must be
$< 1.8\gevcc$.  The calorimeter
cluster must be narrow, and its energy must be consistent with the
total $\tau$ momentum.
Finally, clusters consistent with being electrons or muons
are removed.

\begin{figure}
\epsfig{file=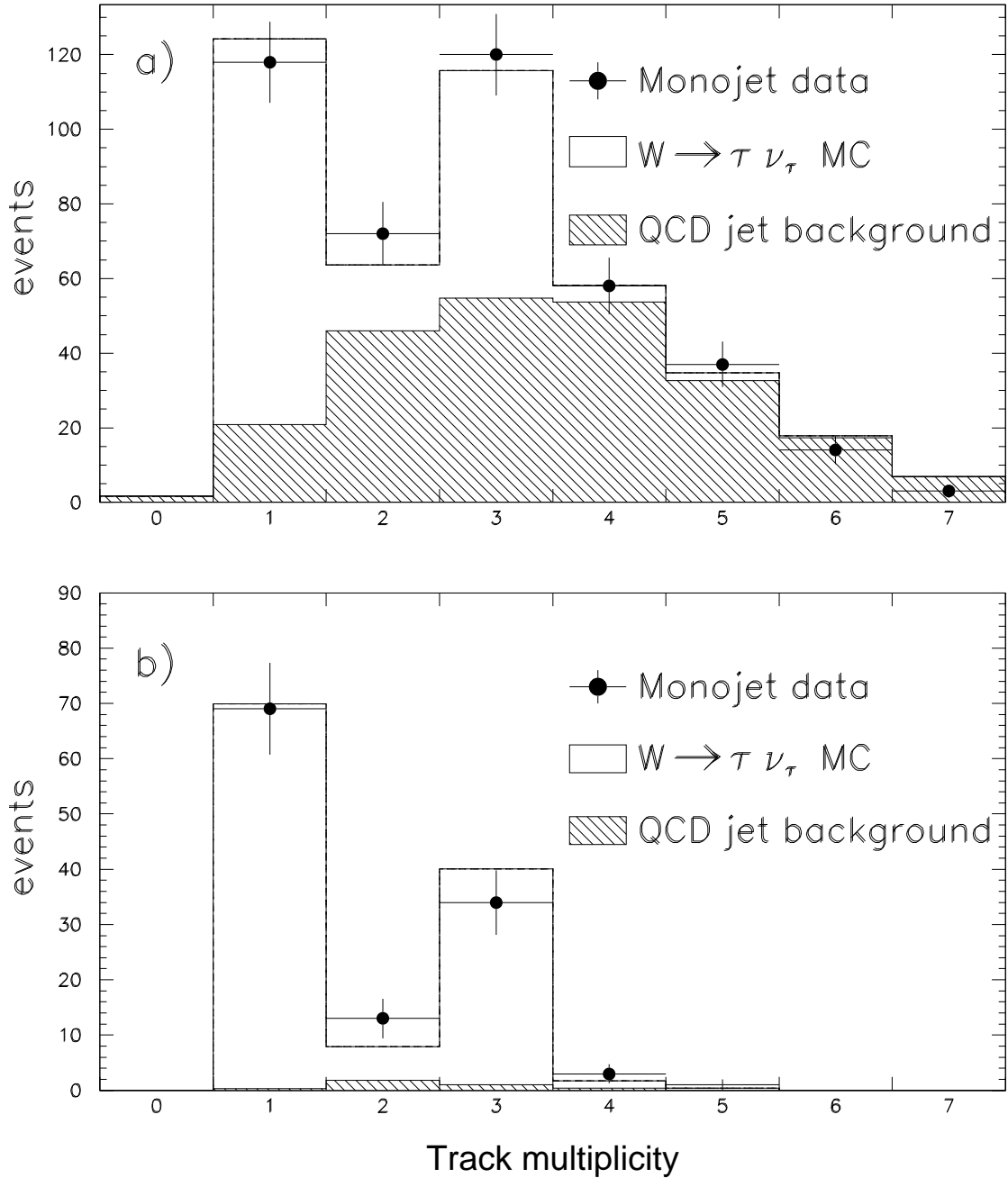,width=\hsize}
\caption{Track multiplicity in the monojet data sample.
 a) No $\tau$ identification requirements.
 b) All calorimeter-based $\tau$ identification requirements,
    except for track multiplicity.
 From Ref.\protect\onlinecite{cdftoptotau97}.}
\label{fig:cdftaudilep1}
\end{figure}

The efficacy of the calorimeter-based selection is demonstrated in
\figref{fig:cdftaudilep1}, which shows the track multiplicity for a
monojet sample which required one jet with $15 < \et < 40\gev$,
$20 < $\met$ < 40\gev$, and no other jet with $\et > 7\gev$.  The
excess in the one and three track bins from $W\ra \tau\nu$ is
apparent; it is also seen that the calorimeter-based selection
drastically reduces the background.  About $45\%$ of hadronic $\tau$
decays satisfy the kinematic requirements; of
these, about $55\%$ are identified by this algorithm.

\subsubsection{Neutrinos}

Neutrinos do not interact in the detector with any significant
probability, and therefore cannot be observed directly.  Instead, their
presence is inferred from an imbalance in the total transverse
momentum in the event.  Because the remnants of the
spectator partons escape down the beam pipe, the longitudinal
component of the neutrino momentum cannot be measured.

The missing transverse energy $\vecmet$ is defined by
\begin{eqnarray}
  \met_x &= - \sum_i E_i \sin\theta_i \cos\phi_i, \\
  \met_y &= - \sum_i E_i \sin\theta_i \sin\phi_i, \nonumber
\end{eqnarray}
where the sums are over all calorimeter cells.  If there are muons in
the final state, then their transverse momenta should also be subtracted
from $\vecmet$.  (Some analyses make use of the $\met$
measured by the calorimeter without correcting for muons.  This quantity will
be denoted by $\metcal$.)  The resolution of $\met$ is usually
parameterized in terms of the total transverse energy in the event (a
scalar sum).  CDF quotes\cite{cdftopprd94} a resolution of
$\sigma(\met) = 0.7 \sqrt{\sum{\et}}$ (units in GeV), while \dzero\
quotes\cite{d0topprd} $\sigma(\met) = 1.08\gev + 0.019 \sum \et$.

\subsubsection{Tagging $b$-Jets}
\label{btagging}

A prominent feature of $\ttbar$ decays is that each event contains
two $b$-quarks.  This is in contrast to the principal backgrounds, in
which heavy flavors are expected to be relatively rare.  Clearly, a
method for identifying, or \emph{tagging}, $b$-jets would be valuable
in separating $\ttbar$~events from background.  The two
strategies developed for doing this, soft-lepton
tagging (SLT) and displaced-vertex tagging (SVX, using the silicon
vertex detector), are discussed below.

\paragraph{$b$-Tagging with Soft Leptons}

\begin{figure}
\epsfig{file=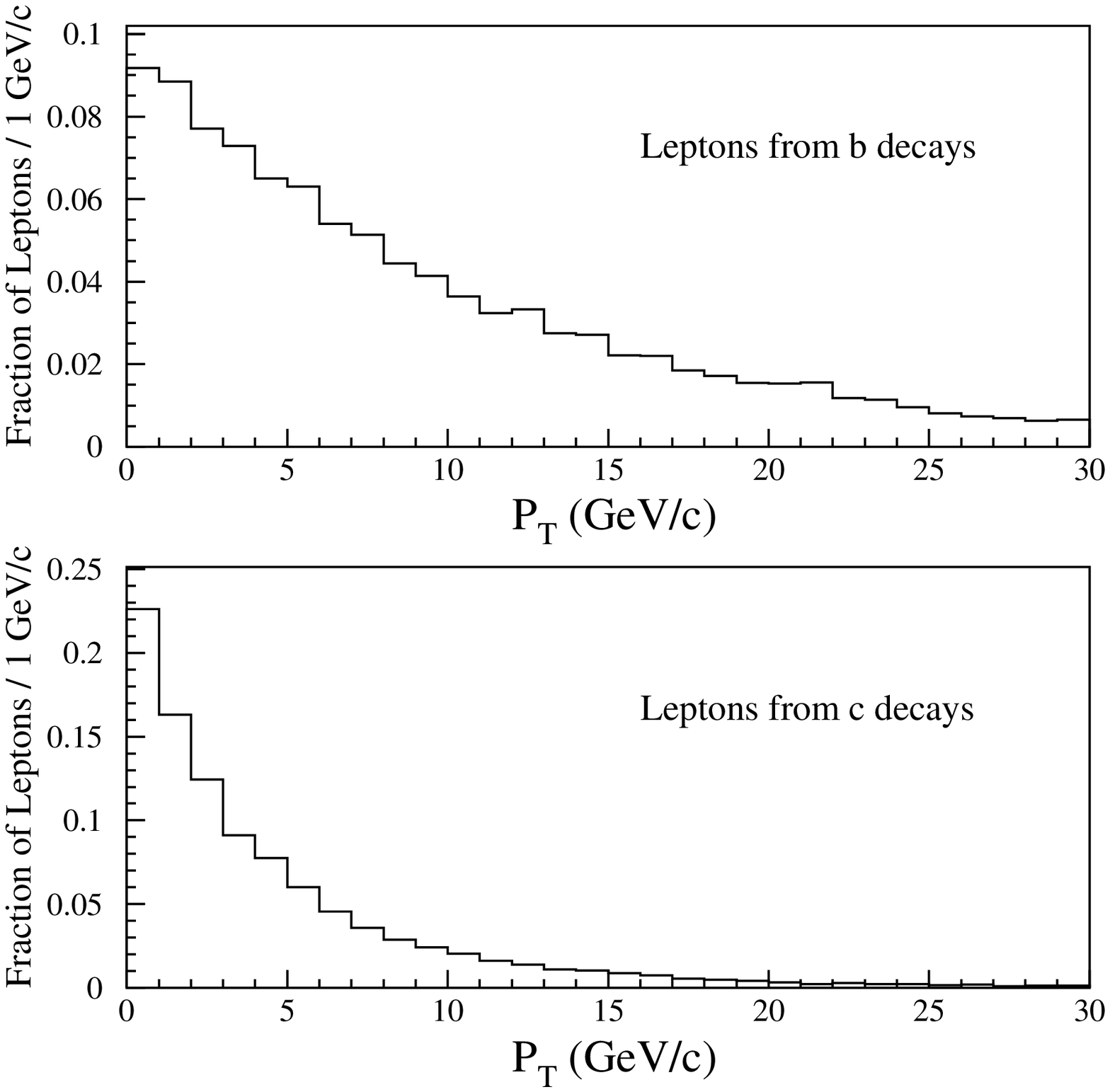,width=\hsize}
\caption{$\pt$ spectra of leptons from decays of $b$- and
  $c$-quarks in $\ttbar$ Monte Carlo events ($m_t = 160\gevcc$).
  From Ref.\protect\onlinecite{cdftopprd94}.}
\label{fig:blepton-pt}
\end{figure}

Approximately $22\%$ of the time, a decaying $b$-quark will yield a
muon, either directly or through a sequential decay via a
$c$-quark.  The branching ratio to electrons is also $22\%$.
One method of $b$-tagging is based on observing such leptons close to
a jet.  These leptons are much softer than the leptons from $W$~boson
decay, typically below about $20\gevc$.  (See \figref{fig:blepton-pt}.)
Because they are soft and
not isolated, their detection efficiencies are significantly lower than
for the high-$\pt$ isolated leptons from $W$~boson decay.

Both experiments tag $b$-jets with this method.  CDF uses both
electron and muon tags, while, in the analyses reviewed here, 
\dzero\ uses only muon tags.  Without a
central magnetic field, electron tagging is less effective at \dzero.
However, even at CDF, the electron tags play a relatively minor role:
the efficiency for electron tagging is about $1/3$ of that for muon
tagging.  Another difference is that CDF selects soft leptons with
$\pt > 2\gevc$, while \dzero\ requires $\pt > 4\gevc$
because muons with lower
momenta have insufficient energy to traverse the
calorimeter and toroid.  
However, \dzero\ has a larger muon acceptance than CDF.
In addition, the fake rate for muon tags is significantly smaller
at \dzero, due both to the large amount of material in the calorimeter
and muon toroid (which reduces hadronic punchthrough) and the small
size of the tracking volume (which reduces background due to
in-flight $\pi/K$ decays).  When all factors are accounted for,
the probability of finding a lepton tag in a $\ttbar$
decay is nearly the same for both experiments, about $20\%$.

\paragraph{$b$-Tagging with Displaced Vertices}
\label{svxtag}

Another method of tagging $b$-jets profits from the relatively
long lifetime of $b$ and $c$ hadrons (about $1\unit{ps}$).
Given the typical boost of $b$-quarks in $\ttbar$ events, this allows
the $b$ hadrons to travel up to several mm before decaying.
Detecting a vertex displaced by this distance is well within the
capabilities of modern silicon microstrip detectors.

\begin{figure}
\epsfig{file=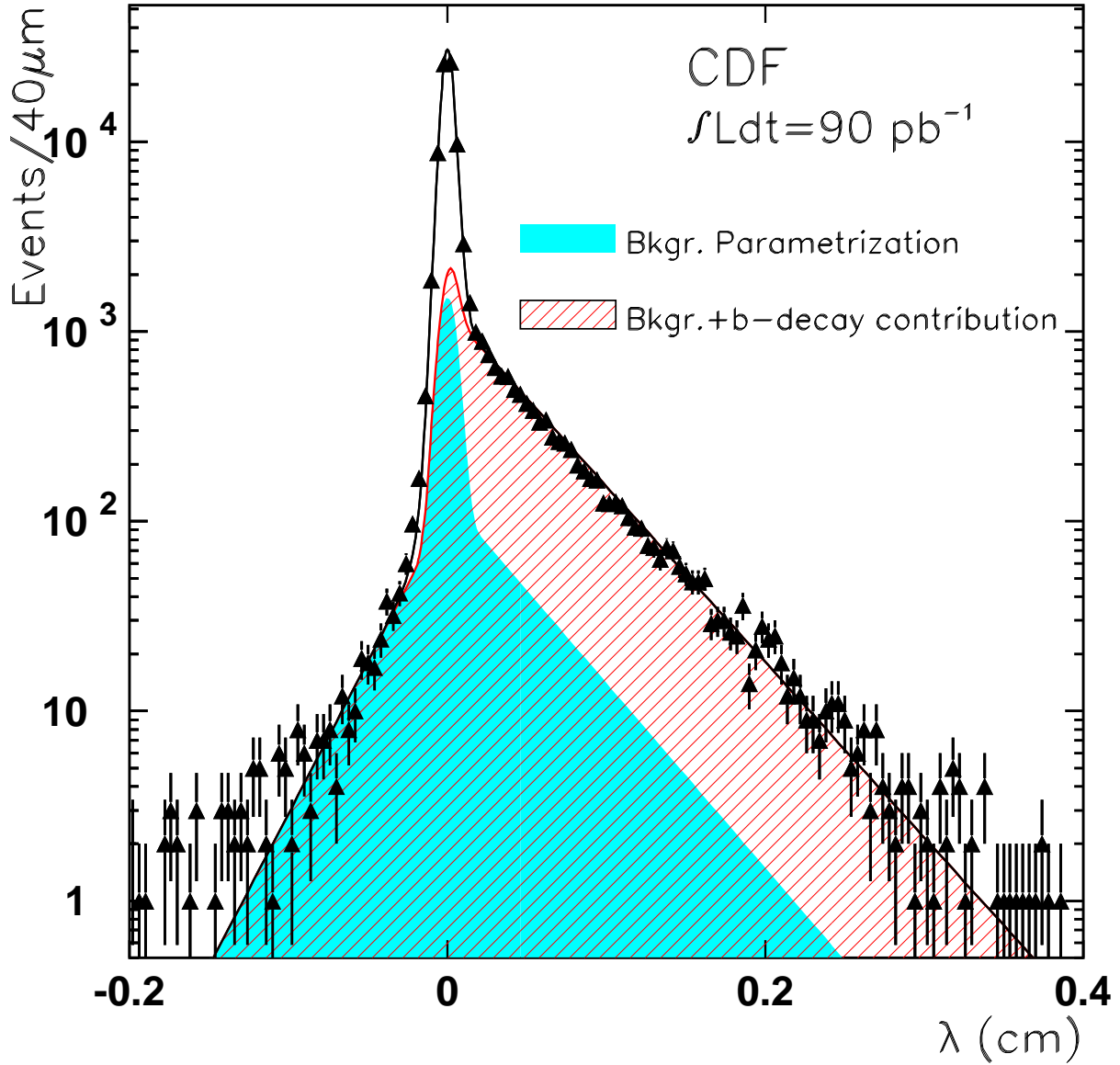,width=\hsize}
\caption{
  The distribution of ``pseudo decay length''
  $\lambda\sim M L_{xy}/ \pt$ from CDF $B\ra J/\psi \ra \mu\mu$ data.
  The points are the data, and the curves are the signal and background
  models.  The heavily shaded area is the background; it is asymmetric
  because of sequential semileptonic $B$ decays in the dimuon sample.
  The lightly shaded area is the signal contribution, and the unshaded
  area is from prompt $J/\psi$ mesons.
  From Ref.~\protect\onlinecite{cdfblifetime}.
}
\label{fig:cdfbtag}
\end{figure}

At present, CDF is the only experiment to have operated a silicon
vertex detector (the SVX) at a hadron collider.\mcite{cdfsvx2,*cdfsvx}
The tagging algorithm works by finding 
combinations of at least two tracks consistent with originating
from a vertex displaced from the primary vertex of the hard interaction.
Such displaced vertices are sometimes called \emph{secondary} vertices.
For each possible secondary vertex, one estimates
the distance ($L_{xy}$) in the transverse plane between that vertex and
the primary one, along with its associated uncertainty ($\sigma_{Lxy}$).  In
order to be accepted as a $b$-tag, a vertex must satisfy the condition
$L_{xy} / \sigma_{Lxy} > 3$.  The sign of $L_{xy}$ is given by the
sign of the dot product between the direction of $L_{xy}$ and the
direction of the vector sum of the momenta of the tracks used.
It is predominantly positive for real $b$-decays; displaced vertices
with negative $L_{xy}$ are due primarily to track mismeasurements.
Jets which contain many mismeasured tracks but no real secondary
vertices are equally likely to have $L_{xy}$ positive or negative,
a fact which is used to measure the background from this source.
Only tags with positive $L_{xy}$ are accepted as
$b$-tags.
The probability of finding at least one displaced vertex
(or SVX) tag in a $\ttbar$ event is $(39\pm3)\%$.\cite{cdfxs98} 
This includes a geometric efficiency of 
$67\%$
caused by the length of the luminous region relative to
the length of the SVX.  See \figref{fig:cdfbtag} for a sample result
from this technique.

\subsection{Characteristics of Signal and Background Events}
\label{signal_background}

We have discussed the decay modes of the top quark and the experimental 
signatures for $\ttbar$ production in \secref{production}. 
We have also outlined how the objects in the final state of a $\ttbar$ decay
are identified and measured in the two detectors.  Here, we shall
discuss briefly the kinematic properties of $\ttbar$ events
in the various decay modes, and how these properties differ for
background processes.

\begin{figure}
\centering
\epsfig{figure=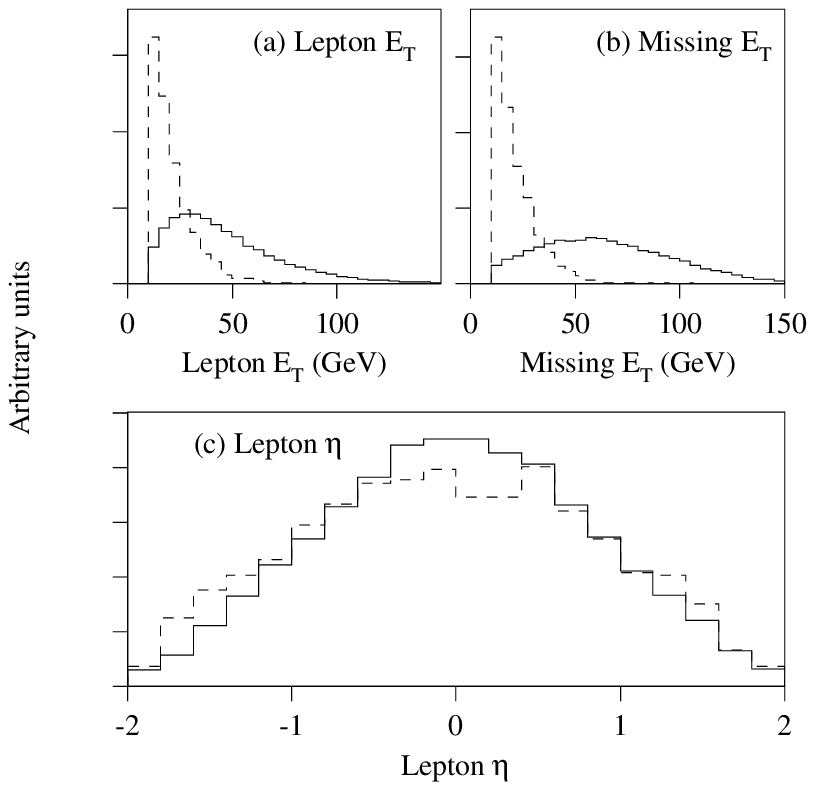}
\caption{Expected distributions for $e\mu$ dilepton events
  of (a) lepton $\pt$ (two entries per
  event), (b) $\met$, and (c) lepton $\eta$ (two entries per event).
  The solid histograms are $\ttbar\ra e\mu + X$
  signal events (generated with \progname{herwig} with
  $m_t=175\gevcc$ for $\ppbar$ collisions at $\sqrt{s}=1.8\tev$).  The
  dashed histograms are $Z+\jets\ra\tau\tau+\jets\ra e\mu+\jets$
  events (also generated with
  \progname{herwig}).  All histograms are normalized to unity.
  We require that events have $\pt^{\ell} > 10\gevc$, $\met > 10\gev$, and
  at least two jets with $\et > 15\gev$ and $|\eta| < 2.0$.
}
\label{fg:dilkin-leps}
\end{figure}

\begin{figure}
\centering
\epsfig{figure=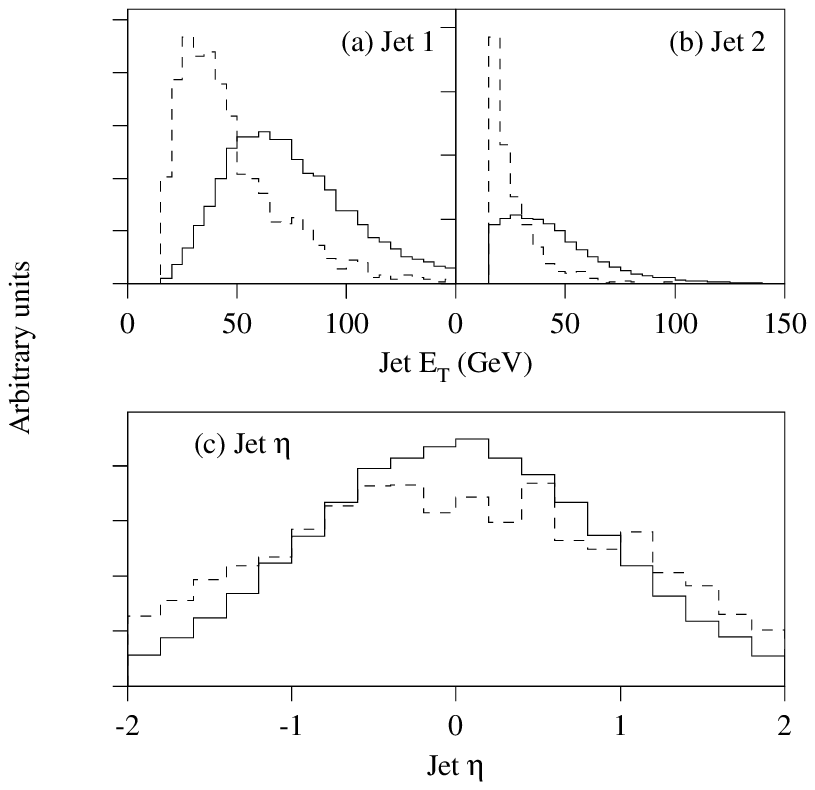}
\caption{Expected distributions for $e\mu$ events of (a and b) the
  transverse energies of the two leading jets and (c) the jet $\eta$
  (two entries per event).  See \figref{fg:dilkin-leps} for further
  details.
}
\label{fg:dilkin-jets}
\end{figure}

\begin{figure}
\centering
\epsfig{figure=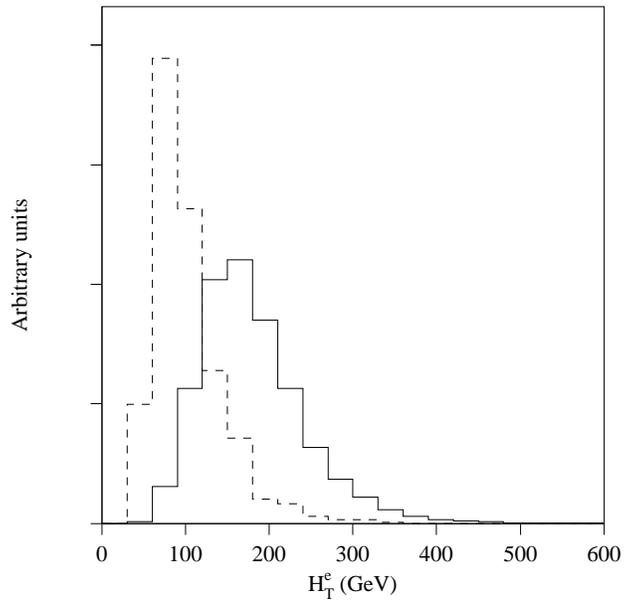}
\caption{Expected distributions for $e\mu$ events of 
  $\htran^e \equiv \sum_{\text{jets}} \et^{\text{jet}} + \et^{e1}$.
  See \figref{fg:dilkin-leps} for further details.
}
\label{fg:dilhte2}
\end{figure}

In the dilepton channels, the two leptons arise from $W$ boson decays,
and therefore tend to be central in pseudorapidity, isolated,
and of
large transverse momenta.  The two $b$-jets also have high transverse energies.
There are two high-$\pt$ neutrinos, so the missing transverse energy
tends to be large as well.  The major background for these channels is from the
Drell-Yan process, which produces isolated lepton pairs in profusion.
Additional jets can arise from initial or final state radiation.  
The Drell-Yan process yields $ee$ and $\mu\mu$ events directly, while
$e\mu$ events can be produced via $\tau\tau$ production and subsequent decay.
The  $ee$ and $\mu\mu$ events at the $Z$ resonance 
can be easily eliminated by rejecting events in which the invariant
mass of the two leptons is consistent with that of the $Z$ boson,
$M_Z$.
Additional rejection of background is obtained by requiring 
at least two jets in the final state.  In Drell-Yan events, the additional
jets are due to gluon radiation; consequently, every additional jet
reduces the cross section by a factor of
$O(1/\alpha_s) \sim 6.7$.  Requiring a large $\met$ further
reduces the background.  In case of
$Z\ra \tau\tau \ra (\ell\nu\nu)(\ell\nu\nu)$, the $Z$~boson cannot be
reconstructed because of the presence of four unobserved neutrinos in the
final state.  However, the leptons in these events
have much smaller $\pt$ than those in $\ttbar$ events.
So, requiring two high-$\pt$ leptons, two or more jets, and large $\met$
will greatly suppress this background.  Other backgrounds which must
be considered are diboson
production ($WW$, $WZ$, $ZZ$),
QCD production of $\bbbar$ (with semileptonic decays of
the $b$-quarks), and QCD events with jets misidentified as leptons.
The distributions of several kinematic
quantities for dilepton events are shown
in \figsref{fg:dilkin-leps}, \ref{fg:dilkin-jets}, and
\ref{fg:dilhte2}.

\begin{figure}
\centering
\epsfig{figure=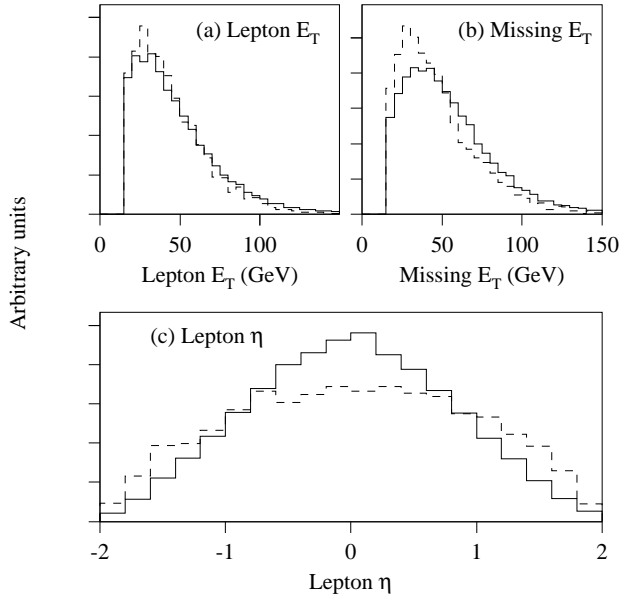}
\caption{Expected distributions for lepton+jets events
  of (a) lepton $\pt$, (b) $\met$, and (c) lepton~$\eta$.
  The solid histograms are $\ttbar$
  signal events (generated with \progname{herwig} with
  $m_t=175\gevcc$ for $\ppbar$ collisions at $\sqrt{s}=1.8\tev$).  The
  dashed histograms are $W+\geq 4~\jets$ events (generated with
  \progname{VECBOS}).  All histograms are normalized to unity.
  We require that events have $\pt^{\ell} > 15\gevc$, $\met > 15\gev$, and
  at least four jets with $\et > 15\gev$ and $|\eta| < 2.0$.
}
\label{fg:ljtkin-leps}
\end{figure}

\begin{figure}
\centering
\epsfig{figure=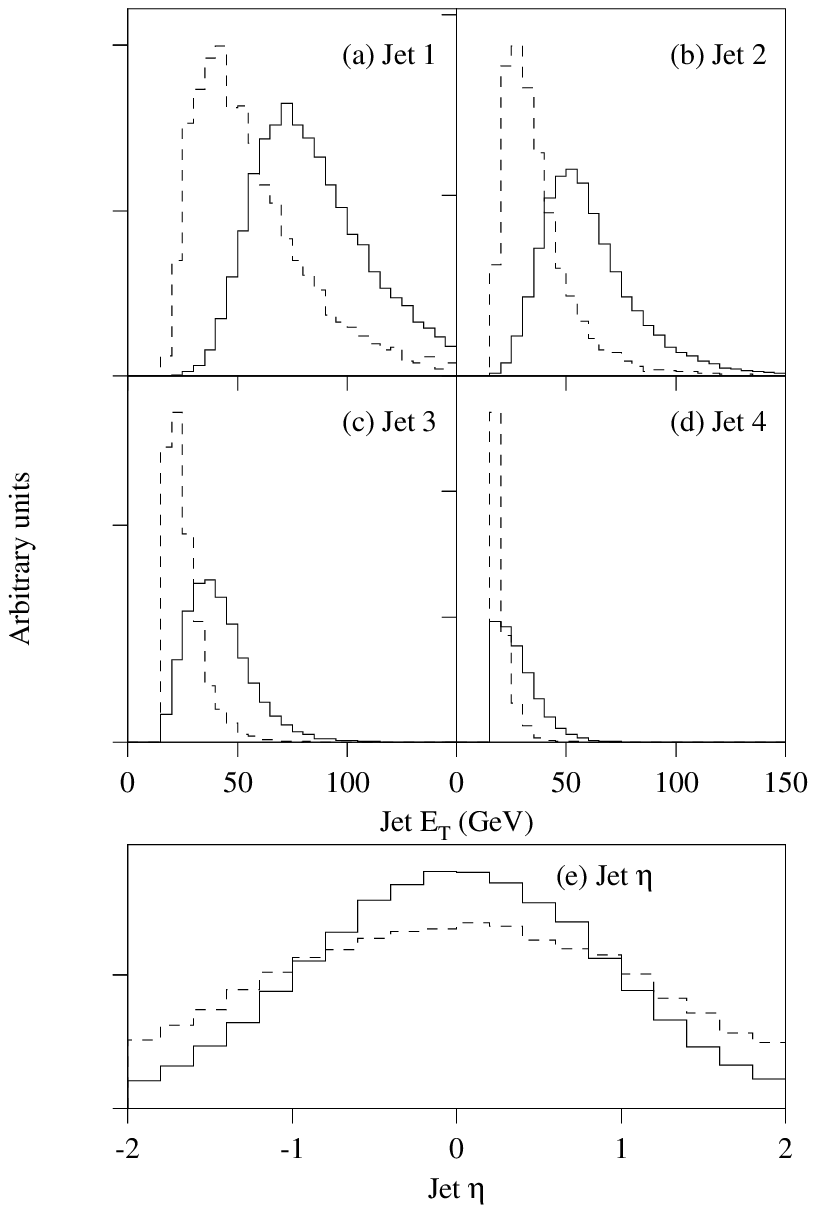,width=0.8\hsize}
\caption{Expected distributions for lepton+jets events of (a--d) the
  transverse energies of the four leading jets and (e) the jet $\eta$
  (four entries per event).  See \figref{fg:ljtkin-leps} for further
  details.
}
\label{fg:ljtkin-jets}
\end{figure}

The dominant background to the lepton+jets channel comes from the
production of
single $W$ bosons in association with jets.  Other backgrounds include
false lepton events with mismeasured $\met$ and/or misidentified $b$-tags, 
$W\bbbar$ and $W\ccbar$ production, and QCD heavy flavor processes.
The distributions of $\et^{\ell}$, $\eta_{\ell}$, and $\met$ for the
signal and for the $W$+jets background are shown in \figref{fg:ljtkin-leps}.
The major differences between the signal
and background processes stem from the number of
jets in the event and the event kinematics.  Ideally, a lepton+jets
$\ttbar$ event has four jets, two of which come from $b$-quarks.
However, as discussed in \secref{jetid}, the number of observed jets 
can be greater or lesser than four.
Therefore, the analyses usually require at least three jets
if a $b$-tagged jet is required in the event,
and at least four jets otherwise.
The jets in background events have lower transverse energies
than those in the signal, and are produced over a wider range of
$\eta$, as is evident from \figref{fg:ljtkin-jets}.

\begin{figure}
\centering
\epsfig{figure=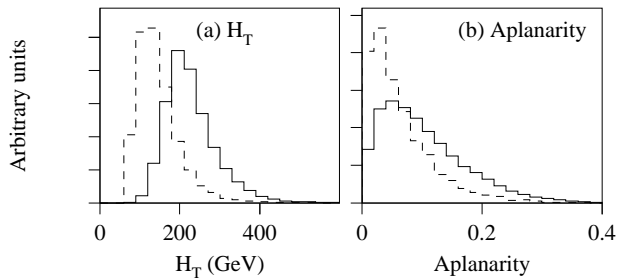}
\caption{Expected distributions for lepton+jets events of
  (a) $\htran$ and (b) aplanarity.
  See \figref{fg:ljtkin-leps} for further
  details.
}
\label{fg:ljtkin-shape}
\end{figure}

Certain variables describing the overall shape of the event provide
powerful means of discriminating lepton+jets signal from
background.\cite{d0topprd,d0xsecprl97}
One such variable is the aplanarity\cite{barger93} $\aplan$, defined as
$2/3$ times the smallest eigenvalue of the total normalized 
momentum tensor in the event.  This is $0.5$ for
spherical events and zero for planar or linear events.  Top~quark
events are expected to be more spherical than radiative QCD processes,
and hence to have larger aplanarities.  Another useful variable is the
sum of the transverse energies of all jets, $\htran$.  This
variable reflects the ``temperature'' of the interaction;
a large $\htran$ is a signature of the
decay of massive objects.\mcite{baer89,*likhoded90}
Distributions of $\htran$ and $\aplan$ are shown
in \figref{fg:ljtkin-shape}.

\section{Discovery of the Top Quark}
\label{discovery}

The pursuit of the top quark began in earnest in the late 1970s, shortly after
the discovery of the $b$-quark at Fermilab.\cite{herb77}
Searches at the $e^+e^-$ colliders PETRA 
\mcite{behrend84,*bartel79a,*bartel79b,*bartel81,*barber79,*barber80,
*adeva83a,*adeva83b,*adeva85,*adeva86,*berger79,*brandelik82,*althoff84a,
*althoff84b}
(1979--84, ${\sqrt s}=\text{12--46.8}\gev$),
TRISTAN
\mcite{sagawa88,*igarashi88,*adachi88,*yoshida87,*abe90}
(1986--90, ${\sqrt s}=61.4\gev$),
and SLC \cite{abrams89} and LEP
\mcite{decamp90,*abreu90,*akrawy90}
(1989--90, ${\sqrt s}\approx M_Z$)
eventually raised the lower limit on the top quark mass to $45.8\gevcc$.
(This limit and all others mentioned in this section
are $95\%$ confidence level results.)  Owing to the pioneering work of  
S. van der Meer, C. Rubbia, and others, at CERN and
elsewhere, high-energy
$\ppbar$ colliders were developed in the 1980s.
The first was the ISR (intersecting storage rings)
at CERN\cite{isrppbar}.  Next came the
S$\ppbar$S, also at CERN.  With $\sqrt s$ up to $630\gev$, this machine
had a beam energy an order of magnitude higher than the ISR.
This was followed by the Fermilab Tevatron, with $\sqrt{s} = 1.8\tev$.
These machines provided much higher center-of-mass energies, 
enabling searches for particles with higher masses. 
 
Searches for the top quark at $\ppbar$ colliders do
not provide direct limits on the mass of the top quark, but rather upper
limits on its production cross section.  These upper limits can be turned
into lower limits on the mass using 
calculations of the production cross section.

In 1984, the UA1 collaboration reported~\cite{arnison84} 
evidence for the production of top quarks with $m_t=40\pm10~\gevcc$.
In a subsequent
analysis, however, with a larger data sample and a more 
thorough evaluation of
backgrounds, the putative signal vanished!\cite{albajar88}  A limit of
$m_t>45\gevcc$ was inferred
from this latter analysis.  The UA1 and UA2 experiments continued
running through 1989, eventually setting limits of
$m_t>60\gevcc$ and  $m_t>69\gevcc$,
respectively.\mcite{albajar90,*akesson90}  Yet, 
even as they were in their last
stretch, the CDF detector came online at the Fermilab Tevatron in 1988
and started recording $\ppbar$ collisions at the
unprecedented ${\sqrt s}=1.8\tev$, racing the CERN collaborations
for evidence of top quark production with $m_t<M_W$.
The CDF collaboration soon set limits of $\mt > 77\gevcc$ from
the $e+\jets$ channel and $\mt > 72\gevcc$
from the $e\mu$ channel.\mcite{cdfejets90,*cdfemu90,*cdfejets91}
Even with a smaller integrated luminosity ($\lumint = 4.4\ipb$ for CDF
\vs\ $\lumint = 7.5\ipb$ for UA2), this limit was already better than
could be achieved at the S$\ppbar$S experiments (because of the higher beam
energy of the Tevatron).  CDF later extended this
analysis, adding the $ee$, $\mu\mu$, and $\mu+\jets$ channels and
using soft-lepton $b$-tagging in the lepton+jets channels, arriving at
a final limit from the 1988--89 run\cite{cdftopprl92,*cdftopprd92}
of $\mt > 91\gevcc$.
Given these limits, the CERN experiments were out of the running;
the search would continue only at the world's highest energy collider,
the Fermilab Tevatron.

Collider operations at the Tevatron resumed in July, 1992, at which time
the CDF detector was joined by the newly-commissioned \dzero\
detector.  Collider running continued through 1996, with the Tevatron
reaching peak luminosities of over $2\times 10^{31}~\lumunits$,
a factor of 10 higher than the previous run, and twenty times
the design luminosity.  The average
luminosity for this period was $1.4\times 10^{31}~\lumunits$.

Using the data from the first period of running (Run 1a),
with $\lumint=13.5\ipb$, the \dzero\
experiment soon set a limit of $\mt > 131\gevcc$ using the $ee$,
$e\mu$, $e+\jets$, and $\mu+\jets$
channels.\cite{d0topprd,d0xsecprl94}
(This limit was later
revised downwards to $128\gevcc$ because of a recalibration 
of the luminosity at \dzero.)
In April 1994, however, the CDF collaboration
claimed~\cite{cdftopprd94,cdftopprl94}
the first evidence for $\ttbar$ production.  With an
integrated luminosity from Run~1a of $19.3\ipb$,
CDF observed twelve candidate events in the dilepton and
lepton+jets channels and estimated a $0.26\%$
probability for the background to fluctuate to at least that many events.
Assuming that the excess was due to $\ttbar$ production, 
the cross section was measured to be $\sigma_{\ttbar}=13.9^{+6.1}_{-4.8}\pb$.
Under the same assumption, CDF also measured the top quark mass
using the $b$-tagged sample, obtaining a result of
$m_t=174\pm 10^{+13}_{-12}\gevcc$.  Meanwhile, the \dzero\
collaboration had reoptimized its analysis to search for high-mass
top~quarks, with $\mt \sim 180\gevcc$.
Nine candidate events were observed,\cite{d0topprd,d0xsecprl95}
compared to $3.8\pm 0.9$
events expected from background. Taking $m_t=180\gevcc$ yielded a
$\ttbar$ production cross section of $8.2 \pm 5.1\pb$.  The
chance of the observed signal being an upward
fluctuation of the background was calculated to be $2.7\%$;
therefore, \dzero\ concluded
that the excess was of insufficient statistical
significance to demonstrate the existence of the top quark.   

\begin{figure}
\centering
\epsfig{file=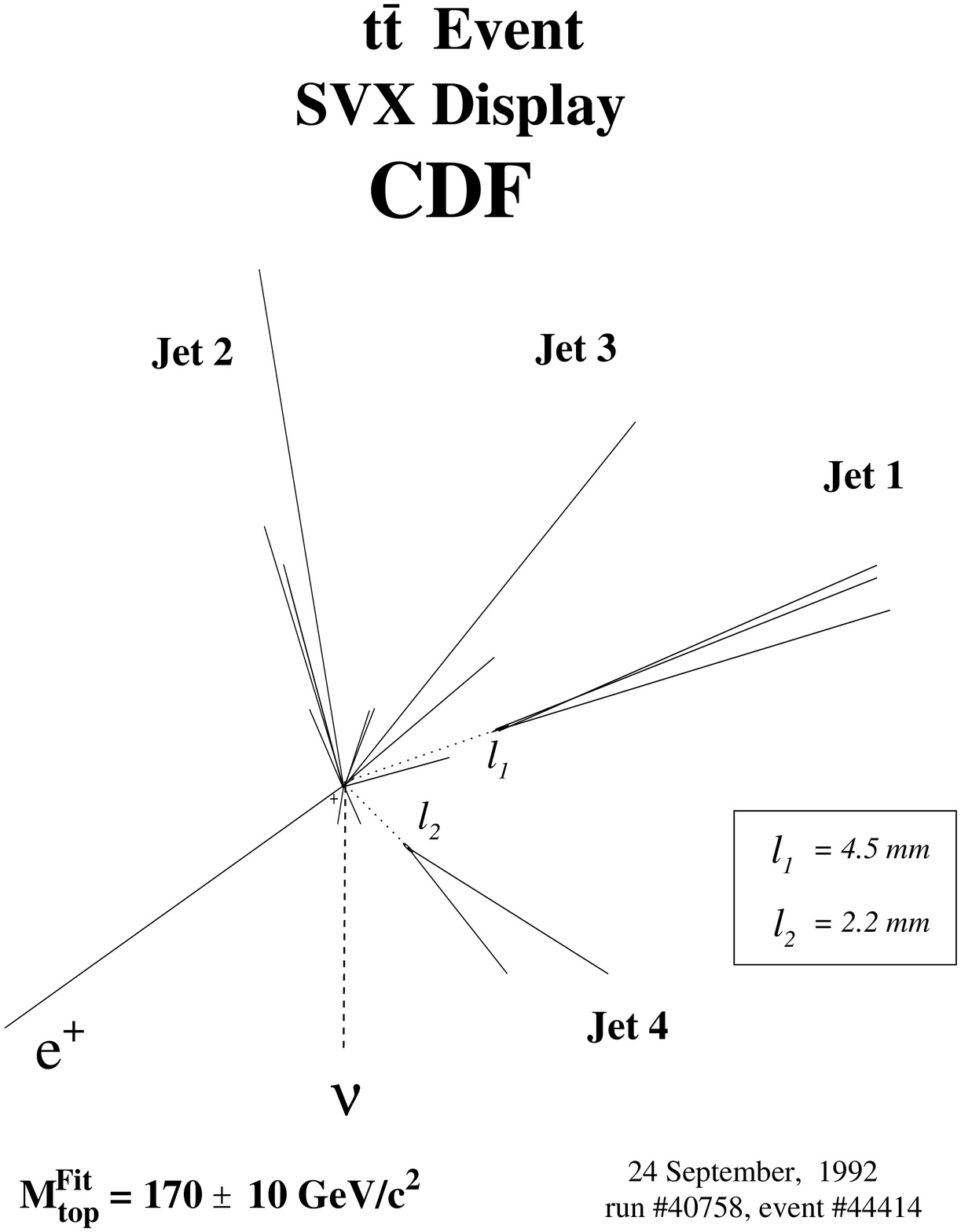,width=0.9\hsize}
\caption{A $\ttbar\rightarrow e+\jets$ candidate event from the
  CDF experiment. 
  Jets 1 and 4 are identified by the SVX detector  
  as $b$-quark decays with vertices displaced from the
  primary event vertex. Jets 2 and 3 could have come from a $W$ boson, and
  the $e^+$ and $\nu$ from the
  other $W$ boson in the event.
  From Ref.~\protect\onlinecite{kim95}.}
\label{fig:cdfljetsevent}
\end{figure}

Run 1a brought several spectacular $\ttbar$ candidate events.
In September, 1992, CDF recorded a
beautiful $\ttbar\rightarrow
(W^{+}b)(W^{-}\bar{b}) \rightarrow (e^{+}\nu b)(q\bar{q} \bar{b})$
candidate.  A display of the
SVX tracks in the event is shown in \figref{fig:cdfljetsevent}.  The
event has an isolated electron, large $\met$, and two jets with
clearly identified displaced vertices (indicative of $b$-quark
decays).  This event will surely find its way into the textbooks as an
ideal top-antitop event!
A kinematic fit of this single event to the $\ttbar$
decay hypothesis yields $m_t=170\pm 10\gevcc$.

\begin{figure}
\epsfig{file=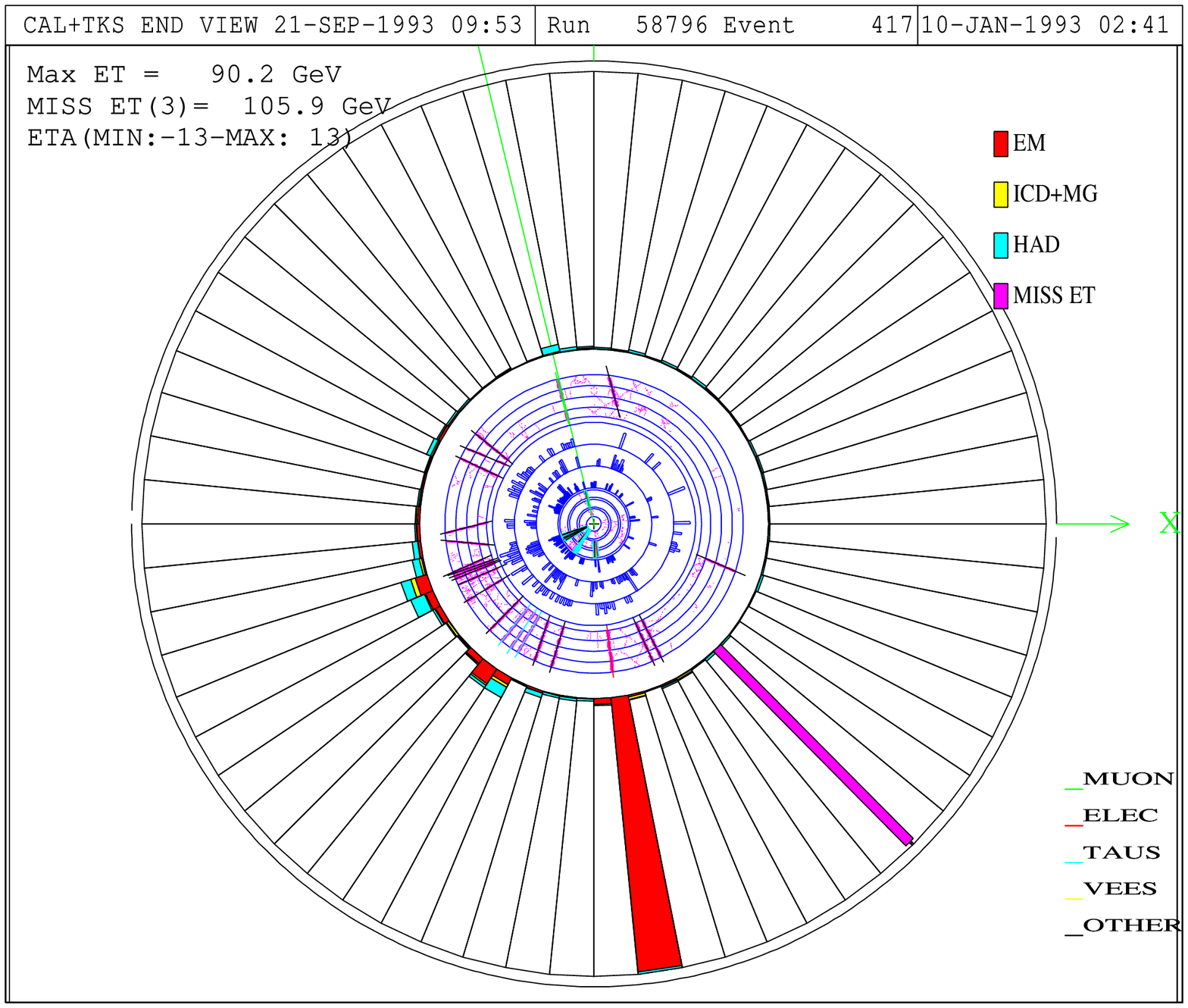,origin=c,angle=270,width=\hsize}
\caption{An $e\mu+\jets$ candidate event from the \dzero\ experiment.
  This is an end view of the tracking detectors and the central
  calorimeter.  The muon is shown by the dotted line near the top, the
  electron is the large EM deposition near the bottom, and the two
  jets are the smaller depositions at about 7 and 8 o'clock.
  The $\met$ is shown by the narrow bar at about 5 o'clock.}
\label{fig:d0event417}
\end{figure}

Another spectacular event is
a dilepton ($e\mu$) event recorded by \dzero\ in January, 1993.  
An event display is shown in \figref{fig:d0event417}.
This event has two high-$\pt$ leptons ($E_T^e=98.8\gev$,
$p_T^\mu=194.6\gevc$), large $\met$ ($100.7\gev$),
and two jets ($E_T^{j1}=26.1\gev$, $E_T^{j2}=23.0\gev$).
A multivariate Fisher discriminant analysis~\cite{pushpadpf} showed that
this event is eighteen times more likely to be a $\ttbar$ event than
a $(Z\rightarrow \tau\tau)+\jets$ event, and ten times more likely to be
$\ttbar$ than $WW+\jets$.

Run 1a continued through June, 1993.  Over the
summer, the accelerator was upgraded, with improvements to the linac
and the installation of electrostatic separators.  In December, collider
operations resumed with Run 1b.
By February, 1995, both experiments had quadrupled their data sets,
and had observed large excesses of events over background
that were fully consistent with the $\ttbar$ hypothesis.  Finally, on
March 2, 1995, the collaborations announced
that the long search was over:
the top~quark had been found.\cite{cdfdiscovery,d0discovery}

CDF, in its $67\ipb$ data set, observed 37 lepton+jets events with
at least one $b$-tag.  In this sample, there were 27 SVX $b$-tags,
compared to $6.7\pm 2.1$ tags expected from background, and 23 SLT
$b$-tags, compared to an expected background of $15.4 \pm 2.0$ tags.  
Six dilepton events were also seen, compared to $1.3 \pm 0.3$ events
expected from background.  The probability for the estimated
background to fluctuate to at least the observed number of events
was calculated to be $1\times 10^{-6}$, corresponding to a $4.8\sigma$
deviation for a Gaussian distribution.  CDF obtained a total $\ttbar$
cross section of $6.8^{+3.6}_{-2.4}\pb$, and a mass for the top~quark of
$m_t=\statsyst{176}{8}{10}\gevcc$.

Simultaneously, \dzero, using approximately $50~\ipb$ of data,
observed 17 events over an estimated background of $3.8
\pm 0.6$ events.  The probability for the background to fluctuate
to at least the measured yield was $2\times 10^{-6}$, equivalent to
$4.6 \sigma$ for a Gaussian distribution.  \dzero\ measured a
top quark mass of $m_t=199^{+19}_{-21}~\text{(stat)}\pm
22~\text{(syst)}\gevcc$ and a $\ttbar$ production cross section of
$6.4 \pm 2.2\pb$.

\section{Measurement of the $\ttbar$ Production Cross Section}
\label{topxs}
\subsection{General Strategy}

Measuring the $\ppbar \rightarrow \ttbar$ production cross section
can be done in several decay modes.
These are categorized as either dilepton, lepton+jets,
or all-jets, as discussed in \secref{production}.  These can be
further subdivided based on the lepton flavors in the
final states, on whether or not $b$-tags are present, and on 
the method used for $b$-tagging.  For each subchannel,
the cross section is given by
\begin{equation}
\sigma_{t\bar{t}} = \frac{(N - B)}{\epsilon \times \lumint},
\label{eq:xsect}
\end{equation}
where $N$ is the number of observed events, $B$ the estimated
background count, $\epsilon$ the total detection efficiency for
$\ttbar$ events in the subchannel, and $\lumint$ the integrated
luminosity of the data set. The total efficiency $\epsilon$
includes the branching fraction, the geometrical
acceptance of the detector, and
the efficiencies for trigger selection,
identification of leptons and jets, and
kinematic selections.  For any given set of kinematic selections,
the signal efficiency depends on the mass of the top~quark; therefore,
the measured cross section depends on
the assumed value of the mass.  For some channels, the
background prediction is based on data samples in which there
is a small
contribution from $\ttbar$ production.  In such analyses, the measured
$\ttbar$ cross section is used to correct the background
prediction iteratively,
until the measured cross section is stable.  
To obtain a final measurement of the cross section, all
channels are combined, taking into account any correlated
uncertainties, such as those on integrated luminosity and on
the signal and background models.

Both collaborations use the \progname{HERWIG}\cite{herwig}
Monte Carlo program as
the primary model for $\ttbar$ events. In addition, CDF uses
\progname{PYTHIA}\cite{pythia}
to assess systematic uncertainties due
to modeling, while \dzero\ uses \progname{ISAJET}.\cite{isajet}  The $W$+jets
background, which is the principal background in the lepton+jets
channels, is modeled using a combination of collider data and
simulations based on the \progname{VECBOS} program.\cite{vecbos}  The other
important background in these channels arises from misidentification
of one or more jets as leptons in multijet events.  The
misidentification rate (called the fake-lepton probability) is obtained
from the data, as are the lepton identification efficiencies and the
$b$-tag probabilities of the background.  Other (smaller) backgrounds
are estimated using a combination of Monte Carlo simulations and
object identification efficiencies measured from data.

\subsection{CDF Analyses}
\label{cdfxs}

The most up-to-date analyses from CDF use data from Run~1a (1992--93) and
Run~1b (1994--95).\cite{cdftoptotau97,cdfxs98,cdfdilep98,cdfalljets}
Since Run~1c was short,
CDF chose not to trigger on top-like events, 
but instead used the run to pursue other specialized studies.
The total integrated luminosity used in the top analyses is
$109 \pm  7\ipb$. If one assumes the predicted $\ttbar$ cross section of
$5.5\pb$ (at $m_t = 175 \gevcc $), about 600 $\ttbar$ events should be 
present in the CDF data.   A subset of  these data ($\lumint
\sim 67\ipb$) supported CDF's observation of the top
quark.\cite{cdfdiscovery}

Using the full data set, CDF has updated analyses in the 
dilepton and 
lepton+jets/b-tag channels and has performed new analyses in the 
$\tau$ dilepton
($e\tau$ and $\mu \tau$) and all-jets
(all hadronic) channels.  The $\tau$ dilepton
channels are of special 
interest  because, if charged Higgs bosons 
$H^{\pm}$ with $m_H^{\pm} < m_t$ exist, they 
could produce an excess of events in these modes via the decay chain 
$ t \ra H^+b \ra \tau^+ \nu_{\tau}b $.

The all-jets final state accounts for $44\%$ of all $\ttbar$
events.  It is therefore both important and interesting to
test this key prediction through an independent observation of
signal in this channel.

CDF calculates $\sigma_{\ttbar}$ assuming $\mt = 175\gevcc$, while \dzero\
uses $\mt = 172\gevcc$. This corresponds to
a typical difference of $\approx 0.3$~pb in extracted cross sections, due to
the dependence of the efficiency on
mass.

\subsubsection{Dilepton Channels}
\label{cdfxs-dilepton}

We discuss first the analyses in the ``standard'' dilepton channels,
$ee$, $e\mu$, and $\mu\mu$,\cite{cdfdilep98} and leave
dilepton channels with identified $\tau$ leptons for later.

\begin{figure}
\epsfig{file=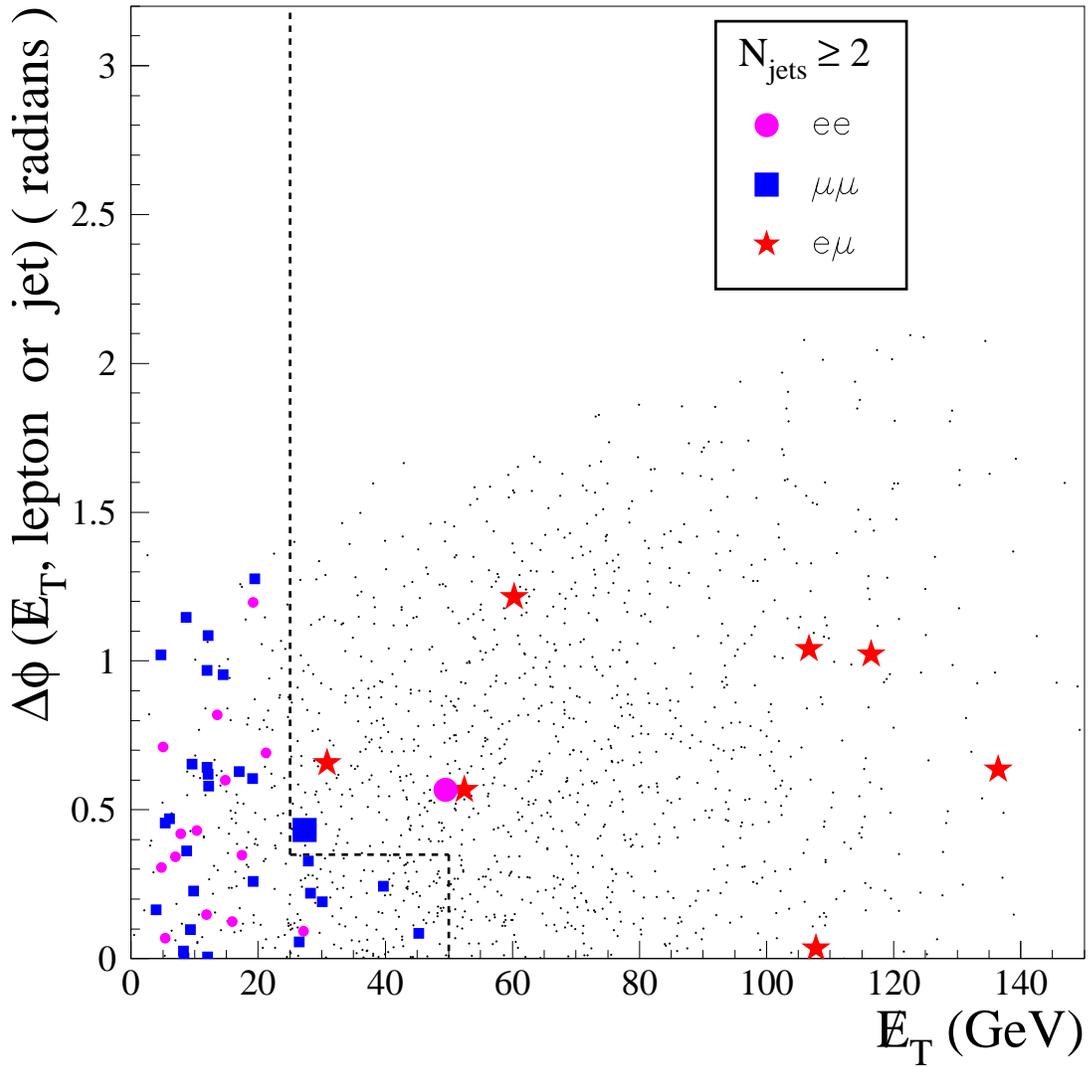,width=\hsize}
\caption{The azimuthal angle $\Delta\phi$ between the $\met$ vector
  and the nearest lepton or jet \vs\ $\met$ in CDF dilepton events.
  The small points show the distribution expected from $\ttbar$ signal
  with $m_t = 175\gevcc$.  The larger symbols represent data. The
  dashed line shows the cuts on $\met$.  From
  Ref.\protect\onlinecite{cdfdilep98}.}
\label{fig:cdfkincuts}
\end{figure}

\begin{figure}
\epsfig{file=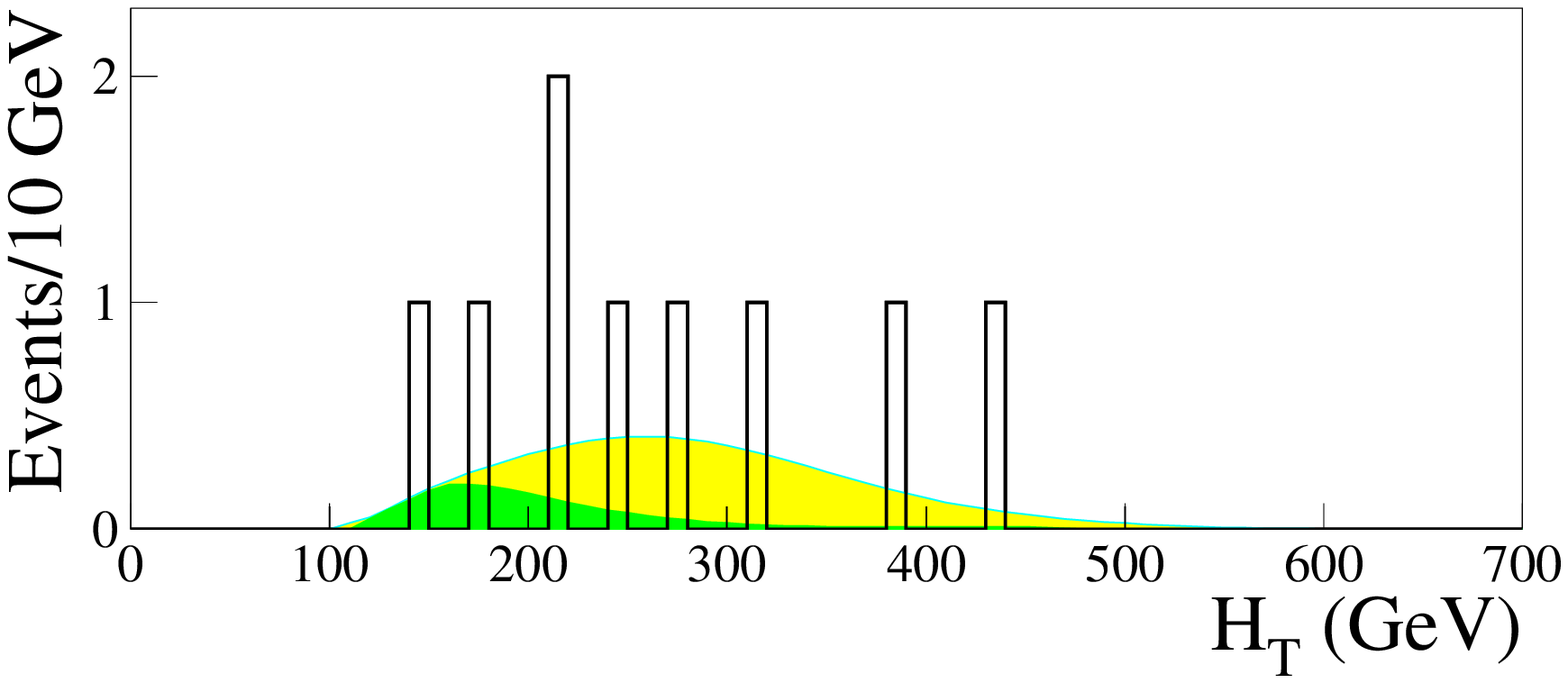,width=\hsize}
\caption{Comparison of $\htran^{\text{all}}$
  for the CDF dilepton candidate events
  (histogram) with the expectation for $\ttbar$ production
  ($m_t=175\gevcc$) plus background (lighter shaded area).  The darker
  shaded area represents background alone.  From
  Ref.\protect\onlinecite{cdfdilep98}.
}
\label{fig:cdfdilepcuts}
\end{figure}

The initial event selection in the standard dilepton analyses requires
the presence of two oppositely charged high-$\pt$ leptons ($e$ or $\mu$),
two or more jets as expected from the $b$-quarks, and large $\met$
as the signature for the neutrinos. The kinematic selection criteria
are shown in \tabref{tab:cdfxscuts}.  Since both leptons in
dilepton $\ttbar$ events come from $W$ boson decays, at least one of
the two leptons is required to be isolated.
The criterion for lepton
isolation is that the transverse energy in the calorimeter in a cone
of $R=0.4$ around the lepton be less than $10\%$ of the lepton's $\et$
(or $\pt$).  The dominant background to the $ee$ and $\mu\mu$ channels
comes from $Z(\ra\ell\ell)+\jets$.  This is largely eliminated by
rejecting events with a dilepton invariant mass $M_{\ell\ell}$ within
a narrow window about the $Z$ boson mass, that is, with
$|M_{\ell\ell}-M_Z| < 15 \gevcc$.  Events containing an isolated
photon with $\et > 10\gev$ consistent with a radiative
decay of a $Z$ boson are removed. The requirement on the $\met$
is tightened (to $\met > 50\gev$) when the $\met$ vector is nearly
collinear with either a lepton or a jet
($\Delta \phi(\met,\text{$\ell$ or jet}) < 20\degree$).
This suppresses
backgrounds from $Z\ra \tau\tau$, where the two $\tau$ leptons (and
hence their decay products) are spatially close when 
the decaying $Z$ boson has high momentum, 
and background from the Drell-Yan continuum where
$\met$ arises from mismeasurements of jet or lepton
energies. The latter process has a very large cross section, and therefore
is an important source of background.  The distribution of
$\Delta\phi($\met$,\text{$\ell$ or jet})$ \vs\ $\met$ is shown in
\figref{fig:cdfkincuts} for events that pass all but the final
selection criterion.
The expected distribution for $\ttbar$ events, calculated with
\progname{herwig} assuming $m_t=175\gevcc$, is superimposed for
comparison.  Nine events --- seven $e\mu$, one $ee$, and one $\mu\mu$
--- survive all the requirements.  The distribution of $\htran^{\text{all}}$
($\equiv \pt^{\ell 1} + \pt^{\ell 2} +
 \met + \et^{\text{jet1}} + \et^{\text{jet2}}$) for candidate events
is compared to expectations for signal and background in
\figref{fig:cdfdilepcuts}.

\begin{table*}
\caption{Basic kinematic requirements in various channels in the CDF cross 
section measurements.}
\label{tab:cdfxscuts}
\begin{center}
\begin{tabular}{lcccc}
 & Standard Dilepton & $\tau$-Dilepton   & $\ell+\jets$/$b$-tag   & All jets\\
 & ($ee$, $e\mu$, $\mu\mu$) &  ($e\tau$, $\mu\tau$)      &        &         \\
\hline
Lepton $p_T (\ugevc)$     & $> 20$  & $\pt^\tau>15$      & $> 20$ &   ---   \\
                          &         & $\pt^{e,\mu}>20$   &        &         \\
Lepton $|\eta |$          & $< 1.0$ & $|\eta^\tau|<1.2$  & $< 1.0$&   ---   \\
                          &         & $|\eta^{e,\mu}|<1.0$ &      &         \\
$\met$ (GeV)              & $> 25$  &  ---               & $> 20$ &   ---   \\
$S_{\met}$ ($\ugev^{1/2}$)&   ---   & $>3$               &  ---   &   ---   \\
Jet $\et$ (GeV)           & $> 10$  & $> 10$             & $>15$  &  $>15$  \\
Jet $|\eta |$             & $< 2.0$ & $< 2.0$            & $<2.0$ &  $<2.0$ \\
Number of jets            & $\geq2$ & $\geq2$            & $\geq3$&  $\geq5$\\
$\htran$ (GeV)            & ---     & ---                & ---    &  $>300$ \\
$\htran^{\text{all}}$(GeV)& ---     & $>180$             & ---    &  ---    \\
 \end{tabular}
\end{center}
\end{table*}

The estimates of the backgrounds from various sources are listed in
\tabref{tab:cdfllback}.  After the initial event selection,
Drell-Yan ($Z/\gamma^* \ra ee, \mu\mu $) production
continues to be the major background for the $ee$ and
$\mu\mu$ channels.  The $e\mu$
channel, the cleanest of the dilepton channels, has background mainly
from decays of $W$ boson pairs ($WW$) and $Z \ra \tau\tau$.  In all
cases, additional jets can arise from gluon radiation, and $\met$ from
either neutrinos or mismeasurement of energies.  CDF also estimates
the background from false signatures due to particle
misidentification, such as a jet or a track faking one of the leptons,
or overestimated $\met$ due to mismeasured muon momenta.  The
backgrounds from Drell-Yan production, fake leptons, and mismeasured
tracks are estimated from the data. Other backgrounds are calculated
using Monte Carlo simulations, which use lepton identification
efficiencies estimated from data.

\begin{table}
\caption{Expected backgrounds to the standard dilepton channels in the
  CDF data, corresponding to $\lumint= 109 \pm 7\ipb$.
  From Ref.~\protect\onlinecite{cdfdilep98}.}
\label{tab:cdfllback}
\begin{center}
\begin{tabular}{lc}
Background type & Expected Number of events \\
& (All dilepton channels) \\
\hline
Drell-Yan                       & $0.61 \pm 0.30$  \\
$Z\ra \tau\tau$                 & $0.59 \pm 0.14$  \\
Fake leptons                    & $0.37 \pm 0.23$  \\
$WW$                            & $0.36 \pm 0.11$  \\
Mismeasured muon tracks         & $0.30 \pm 0.30$  \\
QCD $b\bbar$                    & $0.05 \pm 0.03$  \\
Other (radiative $Z$, $W\bbar$,  
       $WZ$, $ZZ$)              & $0.10 \pm 0.10$  \\
\hline                          
Total                           & $2.40 \pm 0.50$  \\
\end{tabular}
\end{center}
\end{table}

For the acceptance of $\ttbar$ events, CDF takes the average of the
results from \progname{HERWIG} and \progname{PYTHIA}.  The lepton
identification efficiencies are measured to be 91\% for muons and 83\%
for electrons, using $Z \ra \ell^+\ell^-$ data.  (These efficiencies
do not include the geometric acceptance of the detector or the isolation
cuts.)  The overall
efficiency for detection of $\ttbar$ events
(including the branching ratio) is estimated to be
$\epsilon = (0.74 \pm 0.08)\%$~for $m_t = 175\gevcc$. The uncertainty
in the efficiency reflects uncertainties in event modeling (estimated
from the differences between \progname{HERWIG} and \progname{PYTHIA})
and in the simulation of the detector.  Assuming a $\ttbar$ production cross
section of $5.5\pb$, CDF expects to see a total of 4.4
$\ttbar$ events in the standard dilepton channels, of which
$(58 \pm 2)\%$ are expected to be $e\mu$, $(27 \pm 1)\%$ $\mu\mu$, and
$(15 \pm 1)\%$ $e e$.  Both charged leptons directly come from $W$~boson
decays in $(86 \pm 2)\%$ of the observed dilepton $\ttbar$ events. In
the remaining events, one of the leptons comes from the decay chain
$W \ra \tau \ra e/\mu$. The total $\ttbar$ acceptance increases by
35\% as $m_t$ increases from 150 to $200\gevcc$.

With nine dilepton events in the data sample, an estimated background
of $2.4 \pm 0.5$ events, and an overall $\ttbar$ efficiency of
$(0.74\pm 0.08)\%$, the cross section for $\ttbar$ production
for $\mt = 175\gevcc$ is found
to be
$8.2_{-3.4}^{+4.4}\pb$.  Of the nine candidate events, four have one
jet tagged as a $b$-jet by the SVX algorithm; of these four, two are
also tagged by the SLT method. If all nine candidates are assumed to
be from background processes, the number of expected $b$-tagged jets
is only $0.7 \pm 0.2$.

CDF also searches for $\ttbar$ events in the $e\tau$ and $\mu\tau$
decay channels.\cite{cdftoptotau97}   The total branching ratio for these
channels is {$4/81$}, the same
as for all the standard dilepton modes (see \secref{topdecay}). This
could, in principle, double the efficiency for dilepton modes.
That is not the case, however, since only hadronic decays of the  
$\tau$ are used (branching fraction $\approx 64\%$: $50\%$
one-prong and $14\%$ three-prong decays), and the
identification of $\tau$ leptons is far less efficient than of electrons
or muons.

\begin{figure}
\epsfig{file=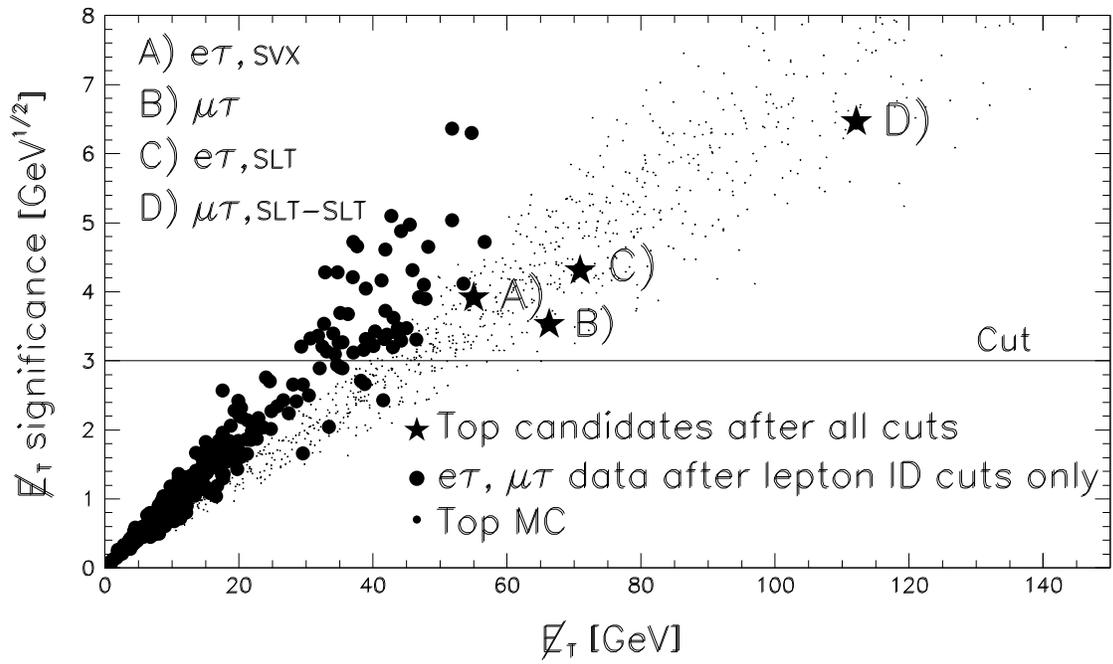,width=\hsize}
\caption{The distribution of $S_{\met}$ \vs\ $\met$ for CDF events
  with a primary lepton and a $\tau$ candidate.
  Three of the four final candidate events (stars) have $b$-tagged
  jets.  From Ref.~\protect\onlinecite{cdftoptotau97}.
}
\label{fig:cdftaudilep2}
\end{figure}

Recall from \secref{tauid} that CDF uses two different $\tau$
identification algorithms, one ``track-based'' and the other
``calorimeter-based.''  Separate cross section analyses are performed
for each of these selections, starting from samples containing events
with $\tau$ candidates along with oppositely charged leptons satisfying
the kinematic cuts summarized in \tabref{tab:cdfxscuts}.
Each event must also
contain at least two jets (since we expect two $b$-quarks) with
$\et > 10\gev$ and $|\eta|<2.0$, and have 
significant $\met$ due to the unobserved
neutrinos. Instead of directly applying a cut on $\met$, CDF applies
a cut on the $\met$ significance, defined as
$S_{\met}=\met/\sqrt{\sum \et}$ for $e\tau$ events, and
$S_{\met} =\met/\sqrt{\sum \et + \pt^\mu}$ for $\mu\tau$ events. Here
$\sum \et$
is the scalar sum of the transverse energies measured in
the calorimeter and provides a measure of the $\met$ resolution. 
The requirement on the $\met$ significance is
$S_{\met} > 3\gev^{1/2}$.
The distributions of $S_{\met}$ \vs\ $\met$ for
the $e\tau$ and $\mu\tau$ data samples as well as for 
$\ttbar$ Monte Carlo events are shown
in \figref{fig:cdftaudilep2}. A cut $\htran^{\text{all}} > 180\gev$ 
is also imposed (where $\htran^{\text{all}} \equiv \pt^{\ell} +
\pt^\tau + \met\ + \sum_{\text{jets}} \et^{\jet}$,
$\ell = e$ or $\mu$).  Using either $\tau$ identification
method, the same four events are left after all
cuts, two $e \tau$ and two $\mu\tau$ (shown as
stars in \figref{fig:cdftaudilep2}). Three of these events are
$b$-tagged, and one has an SLT-SLT double-tag.

The $\tau$ identification and signal efficiencies are estimated using
\progname{PYTHIA} $\ttbar$ events decayed with the \progname{TAUOLA}
package,\cite{tauola} which properly treats the $\tau$ polarization.  The
$\tau$ identification is cross checked using a data sample enriched in $W
\ra \tau\nu_\tau$ decays (see \figref{fig:cdftaudilep1}).  Of all the
selected one-prong events, $19\%$ and $38\%$ are found only by the
track-based and calorimeter-based techniques, respectively, and $43\%$
by both.  The overall signal efficiency is estimated to be
$\epsilon = (\statsyst{0.085}{0.010}{0.012})\%$
for the track-based analysis, and
$\epsilon = (\statsyst{0.134}{0.013}{0.019})\%$ for the
calorimeter-based analysis.

The systematic uncertainty is dominated by uncertainties in
identification efficiencies for the $\tau$ (6\%) and $e/\mu$ (7\%) and
in the hadronic energy scale of the calorimeter (5\%). The uncertainty in
the top quark mass contributes another 6\%.  If one uses the $\ttbar$
production cross section of $\sigma_{\ttbar} = 7.6^{+1.8}_{-1.5}\pb$,
measured by CDF from other decay channels, one expects
$\statsyst{0.7}{0.2}{0.1}$ and $\statsyst{1.1}{0.3}{0.2}$ $\ttbar$
events in the track-based and calorimeter-based analyses,
respectively.  So, if the top quark decays as predicted by the
Standard Model, these channels are not expected to 
improve significantly the acceptance for signal.

The primary background in the $e\tau$ and $\mu\tau$ decay channels
comes from $Z/\gamma^* \ra \tau^+ \tau^- + \jets$ events, where one
$\tau$ decays leptonically and the other hadronically. The other
backgrounds include $WW$ and $WZ$ production and ``fake $\tau$'s'' in
$W+\jets$ events, where a jet is misidentified as a $\tau$.  The fake $\tau$
background is calculated by weighting the $\et$ spectrum of all jets
that could be misidentified as $\tau$'s in the $W+\geq 3~\jets$ sample by
the fake rate, determined by applying the $\tau$ selection criteria to a
multijet event sample.  The background estimates are shown in
\tabref{tab:cdftaubkgd}.  The total background for events with
$N_{\jet} \geq 2$ is estimated to be $1.28 \pm 0.29$ events for the
track-based selection and $2.50 \pm 0.43$ events for the
calorimeter-based selection. 
The measured $\ttbar$ production cross sections, 
based on the four events observed, are
$\sigtop=29.1_{-18.4}^{+26.3}~\text{(stat)}\pm4.7~\text{(syst)}\pb$
for the track-based selection and
$\sigtop=10.2_{-10.2}^{+16.3}~\text{(stat)}\pm1.6~\text{(syst)}\pb$
for the calorimeter-based selection. Unfortunately, 
because of their large uncertainties,
these
measurements do not improve the
precision of the overall cross section measurement. But the analysis
is important because non-standard decays of top, specifically $t\ra
H^+b$ with $H^+\ra \tau^+ \nu_{\tau}$, could have produced an excess
above the Standard Model prediction. No significant excess is
evident.
 
\begin{table*}
\caption{Numbers of events expected and observed for the CDF
  $\tau$-dilepton analyses.  The one-jet columns are background
  dominated; the columns with two or more jets correspond to the signal
  region.  From Ref.~\protect\onlinecite{cdftoptotau97}.}
\label{tab:cdftaubkgd}
\begin{center}
\begin{tabular}{ccccc}
Selection       & \multicolumn{2}{c} {Track-based} & \multicolumn{2}{c} 
{Calorimeter-based} \\ 
$N_{\jet} \geq 10\gev$  & 1  & $\geq2$ & 1  & $\geq2$ \\
\hline
$\tau$ fakes       
            & $0.14\pm0.01$ & $0.25\pm0.02$ & $0.47\pm0.03$ & $0.78\pm0.04$ \\
$Z/\gamma\ra\tau^+ \tau^- $  
            & $0.22\pm0.12$ & $0.89\pm0.28$ & $0.54\pm0.16$ & $1.48\pm0.38$ \\
$WW$, $WZ$  
            & $0.14\pm0.06$ & $0.14\pm0.08$ & $0.20\pm0.09$ & $0.24\pm0.10$ \\
Total Background
            & $0.50\pm0.14$ & $1.28\pm0.29$ & $1.21\pm0.28$ & $2.50\pm0.43$ \\
\hline
Expected from $\ttbar$ 
            & $0.08\pm0.02$ & $0.70\pm0.30$ & $0.13\pm0.03$ & $1.10\pm0.40$ \\
\hline
Observed events 
            & 1             & 4             & 0             & 4   \\
\end{tabular}
\end{center}
\end{table*}

\subsubsection{Lepton+Jets Channels}
\label{cdfxs-lepton}

To measure the $\ttbar$ production cross section in the lepton+jets
channels,\cite{cdfxs98} CDF accumulated data using inclusive electron
and muon triggers that required $\pt^{\ell}>18\gevc$.  A $\met$
trigger was used to compensate for small inefficiencies in the
lepton triggers.

The signal in this channel, $\ttbar\ra
(W^+b)(W^-{\bbar}) \ra (\ell \nu b)(q \qbar' \bbar)$, is subject to
a large background from $W$+jets production.
Since it is hard to identify $W$ bosons that decay hadronically,
these $\ttbar$ events comprise a subset of the $W(\ra\ell\nu)+\jets$
sample.  This sample is used for systematic studies
of the background involving
a $W$ boson, as well as for extracting the signal.

The initial data set consists of high-$\pt$
inclusive lepton events that contain a central ($|\eta| < 1.0$)
isolated electron or muon with $\pt > 20\gevc$.
The lepton isolation $I$ is defined as the  
ratio of the transverse energy in a cone of
$R = 0.4$  around the lepton (excluding the lepton energy) to the
$\pt$ of the lepton; the analysis requires $I< 0.1$.
Events consistent with $Z\ra \ell\ell$ or $Z\ra \ell\ell\gamma$ are
rejected, as in the dilepton analysis.
From this inclusive lepton sample, events with
$\met > 20\gev$ are selected for further study.  Events that pass the dilepton
selection criteria are removed in order to keep the two samples
disjoint. This gives an inclusive sample of $W+\jets$ events.

Nominally, one expects to observe four jets per $\ttbar$ 
event in the lepton+jets
channels.  But, as discussed in \secref{jetid}, the number of jets
may be larger or smaller than this.
Therefore, the analysis requires a minimum of only three jets.
The kinematic selection criteria are
summarized in \tabref{tab:cdfxscuts}. 

CDF uses both the SVX and SLT algorithms for $b$-tagging, with
efficiencies of $(39 \pm 3)\%$ and $(18 \pm 2)\%$, respectively, for
tagging at least one $b$-quark in a $\ttbar$ event with at least three
jets.  These tagging efficiencies are estimated from a $\ttbar$ Monte
Carlo simulation.  After all cuts are applied, the $W + \geq 3~\jets$
sample contains 34 SVX-tagged events containing a total of 42
SVX-tagged jets, and 40 SLT-tagged events with 44 SLT-tagged jets.  Of
these events, 11 are $b$-tagged by both algorithms.
 
The primary background is from $W$+heavy flavor production processes.
For the SVX-tagged sample, CDF estimates this background by using
\progname{vecbos} and \progname{herwig} to predict, as a
function of jet multiplicity, the fraction of $W+\jets$ events that
contain heavy flavor.  This is then combined with the tag probability and
applied to the observed $W+\jets$ events.  Backgrounds from processes
that do not produce a $W$~boson are calculated from the data by
measuring the tagging rate in a sample with low
$\met$ and without isolated leptons, where there are few $W$~bosons.
This is then
used to predict the amount of contamination in the signal region.  The
background due to mistags can be found by assuming that the distribution of
decay lengths from this source is symmetric around
zero. CDF measures the negative half of this distribution from
jet data and uses that to predict the number of events
fluctuating above the cut.  The background expected from
single top production is also calculated, using the latest theoretical
cross sections\cite{heinson97b} and efficiencies derived from Monte
Carlo simulations.  Additional small backgrounds from $WW$, $WZ$, and
$Z \ra \tau \tau$ processes are calculated from Monte Carlo
simulations.

For the SLT-tagged sample, the background is mainly from $W+\jets$
events with fake leptons, $K$ decays in flight, electrons from
photon conversions, and events containing heavy flavor
jets ($W\bbbar$, $W\ccbar$).  These backgrounds are calculated from
jet data by
measuring the tag probability per track as a function of the track
$\pt$. These tag probabilities are then applied to
the tracks in the $W+\jets$ events to estimate the background from
these sources.

The signal efficiency, which includes the trigger, tagging
and lepton identification efficiencies, kinematic and geometric
acceptances, and the branching fraction, 
is estimated using a combination of data and
Monte Carlo events.
For $m_t = 175\gevcc$, the overall efficiency is
$(3.7 \pm 0.3)\%$ with SVX-tagging and
$(1.7 \pm 0.3)\%$ with SLT-tagging.

The main systematic uncertainties arise from the jet
energy scale ($\pm5\%$), modeling of initial and final state gluon
radiation 
$(\pm2\%$ and $\pm 5\%$, respectively), Monte Carlo
generators ($\pm5\%$), detector resolution effects
($\pm2\%$), and the instantaneous  luminosity ($\pm1\%$).

\begin{figure}
\epsfig{file=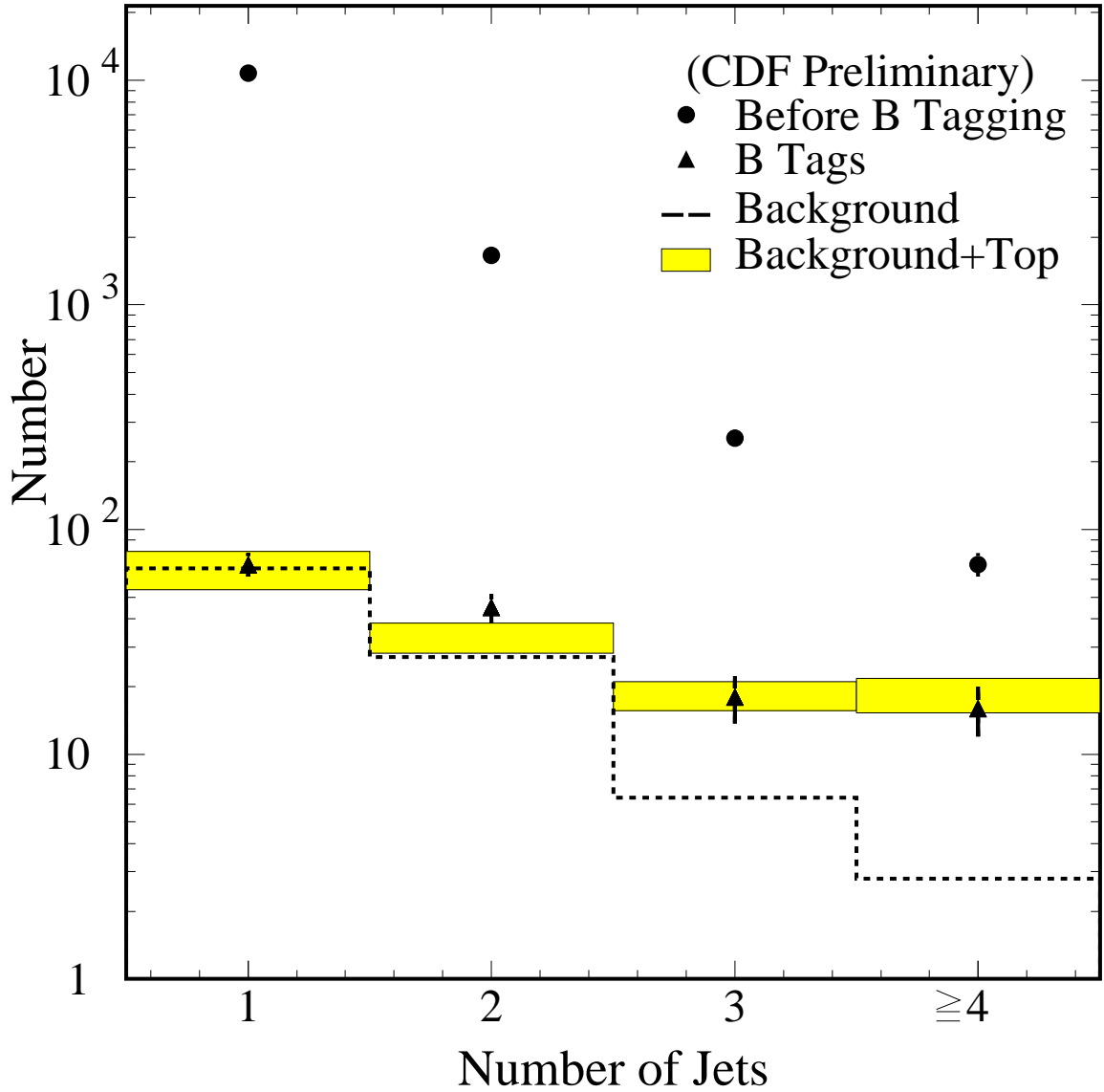,width=\hsize}
\caption{The CDF jet multiplicity distribution of SVX-tagged lepton+jets
  events.  Note the clear excess above background for events
  with three or more jets.  From Ref.~\protect\onlinecite{gerdes97}.}
\label{fig:cdfljets}
\end{figure}

The results of the analysis are summarized in \tabref{tab:cdfxssum};
the SVX-tagged results are also displayed in \figref{fig:cdfljets}.
Note that the SVX-based analysis achieves a signal to background ratio
about five times better than the SLT-based analysis, demonstrating the
clear advantage of $b$-tagging with a silicon vertex detector.
Finally, for $\mt = 175\gevcc$,
CDF obtains a cross section of $6.2^{+2.1}_{-1.7}\pb$ using
the SVX analysis, and $9.2^{+4.3}_{-3.6}\pb$ using the SLT analysis.

\begin{table*}
\caption{Summary of CDF acceptance factors, event yields, backgrounds,
  and  measured cross sections for each analysis channel, for
  $\mt = 175\gevcc$.}
\label{tab:cdfxssum}
\begin{center}
\begin{tabular}{lccccc}
                          & Dilepton
                          & \multicolumn{2}{c} {Lepton+Jets}
                          & \multicolumn{2}{c}{All-Jets} \\
\cline{2-6}
 & & SVX  & SLT     & SVX & 2 SVX  \\
\hline
$\epsilon_{\text{total}}$ & $0.0074\pm0.0008$
                          & $0.037\pm0.005$ 
                          & $0.017\pm0.0003$
                          & $0.044\pm0.010$
                          & $0.030\pm0.010$ \\
Obs. Events               &  9
                          & 34
                          & 40
                          & 187 
                          & 157   \\
Background                & $2.4\pm0.5$
                          & $9.2\pm1.5$
                          & $22.6\pm2.8$
                          & $142\pm12$
                          & $120\pm18$   \\
Expected $\ttbar$         & 4.4
                          & 31.0
                          & 14.3
                          &
                          & \\
$\sigtop$ (pb)            & $8.2_{-3.4}^{+4.4}$
                          & $6.2_{-1.7}^{+2.1}$
                          & $9.2_{-3.6}^{+4.3}$
                          & $9.6_{-3.6}^{+4.4}$
                          & $11.5_{-7.0}^{+7.7}$  \\
\end{tabular}
\end{center}
\end{table*}

\subsubsection{All-Jets Channel}
\label{cdfxs-alljets}

The all-jets channel has the advantage of having a large 
branching fraction ($\approx 44$\%)
and a fully reconstructible final state (since
no high-$\pt$ neutrinos are present).  However, it is subject to an 
overwhelming background from QCD multijet production, with about one thousand
times the rate of $\ttbar$ production.  But, by requiring large transverse
energy in the final state as  well as the presence of $b$-quarks, the
multijet background can be reduced to manageable levels.  

CDF recently reported~\cite{cdfalljets} 
the first observation of $\ttbar$ decays in the 
all-jets channel (referred to as the all hadronic  channel by CDF).  
The collaboration performs 
two analyses, which require either a single or a double $b$-tag,
and, in both cases, finds an excess relative to the estimated background.

The data sample was collected using a special multijet online
trigger, specifically developed for $\ttbar$ decays into the all-jets
final state. This trigger requires four or more clusters of contiguous
towers in the calorimeter, each with
$\et \ge 15\gev$ and a total transverse energy of
$\sum \et^{\jet} \ge125\gev$. 
Offline, jets are reconstructed using a cone algorithm with
$R=0.4$, and jet energy corrections are applied (see
\secref{jetid}).
Selecting events with five or more jets and requiring
$\sum \et \ge 300\gev$ yields 21,840 events.  Events with high-$\pt$
electrons and muons (which would appear in the dilepton and  lepton+jets
channels) are removed.  After this selection, the QCD background still
dominates $\ttbar$
production by a factor of about 110. But when one demands at least one
SVX $b$-tag, the signal to background ratio increases to
$S/B \approx 1/20$, in a sample of $1596$ events.  Thereafter, 
two approaches (referred to as Techniques I and II) are used to
further reduce background.

In Technique I, two new variables are introduced that together prove decisive
in achieving the final $S/B$ ratio of $\approx 1/4$:
\begin{itemize}
\item The \emph{centrality} $C = \htran/\sqrt{\hat{s}}$, where
  $\sqrt{\hat{s}}$ is the invariant mass of the multijet system.
\item The aplanarity $\aplan$, computed from the jet momenta.
\end{itemize}
The following cuts are then made on these variables:
\begin{itemize}
\item $C > 0.75$.
\item $\aplan > -0.0025 H_{T3} + 0.54$; 
  ($H_{T3}=\htran-\et^{\jet1}-\et^{\jet2}$).
\end{itemize}
The motivation for the first cut is the observation that
the jets from
$\ttbar$ decay are
more central than those from QCD production (see \secref{production}).
The \dzero\ collaboration uses
a similar variable, except that the 
$\htran$ is divided by the sum of the jet energies 
(see \secref{d0xs}). The second cut provides a more optimal way to
separate signal from background than that afforded 
by independent cuts on the aplanarity and $\htran$. 
(Cuts of this kind in the ($\aplan, \htran$)-plane were
first used by \dzero.\cite{pushpadpf,pushpapbarp})

After all cuts, 187 events containing 222 $b$-tags remain. This is a
statistically significant excess over the estimated background of
$164.8 \pm 10.8$ $b$-tags.  The background is largely from QCD
production of heavy quark pairs and from fake
$b$-tags.  The probability for finding a $b$-tag in a multijet sample,
using the SVX
algorithm, is parameterized in terms of the jet
$\et$, $\eta$ and track multiplicity, and the event aplanarity.  To
estimate the background, each multijet event is weighted by its tag
probability. The sum of these weights, for the multijet events that
pass all cuts, gives the background estimate. But, because the multijet
sample contains $\ttbar$ events, the background estimate must be
corrected to account for the $\ttbar$ contribution in the multijet
sample.  This procedure is done iteratively.  The tag probabilities are
cross checked by comparing predicted $b$-tagged distributions with
those observed, using an independent multijet sample.

The uncertainty of the background estimate is dominated by systematics
arising from the dependence of the tag probability on the run
conditions (such as instantaneous luminosity), the event shape, and
kinematics.  In \tabref{tab:cdfxssum} we summarize the event yields
and the background estimates for $N_{\jets} \ge 5$.  CDF estimates
the probability of the background fluctuating to at least the
number of observed $b$-tags as ${\cal P}= 1.5 \times 10^{-3}$,
corresponding to three standard deviations for a Gaussian
distribution.

From the number of tagged events and the background estimate, corrected
for the $\ttbar$ content, CDF extracts the number of $\ttbar$ candidates 
to be  $10.4\pm 6.0$ for Run 1a data and $34.7\pm 16.1$ for Run 1b data.
The total signal efficiency is estimated to be $(9.9\pm 1.6)\%$ for 
$m_t=175~\gevcc$.  The uncertainty in the efficiency 
is largely due to
systematic uncertainties in the jet energy scale (9\%),
in modeling of fragmentation (9\%) and gluon radiation (11\%).
The $b$-tagging efficiencies for the two data periods are
$(38\pm 11)\%$ and $(46\pm 5)\%$, respectively.  The $\ttbar$ cross section
is measured to be
$\sigtop=9.6\pm 2.9~\text{(stat)}^{+3.3}_{-2.1}~\text{(syst)}\pb$.

In Technique II, two or more $b$-tags are required in each event.
Requiring the second $b$-tag significantly reduces the acceptance.
Most of this is recovered, however, by removing the centrality and
aplanarity requirements.
CDF observes 157 such
events with $\geq5~\jets$, while background from QCD
heavy flavor production and fake double tags is predicted to be $122.7\pm 13.4$
events.  The number of signal events is extracted from
a likelihood fit of a sum of contributions from
fake double tags, QCD heavy flavor
background, and $\ttbar$ signal, to 
the observed number of events as a function of jet multiplicity.
The number of $\ttbar$
candidates is found to be $5.9\pm 3.9$ for Run~1a and $31.6\pm 16.4$
for Run~1b.  The corresponding numbers of background events are
$21.1\pm 4.5$ and $98.4\pm 17.3$.  The detection efficiency before tagging for
$\ttbar$ events is $(26.3\pm 4.5)\%$ for $m_t=175\gevcc$.  
The estimated
efficiencies for tagging $\geq2$ heavy flavor jets are $(7\pm 6)\%$ and
$(12 \pm 2)\%$ for the two data-taking periods.
The cross section is 
$\sigtop=11.5\pm 5.0~\text{(stat)}^{+5.9}_{-5.0}~\text{(syst)}\pb$.
The probability for the background to produce at least the
observed excess is ${\cal P}= 2.5\times 10^{-2}$, 
which corresponds to two standard deviations for a
Gaussian distribution.

CDF combines the results from the two techniques to calculate an overall
cross section, taking into account the correlations in the
efficiencies for the two techniques as well as the overlap between
the two data samples (for which 34 events are in common).  The
correlation coefficient between the two techniques is estimated to be
$\rho=0.34\pm 0.13$.  For $m_t=175~\gevcc$, the combined measurement
is $\sigtop=10.1\pm1.9~\text{(stat)}^{+4.1}_{-3.1}~\text{(syst)}\pb$.

\subsection{\dzero\ Analyses}
\label{d0xs}

The \dzero\ collaboration has used its 
full
Run~1 data set of $125 \pm 7\ipb$ to measure the $\ttbar$
production cross section.\cite{d0xsecprl97}
Here, we review the \dzero\
measurements for the various decay channels.
These cross sections are calculated for 
$\mt = 172.1\gevcc$.\cite{d0ljtopmassprd}

\subsubsection{Dilepton Channels}

\begin{figure}
\epsfig{figure=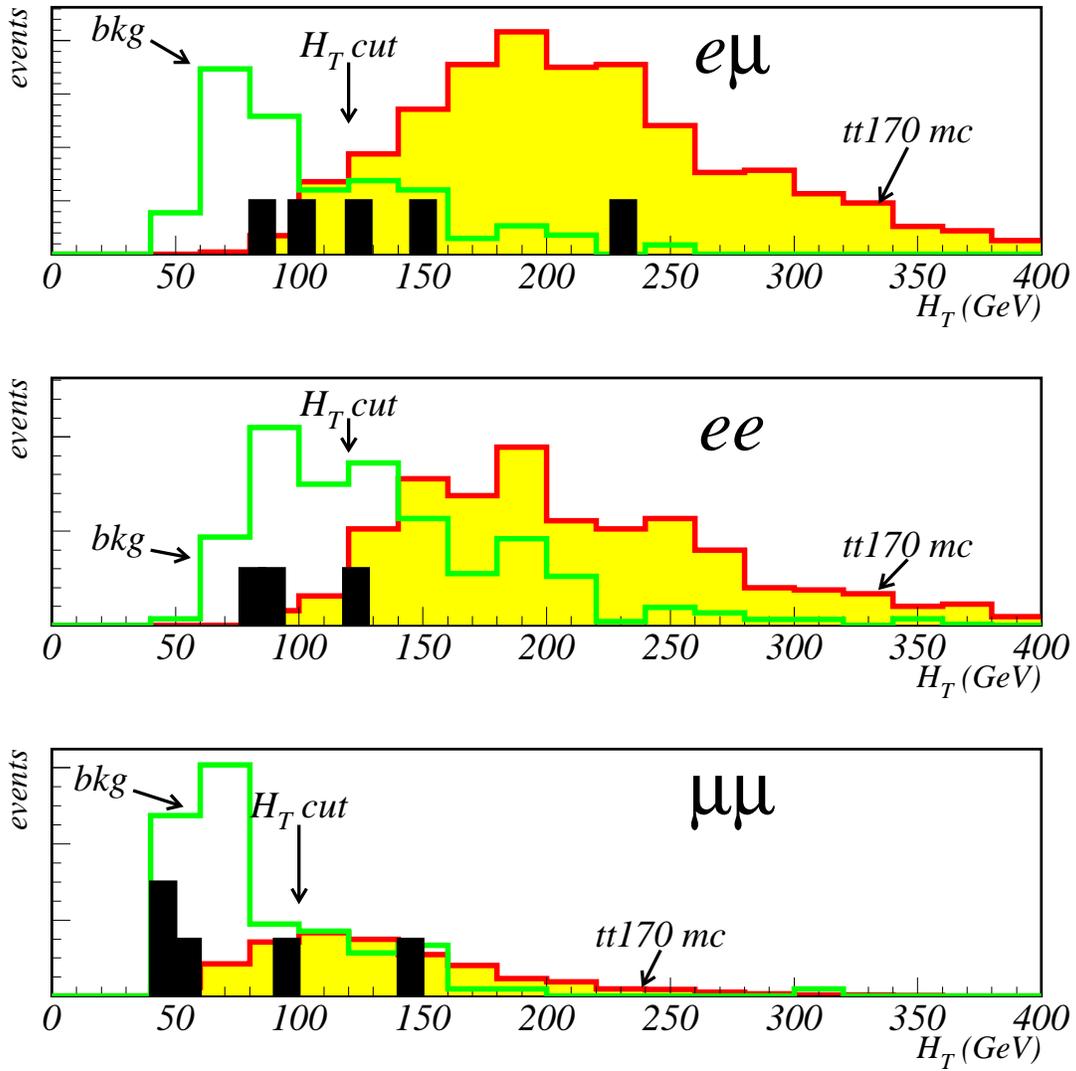,width=\hsize}
\caption{$\htran$ distributions of dilepton events (for $m_t=170\gevcc$) 
  in the \dzero\ analyses.  The location of observed events is shown by 
  dark histograms.
}
\label{fig:d0htll}
\end{figure}

\dzero\ used the standard dilepton channels $e e$, $e\mu$, and
$\mu\mu$ to support 
the discovery in 1995.  Since then, the collaboration has added
the inclusive $e\nu+\jets$ channel. 
The dilepton channels are characterized by two high-$\pt$ leptons,
two high-$\et$ jets, and substantial missing transverse
energy.  The selection criteria are summarized
in \tabref{tab:d0xscuts}.  Note the use of the variable 
$\htran^e \equiv \htran + \et^e$, where $\et^e$ is the transverse energy of
the leading electron. This variable is found to provide better 
discrimination between signal and background than $\htran$ alone.
After additional cuts (discussed below), designed specifically to 
suppress the background from $Z+\jets$ events, the 
sample contains three $e\mu$ events, one $e e$ event, and one $\mu\mu$
event. The distributions of the $\htran$ variable for signal, background,
and candidate events are shown in \figref{fig:d0htll}
(before the cut on  $\htran$).

\dzero\ sought to increase
the $\ttbar$ acceptance by using the ``$e\nu$'' channel to recover some  
events that fail the standard kinematic selection.
This channel requires only a single electron, but has very high
$\met$ and jet cuts.  It is populated by events from the channels
$ee$, 
$e\mu$, and $e+\jets$,  and yields four events that pass the
selection criteria listed in
\tabref{tab:d0xscuts}. We note that
the significance of these data is comparable to that of the dilepton
data; therefore, the $e\nu$ channel makes a useful 
contribution  to the cross section measurement. 

\begin{table}
\caption{Kinematic event selection criteria for various $\ttbar$ 
  decay channels applied in the cross
  section analysis by \dzero.  All energies are in GeV, $\eta$ is the 
  pseudorapidity, $\htran=\sum \etjet$ with $\etjet>15\gev$, and 
  $\htran^e=\htran+\et^e$. }
\label{tab:d0xscuts}
\begin{center}
\begin{tabular}{lcccc}
           & Dilepton  & $\ell +\jets$  & $\ell +\jets/\mu$   & $e\nu$  \\
\hline
Lepton $\pt$  &   $> 15$        &  $> 20$       &  $> 20$       & $> 20$  \\
              &   $> 20$ ($e e$)   &            &               &         \\
Electron $|\eta|$ & $< 2.5$     &  $< 2.0$      &  $< 2.0$      & $< 1.1$ \\
Muon $|\eta|$     & $< 1.7$     &  $< 1.7$      &  $< 1.7$      & ---     \\
\met          & $> 20$ ($e\mu$) &  $> 25$ ($e$) &  $> 20$       & $> 50$  \\
              & $> 25$ ($e e$)  &  $> 20$ ($\mu$)&              &         \\
Jet $\et$     & $> 20$          &  $> 15$       &  $> 20$       & $> 30$  \\
Jet $|\eta|$  & $< 2.5$         &  $< 2.0$      &  $< 2.0$      & $< 2.0$  \\
Number of jets& $\geq 2$        &  $\geq 4$     &  $\geq 3$     & $\geq 2$ \\
$\htran^e$    & $ > 120$ ($e e$,$e\mu$)&  ---   &  ---          & ---    \\
$\htran$      & $> 100$ ($\mu\mu$)&  $> 180$    &  $> 110$      & ---     \\
$\aplan$      & ---             &  $> 0.065$    &  $> 0.040$    & ---     \\
$\et^L$       & ---             &  $> 60$       &  ---          & ---     \\
$|\eta^W|$    & ---             &  $< 2.0$      &  ---          & ---     \\
Tag muon      & ---             &  veto         &  $p_T$ $> 4$  & ---      \\
 &            &                 & $\Delta {\cal R}(\mu,\jet)$  $< 0.5$&    \\
$M_T^{e\nu}$  & ---             &  ---          &  ---          & $> 115$  \\
\end{tabular}
\end{center}
\end{table}

The main backgrounds in the dilepton channels are from $Z$~boson,
Drell-Yan, and diboson processes. 
In the $e e$ channel, the large background
from $Z \rightarrow e e$ is reduced by tightening the $\met$
cut from $25\gev$ to $40\gev$ 
if the dielectron mass lies within $12\gev$ of the mass of the $Z$
boson.  For the $\mu\mu$ channel, the $Z+\jets$ background is also dominant.
Because the $\mu\mu$ and $e e$ channels
have identical kinematics one could, in principle,
apply the same cuts in both cases.
However,  \dzero\ does not measure the momenta of muons
as well as those of electrons.
Consequently, the missing transverse energy is more
susceptible to mismeasurements of muon momenta, and is therefore
less useful as a variable to discriminate
signal from background.   Instead, events are fit
to the $Z (\rightarrow \mu\mu)+\jets$ hypothesis;
events with acceptable fits ($\text{Prob}(\chisq) > 1$\%) are rejected.
The $e\mu$ channel, by contrast, is particularly clean: with the 
selection criteria
given in \tabref{tab:d0xscuts}, \dzero\ obtains 
a signal to background ratio of
$10:1$. Half the background comes from
$Z \rightarrow \tau \tau$ events. 

The $W (\rightarrow e\nu)+ \jets$ background, the main background in
the $e\nu$ channel, is very effectively reduced by the application of
a large transverse mass requirement: $M_T^{e\nu} > 115\gev$,
where
$M_T^{e\nu} \equiv \sqrt{2|\pt^{e}||\pt^{\nu}|- 2p_T^{e}\cdot\pt^{\nu}}$.
The size of the
background is estimated from data: one scales 
down the number of $W + \geq 2~\jets$ events
by the efficiency of the transverse mass cut, the latter
determined from a simulation of $W + \geq 2~\jets$ events using the
\progname{VECBOS} program.

The systematic uncertainties mainly arise from the 
uncertain knowledge of the jet energy scale 
and Monte Carlo modeling, and
are estimated to be between 10 and 15\%.  The $\mu\mu$ channel has an
additional uncertainty of $10\%$, 
due to the $Z\ra \mu\mu$ kinematic fit.
From these channels,
\dzero\ measures $\sigtop(\mt=172.1\gevcc) = 6.4 \pm 3.3\pb$.

\subsubsection{Lepton+Jets Channels}
\label{d0ljetxs}

The lepton + jets channels include the $e+\jets$ and 
$\mu+\jets$ subchannels.
These events are divided into two disjoint
subsamples:
untagged ($\ell+\jets$) and (SLT) tagged ($\ell+\jets/\mu$). 
The major background for both of these subsamples is from 
$W+\jets$ events.  There is also a smaller component from QCD
multijet events, in which one jet fakes a lepton
signature.

The cuts imposed for the untagged $\ell+\jets$ channel are summarized
in \tabref{tab:d0xscuts}, the basic requirements being a high-$\pt$
central lepton, large $\met$, four jets, and no $b$-tag.  For the
first half of Run~1b, the muon acceptance was restricted to
$|\eta| < 1.0$ because of radiation damage in the forward muon chambers.
For Run~1a and the second half of Run~1b (after the damage was
repaired), the acceptance extended out to $|\eta| < 1.7$.
The variable $\eta^W$ shown in the table is the pseudorapidity of the
$W$-vector formed by combining the lepton and the $\met$.  The
longitudinal component is found by constraining the lepton-neutrino
pair to the $W$~boson mass.  A cut $|\eta^W| < 2.0$ is imposed
because \dzero\ observes that \progname{vecbos}, which is used to
model the kinematics of the $W+\jets$ background, does not agree well
with data in the forward region.\cite{d0ljtopmassprd}  The cut has
very little effect on the top quark signal, however.  The variable
$E_T^L$ is defined as the scalar sum of the lepton momentum and the
$\met$.  It is useful mainly for rejecting QCD multijets background;
again, this cut does not have much effect on the $\ttbar$ signal.

\begin{figure}
\centering
\epsfig{figure=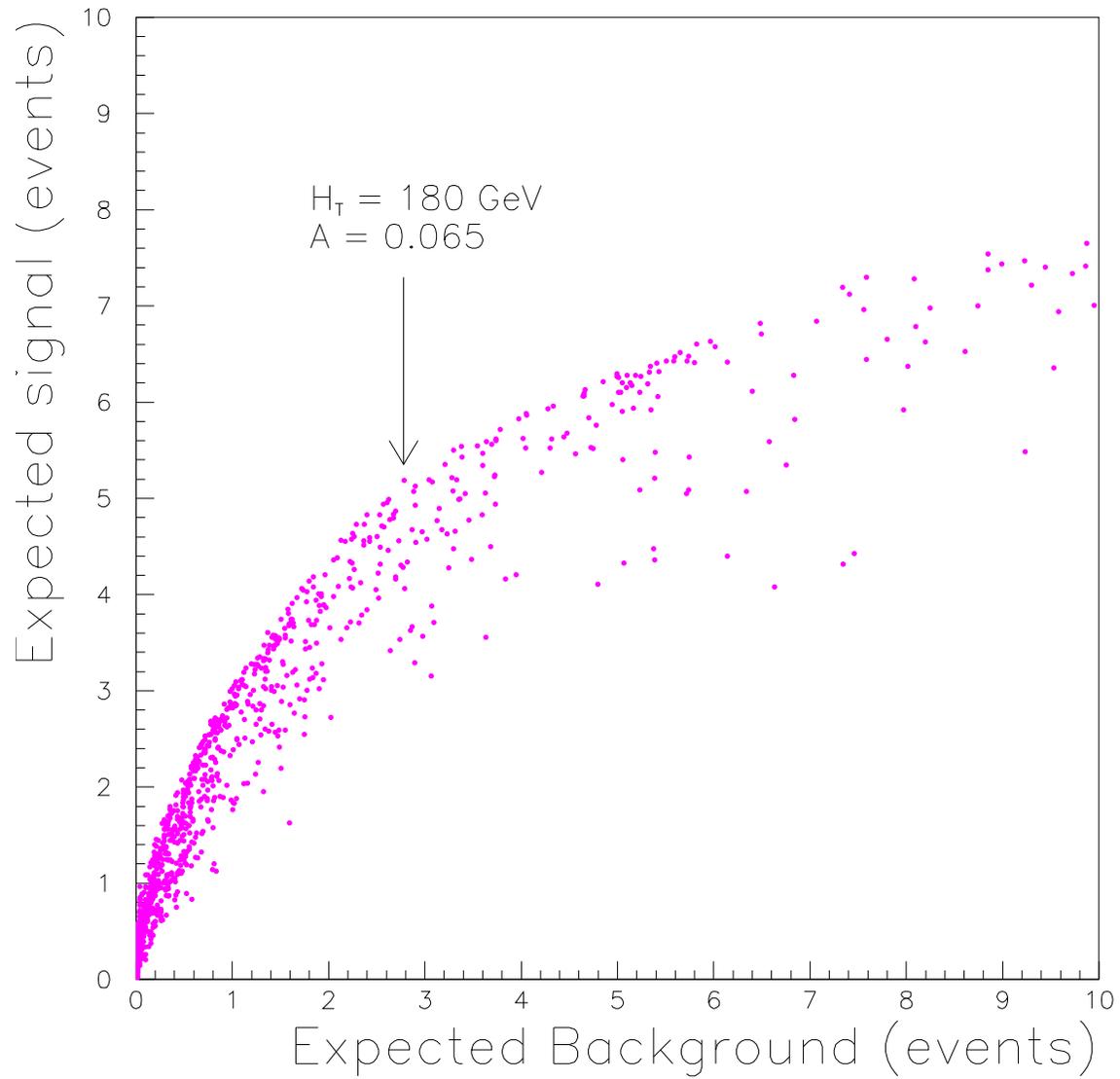,width=\hsize}
\caption{Expected signal (for $\mt = 180\gevcc$, $e+\jets$ channel)
  versus background yields in $77\ipb$
  for various cuts on $\aplan$ and $\htran$.
  The arrow identifies the optimal cut found by the
  random grid search method.}
\label{fig:d0rgs}
\end{figure}

\begin{figure}
\centering
\epsfig{figure=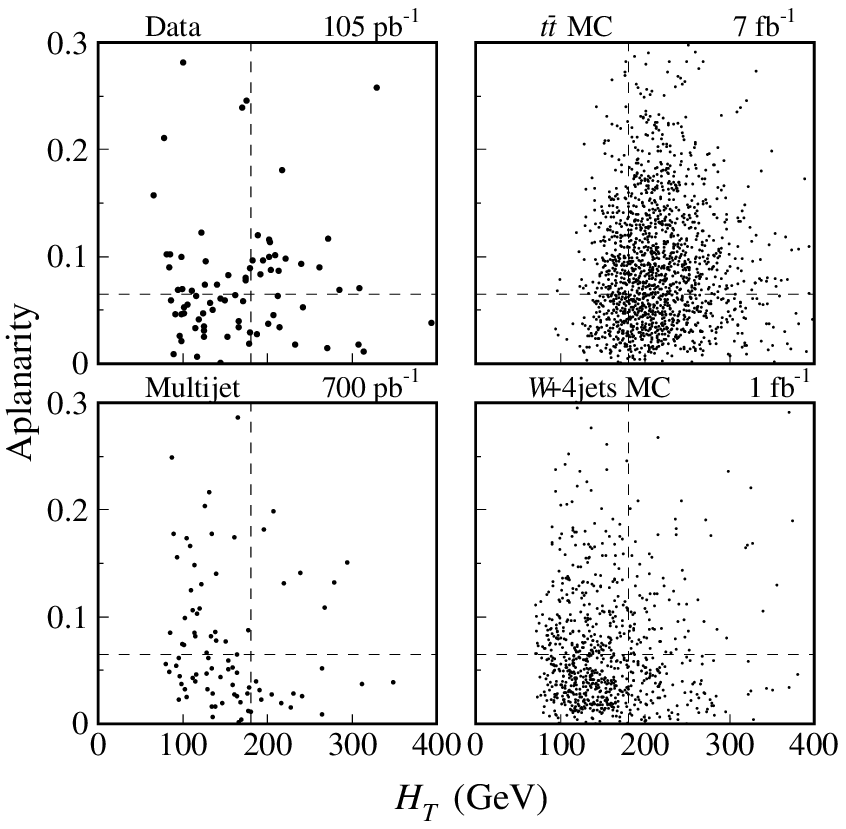}
\caption{Distributions the in $(\aplan, \htran)$ plane for 
  data (top left),
  $\ttbar$ MC (top right), multijet background (bottom left), and
  $W+\jets$ \progname{VECBOS} MC (bottom right).
  From Ref.~\protect\onlinecite{d0xsecprl97}.
}
\label{fig:d0aht}
\end{figure}

\dzero\ also applies cuts in the $(\aplan, \htran)$ plane.
The cut values are chosen through an exhaustive search of possible
cuts in order to optimize the expected precision of the cross section
measurement.  The search method used by \dzero, called
the \emph{random grid search},\cite{rgs} was a key development as it 
provides a highly efficient way to search $n$-dimensional parameter spaces.
It differs from a conventional grid search in 
that the grid of cuts is determined
by the expected distribution of the signal; hence the term ``random'' grid.
The results of the search, along with the optimal cuts, are
shown in \figref{fig:d0rgs}. 
The distribution of events in the $(\aplan, \htran)$ plane
is shown in \figref{fig:d0aht}.
Nineteen events survive all the cuts.

The QCD multijets background is estimated from data by first measuring
the probability for a jet to fake a lepton signature, and then directly
applying these probabilities to the QCD multijets data.
The dominant $W+\jets$ background is estimated using the technique of
``Berends scaling.''  This is based on the observation that the number
of $W+\jets$ events falls off exponentially with the number
of jets.\cite{d0topprd,vecbos}  Thus, after subtracting the
contribution from QCD multijets background, \dzero\ extrapolates from
the number of $\ell + \jets$ events observed at low multiplicities to
estimate the number of $W+\jets$ events with at least four jets.  The
efficiency for these $W+\jets$ events to pass the $(\aplan, \htran)$
requirement
is estimated using \progname{vecbos}.
By this procedure, \dzero\ estimates the
background to be $8.7 \pm 1.7$ events.  About $15\%$ of this
background is due to QCD multijet events; the rest is from $W+\jets$.

\dzero\ finds that the dominant systematic uncertainty of 15\%
is due to the differences 
in the $(\aplan, \htran)$ distributions between data 
and the corresponding simulations
of $W+2~\jet$ and $W+3~\jet$ events.  The uncertainty due to possible
departures from the Berends scaling law is found to
be 10\%. The jet energy scale
uncertainty, however, is only 5\%. With 19 candidate events and
a background of $8.7 \pm 1.7$ events, the measured cross section 
for the $\ell+\jets$ channel is
$\sigtop(\mt=172.1\gevcc) = 4.1 \pm 2.1\pb$.

The cuts for the tagged channels $\ell+\jets/\mu$ are summarized
in \tabref{tab:d0xscuts}.  Note that for these channels,
only three jets are required, as opposed to four for the untagged
channels.  The $(\aplan, \htran)$ cuts are also looser.
This is to compensate for the loss of acceptance caused by 
requiring events to be tagged.  To reject QCD multijets background,
\dzero\ makes the cuts:\cite{d0topprd}
\begin{itemize}
\item $e+\jets/\mu$ channel: $\met > 35\gev$ if
  $\Delta\phi(\met,\mu) < 25\degree$.
\item $\mu+\jets/\mu$ channel: $\Delta\phi(\met,\mu) < 170\degree$
  and $|\Delta\phi(\met,\mu) - 90\degree|/90\degree < \met/(45\gev)$.
\end{itemize}
In addition, in the $\mu+\jets/\mu$ channel, \dzero\ rejects $Z\ra\mu\mu$
background using a kinematic fit to the $Z$~boson decay hypothesis.
With these cuts, eleven candidates survive.

The $W+\jets$ background is computed by starting with the assumption
that the heavy flavor content
of $W+\jets$ events is the same as that in multijet events. 
Therefore, from the
tag probability per jet $p_j$,
as determined from a multijet sample (and parameterized in terms of
$\et$ and~$\eta$), \dzero\
determines the tagged $W+\jets$ background by
weighting a sample of background-dominated
untagged events 
by the per-event tag probability $\sum_j p_j$, where the
sum runs over all jets.

\begin{figure}
\centering
\epsfig{figure=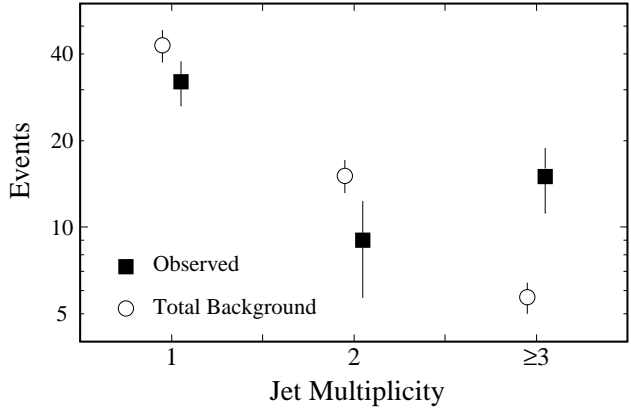}
\caption{Jet multiplicity of $b$-tagged lepton+jets events
  before $\aplan$ and $\htran$ cuts.  Square points represent data and
  the circles the expected background.  From
  Ref.~\protect\onlinecite{d0xsecprl97}.}
\label{fig:d0xsljmult}
\end{figure}

\Figref{fig:d0xsljmult} compares the observed 
numbers of events for different
jet multiplicities to the background prediction. 
The agreement for the one and two jet
samples is good, while for
three or more jets there is a clear excess of events.  The estimated
background in the $\ge 3~\jet$ signal region is
$2.4 \pm 0.5$ events, including a $10\%$ systematic uncertainty
due to the tagging procedure.  The measured cross section for these
channels is 
$\sigtop(\mt=172.1\gevcc) = 8.3 \pm 3.5\pb$.

\subsubsection{All-Jets Channel}

The all-jets channel, as we noted earlier, suffers from a huge QCD background.
It is a channel for which $b$-tagging is particularly 
effective and \dzero, like CDF, takes advantage of this.
However, since \dzero\ cannot tag $b$-jets nearly as efficiently as CDF, 
extensive use is made of multivariate methods, in order to
exploit other, more subtle, differences between the signal and the
background.  The resulting analysis\cite{amos98,klima98,d0alljetsprd} is
rather complex; therefore, we focus here on only the
key ideas.  

The measurement of the $\ttbar$ production cross section in the all
jets channel is based on
an integrated luminosity of $110 \ipb$.  The trigger used is sensitive
to events with many jets and large $\htran$.  The initial selection
cuts are:
\begin{itemize}
\item No isolated leptons.  This ensures that there is no overlap with
  the other channels.
\item $\htran > 115 \gev$ (using jets with $R = 0.5$, $\et > 15 \gev$,
  and $|\eta| < 2.5$).
\item At least six $R = 0.3$ cone jets and no more than eight
  $R = 0.5$ cone jets (with $E_T > 10 \gev$ and $|\eta| < 2.5$).
  Both of these requirements improve the signal/background ratio: the
  lower bound rejects only $14\%$ of the signal while removing $36\%$
  of the background, and the upper bound loses $5\%$ of the signal but
  $13\%$ of the background.
\end{itemize}
Additional cuts are applied to remove events with noise from the Main
Ring.  After these cuts, about 280,000 events remain.  At this stage, 
the signal to background ratio is about $1:1000$.  Requiring an SLT
$b$-tag increases the signal/background by about an order of magnitude,
leaving 6000 events.

To make further progress, \dzero\ performs a multivariate analysis using
thirteen variables, described briefly in \tabref{tab:d0alljvar}.
The principal ones are $H_T$,
$H_{T3} \equiv H_T - E_T^{\jet1} - E_T^{\jet2}$,
the average jet count $N_{\jets}^{A}$, the aplanarity
${\cal A}$, the centrality $C \equiv H_T/\sum_{\jets} E^{\jet}$,
and the transverse momentum of the muon $\pt^\mu$.
A particularly powerful (and unusual) 
variable is the average jet count, defined by
\begin{equation}
  N_{\jets}^A = {\int_{15\gev}^{55\gev} E_T N(> E_T)\; dE_T \Bigg/
  \int_{15\gev}^{55\gev} E_T\; dE_T},
\end{equation}
where $N( > E_T)$ is the number of jets with transverse energy greater
than $E_T$. This variable, inspired by the work of Tkachov,\cite{tkachov95}
is interesting in that it  assigns a nonintegral ``number of jets'' to
an event. 

\begin{table}
\caption{List of variables used by \dzero\
in its all-jets cross section measurement. The $Q_i$ are 
the eigenvalues (in increasing size)
of the normalized momentum tensor 
$M_{ab} = \sum_j P_{ja} P_{jb}/\sum_j |P_j|^2$ of the jets in the
event; $a, b$ run over the $x,y,z$ components of the jet momenta $P_j$.}
\label{tab:d0alljvar}
\begin{center}
\begin{tabular}{ll}
Variable  & Description \\ 
\hline
$H_T$           & Total jet transverse energy \\        
$\sqrt{\hat s}$ & Total $\ttbar$ center of mass energy  \\
$H_{T3}$        & Transverse energy of the non-leading jets     \\
$N_{\jets}^A$   & Average jet count (see text for details)      \\
$\aplan$        &       Aplanarity ($\frac{3}{2}Q_1$)   \\
${\cal S}$      & Sphericity ($\frac{3}{2}(Q_1+Q_2)$)           \\
${\cal C}$      & Centrality ($H_T/\sum_{\jets} E^{\jet}$)       \\ 
\hline
$\et^{\jet1}/H_T$
                & Transverse energy fraction of leading jet     \\
$\sqrt{\et^{\jet5}\et^{\jet6}}$
                & Geometric mean of $\et$ of the fifth and sixth jets    \\
$\brocket{\eta^2}$
                & Mean square rapidity of all jets \\
\hline
$p_T^\mu$       & Transverse momentum of the tag muon   \\
$\cal F$        & Fisher discriminant: A   measure of the difference \\
                & in jet widths between signal and background \\
$\cal M$        & Mass likelihood: A measure of the degree to which \\
                & jet invariant masses satisfy the constraints that there \\
                & be two $W$ bosons and that the two top quarks have \\
                & equal mass. \\
\end{tabular}
\end{center}
\end{table}

These thirteen variables are combined using feed-forward
neural networks into a single discriminant $NN_2$, the value of which
lies between zero and unity. This is done in three steps:
\begin{itemize}
\item $NN_0$ --- a network is trained with the first seven variables
  listed in \tabref{tab:d0alljvar}, using all events; this
  network is used to select  better
  event samples for training the subsequent networks.

\item $NN_1$ --- a network is trained with
  the first ten variables of \tabref{tab:d0alljvar},
  having twenty hidden nodes and one output node, using events which
  satisfy the cut $NN_0 > 0.3$.

\item $NN_2$ --- a final network is trained, using events with
  $NN_0 > 0.3$, with four input nodes, eight hidden nodes and one output
  node; the inputs are
  the output of $NN_1$ plus the remaining three variables. 
\end{itemize}

Since the data are predominantly background, \dzero\ uses
\emph{untagged} events as the background sample in the
training, and  \emph{tagged} \progname{HERWIG} 
$\ttbar$ Monte Carlo events (with $m_t = 180 \gevcc$) for the signal.
There is, however, a complication. Untagged
events, by definition, do not have muons; but the latter
are nonetheless needed to train $NN_2$! The solution adopted is to
generate Monte Carlo muons according to the measured muon $\pt$ spectrum
and to add them to the untagged sample, thereby transforming the
background sample
into a tagged sample. One might worry that
this procedure would fail because of correlations between
the muon $\pt$ and the remaining variables. If correlations exist, the
incoherent addition of  Monte Carlo muons to untagged events would
not be expected to reproduce the characteristics of tagged events.

\begin{figure}
\centering
\epsfig{figure=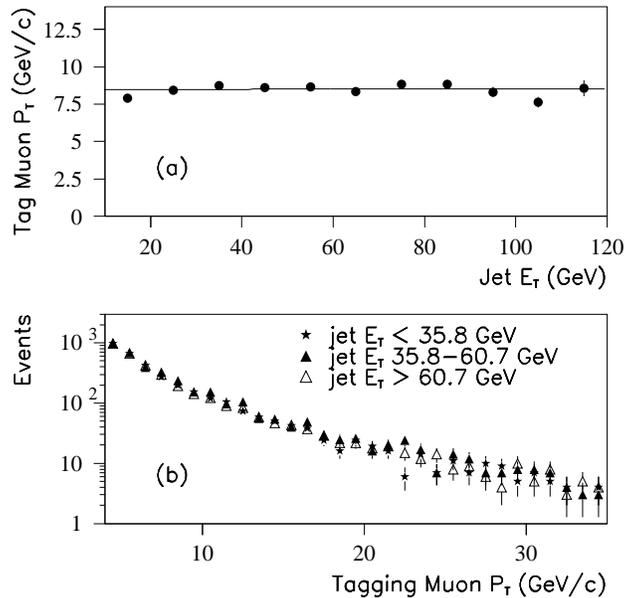}
\caption{
  (a) The mean muon $\pt$ versus jet $\et$, as observed by \dzero\
  in their all-jets cross section analysis. We see that the muon $\pt$ is
  independent of the jet $\et$. (b) The observed muon $\pt$ distribution
  for three jet $\et$ ranges (normalized to the same number of
  events).
  From Ref.~\protect\onlinecite{d0alljetsprd}.
  }
\label{fig:d0alljmupt}
\end{figure}

However, the procedure works.  As
shown in \figref{fig:d0alljmupt}, the muon
$\pt$ is uncorrelated with the $\et$ of the tagged jet, provided that the
tagged
jet is \emph{not}
corrected for the presence of the muon. This absence of correlation between
muon $\pt$ and jet $\et$ precludes the existence
of correlation between the muon $\pt$ and the
rest of the event.

\begin{figure}
\centering
\epsfig{figure=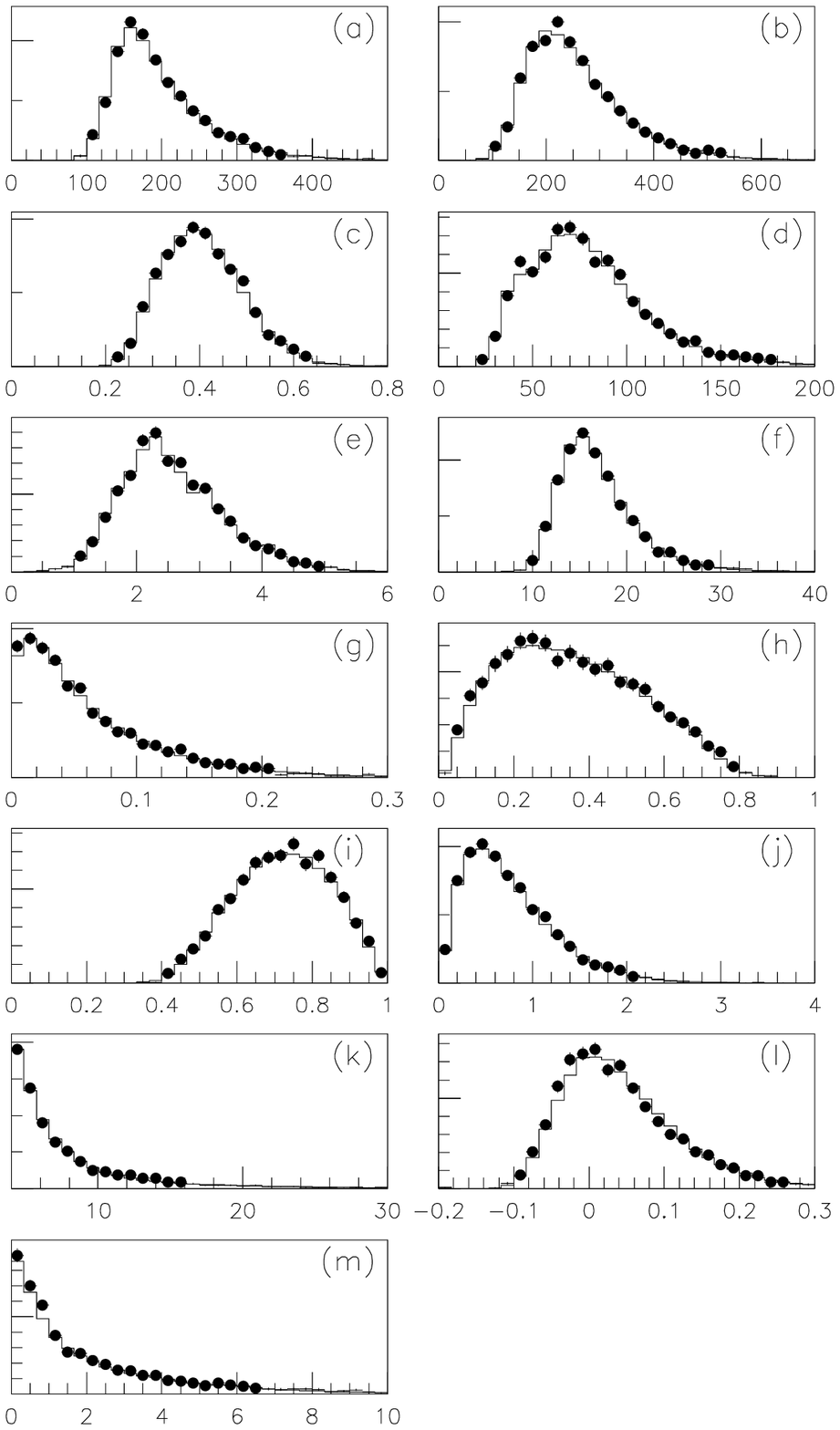,width=11.9cm}
\caption{
  A comparison of \dzero's absolute predictions (histograms)
  for the distributions of the
  thirteen variables used in the all-jets cross section
  measurement: (a) $\htran$ (GeV), (b) $\sqrt{\hat s}$ (GeV),
  (c) $\et^{\jet1}/H_T$, (d) $H_{T3}$ (GeV), (e) $N_{\jets}^A$,
  (f) $\sqrt{\et^{\jet5}\et^{\jet6}}$ (GeV),
  (g) $\aplan$, (h) ${\cal S}$, (i) ${\cal C}$,
  (j) $\brocket{\eta^2}$, (k) $p_T^\mu$,
  (l) $\cal F$, and (m) $\cal M$.
  The points represent the observed
  data in a sample of 3853 tagged events. 
  From Ref.~\protect\onlinecite{d0alljetsprd}.
}
\label{fig:d0alljtagpred}
\end{figure}

The background to the tagged data is estimated by weighting the untagged
data (with the Monte Carlo muons) by the tag probability per event. 
\Figref{fig:d0alljtagpred} illustrates 
how well the weighting procedure works.
The figure shows an absolute comparison
of the tagged events to the predictions for the thirteen variables.
For all variables, the agreement is seen to be
good.

\begin{figure}
\centering
\epsfig{figure=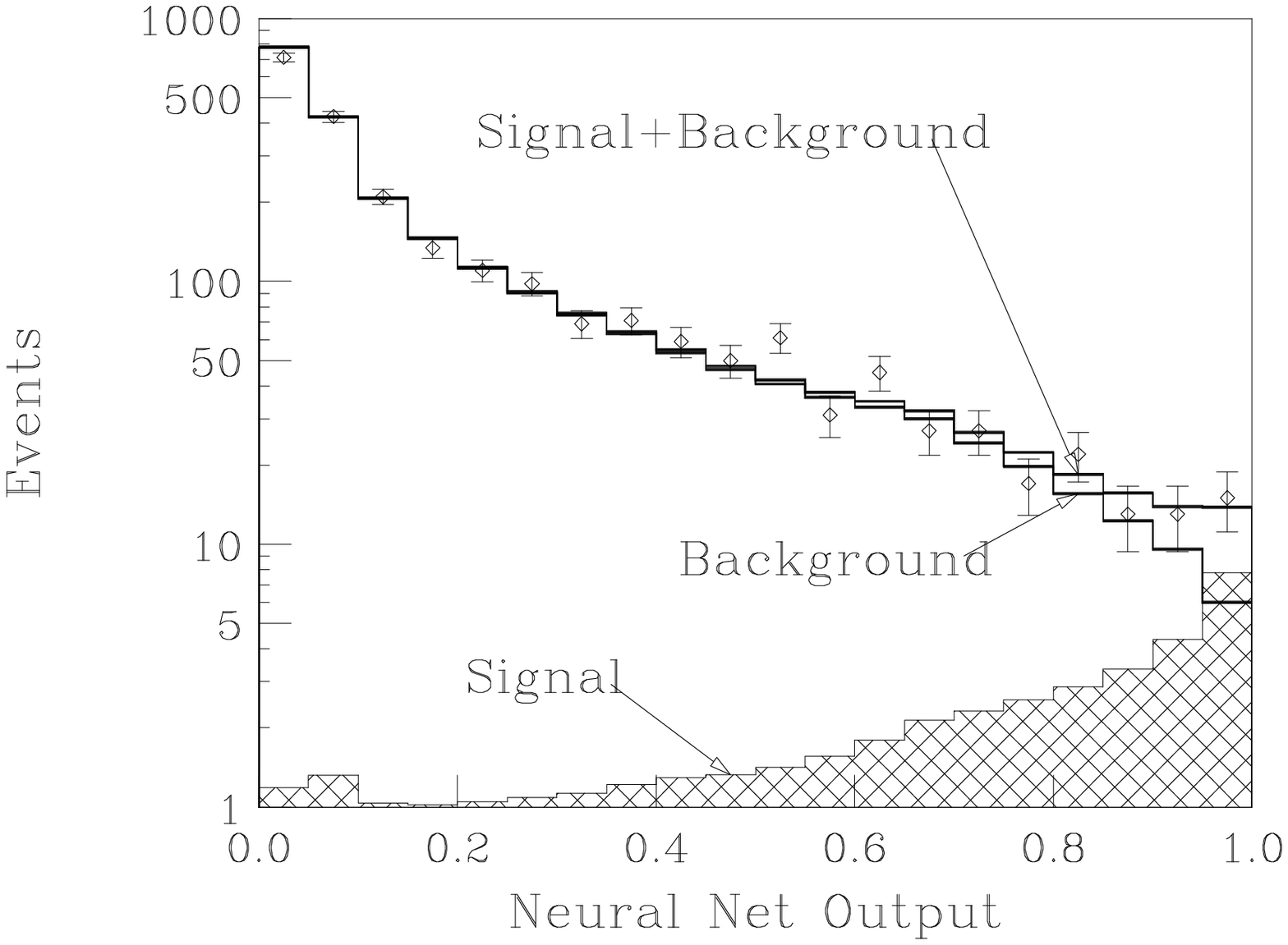,width=\hsize}
\caption{
  The distribution of the final neural network output ($NN_2$) in the
  \dzero\ all-jets cross section analysis showing the results of a fit of
  the observed tagged distribution to the predictions for signal and
  background, with $m_t = 180 \gevcc$. 
  Only statistical errors are shown, and events with $NN_2 < 0.02$ are
  not plotted.  The fit has a $\chi^2$ per degree of freedom
  of 16.9/17.
  From Ref.~\protect\onlinecite{d0alljetsprd}.
}
\label{fig:d0alljfit}
\end{figure}

To extract the cross section, \dzero\ fits the observed $NN_2$
distribution for tagged events to the expectations for signal and
background.  The fit is performed for the 2207 tagged
events that pass the cut $NN_2 > 0.02$.  The results of the fit (for
$m_t = 180 \gevcc$) are shown in \figref{fig:d0alljfit}.  This gives a
cross section for $m_t = 180 \gevcc$ 
of $6.3\pm 2.5\ \text{(stat)}\pb$.  The major systematic
uncertainties in this measurement are in the background
tag probability (7\%), the $\ttbar$ tag probability (7\%), the modeling
of the muon $\pt$ spectrum, and 
the jet energy scale (9\%).
After interpolating to the measured top quark mass,
the final result is
$\sigtop(\mt = 172.1\gevcc) = \statsyst{7.1}{2.8}{1.5}\pb$.  

As a cross check, \dzero\ also estimates the cross section using the
usual counting method.  Here, a cut $NN_2 > 0.85$ is imposed.  This
yields 41 observed events, with an estimated background of $24\pm 2.4$
events.  The signal efficiency at $\mt = 180\gevcc$ is
$0.022\pm 0.0037$, yielding a cross section of
$\statsyst{6.5}{2.6}{1.4}\pb$, in good agreement with that found by
the fit.
If the cut value is reoptimized to maximize the expected
significance (using Monte Carlo), the result is $NN_2 > 0.94$,
yielding 18~observed
events and an estimated background of $6.9\pm 0.9$ events.
The chance of this background fluctuating to give at
least the observed signal is $6\times 10^{-4}$, or 
a $3.2\sigma$ deviation for a Gaussian.

\begin{table}
\caption{Summary of \dzero\ cross section measurements,
  for $\mt = 172.1\gevcc$.}
\label{tab:d0xssum}
\begin{center}
\begin{tabular}{lcccc}
Channel  & $\epsilon \times \text{BR}(\%)$ &Data & Background & $\sigtop$ (pb)
        \\
\hline
$\ell\ell$ (with $e\nu$) & $0.91\pm 0.17$ & 9 & $ 2.6\pm 0.6$ & $6.4\pm 3.3$ \\
$\ell+\jets$ (untagged)  & $2.28\pm 0.46$ &19 & $ 8.7\pm 1.7$ & $4.1\pm 2.1$ \\
$\ell+\jets/\mu$         & $0.96\pm 0.15$ &11 & $ 2.4\pm 0.5$ & $8.3\pm 3.5$ \\
\hline
$\ell\ell$ and $\ell+\jets$
                         & $4.14\pm 0.69$ &39 & $13.7\pm 2.2$ & $5.6\pm 1.8$ \\
\hline
All-jets ($NN_0 > 0.85$, $\mt=180\gevcc$)
                         & $2.2\pm 0.4$   &41 & $24\pm 2.4$   & $6.5\pm 3.0$ \\
\hline
All-jets (from fit)      &                &   &               & $7.1\pm 3.2$ \\
\hline
Total                    &                &   &               & $5.9\pm 1.7$ \\
\end{tabular}
\end{center}
\end{table}

\subsection{Summary of Cross Section Measurements}
\label{summaryxs}

\begin{figure}
\centering
\epsfig{file=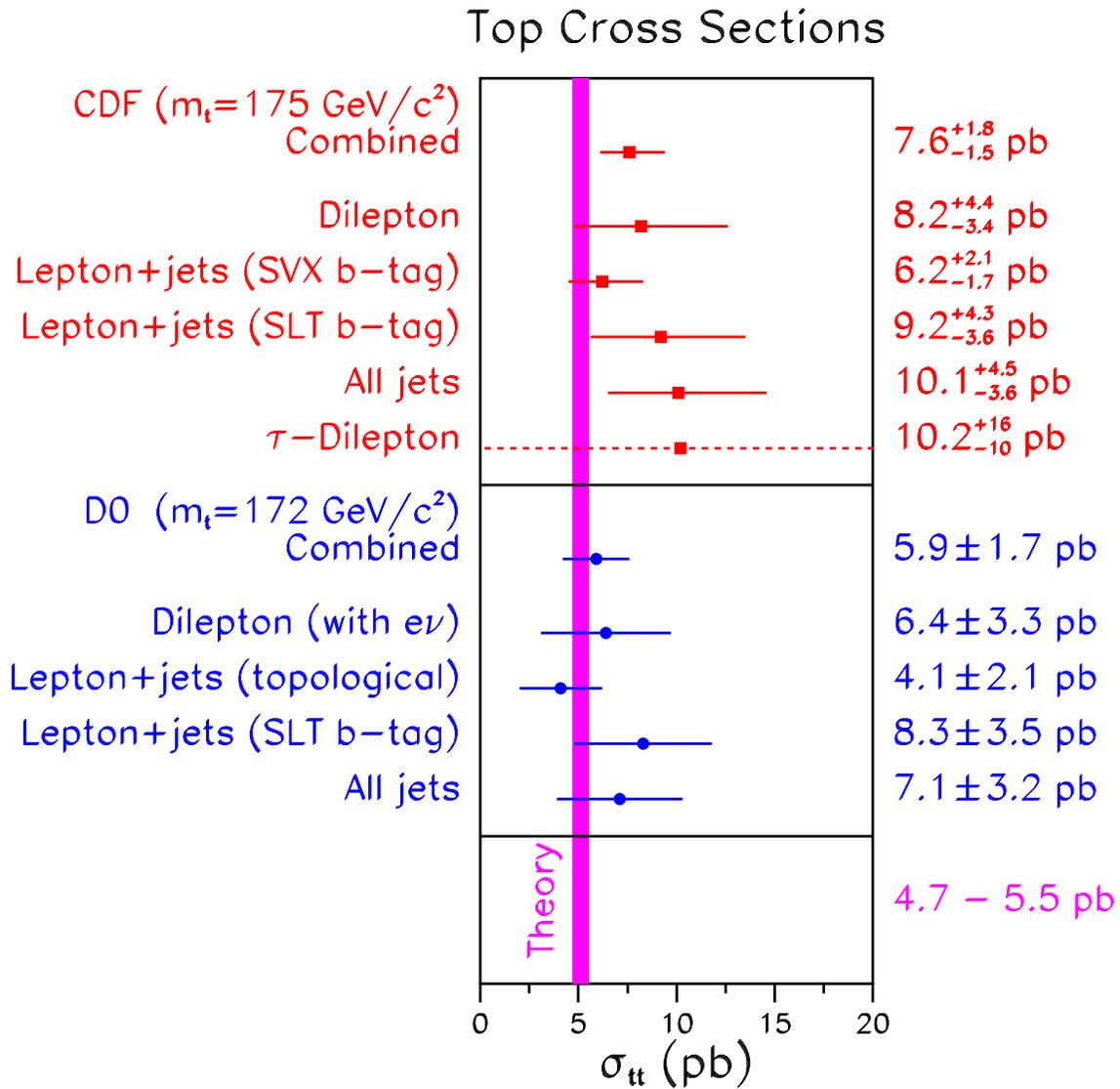,width=\hsize}
\caption{The $\ttbar$ production cross section measured in the
  channels studied  by CDF and \dzero.  
  Also shown is the range of predictions from various theoretical
  calculations.}
\label{fig:item_cs_cdf_d0}
\end{figure}

The results of the CDF and \dzero\ cross section analyses 
are summarized in \tabsref{tab:cdfxssum} and~\ref{tab:d0xssum},
respectively, and graphically in \figref{fig:item_cs_cdf_d0}.

\begin{figure}
\centering
\epsfig{file=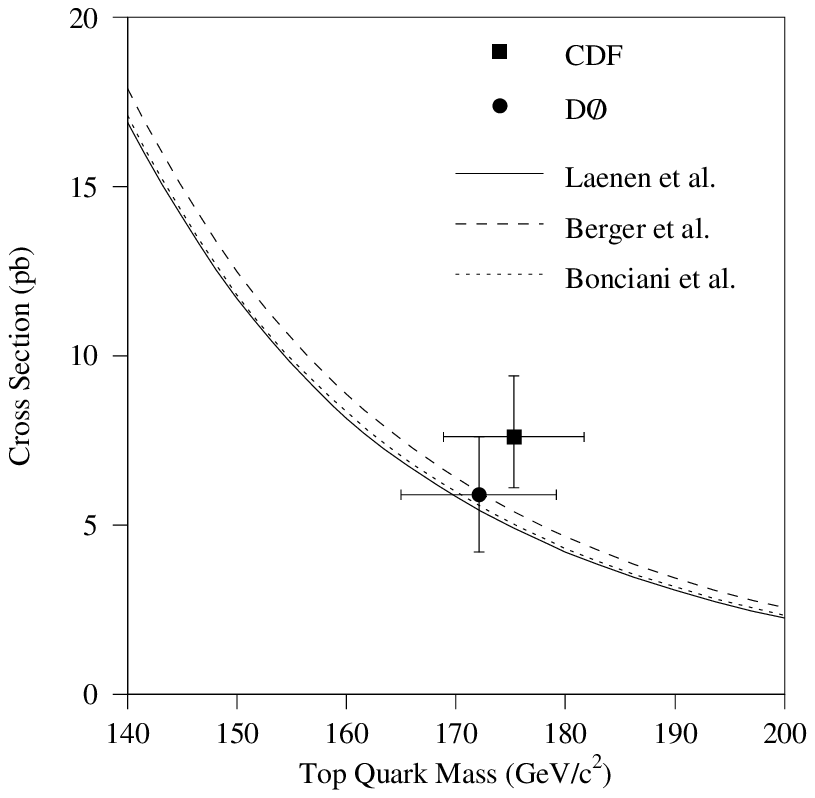}
\caption{The $\ttbar$ production cross sections as measured by CDF and
  \dzero.  Also shown are various theoretical calculations.%
  \protect\cite{laenen94,berger95,bonciani98}
  }
\label{fig:top_xsec_1997}
\end{figure}

Excluding the all-jets channel,
\dzero\ observes a total of 39 events with a background of $13.7 \pm 2.2$
events. 
For a top quark mass of $172.1\gevcc$
this gives a measured cross section of $5.6 \pm 1.8\pb$.
If the measurement from the all-jets channel
is included, the result changes 
to $5.9\pm 1.7\pb$.
The measurement from CDF, which includes
all channels except the $\tau$-dilepton channels,
is $\sigtop=7.6^{+1.8}_{-1.5}\pb$ for a top quark mass
of $175\gevcc$.
Both results are presented graphically 
in \figref{fig:top_xsec_1997}, 
which compares the measurements to several predictions.
Although the cross sections obtained by CDF are typically higher than those
from \dzero, both sets of results are consistent with each other and with
predicted values. The results are also consistent with the expected branching
ratios.

\section{Measurement of the Top Quark Mass}
\label{topmass}

Early
direct measurements of the
top quark mass were published in the papers 
announcing the observation of the top
quark in 1995.\mcite{cdfdiscovery,d0discovery,wimpenny96,*franklin97}
Since that discovery, 
the two collaborations have mounted prodigious efforts
to measure the top quark mass 
as precisely as possible.
It is to these results that we now turn. After
sketching the analysis strategies used by the 
two collaborations,
we describe in greater detail interesting aspects
of the analyses of CDF and
\dzero. We end with
a summary of the results.

\subsection{General Strategy}
\label{topmass-general-strategy}

Both
collaborations use two different strategies 
to measure the mass of the top quark.
The first involves kinematically fitting the
events to a $\ttbar$ 
hypothesis and then extracting from the fits
an estimate of the mass of the top~quark.
The second strategy is based on constructing
mass-dependent variables that do not require a 
kinematic fit.

In the first strategy,
one tries to identify the
observed objects (jets, leptons, and missing transverse
energy) with those at the partonic level. 
To the degree that such an identification is possible,
one can perform a kinematic fit of the observed event
to the $\ttbar$ decay hypothesis.  When the mass of the top quark is
extracted from such a fit, it is commonly called
the \emph{fitted mass}.  Because the $\ttbar$ decay hypothesis does
not always hold true, this is not the same as the true top quark mass;
nevertheless, it is a well-defined observable, the mean of which depends on
the mass of the top quark. 
This defines one class of mass-dependent observables or variables.   

The principal 
advantage of a kinematic fit is that, generally, the resulting
variables have the greatest sensitivity to the mass of the top~quark.
But such fits are not always possible; this is so, 
for example, if there are fewer observed jets than partons.

The alternative strategy eschews the complexities of a kinematic fit, and
instead
uses one or more kinematic quantities that are known to be mass-dependent. 
An example
of such a variable is $\htran = \sum \et^{\jet}$, 
the scalar sum of the jet transverse
energies.  The disadvantage of this approach is that 
variables such as $\htran$ tend to have much broader
distributions than those of the fitted mass. On the other hand, 
this approach is simple and direct (compared with
kinematic fitting), and it can be used for all events.
Moreover, it may ultimately be
the more satisfactory strategy for the following reason.
As the accuracy of the mass measurements 
improve, we shall reach a point when the quantum nature of partonic
decay can no longer
be neglected and our classical one-to-one identification 
of a jet with a parton will break
down.\cite{giele97}
When this point is reached, it may be necessary to use observables that
do not require forcing a one-to-one map back to the partonic level.
Ultimately, the limitation will be the sensitivity of the mass
to the QCD modeling of the final state.

Each strategy furnishes a mass-dependent variable $x$ per event, the
distribution
of which, $f(x)$, depends upon the mass of the top quark $m_t$ if the event
is a top quark event.  The functions $f_s(x|m_t)$
and $f_b(x)$ pertaining to the signal and background, respectively, 
are calculated
using detailed Monte Carlo simulations of the signal and background reactions.
The functions $f_s(x|m_t)$ and $f_b(x)$ are sometimes 
referred to as \emph{templates}.
Given a set of templates and a set of
measurements $(x_1,\cdots,x_n)$, there are a variety of standard methods
available,
most commonly \emph{maximum likelihood},\cite{MaxLikelihood}
to extract a single estimate of the
parameter of interest, here the mass $m_t$.

Much of
the effort in these analyses is devoted to
constructing reliable templates. This is a delicate matter because of the
necessary reliance on complicated Monte Carlo simulations to compute
them.  The key issue, of course, is to establish
that the templates, so obtained, agree with the data.
For the background, this is less of an
issue because the data, being largely background, 
afford a direct check of the correctness of those templates 
in  regions of parameter space that are relatively free of signal.
However, for the signal, 
the verification of the templates is, necessarily, more
problematic because the signal events are generally less
numerous. It is therefore notable that
in spite of these difficulties, both collaborations have been
able to produce very precise and consistent results.

At a fundamental level, there is really no difference
between the two strategies we have outlined: the result of the kinematic
fit \emph{is} a mass-dependent kinematic quantity, albeit a particularly
complicated one.  Nevertheless, the practical details of carrying out
the analyses are sufficiently different that the distinction is
worth making.

In the following sections
we describe how these strategies have been realized by
each
collaboration. We begin with the analyses of CDF, followed by those of \dzero.

\subsection{CDF Analyses}
\label{cdfmass}

The CDF collaboration has measured the top quark mass
in all of the dilepton, lepton+jets, and all jets channels, 
using their full Run~1 data sample.

\subsubsection{Dilepton Channels}
\label{cdfmass-dileptons}

In this analysis,\cite{cdfdilep98} CDF considers the channels $e\mu$, 
$\mu \mu$, and $e e$, in which both 
$W$ bosons decay leptonically.  The mass analysis starts with the same
data sample and cuts as used in the dilepton cross section analysis
(see \secref{cdfxs} and \tabref{tab:cdfxscuts}).
To improve the signal/background ratio beyond the level used for
the cross section analysis, CDF imposes 
the cut $\htran^{\text{all}} > 170\gev$, where 
$\htran^{\text{all}}
 = \pt^{\ell1} + \pt^{\ell2} + \et^{\jet1} + \et^{\jet2} + \met$ 
is the sum of the transverse momenta of
the two leptons, the transverse energies of the two highest-$\et$
jets, and the missing transverse energy. 
The power of such variables to discriminate
signal from background was recognized
early on,\mcite{baer89,*likhoded90}
and was first applied to good effect by \dzero.\cite{d0topprd,d0xsecprl95}
With this final cut, the sample is reduced to eight events 
with an estimated background
of $1.3 \pm 0.3$ events. The CDF collaboration then estimates the top
quark mass using two different methods.

\begin{figure}
\centering
\epsfig{file=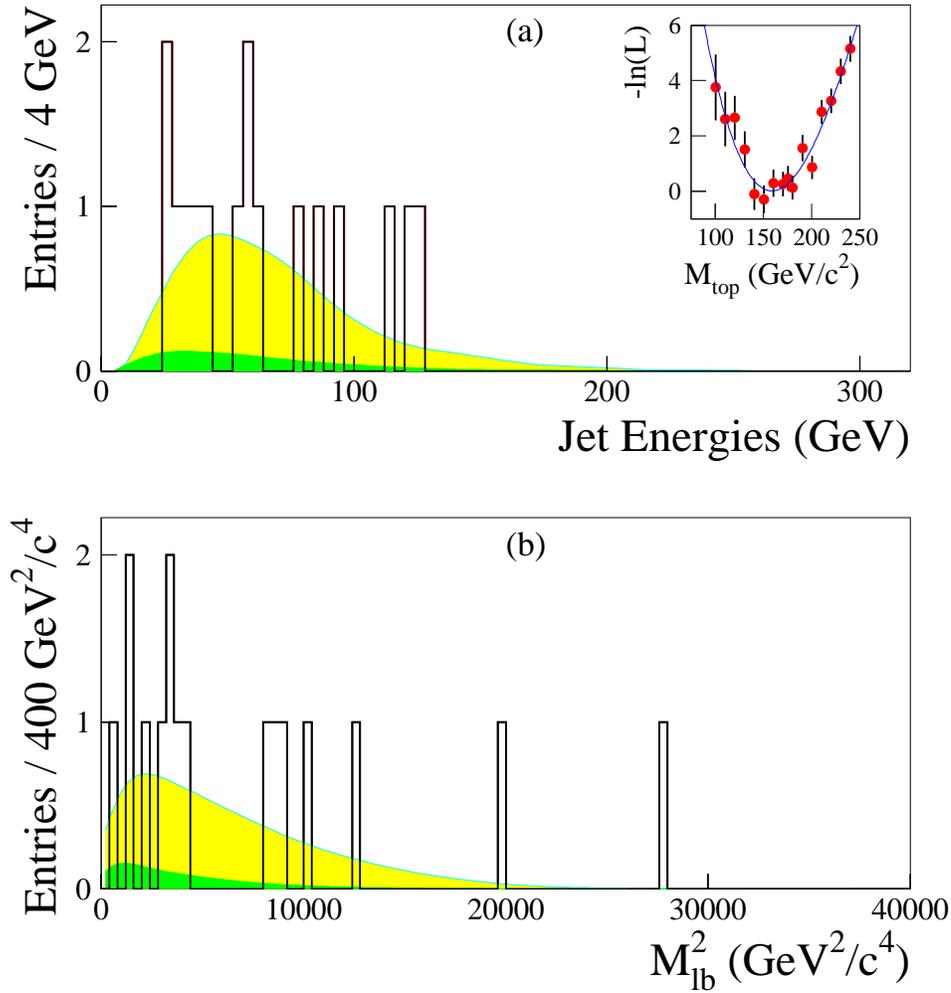,width=0.9\hsize}
\caption{(a) The energy distribution of the
  two highest $\et$ jets for the CDF dilepton events (histogram, two
  entries per event).
  The lightly shaded curve is the prediction for $\ttbar$ signal
  ($\mt = 160\gevcc$) plus background, and the heavily shaded curve is
  the prediction for background alone.  The inset shows
  the $-\ln({\cal L})$ fit as a function of $m_t$.
  (b) The distribution of $\Mlbminsq$ (histogram, two entries
  per event).
  The shaded curves are as in (a).
  From Ref.~\protect\onlinecite{cdfdilep98}.
}
\label{fig:cdfllmass}
\end{figure}

Both methods are a realization of the second strategy,
being based on relatively simple mass-dependent kinematic variables.
For the first method, CDF uses
the observation that the energies of the $b$-quarks are sensitive
to the mass of the top quark, $m_t$. Indeed, it is found 
that the mean energy of
the two highest
$\et$ jets depends linearly on $m_t$, 
with a slope of about 0.5.  From Monte Carlo simulations,
CDF obtains a set of
templates $f(E|m_t)$ describing the observed jet energy distribution, 
one for each top quark mass considered in the range 100 to $240 \gevcc$. 
Templates are also constructed for the backgrounds. CDF
then performs
a maximum likelihood fit of the
observed distribution of jet energies to a mixture 
of $\ttbar$ and background events,
with the background count
constrained to $1.3 \pm 0.3$ events. For each
assumed top quark mass $m$,
$-\ln\left({\cal L}(m)\right)$, where ${\cal L}(m)$ 
is the value of the likelihood at that mass, is calculated.
By fitting a third order
polynomial
to the negative logarithms, and using the position of
its minimum as an estimate of
the mass, CDF obtains
the mass estimate 
$\statsyst{159}{23}{11}\gevcc$. 
Following standard practice, CDF defines
the statistical error by applying the ``0.5-rule'': the error 
is the mass difference, relative to the value at the minimum ($159 \gevcc$), 
that increases $-\ln({\cal L})$ by 0.5. 
\Figref{fig:cdfllmass}a
shows the jet energy distribution and the resulting log-likelihood
function.

The second method exploits the relationship between 
$m_t$ and the invariant mass $\Mlb$ 
of the lepton and the associated
$b$-quark.
Since the top quark decays to a $b$-quark and a $W$~boson, we can write
$m_t^2 = m_b^2 + M_W^2 + 2 E_b M_W$,
where $E_b$ is the energy of the $b$-quark in the rest frame
of the $W$~boson. 
The $b$-quark's energy $E_b$, which is \emph{constant} 
in the $W$~boson rest frame,
can be calculated from the mean invariant mass $\brocket{\Mlb}$ 
and 
opening angle $\brocket{\cos\theta_{{\ell}b}}$ between the $b$-quark
and the lepton:
$\brocket{\Mlbsq} =  E_b M_W (1 - \brocket{\cos\theta_{{\ell}b}})$.
To lowest order in the
Standard Model, $\brocket{\cos\theta_{{\ell}b}} = 
M_W^2/(m_t^2 + 2 M_W^2)$, 
which leads
to $m_t^2 = \brocket{\Mlbsq} + 
\sqrt{M_W^4 + 4 M_W^2 \brocket{\Mlbsq} + \brocket{\Mlbsq}^2}$.  
Given this relationship, a measurement of
$\brocket{\Mlbsq}$ is tantamount to a direct measurement of $m_t$.
(For an interesting variation on this theme, see Ref.~\onlinecite{prosper95}.)

Unfortunately, there is a difficulty: a priori, we do
not know which jet should be paired with which lepton; whether
$(\ell^+j_1, \ell^-j_2)$ is the correct $(\ell^+b, \ell^-\bbar)$ assignment or
$(\ell^+j_2, \ell^-j_1)$.  In addition, since $\Mlb$ is defined by a
quadratic equation, there are two solutions to be considered
for each jet configuration.  CDF resolves the first of these
difficulties by selecting the jet configuration with the smaller sum
of invariant masses. This selection is correct
55\% to 75\% of the time,
depending upon the top quark mass.
The values of $\Mlb$ selected in
this manner are
denoted $\Mlbmin$. 
The second difficulty is dealt
with by retaining both solutions, and plotting each event twice.
\Figref{fig:cdfllmass}b shows the resulting
$\Mlbmin$ distribution for the eight events in
the sample, along with
the signal and background predictions.
Of course, 
owing to selection biases and jet energy mismeasurements,
$\brocket{\Mlbmin}$ is not the same as 
$\brocket{\Mlb}$; however, from Monte Carlo
studies, CDF finds that the two
quantities are
linearly related. The result from this method is 
$m_t = \statsyst{163}{20}{9}\gevcc$.
Combining this with
the previous result, taking into account their mutual correlation,
gives $m_t = \statsyst{161}{17}{10}\gevcc$.

For both methods, the major sources of systematic uncertainties are
the knowledge of the jet energy scale and the shape of the
background distributions. 

At a recent conference in Vancouver (ICHEP98), CDF presented
an updated dilepton channel mass measurement,\mcite{roser,*yao98}
using a method similar to the \dzero\ $\nu$WT method
(see \secref{d0-dilepton-mass}).  The result of this analysis
was $m_t = \statsyst{167.4}{10.3}{4.8}\gevcc$.  (The details of
this analysis were not available in time for inclusion in this review.)

\subsubsection{Lepton+Jets Channels}
\label{cdfmass-lepton}

These channels provide CDF with its 
most precise measurement of the top quark
mass.\cite{cdftopmass98}
This is due to both the relatively large branching ratios for
these channels ($\approx 30\%$) and the large signal/background ratio
that can be achieved through $b$-tagging (see \secref{btagging}).
The event selection used is similar to that used for the cross section
analysis (see \tabref{tab:cdfxscuts}), except that four jets are
required, and events without a $b$-tag are accepted.  But in order to
increase the efficiency for the four jet requirement, the cuts on the
fourth jet are loosened to $\et > 8\gev$ and $|\eta| < 2.4$, provided
that one of the four leading jets is $b$-tagged with either the SLT or
the SVX algorithm.  (For untagged events, all four jets must satisfy
$\et > 15\gev$ and $|\eta| < 2.0$.)
This selection yields a sample of 83 events.

This analysis follows the strategy of kinematic fitting.
The observed event is fit to the hypothesis 
$\ppbar \rightarrow \ttbar + X \ra (\ell \nu b) (b q \qbar) + X$,
with the four leading  jets assumed to map onto the four quarks.
The measured variables are the three-momenta of the lepton and the
four leading  jets, plus the vector sum of all remaining transverse
energy (excluding the lepton and the leading jets).  This latter
vector gives the transverse momentum of the recoiling system~$X$.
The momentum of the neutrino is not measured, yielding three unknowns.
The neutrino and lepton are assumed to be massless, $b$-quarks are
assigned a mass of $5\gevcc$, and light quarks are assigned a mass of
$0.5\gevcc$.
Five constraints can then be imposed:
\begin{itemize}
\item The transverse momentum components of 
  the $\ttbar$ + $X$ system are zero.
\item The invariant mass of the lepton and neutrino equals $M_W$.
\item The invariant mass of the $q$ and the $\qbar$  equals $M_W$.
\item The mass of the $t$ equals that of the $\tbar$.
\end{itemize}
Since there are two more constraints than unknowns, this is 
a 2C fit. It is solved by standard $\chisq$-minimization
techniques, resulting, for each
event, in a
fitted mass $\mfit$ and a $\chisq$ value that is taken 
as a measure of how well
the event satisfies the $\ttbar$ hypothesis.

The energy of
each jet is adjusted to
match as closely as possible the energy of the parton from which the jet is
presumed to have originated,%
\footnote{Note, however, that 
this procedure is not strictly necessary provided that
the Monte Carlo simulation offers a faithful rendition of
the characteristics of the data, the most critical 
being that the energy scale of the simulation agrees with that
of the data. Once this has been achieved, the only requirement is that
the \emph{same} kinematic fitting algorithm be applied both to real and
Monte Carlo events.  The rescaling may, however, improve the
resolution slightly because of the imposition of the $M_W$ constraints.}
starting with corrections of the sort outlined in \secref{jetid}.
In addition, CDF applies energy corrections derived by directly
comparing, in \progname{herwig} Monte Carlo samples, the energies of
partons with the reconstructed jets.  These corrections
depend on parton flavor; hadronically decaying $b$-quarks,
leptonically decaying $b$-quarks, and light quarks are all corrected
differently.
Clearly,
these corrections depend critically
on the ability of \progname{HERWIG}
to model the fragmentation properties of jets.
It is therefore pertinent to ask how well these
properties are modeled. The answer is, surprisingly well, given the
phenomenological nature of the fragmentation model. 
Detailed studies of the energy flow
within jets (see, e.g., Ref.\onlinecite{d0jetshape}) confirm
the ability of \progname{HERWIG} to model jets accurately.

In the lepton+jets 
channels, there is considerably more ambiguity in assigning
the jets to the partons than is the case for the dilepton channels. 
There are twelve possible configurations, or assignments of the four
leading jets to the four partons.
Moreover, one must choose an initial value for the longitudinal
component of the neutrino momentum.  Using one of the constraints,
this quantity can be found to within a two-fold ambiguity.  To guard
against finding a local minimum, the fit is tried with both starting
points, giving a total of up to twenty-four kinematic fits per event.
However, jets that are tagged are only assigned to $b$-quarks, 
thus
reducing the number of configurations.
The configuration with the smallest $\chisq$ is chosen, 
but events whose smallest
$\chisq$ value exceeds 10 are rejected. 
This $\chisq$ cut reduces the sample from 83 to 76 events.

The $\chisq$ variable, unfortunately,
is not very effective at selecting the correct configuration (see the
discussion in \secref{d0-ljets-mass}). 
Consequently, although the standard deviation of the $\mfit$
distribution would be only $\sim 13 \gevcc$
(for a top quark mass of $175 \gevcc$) 
were
the selection procedure perfect, CDF finds that, in practice, 
the standard deviation of the
$\mfit$ distribution is approximately double that value. 

The CDF collaboration 
uses a maximum-likelihood method to extract an estimate of 
the top quark mass
from the sample of 76 events. The likelihood function has the form
\begin{equation}
        {\cal L} = 
{\cal L}_{\text{shape}}(D|m_t, x_b, a) \times 
{\cal L}_{\text{backgr}}(x_b|x_b^0) \times 
{\cal L}_{\text{param}}(a|a^0),
\label{eq:cdfmasslike}
\end{equation}
where
\begin{equation}
        {\cal L}_{\text{shape}}(D|m_t, x_b, a) = \prod_{i=1}^{N}
[(1 - x_b) f_s(m_i|m_t,a) + x_b f_b(m_i|a)]
\end{equation}
is the joint probability density assigned to the sample of $N$ fitted masses 
$D \equiv (m_1,\ldots,m_N)$;
$f_s(m_i|m_t,a)$ is the probability density to observe the fitted
mass $m_i$ 
given a top quark mass
$m_t$, assuming that the $i$th event is signal and $f_b(m_i|a)$ 
is the probability
density to observe $m_i$ assuming the event is background; $x_b$ is the
background fraction; and $a$ denotes the set of parameters that
determine the shape of the functions $f_s(m_i|m_t,a)$ and $f_b(m_i|a)$.
These functions (templates)
are derived from Monte Carlo calculations
(based on \progname{HERWIG} 
for top events and \progname{VECBOS} for the
$W$+jets background). In previous work,\cite{cdfdiscovery} CDF used a 
discrete set  of templates. However, the collaboration now finds
that smaller systematic uncertainties result from deriving from
the discrete set
a smooth parameterization of $f_s(m_i|m_t,a)$,
both
in $m_i$ and in $m_t$.
(Amusingly, \dzero, which earlier used
parameterized templates, now finds it better to use a discrete set!)

These functions contain parameters $a$, other
than $m_t$,
that define the shape of the functions. The
parameters have estimates $a^0$ and uncertainties that are encoded in the
likelihood function ${\cal L}_{\text{param}}$. This likelihood function
constrains the shape of the templates in ${\cal L}_{\text{shape}}$ to
vary within their uncertainties. The shape parameter
uncertainties reflect simply the finite statistics of the Monte Carlo event
samples that were used to generate the discrete set of templates. 

The background fraction $x_b$ is constrained within its uncertainty to
the independently measured 
background fraction
$x_b^0$ by the
background likelihood function ${\cal L}_{\text{backgr}}(x_b|x_b^0)$.
One can think of ${\cal L}_{\text{backgr}}$ and 
${\cal L}_{\text{param}}$ as prior probabilities 
with respect to the likelihood function
${\cal L}_{\text{shape}}$. They encode the information $(a^0, x_b^0)$
that is known about the background
fraction and the shape of the templates.

In statistical analysis, it is sometimes advantageous to use stratified 
sampling;\cite{FredJames} that is,
to divide a sample into two or more subsamples. 
This is especially true when the 
characteristics vary widely from one subsample to another. 
In order to make optimal
use of all the available information, 
the event sample is divided into 
disjoint subsamples. 
Each subsample is assigned its own likelihood function in
accordance with \eqref{eq:cdfmasslike}, the product of which
is then maximized to obtain
the final estimate of the top quark mass and its uncertainty. 
From Monte Carlo studies, CDF
concludes that four subsamples give the optimum partition: 
\begin{itemize}
\item events with a single SVX tag,
\item events with two SVX tags,
\item events with an SLT tag but no SVX tag, and
\item events with no tag but with all four leading 
  jets satisfying $\et > 15\gev$ and $|\eta| \leq 2$.
\end{itemize}

The expected background fraction $x_b^0$ in each subsample
is derived from the $W + \geq 3~\jets$ background estimate
used in the $\ttbar$ cross section
measurement (see \secref{cdfxs}). To estimate the
background in
the subsamples, one needs the efficiencies of the additional cuts,
including the fourth jet requirement and the $\chisq$ cut.
These efficiencies are 
determined from Monte Carlo studies. Using the tagging efficiencies
from the cross section analysis and the background rates for the
subsample, CDF writes the total
mean number of events $n$ in each subsample
as a function of the mean number of signal
and background events $S$ and $B$, respectively, in
the {\it combined} sample; that is, $n = \epsilon_s S + \epsilon_b B$, where
$\epsilon_s$ and $\epsilon_b$ are known fractions, derived
from the Monte Carlo studies. The parameters
$S$ and $B$ are then estimated by maximizing a multinomial likelihood
function, based on the measured event count $c_i$ in subsample $i$ and
its mean count $n_i$. This leads to the background
fractions $x_{bi}^0 = \epsilon_b B/n_i$, listed
in \tabref{tab:cdfljmass}. 

It is found that about 67\% of the background is due to $W+\jets$, 20\%
is due to
multijet events (in which a jet has been misidentified as a lepton) and
$\bbbar$ events with the 
$b$-hadron decaying semileptonically, and the remaining
13\% comes from $(Z\ra\ell\ell)$+jets events,
diboson events ($WW$, $WZ$, and $ZZ$), and single top. 

As noted above, a key issue is the construction of reliable templates. For
the signal templates, one can judge their reliability 
only after the mass distributions have been fitted. If the fit is
good, this provides \emph{ex post facto} evidence of their soundness. The
reliability of the background templates is generally addressed by
comparing Monte Carlo simulations 
with data
in regions that are largely devoid of signal.  If the
templates agree well in those regions, then one has some confidence 
that they are probably adequate in the signal-enhanced regions. 
These cross checks are not rigorous:
the fact that the templates
are modeled well in the background-dominated regions does not
guarantee that they are also modeled well
in the regions
of interest. This observation applies, of course, not only to 
mass templates but to \emph{any}
distribution based on Monte Carlo simulation. 
However, the quoted uncertainties
account for possible mismodeling of the templates.
Both CDF and \dzero\ confirm the reliability of the templates in this
manner.

\begin{figure}
\centering
\epsfig{file=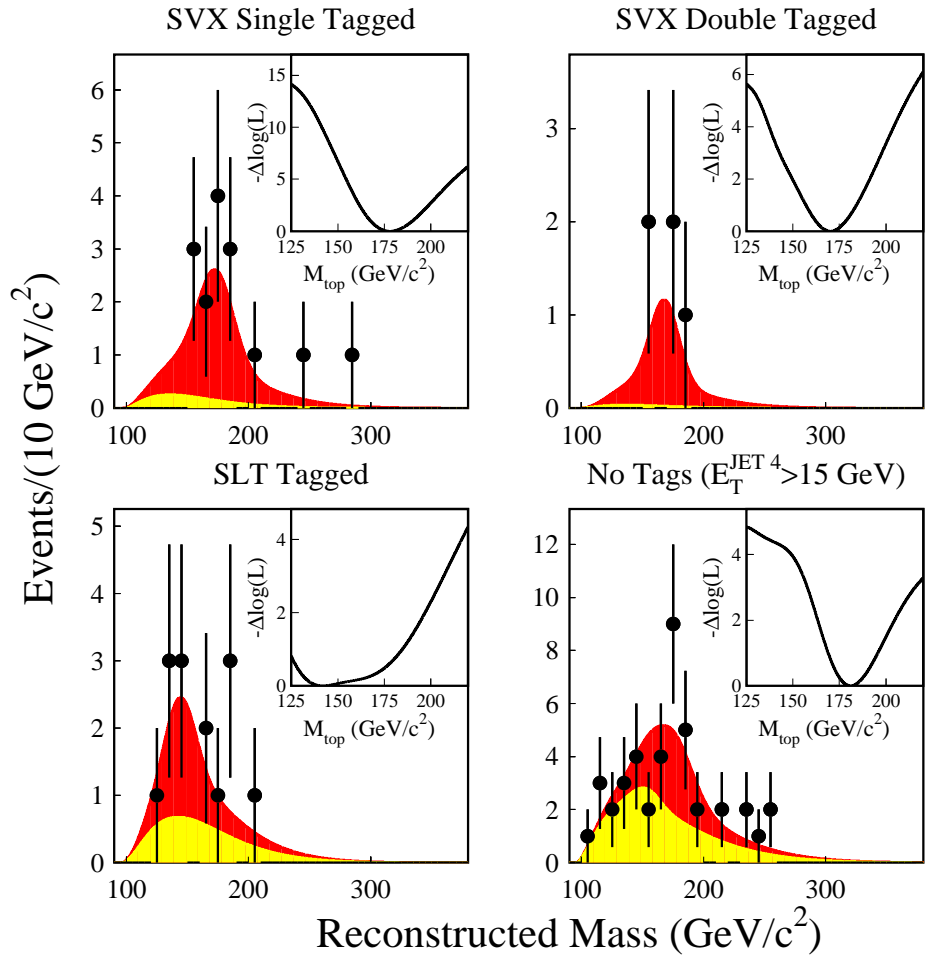,width=0.9\hsize}
\caption{The CDF top quark mass distributions for the four lepton+jets 
  subsamples.  The points are the data, the heavily shaded area is the top
  signal + background combination resulting from the fit, and the
  lightly shaded area is background alone.  The insets show the negative log
  likelihood \vs\ $\mt$ for each subset.
  From Ref.~\protect\onlinecite{cdftopmass98}.
}
\label{fig:cdfljmass1}
\end{figure}

\begin{figure}
\centering
\epsfig{file=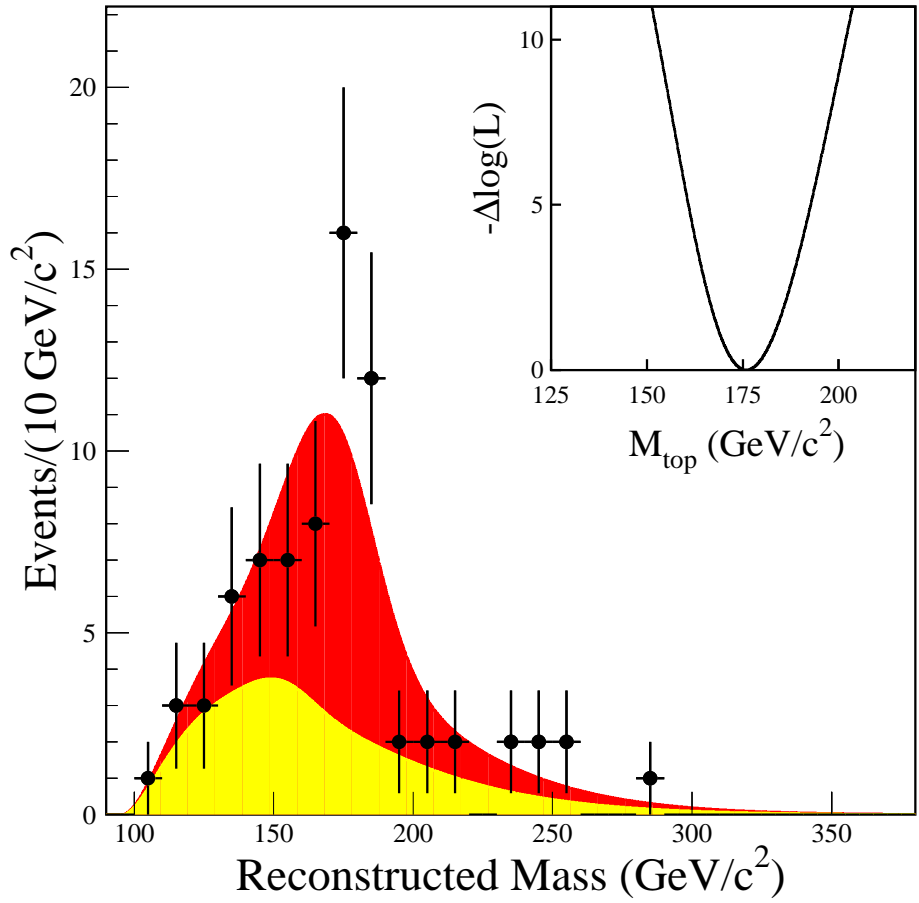,width=0.9\hsize}
\caption{The CDF top quark mass distribution for all four subsamples
  combined.
  The points are the data, the heavily shaded area is the top
  signal + background combination resulting from the fit, and the
  lightly shaded area is background alone.  The inset shows the negative log
  likelihood \vs\ $\mt$.
  From Ref.~\protect\onlinecite{cdftopmass98}.
}
\label{fig:cdfljmass2}
\end{figure}

Each subsample provides an independent mass measurement, as listed in 
\tabref{tab:cdfljmass}. 
The fitted mass distributions and corresponding log-likelihood curves
for each are shown
in \figref{fig:cdfljmass1}. The combined mass distribution,
along with the total log-likelihood, is shown in
 \figref{fig:cdfljmass2}.
From this likelihood curve, CDF measures a top quark mass
of $175.9 \pm 4.8\gevcc$, where the uncertainty is
defined by the usual ``0.5-rule'' (described above). 
The dominant systematic uncertainty,
of $4.4 \gevcc$, is from the jet energy scale;
see \tabref{tab:cdf-ljetmass-systerr}.  A very
important goal for the next round of experiments
is to  substantially reduce this uncertainty. This will
be no
easy task; an enormous effort has already been expended to
achieve the present small uncertainty.

\begin{table}
\caption{Subsamples used in the CDF lepton+jets top quark mass
  measurements.
  $N_{\text{obs}}$ is the the number of observed
  events, $x_b$ is the expected background fraction, and $\mt$
  is the measured top quark mass.  Uncertainties in $\mt$
  are statistical only.  From Ref.~\protect\onlinecite{cdftopmass98}.
  }
\label{tab:cdfljmass}
\begin{tabular}{cccc}
&&$x_b$&Measured $m_t$ \\
Subsample& $N_{\text{obs}}$ & (\%) & ($\ugevcc$)\\
\hline
SVX double tag  &       5       &  $ 5\pm 3$ & $170.1\pm9.3$ \\
SVX single tag  &       15      &  $13\pm 5$ & $178.0\pm7.9$ \\
SLT tag (no SVX)&       14      &  $40\pm 9$ & $142.0^{+33}_{-14}$ \\
No tag          &       42      &  $56\pm15$ & $181.0\pm9.0$ \\
\end{tabular}
\end{table}

\begin{table}
\caption{Systematic uncertainties in CDF's $m_t$ measurement
  in the lepton + jets channels.
  From Ref.~\protect\onlinecite{cdftopmass98}.
}
\label{tab:cdf-ljetmass-systerr}
\begin{tabular}{ld}
Source & Value ($\ugevcc$) \\
\hline
Jet energy measurement            & 4.4 \\
Initial and final state radiation & 1.8 \\
Shape of background spectrum      & 1.3 \\
$b$ tag bias                      & 0.4 \\
Parton distribution functions     & 0.3 \\
\hline
Total                             & 4.9 \\
\end{tabular}
\end{table}

\subsubsection{All Jets Channel}

CDF has also presented a top quark mass measurement in the challenging
all-jets channel,\cite{cdfalljets} in which both $W$~bosons decay
hadronically.  The event sample is based on the kinematic cuts of
technique~I of the cross section measurement (see
\secref{cdfxs-alljets}), except that six jets are required and 
the cut on $\htran = \sum \etjet$ is relaxed from $300 \gev$ to
$\htran > 200\gev$.  The latter change is to reduce the bias
in the mass distributions due to threshold effects.  Before requiring
a $b$-tag, 1121 events survive these cuts.  After tagging, 136~events
remain.

\begin{figure}
\centering
\epsfig{file=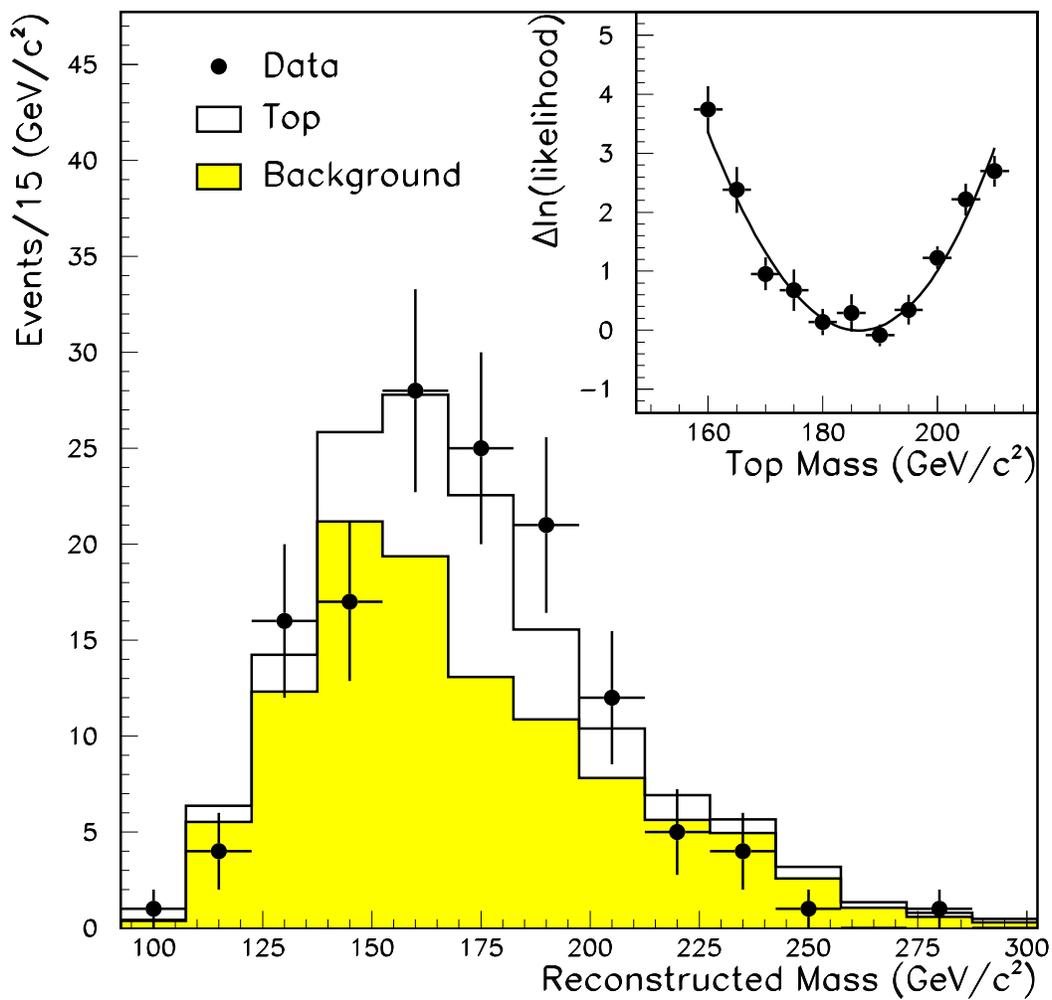,width=\hsize}
\caption[]{The CDF fitted mass distribution for the all jets sample.
  The points are the data, the shaded histogram is the background, and
  the open histogram is the sum of $\ttbar$ signal with
  $\mt = 175\gevcc$ and background.  The inset shows the
  log-likelihood curve.
  From Ref.~\protect\onlinecite{cdftopmass98}.
}
\label{fig:cdfalljmass}
\end{figure}

CDF obtains a fitted mass for each event using a kinematic fit to
the hypothesis
$\ttbar \ra (bW^+) (\bbar W^-) \ra (b\qqbar) (\bbar\qqbar)$.  One can
parameterize these events so that there are thirteen unknowns (the
top quark mass and the three-momenta of both top quarks and $W$~bosons)
and sixteen constraints (four-momentum conservation at both top quark
and $W$~boson decay vertices).  Thus, a 3C fit is possible.
As in the kinematic fit in the lepton + jets channel, CDF tries 
all possible assignments of jets to
the quarks which are consistent with assigning the tagged jet as a
$b$-jet.  The fit which gives the smallest $\chisq$ is
retained.  \Figref{fig:cdfalljmass} shows the resulting three-jet
fitted mass distribution.

A major challenge for this analysis is the background estimation.
This is because not only is the background large, but its fitted mass
distribution is not that different in shape from the signal (see
\figref{fig:cdfalljmass}).  Therefore, an error in the background
normalization could result in a bias in the mass measurement.  CDF
estimates the background using a technique similar to that used for
the lepton+jets and all-jets cross section measurements: each jet in
the sample of 1121 untagged events is weighted by the probability for
a jet in multijet events to be tagged.  This tag probability\footnote{%
The appropriate parameterization of the tag probability has been
 a source of seemingly
endless discussion within CDF and \dzero. Several different parameterizations
have been tried, whose forms
are largely the outcome of a judicious use of intuition and 
experimentation. The chief difficulty
is to understand whether the choice of parameters, in terms of 
which the tag probability is parameterized, is complete 
in the sense that they provide
a full account of the variation of the tag probability from event
to event.
Yet, in spite of this and other difficulties, 
the tag probability functions work
remarkably well in practice.}
is
parameterized in terms of the jet $\et$, $\eta$, and track
multiplicity and the event aplanarity.  This technique yields an
estimated 
background of $108\pm 9$ events.  The shape of the fitted mass
distribution is taken directly from the untagged sample.

The final mass estimate is extracted from the fitted mass distribution
using a maximum likelihood fit based on discrete templates (similar to
what was used in the discovery paper\cite{cdfdiscovery}).  The result
is $186\pm 10\gevcc$ (see \figref{fig:cdfalljmass}).
In the published reference,\cite{cdfalljets} CDF quotes a
systematic uncertainty due to the modeling of gluon radiation
and fragmentation effects of $\pm 8.6\gevcc$.  However, this
uncertainty was recently revised downwards\mcite{roser,*yao98}
to $\pm 3.0\gevcc$.
The uncertainty due to the jet energy scale is $\pm 5.4\gevcc$.
Other systematic uncertainties are due to
the fitting method ($\pm 5.2\gevcc$) and
the background estimation method ($\pm 1.7\gevcc$). 
When these are combined in quadrature,
one obtains a total systematic uncertainty
of $\pm 8.2\gevcc$.

\subsection{\dzero\ Analyses}

The \dzero\ collaboration has measured the top quark mass
in both the dilepton and lepton+jets channels.

\subsubsection{Dilepton Channel}
\label{d0-dilepton-mass}
 
A dilepton $\ttbar$ event is
kinematically underconstrained due to the two unobserved neutrinos. This
would seem to preclude the use of kinematic fits as a means to
estimate the top quark mass. However,
Kondo~\mcite{kondo88,*kondo91,*kondo93} and
Dalitz
and Goldstein~\mcite{dalitz92a,*dalitz92b}
have shown that it is possible to estimate
the mass from 
an ensemble of
kinematic fits by making use of additional information concerning the
likelihood of given final states.
These suggestions were pursued vigorously 
by \dzero\ and led to a 
successful measurement of the top quark mass
in these channels.\cite{d0llmass,d0llmassprd}

The \dzero\ collaboration 
has five dilepton candidate events (3 $e\mu$, 1 $e e$, and 1 $\mu\mu$) 
after applying the
selection criteria used in the cross section measurement. An additional 
$e e$ event is selected by relaxing the track requirement on one of the
electrons in the presence of  a semileptonic (soft muon) $b$-tag. These
six events are used to measure
the top quark mass.\cite{d0llmass,d0llmassprd}  The
expected backgrounds are $0.21\pm0.16$, $0.47\pm0.09$, and $0.73\pm0.25$
events in the $e\mu$,  $e e$, and $\mu\mu$ channels, respectively.

The invariant mass constraints $m(\ell_1\nu_1)=
m(\ell_2\nu_2)=M_W$ and $m(\ell_1\nu_1 b_1)=m(\ell_2\nu_2 b_2)$ depend on the 
unknown neutrino four-vectors, $\nu_1$ and $\nu_2$.
But,
given a value for the top quark mass $m_t$, the system can be solved to within
a four-fold ambiguity. The crucial observation
is that not all hypotheses about the value
of the top quark mass are equally probable: some final-state kinematic
configurations are more likely than others, and so some top quark
masses are preferred over others.
Therefore, if one could assign such probabilities
given the
measured momenta, it would be possible 
to estimate the top quark mass.  For an assumed value of $m_t$, \dzero\ 
assigns a
probability $p(D|m_t)$ to the measured momenta $D$ and uses Bayes'
theorem,\cite{bayes}
\begin{equation}
p(m_t|D) = \frac{p(D|m_t)p(m_t)}{\int_0^{\infty}p(D|m_t')p(m_t')\; dm_t'},
\label{eq:bayes}
\end{equation}
to compute the probability $p(m_t|D)$ of $m_t$ given the fourteen
measured parameters comprising the data $D$.
The expression $p(m_t)$
is the prior probability for $\mt$.

For each value 
of $m_t$,
there are  up to four
possible solutions per jet-to-parton assignment, or configuration. 
For each solution 
and configuration, a likelihood $L$ is computed. These likelihoods are
summed over all solutions and configurations
to arrive at the overall likelihood $p(D|m_t)$. \dzero\ chooses
the prior probability for $m_t$ to be flat, so that
\eqref{eq:bayes} reduces to $p(m_t|D) \propto p(D|m_t)$.

In principle, the likelihood can be computed analytically, using
\begin{equation}
  p(D|m_t) \propto \int f(x) f(\bar x) |{\cal M}|^2 r(D|{\bf v})
    \delta^4\; d^{18}{\bf v}.
\end{equation}
Here, ${\bf v}$ is the set of eighteen parameters needed to completely
specify the kinematics of the event, ${\cal M}$ is the matrix element
for the process
$\qqbar \ra \ttbar + X \ra \ell^+\nu b \ell^-\bar\nu\bbar$, and
$f(x)$ and $f(\bar x)$ are the parton densities for quarks and
antiquarks of momentum fractions $x$ and $\bar x$ in the proton and
antiproton, respectively.  The detector resolution function
$r(D|{\bf v})$ gives the probability to observe the data $D$ given the
true kinematic configuration ${\bf v}$.  The four-dimensional delta
function $\delta^4$ enforces the kinematic constraints:
\begin{equation}
  \delta^4 = \delta(m(\ell^+\nu) - M_W)
             \delta(m(\ell^-\bar\nu) - M_W)
             \delta(m(\ell^+\nu b) - \mt)
             \delta(m(\ell^-\bar\nu \bbar) - \mt).
\end{equation}

Unfortunately, it is not feasible to evaluate this expression for the
large number of Monte Carlo events which are used to construct the
templates.  Therefore, \dzero\ computes only much simplified
approximations to this full likelihood.  Two
different methods are employed for doing so, as discussed below.

The first method, called \emph{matrix element weighting} (${\cal M}$WT),
is a modified version of the procedure suggested by
Dalitz
and Goldstein.\mcite{dalitz92a,*dalitz92b}
Here, the sum of the transverse momenta of the two neutrinos is
required to be equal to the measured $\met$. The system 
is then solved for the neutrino momenta, and hence the top and
antitop four-vectors. The likelihood
for a given solution and configuration $j$ is taken to be
\begin{equation}
L(D|m_t,j) \propto f(x)f(\bar{x})P(E_{{\ell}1}^{CM}|m_t)P(E_{{\ell}2}^{CM}|m_t),
\end{equation}
where $f(x)$ and $f(\bar{x})$ are again the parton distribution
functions (\dzero\ uses CTEQ3M\cite{cteq95}) and
$P(E_{\ell}^{CM}|m_t)$ is the probability density for the lepton energy
in the rest frame of the top quark, $E_{\ell}^{CM}$.

In the second method, called the \emph{neutrino weighting}
($\nu$WT) method, for each configuration $j$,
the expected phase space of neutrino pseudorapidity in
$\ttbar$ events (at a given $m_t$) is divided into
elements with equal phase space weight. 
For each pair of neutrino $\eta$ values, a solution is sought.
If a solution is found, the likelihood $L(D|m_t,j)$ is
assigned based on the degree to which the
sum of the neutrino transverse momenta
agrees with the $\met$ in the
event. For this calculation, 
only the smearing due to the underlying event is taken into account. 
For both methods, one sums $L$ over all solutions and
configurations to obtain $p(D|m_t)$, and hence $p(m_t|D)$.

So far, we have not accounted for the detector resolution function,
$r(D|d)$, where $d$ are the true values of the momenta.
(The information in $d$ is a subset of that in ${\bf v}$.)
To include the effects of $r$, \dzero\ 
uses the \emph{true}, rather than
the measured, momenta to solve the system. 
But since the true momenta are unknown,
it is necessary to consider all
possible values of the true momenta $d$, constrained (probabilistically) 
by the measured momenta $D$, which of course are fixed. 
(In Ref.~\onlinecite{d0llmass}, it is
stated that the measured momenta are ``smeared,'' but this we regard as 
physicists' argot for the statement we have just made.)
The measured
momenta $D$ and the true momenta $d$ are related via
the experimental resolution function $r(D|d)$, which modifies
the relationship between $p(D|m_t)$ and $L(d|m_t,j)$ (now written
in terms of the true momenta $d$) thus:
\begin{equation}
p(D|m_t) \propto \int_{d} r(D|d) \sum_j L(d|m_t,j),
\label{eq:pdmt}
\end{equation}
where the integration is over all possible momenta. 
\dzero\ models its jet resolution function 
by  a double Gaussian, which accounts for
the inherent energy resolution of the \dzero\ calorimeter plus the 
effects
of large angle gluon radiation.  The $\met$ is computed from the
true momenta plus the contribution from the underlying event. The latter
is included by fluctuating each component of $\met$ 
with a Gaussian, with a standard deviation of $4\gev$. 

Because of initial or
final state radiation (ISR or FSR), an event may have more
than two jets.  In that case, \dzero\ considers all
possible interpretations of the jets, merging jets classified
as FSR with the appropriate $b$-jet. Each interpretation is assigned
a weight, which is the product of the
weights assigned to each jet. Therefore, for events with three or more
jets, 
the sum in \eqref{eq:pdmt} becomes a weighted sum.
The
weight assigned to an ISR jet is $\exp(-\et \sin\theta /(25\gev))$, where
$\theta$ is the polar angle of the ISR jet.  For an
FSR jet coming from a $b$-quark, which together
with the $b$-jet form an invariant mass $m$,
the weight assigned is $\exp(-m/(20\gevcc))$.

\begin{figure}
\centering
\epsfig{file=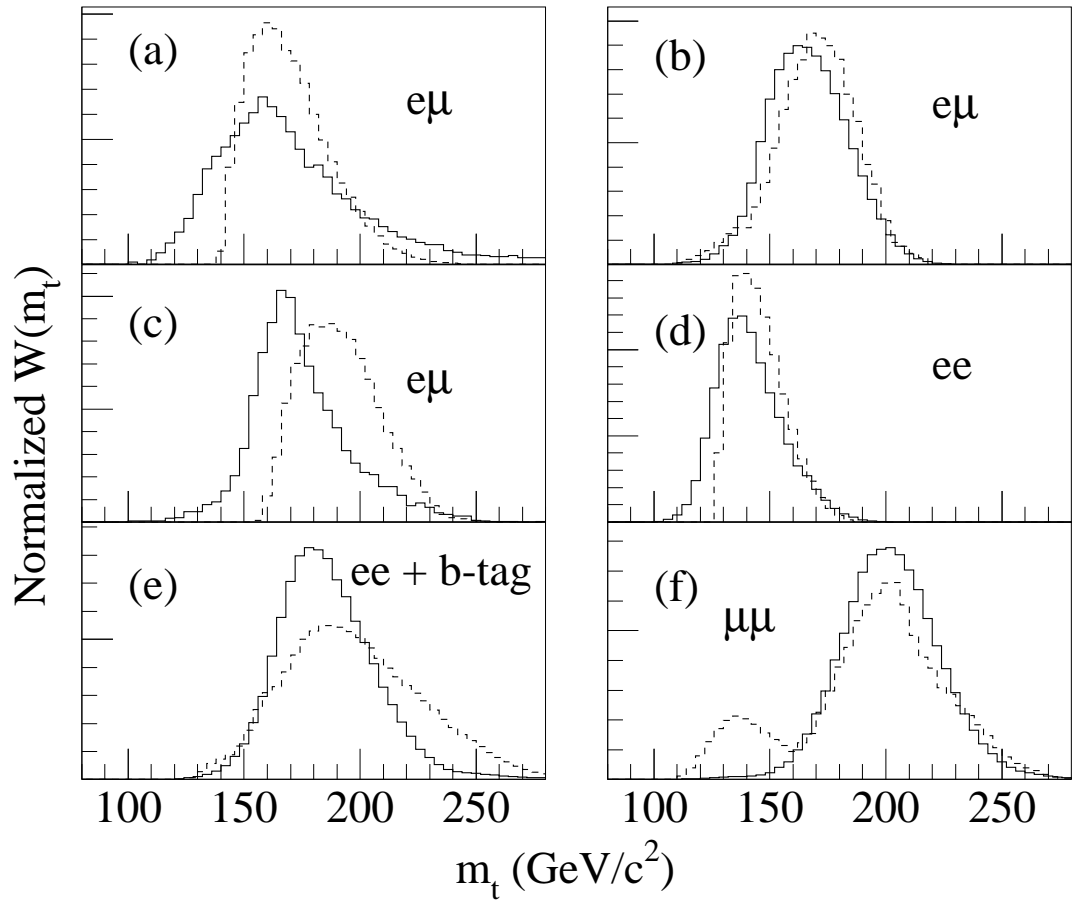,width=\hsize}
\caption{Distributions of $p(D|\mt)$ for the six \dzero\ dilepton
  candidates, using the ${\cal M}$WT (dashed) and $\nu$WT (solid)
  methods.
  From Ref.~\protect\onlinecite{d0llmass}.
}
\label{fig:dilepfig1}
\end{figure}

The normalized distributions of $p(D|m_t)$
for the six candidate
events are shown in \figref{fig:dilepfig1}.  If the background in this
sample of six events were negligible and no approximations were made
in computing $p(D|m_t)$, one could
simply form the product
of the six distributions and use the position of 
its peak or its mean as an estimate of the
top quark mass. However, the background cannot be neglected, and the
effects of the approximations must be taken into account.
Therefore, 
\dzero\ performs a maximum likelihood
fit to a sum of signal and background templates.  The signal templates
are constructed from Monte Carlo simulations, the background templates
from a combination of Monte Carlo and data.

\begin{figure}
\centering
\epsfig{file=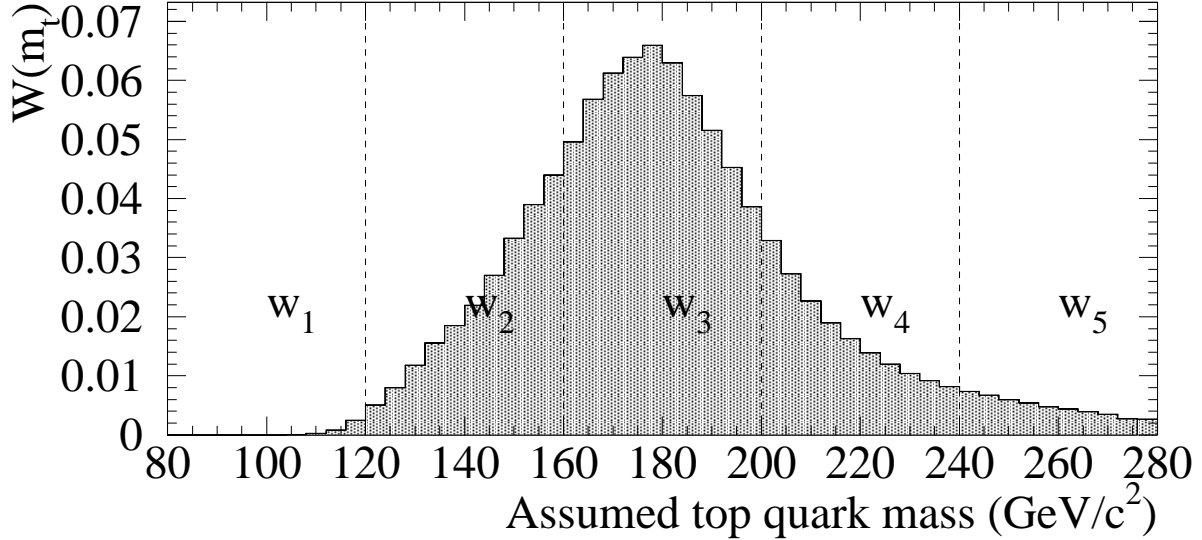,width=\hsize}
\caption{The likelihood curve for a typical Monte Carlo $\ttbar$
  event, normalized to unity.  The vertical lines show the five bins
  into which the likelihood is divided.
  From Ref.~\protect\onlinecite{d0llmassprd}.
}
\label{fig:d0-dilep-binning}
\end{figure}

To make use of the information contained in the shape of the
distributions $p(m_t|D)$, the latter are normalized and divided into
five bins of width $40\gevcc$ (see \figref{fig:d0-dilep-binning}).
For each event $i$, the integrated content of each of the first four
bins form the components of a four-dimensional vector
$\bar{w}_i$. Only four bins are used because the content of the fifth
bin is determined by the normalization condition, and is therefore not
independent.
The likelihood function is then
\begin{equation}
{\cal L}(\bar{w}|m_t,n_s,n_b) = G(\bar{n_b}|n_b)\;P(N|n_s+n_b) \prod_i^N
\frac{n_sf_s(\bar{w}_i|m_t)+n_b f_b(\bar{w}_i)}{n_s+n_b},
\end{equation}
where $n_s$ and $n_b$ are the mean signal and background counts,
$G(\bar{n_b}|n_b)$ is a Gaussian constraint on $n_b$, 
$P(N|n_s+n_b)$ is a Poisson
constraint on $(n_s+n_b)$  to the observed sample size $N = 6$, and
$f_s$ and $f_b$ are four-dimensional probability density templates for the
signal and background, respectively.  These templates are approximated using a
multivariate probability density method.\cite{holmstroem95}
The maximum
likelihood estimate of $m_t$ and its error are determined by a
quadratic fit to the $-\ln \cal{L}$ curve, using nine points about the minimum.

\begin{figure}
\hbox to \hsize{%
\begin{minipage}[b]{.46\linewidth}
\epsfig{figure=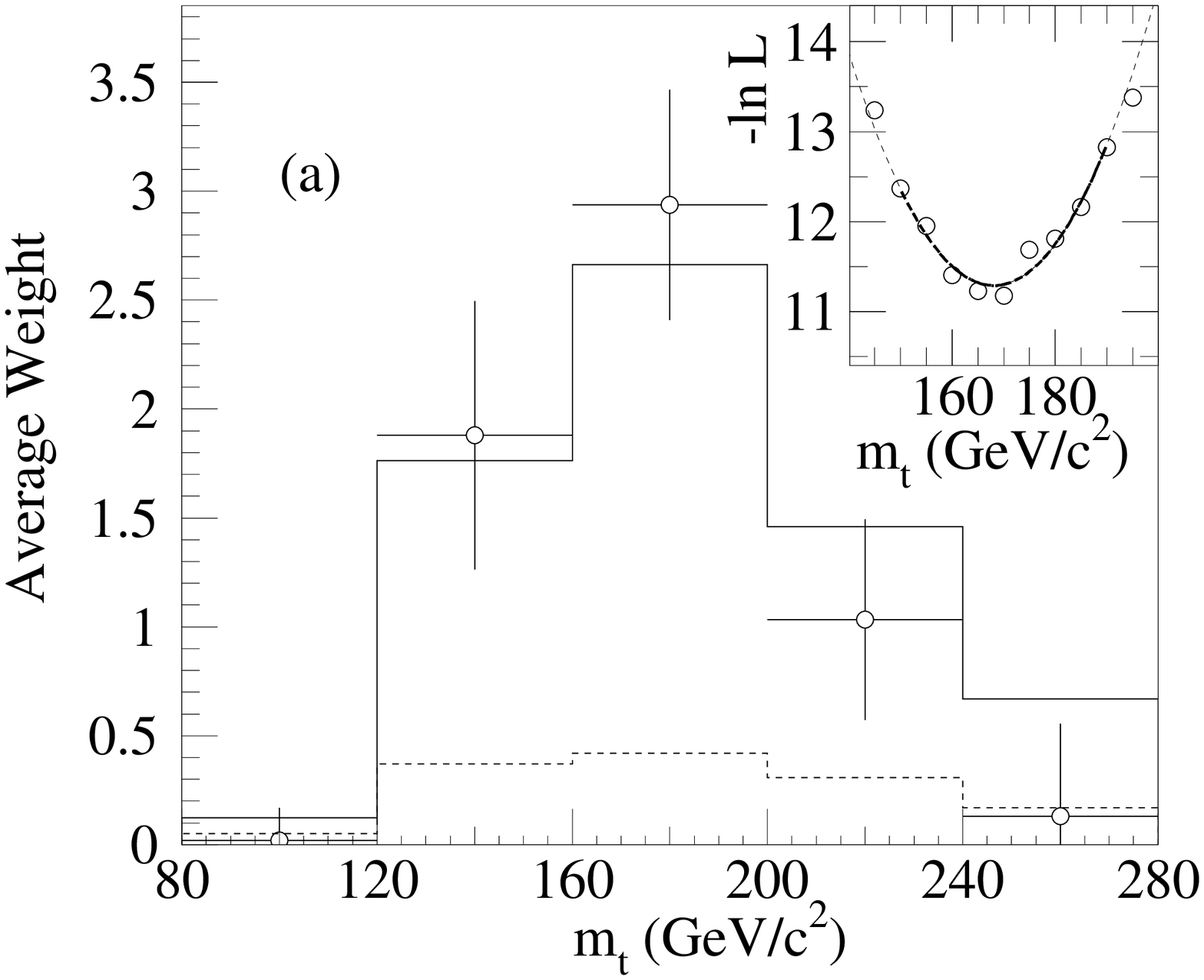,width=\hsize}
\end{minipage}
\hfill
\begin{minipage}[b]{.46\linewidth}
\epsfig{figure=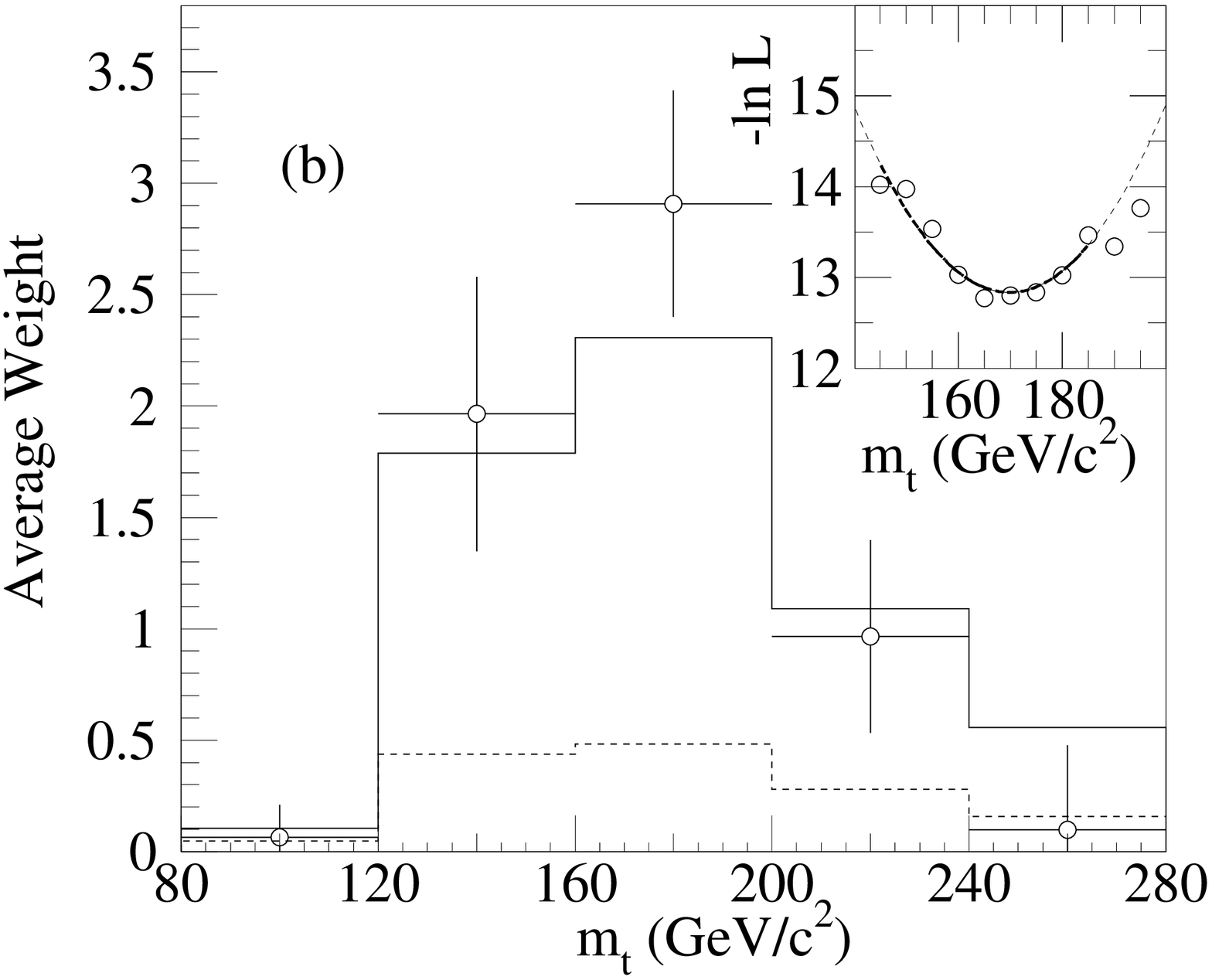,width=\hsize}
\end{minipage}}
\caption{The sum of the normalized candidate likelihoods for the
  ${\cal M}$WT (left) and $\nu$WT (right) analyses, compared to the
  results of the maximum likelihood fit.  The solid histogram is the
  best-fit $\ttbar$ signal plus background, and the dotted histogram
  is background alone.  The insets show the $-\ln {\cal L}$ as a
  function of $\mt$.
  From Ref.~\protect\onlinecite{d0llmass}.
}
\label{fig:dilepfig2}
\end{figure}

The sums of the weights for the six candidate events in the
five mass bins are compared with the
signal and background expectations in \figref{fig:dilepfig2}.  The
insets show the $-\ln \cal{L}$ distributions from which the 
estimates
$m_t = 168.2\pm12.4\gevcc$ ($\cal M$WT) and $m_t = 170.0\pm14.8\gevcc$
($\nu$WT) are extracted.  \Tabref{tab:dileptab1}
shows the results of fits to subsamples of the data.
As shown in \tabref{tab:dilepsys},
the total systematic uncertainty is estimated to be $3.6\gevcc$,
dominated by the jet energy scale uncertainty.
\dzero\ combines the results from the $\cal M$WT
and $\nu$WT analyses, taking into account the $77\%$ correlation
between them, to obtain
$m_t = \statsyst{168.4}{12.3}{3.6}\gevcc$.

\begin{table}
\caption{Summary of \dzero's dilepton $m_t$ measurements for full
  and partial data sets.  Uncertainties are statistical only.
  From Ref.~\protect\onlinecite{d0llmass}.
}
\label{tab:dileptab1}
\begin{tabular}{lcc} 
 Channels Fit             &  ${\cal M}$WT ($\ugevcc$) & $\nu$WT ($\ugevcc$) \\
\hline
$e\mu$ + $e e$ + $\mu\mu$  & $168.2 \pm 12.4$          & $170.0 \pm 14.8$ \\
$e\mu$ + $e e$             & $168.0 \pm 12.7$          & $173.3 \pm 14.0$ \\
$e\mu$                     & $173.1 \pm 13.3$          & $170.1 \pm 14.5$ \\
\end{tabular}
\end{table}

\begin{table}
\caption{Systematic errors for \dzero's measurement of $m_t$
  in the dilepton channels.
  From Ref.~\protect\onlinecite{d0llmass}.
}
\label{tab:dilepsys}
\begin{tabular}{lc} 
 Source   & Error ($\ugevcc$)  \\
\hline
Jet energy scale  &   2.4\\
Signal model          &  1.8 \\
Multiple interactions  & 1.3  \\
Background model      &  1.1  \\
Likelihood fit        &  1.1 \\
\hline
Total                 &  3.6  \\
\end{tabular}
\end{table}

\subsubsection{Lepton+Jets Channels}
\label{d0-ljets-mass}

The distributions of fitted mass ($\mfit$) for signal
and background can overlap significantly, depending on the mass of
the top quark. Therefore, 
in order to extract the top quark mass reliably from an
$\mfit$ distribution, it is necessary either to 
suppress the background sufficiently or to make
optimal use of available information. 
Requiring the presence of
$b$-jets, as we have seen, 
is a simple and effective means of enhancing the signal to
background ratio. However, 
\dzero\ uses only soft-lepton tagging,
for which the tagging efficiency
is only $\approx 20\%$ in $\text{lepton}+\geq 4~\jets$ events,
compared to $\approx53\%$
at CDF, which has the ability to tag $b$-jets with the SVX.

It is therefore noteworthy that 
in spite of  \dzero's relatively poorer $b$-tagging, it has managed to
measure the top quark mass with a precision approaching that of
\hbox{CDF}.\cite{d0ljtopmassprd,d0ljmassprl}
\dzero\ achieves this by using multivariate
techniques\cite{pushpadpf,pushpapbarp,strovink95}
 for separating signal and background while minimizing the
correlation of the selection with the top quark mass.
Two multivariate
methods are used to compute, approximately,
a signal probability $p(\text{top}|D)$ for each
event, given data $D$. 
A likelihood fit of the data to a discrete set of signal and background
models in the $(p(\text{top}|D), \mfit)$ plane is performed to extract the
top quark mass. 

The event selection is based on that used for the cross section
measurement (see \secref{d0ljetxs}), with the following differences:
\begin{itemize}
\item Four jets (with $\et > 15\gev$ and $|\eta| < 2.0$) are always
  required.
\item The aplanarity and $\htran$ cuts are not applied.
\item Events containing high-$\pt$ photons or more than one high-$\pt$
  electron are rejected.
\end{itemize}
These cuts select 91~events, of which seven are tagged.  The dominant
background is $W+\jets$, and about $20\%$ of the background comes from
QCD multijets with a fake lepton.

The events are
analyzed using a kinematic fit to the
$\ttbar \ra \ell\nu b\bbar \qqbar$ hypothesis.  As in the CDF
lepton+jets mass analysis, the jets are first corrected to match the
partons, and the fit is then tried for all possible jet configurations
consistent with the $b$-tagging information.  For each fit, this
gives a fitted mass $\mfit$ and a fit $\chisq$.  The jet configuration
with the lowest $\chisq$ is retained, and events are required to have
this minimum $\chisq < 10$.  After this final cut, 77 events survive,
of which five are tagged.

\begin{figure}
\centering
\epsfig{file=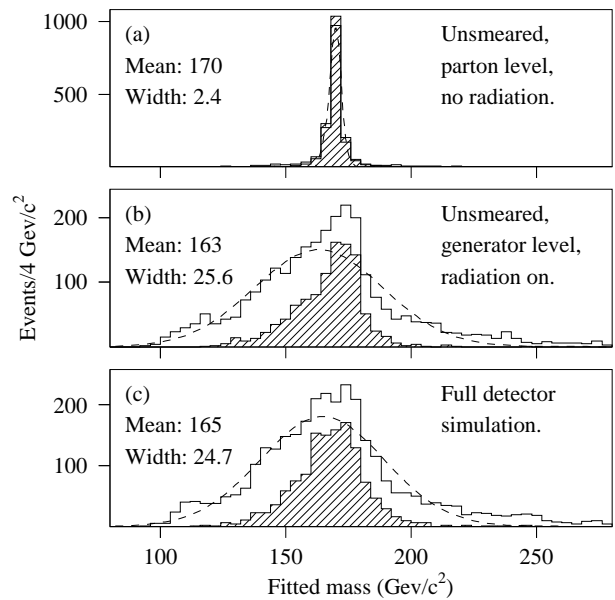}
\caption{Monte Carlo distributions of fitted mass for
  \progname{herwig} $\ttbar$ events with $m_t = 170 \gevcc$, showing
  how radiation and detector resolution affect 
  the width of the mass distribution.
  The hatched histograms show the results when
  the correct jet configuration is used (regardless of whether or not
  it has the lowest $\chisq$).  The means and widths are from the Gaussian
  fits shown by the dashed lines.
  (a)~Using \progname{herwig} partons directly, no radiation or
  detector effects.
  (b)~Using final state Monte Carlo particles clustered into cones.
  Includes the effects of radiation, but not detector effects.
  (c)~After full detector simulation and reconstruction.
  From Ref.~\protect\onlinecite{d0ljtopmassprd}.
}
\label{fig:masstest}
\end{figure}

The kinematic fit is tested on
$\ttbar$ Monte Carlo samples both with and without QCD evolution and gluon
radiation, together with the simulation of detector effects. 
The $\mfit$ distributions
from these tests are shown in \figref{fig:masstest}. The $\mfit$
distribution at the partonic level, without QCD evolution, 
shows a very sharp peak (with $\sim2.4\gevcc$ width). About 80\% of the
time the configuration with the lowest $\chisq$ is the correct one. The width
of the peak is mainly due to the widths of $W$ bosons. Like CDF, \dzero\
finds that, after
QCD evolution and fragmentation effects are included, the width of the $\mfit$
distribution grows to $\sim 26\gev$. (The jets are reconstructed by
clustering particles in a cone of width $R=0.5$.) With these
effects included,
the correct configuration is selected only
about 40\% of the time. The reduction in the selection
efficiency is due to the
confusion arising from the splitting and merging of jets, the jet
finding efficiencies, and increased jet combinatorics. 
Including detector effects does not change the width appreciably
(see \figref{fig:masstest}c).  It can therefore be concluded
that the primary contribution to the width of the mass 
distribution is 
from the extra gluon radiation and jet combinatoric
effects, rather than from the detector resolution.

For the cross section analysis, much of the discrimination between
signal and background is achieved through the use of the variable
$\htran$.  However, this variable
is not usable for the mass measurement because it
is highly correlated with the top quark mass.  Therefore, making
significant cuts on $\htran$ reduces the sensitivity to $\mt$.
In light of this, \dzero\ embarked on an extensive search for variables
that provide good discrimination between signal and background and
are  weakly correlated with $\mfit$.  This resulted in the
following four variables:
\begin{itemize}
\item $x_1\equiv\met$.
\item $x_2\equiv\cal A$, the event aplanarity, as defined earlier.
\item $x_3\equiv (H_T - E_T^{\jet1})/H_z$, where $H_T=\sum E_T$ of all
  selected jets and $H_z \equiv \sum|E_z|$ of all objects in the event
  (lepton, neutrino, and the jets), $E_z$ being the momentum component of
  the object in the beam direction.  This variable 
  measures the centrality of the event.
\item $x_4\equiv \Delta R_{jj}^{\min}\cdot E_T^{\min}/(\et^\ell + \met)$,
  where $\Delta R_{jj}^{\min}$ is the minimum $\Delta R$ of the six
  pairs of four jets and $E_T^{\min}$ is the smaller jet $E_T$ from
  the minimum $\Delta R$ pair.  This variable measures the extent to
  which the jets are clustered together.
\end{itemize}

\begin{figure}
\centering
\epsfig{figure=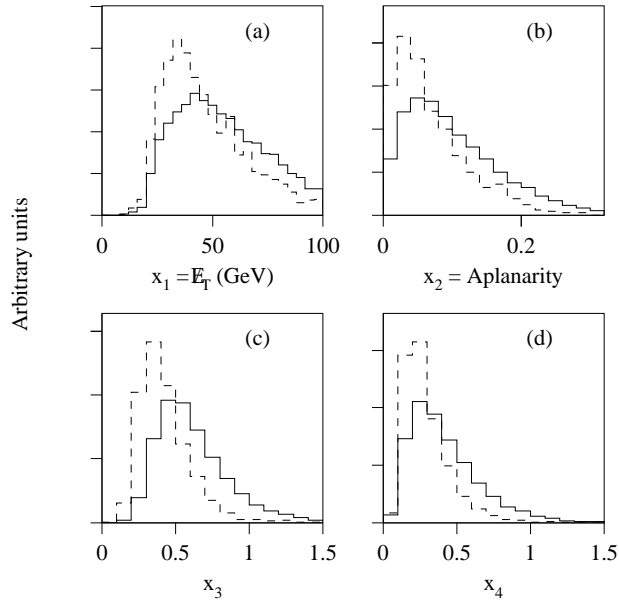}
\caption{\dzero\
  distributions of the discriminant variables $x_1$,~$x_2$,~$x_3$, and
  $x_4$ (see text for
  definitions) for signal (solid histograms) and for background (dashed
  histograms).  All histograms are normalized to unity.}
\label{fig:d0ljvars1}
\end{figure}

\begin{figure}
\centering
\vbox{%
\epsfig{figure=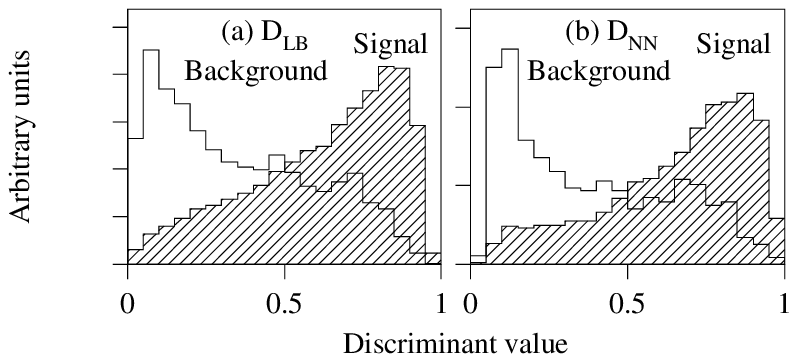}
\epsfig{figure=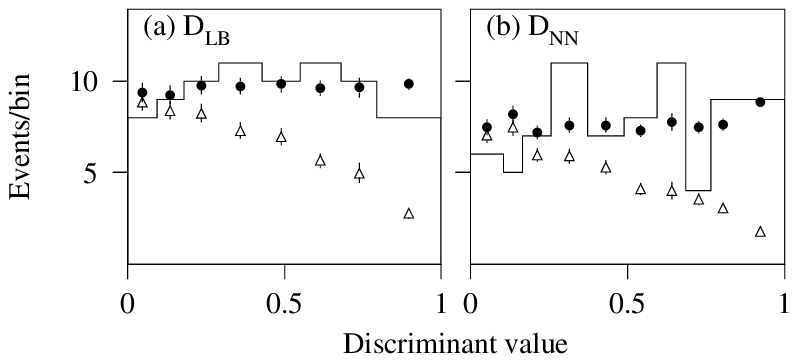}
}%
\caption{Top: \dzero\ distributions of $\DLB$ and $\DNN$ for
  signal (hatched, $\mt = 175\gevcc$) and background (open).  The
  histograms are normalized to unity.
  Bottom: The distributions of $\DLB$ and $\DNN$ for data (histogram),
  compared to the predicted signal plus background (filled circles)
  and background alone (open triangles).  The normalization is from
  the result of the likelihood fit.
  From Ref.~\protect\onlinecite{d0ljtopmassprd}.
}
\label{fig:d0ljdiscrim}
\end{figure}

The distributions of these variables for signal and background are
plotted in \figref{fig:d0ljvars1}.
We see that,
on average, the signal events have larger values of the variables than
do the
background events. But, we also note that, while each variable provides some
discrimination, none is decisive; one might
anticipate that direct cuts on the variables may not be the most effective
way to use them. It proves to be
more effective to treat the variables
collectively and cut on a single multivariate discriminant ${\cal D}(x) =
s(x)/[s(x)+b(x)]$, where $s(x)$ and  $b(x)$ are the signal
and background densities.  Two multivariate methods are used to
approximate  ${\cal D}(x)$: 
(1)~a log-likelihood technique, referred to as the \emph{low bias} (LB) method
(due to the small correlation of the discriminant with the fitted mass),
and (2)~a feed-forward neural network (NN).  In the LB method, \dzero\
parameterizes the
ratios $L_i (x_i) = s_i(x_i)/b_i(x_i)$, where $s_i$ and $b_i$ are
the signal and background densities for variable $x_i$.
Then one computes $L = \prod_i L_i^{w_i}$ and 
$\DLB =L/(1+L)$ for each event. The weights $w_i$ are
adjusted to
minimize the correlation between the discriminant and the fitted
mass.  
The NN method, by construction, takes into account all correlations
between the variables used.
A three layer feed-forward neural network, with four
input nodes, five hidden nodes, and one output node is trained on
samples of Monte Carlo $\ttbar$  ($m_t = 170\gevcc$)
and background events.
The neural network is trained using the back-propagation algorithm,
choosing the output to be unity for the signal and zero for the background. 
For a given event, the network output  $\DNN$
directly approximates the ratio
$s(x)/[s(x)+b(x)]$. \Figref{fig:d0ljdiscrim} shows that
$\DLB$ and $\DNN$ are distributed as predicted and provide
comparable discrimination. As expected, the signal
peaks near unity and the
background near zero.

\begin{figure}
\centering
\epsfig{file=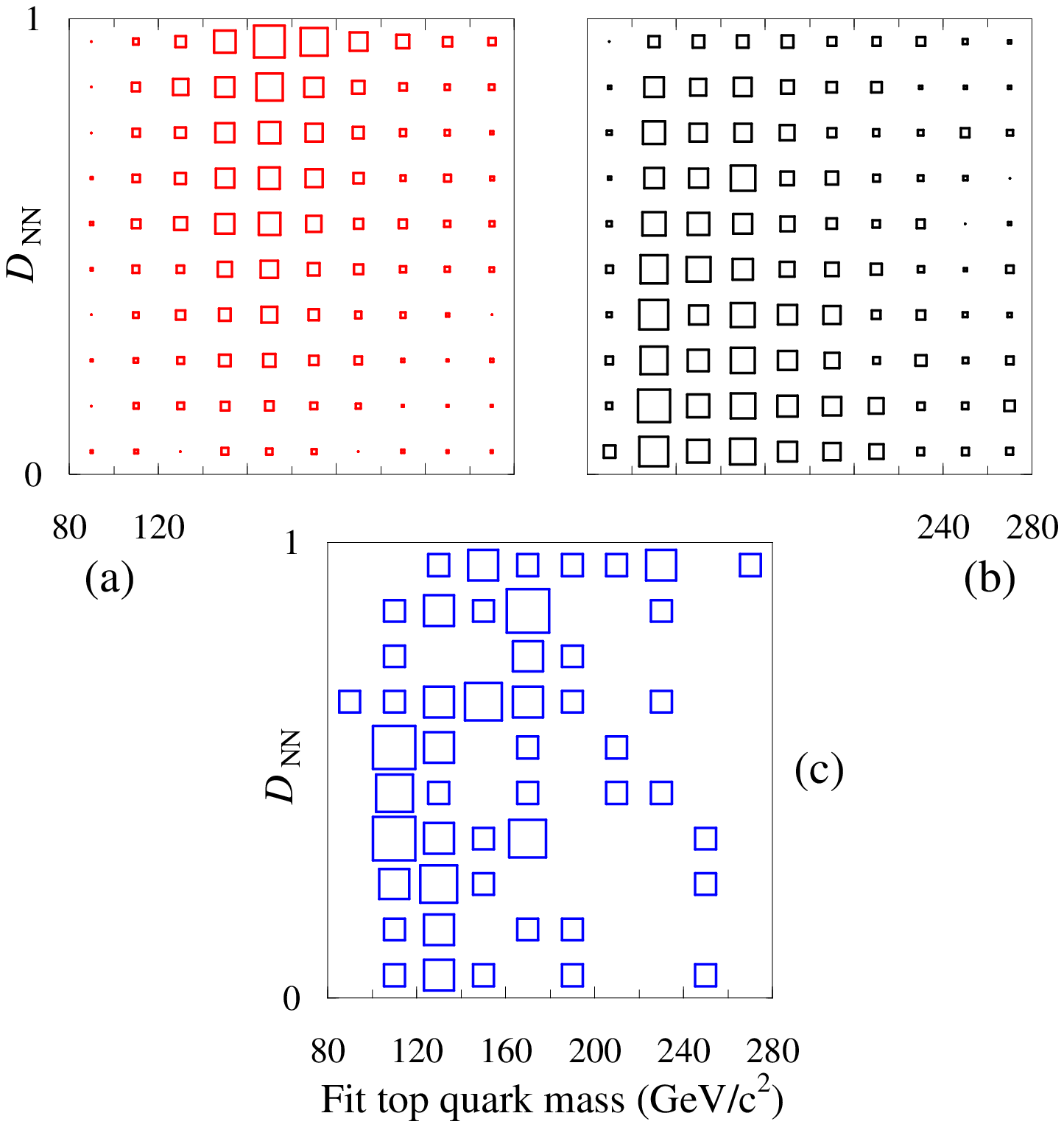}
\caption{Events per bin ($\propto$ areas of boxes) in the
  $(\DNN, \mfit)$ plane for the \dzero\ NN top mass analysis.
  (a)~Expected top quark signal ($\mt = 172\gevcc$).
  (b)~Expected background.
  (c)~Data.
  From Ref.~\protect\onlinecite{d0ljmassprl}.}
\label{fig:d0ljbox}
\end{figure}

\dzero\ carries out two analyses, based on the $\DLB$ and $\DNN$
discriminants.  The data are  binned  in the
$(\calD_{\text{LB/NN}}, \mfit)$ space.  Both analyses use the same binning for
$\mfit$ (20 bins, over the range 80--$280\gevcc$),  but differ in how
the discriminant is binned.  For the $\DLB$ analysis, the events are
split into two bins, depending on whether or not they pass the
``LB cut,'' defined by $\DLB > 0.43$ and
$\htran - \et^{\jet1} > 90\gevcc$.  (The latter condition removes
mostly background; it does not affect the signal much.)  For the
NN analysis, the events are split into ten discriminant bins, as
illustrated in \figref{fig:d0ljbox}. 

\begin{figure}
\centering
\epsfig{file=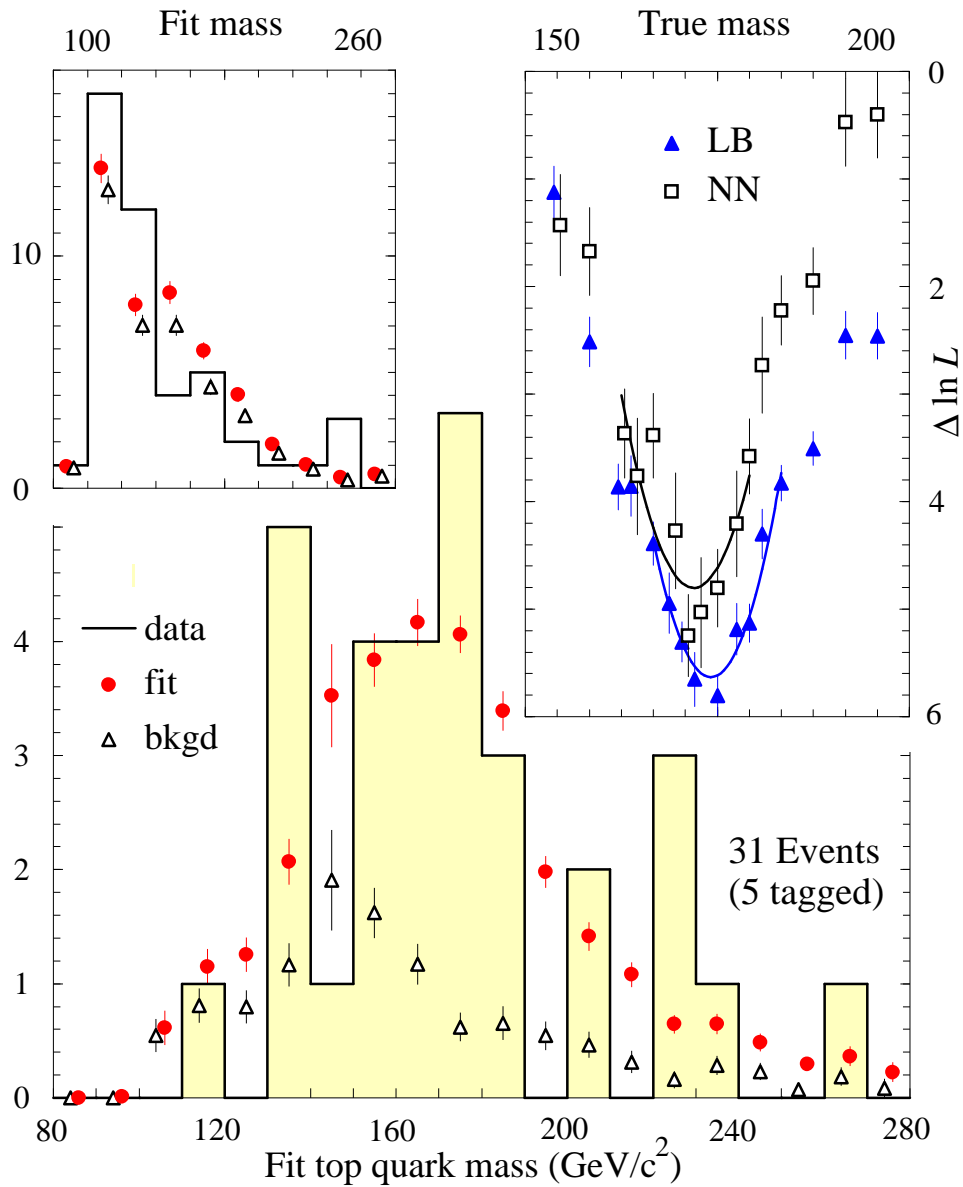}
\caption{Results from the \dzero\ lepton+jets mass analysis.
  Bottom:
  The fitted mass distribution for events which pass the LB cut (a
  signal-rich sample).
  Upper left:
  The same, for events which fail the LB cut (a
  background-dominated sample).
  Upper right:
  The relative $-\ln {\cal L}$ functions for the two methods.
  From Ref.~\protect\onlinecite{d0ljmassprl}.
}
\label{fig:d0ljmass}
\end{figure}

For each method, a top quark mass is extracted from the binned data
using a likelihood fit to a discrete set of signal and background
templates.  The likelihood function, based on
Poisson statistics, is derived by
a Bayesian method.\cite{bayesplb}
In this fit, the number of background events is left unconstrained.
\Figref{fig:d0ljmass} shows the results of the fits and the 
negative log-likelihoods, as functions of the top quark mass.
For the LB method, the fit yields
$m_t= 174.0 \pm 5.6\ \text{(stat)}\gevcc$ and the number of background
events $n_b = 53.2_{-9.3}^{+10.7}$ (out of 77).
For the NN method, the corresponding results are
$m_t= 171.3 \pm 6.0\ \text{(stat)}\gevcc$ and
$n_b = 48.2_{-8.7}^{+11.4}$.
The total systematic uncertainty is estimated to be $5.5\gevcc$
(see \tabref{tab:d0ljmasssyst}),
of which $4.0\gevcc$ comes from the jet energy scale uncertainty and
$3.1\gevcc$ from event modeling (including the modeling of initial and
final state radiation in $\ttbar$ events).

\begin{table}
\caption{Systematic uncertainties in the \dzero\ lepton+jets
  top quark mass measurement.}
\label{tab:d0ljmasssyst}
\begin{tabular}{ld}
Jet energy scale                         &  4.0$\gevcc$   \\
Event modeling&\\
\ \ $t\tbar$ signal                      &  1.9$\gevcc$   \\
\ \ Background                           &  2.5$\gevcc$   \\
Noise/Multiple Interactions              &  1.3$\gevcc$   \\
Monte Carlo statistics                   &  0.8$5\gevcc$  \\
Fit method                               &  1.3$\gevcc$   \\
\tableline
Total                                    &  5.5$\gevcc$   \\
\end{tabular}
\end{table}

\dzero\ combines 
the LB and NN results, taking into account their mutual correlation
of $88\%$, to get an overall measurement of
$m_t = \statsyst{173.3}{5.6}{5.5}\gevcc$. For \dzero,  this
result shows 
a marked improvement in precision with respect to the measurement
published at the time of the discovery. 
This bodes well for the measurement of the top quark mass
in Run 2 by both collaborations.

\subsection{Summary of Mass Measurements}

CDF observes 8 dilepton events with an estimated
background of $1.3 \pm 0.3$ events, and
obtains a mass of $m_t = \statsyst{167.4}{10.3}{4.8}\gevcc$.
In the lepton+jets channel, 76 events are observed, of which 15 are
singly tagged and 5 doubly tagged with the SVX. The mass obtained
from this sample is $\statsyst{175.9}{4.8}{4.9}\gevcc$.  
The 136-event all-jets sample, with a background of $108 \pm 9$ events,
yields $m_t = \statsyst{186}{10}{8}\gevcc$.  Combining all the CDF
measurements yields
$m_t = 175.3\pm 6.4\gevcc$.

\dzero\ observes 6 dilepton events with a background of $1.4 \pm 0.3$ events,
and measures a mass of
$m_t = \statsyst{168.4}{12.3}{3.6}\gevcc$.
In the lepton+jets channel, 77 events are observed
(of which 5 are tagged), with an estimated  background of 
$52 \pm 9$ events. This
gives $m_t = \statsyst{173.3}{5.6}{5.5}\gevcc$. When one 
combines the \dzero\ dilepton and lepton+jets results, one obtains
$m_t = \statsyst{172.1}{5.2}{4.9}\gevcc$.

\begin{figure}
\centering
\epsfig{file=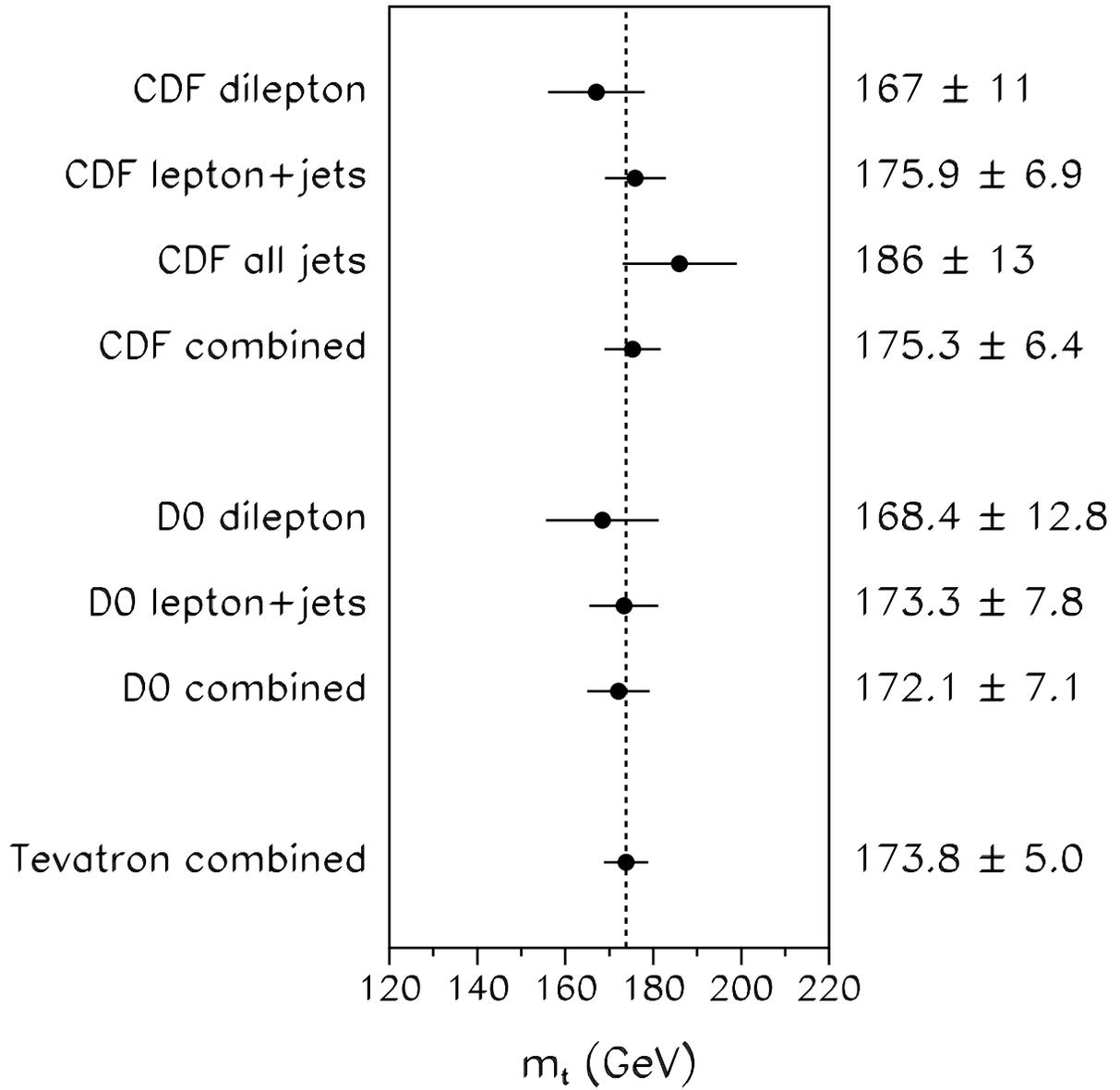,width=\hsize}
\caption{Summary of all the top quark mass measurements from CDF and \dzero.}
\label{fig:item_mass}
\end{figure}

Combining all five measurements
yields\cite{boaz} $m_t = 173.8 \pm 5.0 \gevcc$. 
The measurements, summarized in
\figref{fig:item_mass}, are in striking agreement 
with each other as well as with the indirect measurements 
from precision electroweak data. It is
remarkable that the top quark mass is now known to
a precision (3\%) that is far 
better than what was anticipated a few years ago! 

\section{Other Studies}
\label{other}

\subsection{Measurement of $\Vtb$}

The analyses discussed so far have assumed that the branching
fraction of the decay $t \rightarrow Wb$ is essentially $100\%$.
Clearly, we would like to be able to test this assumption.

The mixing between the three quark generations is described by the
Cabibbo-Kobayashi-Maskawa (CKM) matrix:\mcite{cabibbo63,*km73}
\begin{equation}
\pmatrix{d'\cr s'\cr b'\cr} = 
  \pmatrix{V_{ud} & V_{us} & V_{ub} \cr
           V_{cd} & V_{cs} & V_{cb} \cr
           V_{td} & V_{ts} & V_{tb} \cr}
  \pmatrix{d\cr s\cr b\cr},
\label{eq:ckmdef}
\end{equation}
where the unprimed letters denote mass eigenstates and the primed
ones denote weak eigenstates.  The ratio $R$ of the branching fraction
for a top~quark decaying to a $b$-quark to that of a top~quark decaying
to any down-type quark $q$ can be written as
\begin{equation}
R = {B(t\rightarrow Wb) \over B(t\rightarrow Wq)} =
  {\Vtb^2 \over \Vtd^2 + \Vts^2 + \Vtb^2}
\label{eq:vtbrdef}
\end{equation}
(assuming that top quark decays to non-$W$ final states are negligible).
In the Standard Model with three generations, the CKM matrix \eqref{eq:ckmdef}
must be unitary.  With this constraint, current measurements imply
that $0.9991 < \Vtb < 0.9994$ ($90\%$ confidence level).\cite{pdg96}
If, however, the assumption of three generations is not made (while
maintaining the unitarity of the expanded mixing matrix), then the
$90\%$ confidence level interval for $\Vtb$ opens up to
$0.05 < \Vtb < 0.9994$.  Removing the unitarity assumption leaves
$\Vtb$ unconstrained.  Therefore, if a measurement of $\Vtb$ were to yield
a result significantly different from unity, this would be indicative of
new physics beyond the Standard Model (such as the presence
of a fourth generation).

\begin{table}
\caption{Data for the CDF $\Vtb$ analysis from the lepton+jets (W4J)
  and dilepton (DIL) samples.  From Ref.~\protect\onlinecite{tartarelli97}.}
\begin{tabular}{cccc}
\multicolumn{2}{c}{Bin definition} & \multicolumn{2}{c}{Sample} \\
\hline
SVX tags& SLT tags & W4J & DIL \\
\hline
none & none & 126 &   6 \\
none &  one &  14 & --- \\
 one &  --- &  18 &   3 \\
 two &  --- &   5 &   0 \\
\hline
\multicolumn{2}{c}{Total} & 163 & 9 \\
\end{tabular}
\label{tb:vtbdata}
\end{table}

The CDF collaboration has made a preliminary measurement of $R$
using their combined lepton+jets
and dilepton data samples.\cite{tartarelli97,heinson97a}  As shown in
\tabref{tb:vtbdata}, CDF divides the data into several disjoint
bins, depending on the observed $b$-tags.  For each bin~$i$, the number of
events expected $n_i$ can be calculated as a function of the total numbers
of signal and background events expected in the samples, the
$b$-tagging efficiencies and fake rates, and $R$.
These expectations $n_i$ are compared to the number of events
actually observed in each bin $N_i$ using a likelihood function:
\begin{equation}
{\cal L} = \prod_i P(N_i|n_i (\vec x)) \prod_j G(x_j| \bar x_j, \sigma_j),
\end{equation}
where $P$ is the Poisson distribution
$P(N|n) \equiv e^{-n} n^N / N!$.  The quantities $x_j$ are other
variables on which the predictions $n_i$ depend, each with a measured
value~$\bar x_j$ and uncertainty~$\sigma_j$.  They are convolved into the
likelihood with a Gaussian function
$G(x| \bar x, \sigma) \propto \exp [(x - \bar x)^2 / 2\sigma^2]$.
A maximum likelihood fit is then performed to extract $R$ and the
quantities $x_j$.  For this fit, the number of top~quarks in the
samples is taken to be a free parameter, so that the result will not
depend on the $\ttbar$ production cross section.

\begin{figure}
\epsfig{file=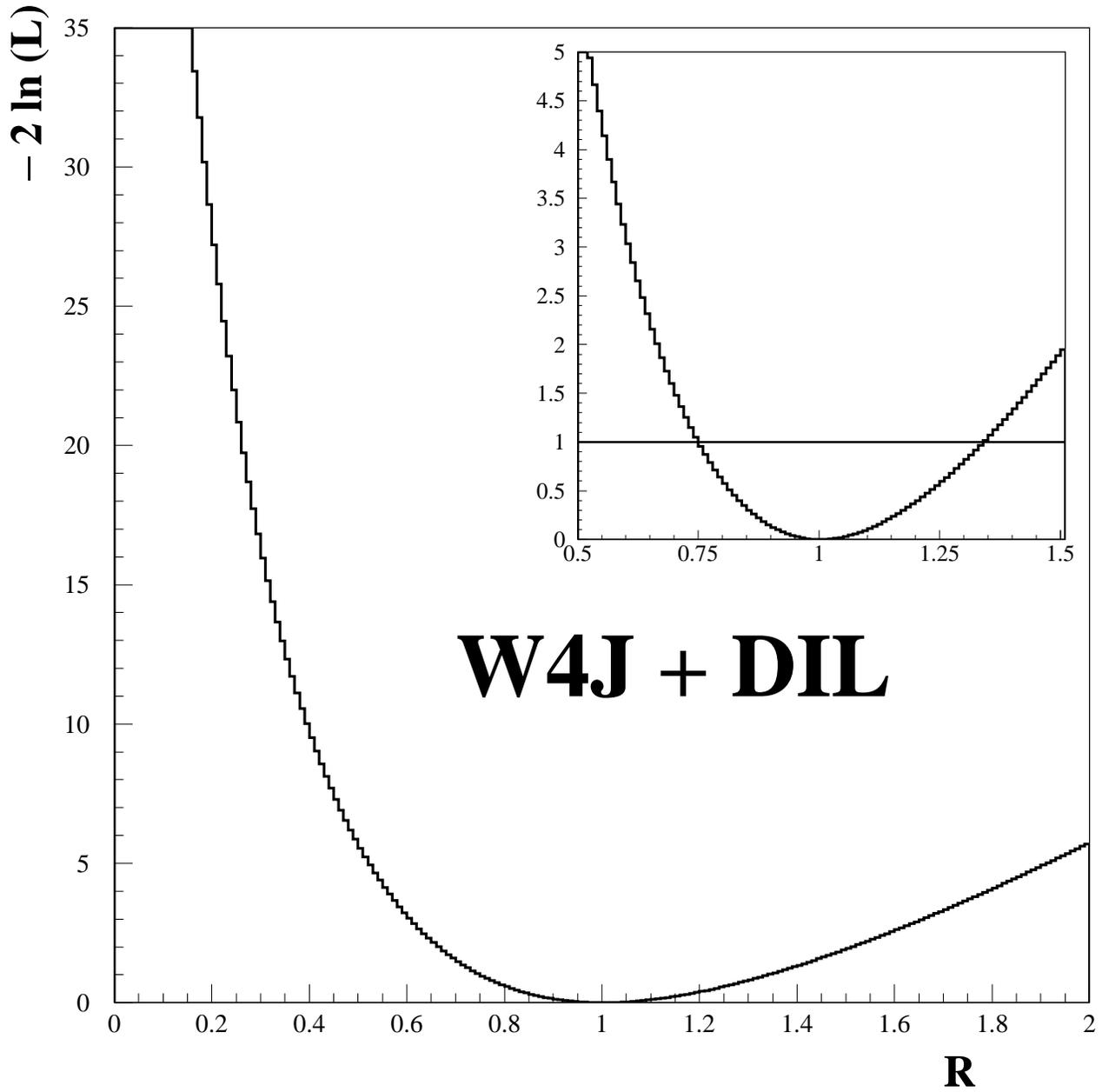,width=\hsize}
\caption{Variation of ${\cal L}$ as a function of $R$ (two different
  scales) for the CDF $\Vtb$ measurement.
  From Ref.~\protect\onlinecite{tartarelli97}.
}
\label{fg:vtblikelihood}
\end{figure}

The resulting likelihood is shown in \figref{fg:vtblikelihood}.  The
fit yields $R = 0.99\pm 0.29$.  The uncertainty includes both statistical
and systematic uncertainties, but is dominated by the statistical
component.  The $90\%$ and $95\%$ confidence limits for $R$ are,
respectively, $R > 0.64$ and $R > 0.58$.

If three-generation unitarity is assumed, then the denominator
in \eqref{eq:vtbrdef} is unity.  The result then corresponds to
$\Vtb = 0.99\pm 0.15$, or $\Vtb > 0.80$ and $\Vtb > 0.76$ at the
$90\%$ and $95\%$ confidence levels, respectively.  If the unitarity
condition is relaxed, no statement can be made about $\Vtb$ without
further assumptions.  CDF sets $\Vtd = 0.009$ and $\Vts = 0.04$, from
the midpoint of their $90\%$ confidence levels determined with the
unitarity assumption. The $90\%$ and $95\%$ confidence limits on
$\Vtb$ are then $\Vtb > 0.055$ and $\Vtb > 0.048$, respectively.

The dominant component of the systematic uncertainty on $R$ is the $b$-tagging
efficiency.  This is measured from the data with a precision
limited by the available statistics of the control samples.
Therefore, the total uncertainty on $R$ should decrease as $1/\sqrt{N}$ as
statistics are increased.  For a $2\ifb$ Run~2 at the Tevatron,
the achievable precision for the $R$ measurement\cite{tev2k} should be about
$2\%$, corresponding to a $95\%$ CL limit of $\Vtb > 0.20$.

An alternate technique which should become possible in Run~2 is to
measure $\Vtb$ from single top production.\cite{heinson97a}
This will be discussed further in \secref{singletop}.
We also note that it is possible to obtain indirect information on
$\Vtb$ from the precision measurements which have been made in the
electroweak sector.  A recent report~\cite{swain97} analyzes the $\Vtb$
dependence of the single-loop corrections to the $Zb\bbar$ vertex.
From a combined analysis using data from LEP, SLC, Tevatron, and
neutrino scattering experiments, the authors derive
$\Vtb = 0.77^{+0.18}_{-0.24}$, independent of unitarity assumptions.

\subsection{Flavor-Changing Neutral Current Decays of the Top Quark}

Flavor-changing neutral current (FCNC) decays can be used to probe for
new physics at mass scales which are otherwise inaccessible.  For
example, the absence of the FCNC decay $K^0_L \rightarrow \mu^+ \mu^-$
was indicative of the presence of the charm quark, even though the
charm quark is several times heavier than the kaon.\cite{gim}  For the top
quark, the SM rates for FCNC decays are very small,\mcite{eilam91,*mele98}
with the branching
fractions $B(t\rightarrow cZ)$ and $B(t\rightarrow c\gamma) \sim
10^{-13} \text{--} 10^{-12}$, $B(t\rightarrow cg) \sim 10^{-10}$,
and $B(t\rightarrow cH) \sim 10^{-14} \text{--} 10^{-13}$.
Any observation of these decays at future experiments
would therefore be evidence for new physics beyond the standard
model.

Because of its large mass, the top quark provides a unique system in which
to look for FCNC contributions. In many models, FCNC terms are
suppressed by ratios of the quark masses to a weak interaction mass
scale.\cite{fritzsch89}  Various models have been proposed in which
the FCNC decays of the top quark are enhanced.  These include models with two
Higgs doublets,\mcite{eilam91,*atwood95,*luke93,*diaz93,*grzadkowski91}
supersymmetry,\mcite{agashe95,*yang94,*li94,yang97} and exotic
fermions.\cite{buchmuller89}  These models typically predict
increases of the FCNC decay rates of 3--4 orders of magnitude,
which are still too small to be seen at any experiments planned in the
near future.  However, in supersymmetric models with baryon number
nonconservation,\cite{yang97} $B(t\ra c\gamma)$ can be as large as
$\approx 2\times 10^{-5}$, which could be observable at
the upgraded Tevatron or the Large Hadron Collider, under construction
at CERN.

Of particular interest are the decays $t\rightarrow c\gamma$ 
and $t\rightarrow cZ$, as they would yield the most distinctive final 
states.\mcite{han96,*han95}
Indirect limits on these branching
fractions may be derived from lower energy data; the results are
$B(t\rightarrow c\gamma) < 1\text{--}2\times 10^{-3}$
and $B(t\rightarrow cZ) < 0.04$.

The CDF collaboration has searched for these decay
modes\cite{cdffcnc97} using their full Run~1 sample of
$\approx 110\ipb$.
The branching ratios are measured relative to a normalization sample,
consisting of the tagged lepton+jets $\ttbar$ candidates.
For the $t\rightarrow q\gamma$ mode, CDF uses two different
cut definitions, depending on whether the $W$~boson (from the normally
decaying top quark) decayed leptonically or hadronically.  For the
leptonic case, the requirements are a central lepton (either an electron or
muon) with $\pt > 20\gevc$, $\met > 20\gev$, a photon with
$\pt > 20\gevc$, and at least two jets with $\et > 15\gev$.  A photon
is identified from a cluster of energy in the electromagnetic
calorimeter with no matching charged track.  For the hadronic case,
CDF requires a photon with $\pt > 50\gevc$ and at least four jets with
$\et > 15\gev$, one of which must contain an SVX $b$-tag.  In both
cases, the invariant mass of one jet and the
photon is required to lie in the range 140--$210\gevcc$, 
consistent with the mass of
the top quark.  For the hadronic case, the tagged jet must be one of the
remaining jets, and these remaining jets must satisfy
$\sum \et > 140\gev$, consistent with the decay of a second top
quark.  Acceptances relative to the SM decay mode are calculated
using \progname{isajet} and a parametric simulation of
the CDF detector; the leptonic and hadronic cases constitute $60\%$ and
$40\%$ of the acceptance, respectively.  The expected backgrounds are
less than half an event in each channel.

A single event is observed in the leptonic channel and none in the
hadronic channel.  The event is kinematically consistent with
a $\tbar\rightarrow W^- \bbar \gamma$ decay, although the photon
momentum ($\pt = 88\gevc$) is quite large for this decay.  CDF does
not subtract the expected background, and thus the single observed
event gives a branching fraction limit ($95\%$ CL) of:
\begin{equation}
  B(t\rightarrow u\gamma) + B(t\rightarrow c\gamma) < 3.2\%.
\end{equation}
The systematic uncertainties in the acceptance calculation are dominated by
the statistical uncertainty on the number of events in the
normalization sample (34~candidates), followed by the uncertainty in
the $b$-tagging efficiency.

For the $t\rightarrow qZ$ mode, CDF searches for the channel in which the
$Z$~boson decays into a pair of leptons (electrons or muons) and the other
top quark decays into three jets.  The selection requirements are four
jets with $\et > 20\gev$ and $|\eta| < 2.4$ and a pair of
opposite-charge, same-flavor leptons with invariant mass in the range
$75$--$105\gevcc$.  Acceptances are calculated as before; the total
expected background is 1.2 events, with the major components being
$Z+\text{multijet}$ production and $\ttbar$ dilepton
decays where the lepton pair invariant mass is close to the $Z$~boson
mass.

A single $Z\rightarrow\mu\mu$ event is observed, with event kinematics
that are more consistent with $Z+\text{multijets}$ than with an FCNC
top quark decay.  Again, no background subtraction is performed, and
the branching fraction limit ($95\%$ CL) is
\begin{equation}
  B(t\rightarrow uZ) + B(t\rightarrow cZ) < 33\%.
\end{equation}
This search is less sensitive than the one for the
$t\rightarrow q\gamma$ mode because of the small branching fraction of
$Z$~bosons into charged leptons.  The sources of systematic uncertainty are
similar.

These limits can be improved during future Tevatron runs.
For a $2\ifb$ Run~2, limits of
$B(t\rightarrow q\gamma) < 3\times 10^{-3}$ and
$B(t\rightarrow qZ) < 0.02$ are probably achievable.\cite{tev2k}

\subsection{Top Quark Decays to Charged Higgs}

The minimal Standard Model contains a single complex Higgs doublet,
giving rise to a single physical neutral Higgs scalar $H^0$.
However, the form of the Higgs sector is not presently constrained by
experiment, so it is interesting to consider models with an expanded
Higgs sector.  In particular, supersymmetric models require two Higgs
doublets, one coupling to the up-type quarks and neutrinos, and the
other coupling to the down-type quarks and charged
leptons.\cite{hhguide}  In such theories, there are five physical
Higgs bosons: two neutral scalars $H^0$ and $h^0$, a neutral
pseudoscalar $A^0$, and a pair of charged scalars $H^\pm$.  The
relevant free parameters are the masses of the Higgs bosons and
$\tan\beta$, the ratio of the vacuum expectation values of the neutral
components of the two doublets.

If $\mch < \mt - \mb$, then the decay $t\rightarrow H^+ b$ can
occur, competing with $t\rightarrow W^+ b$.  The branching ratio
is symmetric in $\log\tan\beta$ with a
minimum at $\tan\beta = \sqrt{\mt / \mb}$.  At the minimum,
$B(t\rightarrow W^+ b) \approx 100\%$, but the situation reverses for
$\tan\beta \rightarrow 0$ or $\tan\beta \rightarrow \infty$, where
$B(t\rightarrow H^+ b) \approx 100\%$.  See \figref{fg:higgsbrs}.
The principal decays of $H^+$ are expected to be
$H^+ \rightarrow c\bar s$, dominant at low $\tan\beta$, and
$H^+ \rightarrow \tau^+ \nu$, dominant at high $\tan\beta$.
In addition, a recent paper~\cite{ma98} has pointed out that
the virtual $t$~quark decay $H^+ \rightarrow b\bbar W^+$
is important for $\mch \gtrsim 135\gevcc$ and small $\tan\beta$.
This is also illustrated in \figref{fg:higgsbrs}.

\begin{figure}
\epsfig{file=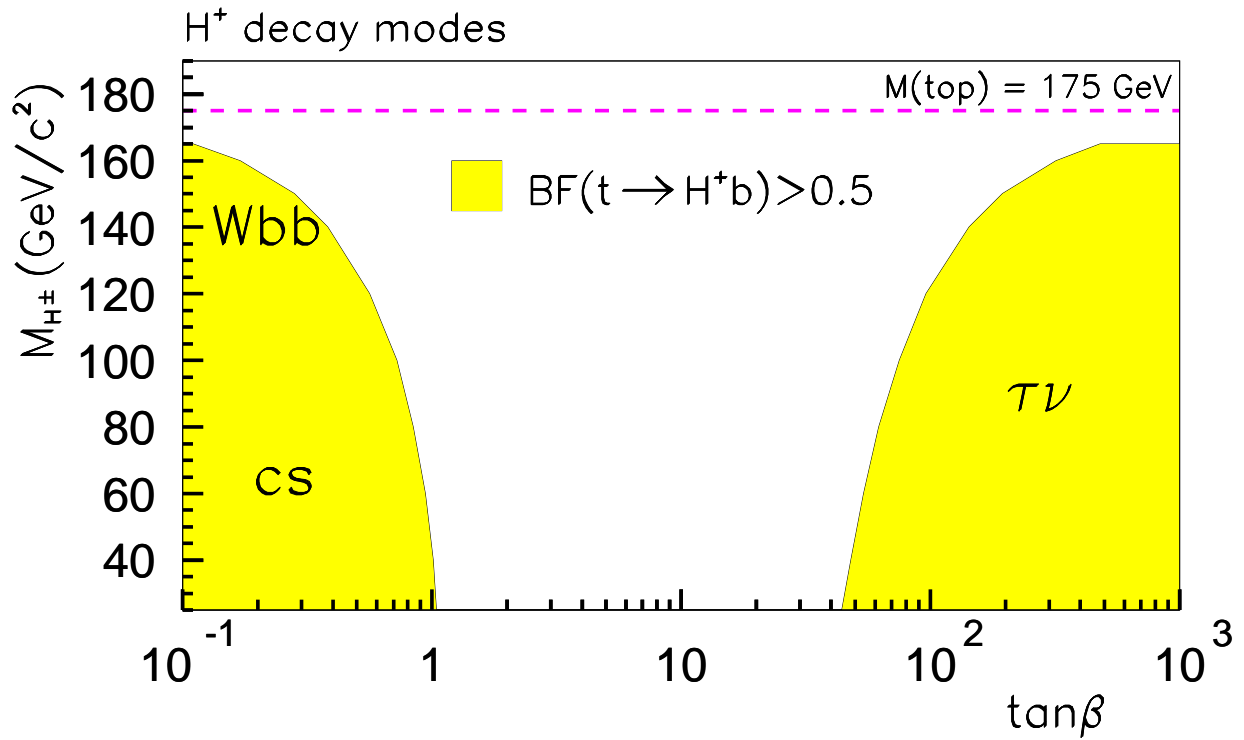,width=\hsize}
\caption{Regions in the $(\mch,\tan\beta)$ plane where the branching
  ratio $B(t\rightarrow H^+ b) > 50\%$.  At low $\tan\beta$,
  $H^+$ decays to $c\bar s$ and $b\bbar W^+$, with the latter mode
  dominating for $\mch > 135\gevcc$.  At high $\tan\beta$,
  $H^+$ decays to $\tau^+ \nu$.
  From Ref.~\protect\onlinecite{bevensee98}.
}
\label{fg:higgsbrs}
\end{figure}

It was also recently realized that there are problems with excluding
very high and low values of $\tan\beta$.  This is because the coupling
of $H^+$ to fermions becomes arbitrarily large in these limits,
causing the perturbative calculations of the branching ratios to break
down.  The widths of the Higgs boson and the top quark also become
very large ($10$--$20\gevcc$) in these regions.
The $\tan\beta$ range over which the coupling is perturbative
is approximately $0.2 < \tan\beta < 170$.

Some early limits on the charged Higgs mass are summarized
in Ref.~\onlinecite{barger90}.  At present, the best limit
independent of $\tan\beta$
is from LEP 2,\mcite{delphihiggs97,*opalhiggs97,*alephhiggs97}
$\mch > 54.5\gevcc$ (95\% CL), extending the LEP 1
limit\mcite{opalhiggs96,*delphihiggs94,*l3higgs92,*alephsearches92}
of $\mch > 44.1\gevcc$.
(Preliminary results presented at the 1998 Moriond conference\cite{dejong98}
further extend this to $\mch > 56.7\gevcc$.)
Limits at high $\tan\beta$ have been derived at
the S$p\pbar\,$S~\mcite{ua1higgs91,*ua2higgs92} and from earlier Tevatron
runs.\mcite{cdfhiggs96,*cdfhiggs94}
A limit based on an independent analysis of CDF data is described in
Ref.~\onlinecite{guchait97}, and an
analysis of $\tau$~lepton decays which yields a 
limit on $\mch$ at large $\tan\beta$
is presented in Ref.~\onlinecite{stahl97}.
In addition, the CLEO collaboration, analyzing
$b\rightarrow s\gamma$ decays and assuming a two-Higgs doublet
extension of the Standard Model, derives a limit\cite{cleopenguin95} of
$\mch > [244 + 63/(\tan\beta)^{1.3}]\gevcc$.
However, models with additional structure, such as supersymmetry, can
evade this limit.\cite{goto96}

The CDF collaboration performs a direct search\cite{cdfhiggs97} for
$t\rightarrow H^+ b$ in the large $\tan\beta$ region by looking for
$H^+ \rightarrow \tau^+ \nu$,
considering $\ttbar$ final states where
either one or both top quarks decay via a Higgs boson.
CDF requires one $\tau$~lepton with $\pt > 10\gevc$,
two jets with $\et > 10\gev$, and one or more
additional objects, either leptons (including $\tau$) or jets, with
$\pt > 10\gevc$.  At
least one of the jets must have an SVX $b$-tag.  If the charged Higgs boson
mass is close to the top quark mass, however,
the $b$-jets may fall below the $\et$
threshold.  To regain acceptance for this case, CDF also considers a second
(``$\tau\tau$'') final state, with no jet requirements but with
two $\tau$~leptons with $\pt > 30\gevc$.  In addition, the two
$\tau$~leptons must not be opposite in azimuth
($\Delta\phi_{\tau\tau} < 160\degree$).  The $\tau$ identification
algorithm used is similar to the calorimeter-based selection described
in \secref{tauid}.

For both final states, it is required that $\met > 30\gev$ and
$\Delta\phi/(1\degree) + \met/(1\gev) > 60$,
where $\Delta\phi$ is the distance in azimuth between the $\vecmet$
and the closest other object.  This cut removes events with
mismeasured $\met$.  In addition, any event containing
a $e^+e^-$ or $\mu^+\mu^-$ pair with mass between 75 and $105\gevcc$
is removed, in order to suppress background from $Z$~boson production.

In $100\ipb$ of data, CDF finds seven candidates passing the above
cuts (six $\tau jjj$ events, one $\tau jje$ event, and no $\tau\tau$
events).  The dominant source of background is fake $\tau$~leptons,
estimated from jet data to be $5.4\pm 1.5$ events for both channels.
Additional background sources, which can produce real $\tau$~leptons,
include electroweak $W/Z + \text{multijet}$ processes and diboson
production.  These backgrounds are estimated using Monte Carlo simulations;
the results are $1.9\pm 1.3$ and $0.08\pm 0.06$ events, respectively.
The
total background for both channels is thus $7.4\pm 2.0$ events,
consistent with observation.

Acceptances are computed using \progname{isajet}.  The resulting
exclusion contour is shown in
\figref{fg:cdfhiggs} (left) for two different assumed top quark
production
cross sections.  The lower of the two, $\sigma_{\ttbar} = 5.0\pb$,
corresponds to the theoretical prediction for a $\mt = 175\gevcc$ top
quark,\cite{laenen94} and the second is $50\%$ larger than that.
For large $\tan\beta$, CDF excludes a charged Higgs boson
with $\mch < 147\gevcc$
at the 95\% confidence level.  This limit can be extended by
requiring that the product
$\sigma_{\ttbar} B(\ttbar\rightarrow W^+bW^-\bbar)$ be consistent with
the (then current) CDF measured cross section\cite{cdfdiscovery} of
$6.8^{+3.6}_{-2.4}\pb$.  The result is
shown in \figref{fg:cdfhiggs} (right).  Note that these limits may not
be reliable for the nonperturbative region $\tan\beta > 170$.

\begin{figure}
\hbox to \hsize{%
\begin{minipage}[b]{.46\linewidth}
\epsfig{figure=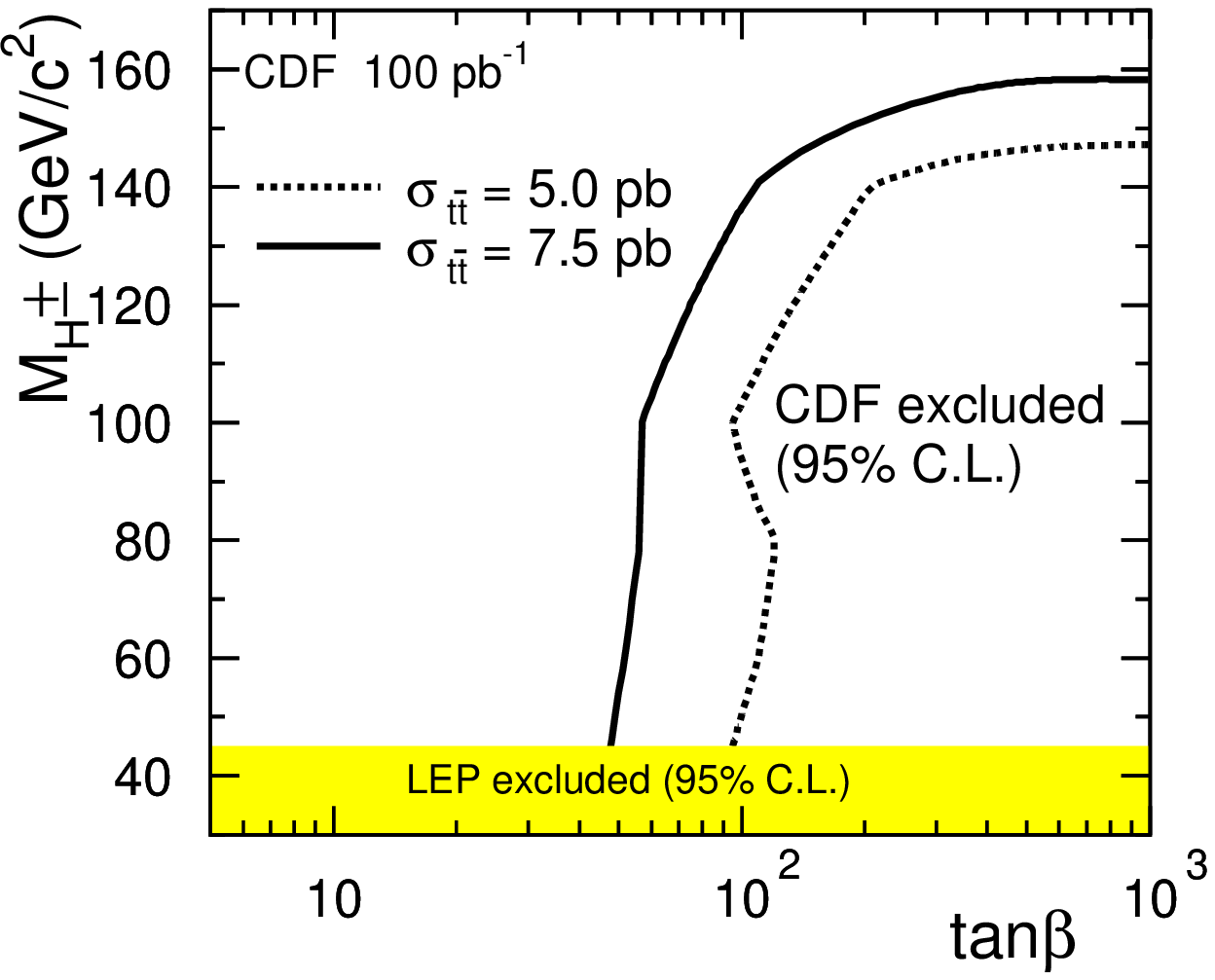,width=\hsize}
\end{minipage}
\hfill
\begin{minipage}[b]{.46\linewidth}
\epsfig{figure=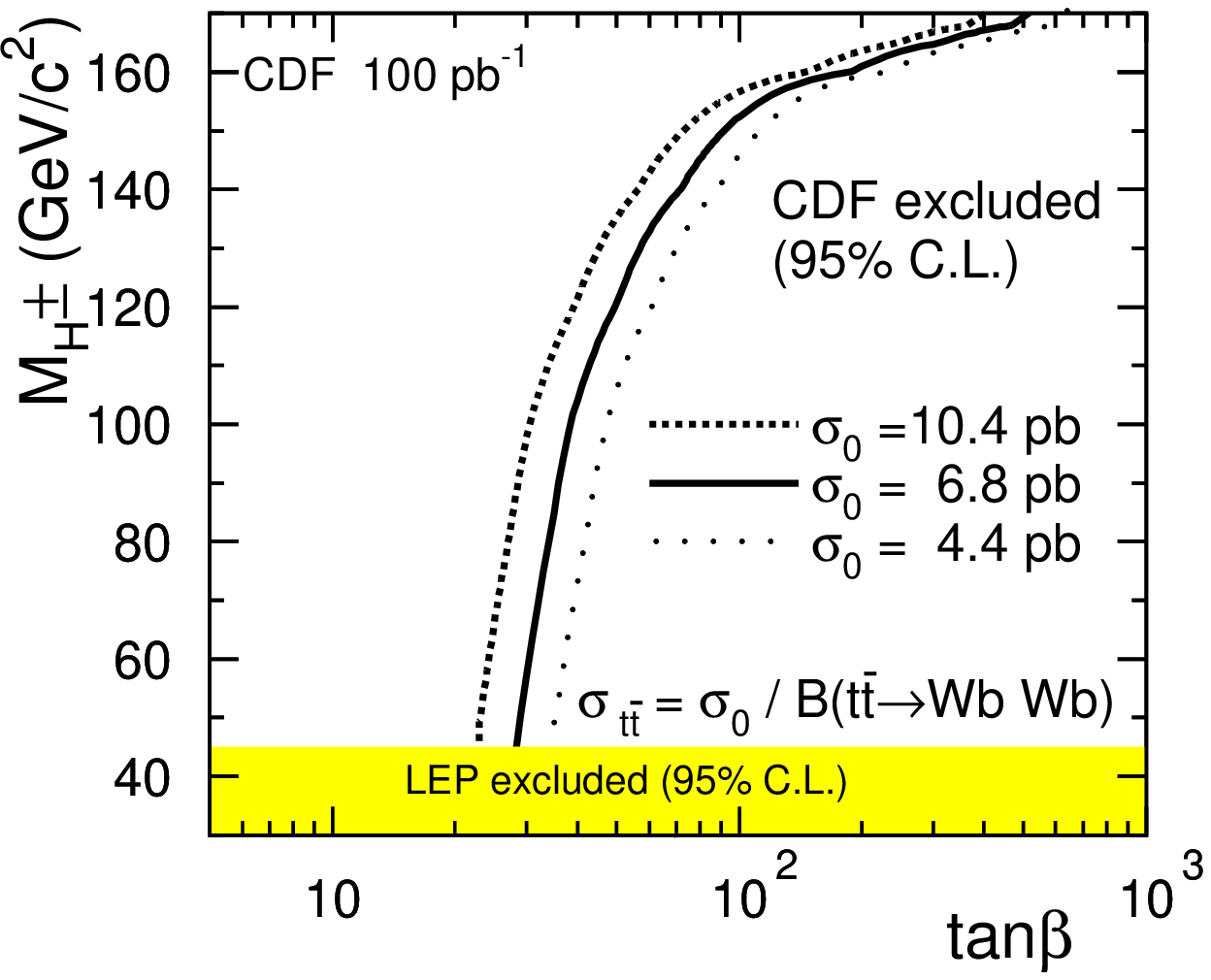,width=\hsize}
\end{minipage}}
\caption{CDF charged Higgs exclusion region for $\mt = 175\gevcc$.
  The plot on the right adds the additional constraint
  $\sigma_{\ttbar} = \sigma_0 / B(\ttbar\rightarrow W^+bW^-\bbar)$.
  Note that these limits may not be reliable for the
  nonperturbative region $\tan\beta > 170$.
  From Ref.~\protect\onlinecite{cdfhiggs97}.
}
\label{fg:cdfhiggs}
\end{figure}

A second analysis strategy is the ``indirect''
search, which uses the results of the $\ttbar$ cross section and top quark
mass measurements.  It is based on the observation
that the standard $\ttbar$
selection cuts are less efficient
for $t\rightarrow H^+b$ than they are for $t\rightarrow Wb$.  Therefore, if
$t\rightarrow H^+b$ were to occur at a significant rate,
there would be a shortfall in the
measured cross section relative to the SM calculation.  As an
example, consider $\mch = 70\gevcc$ and $\tan\beta = 0.4$ (with
$\mt = 175\gevcc$ and $\sigma_{\ttbar} = 5\pb$).  With these
parameters, one has $B(t\rightarrow H^+b) \approx 100\%$ and also
$B(H^+\rightarrow c\bar s) \approx 100\%$.  Therefore, $\ttbar$ pairs will
decay nearly always into a six-jet final state, containing neither a
high-$\pt$ lepton nor any significant $\met$.  The acceptance of the
lepton+jets analyses for these events is practically zero, so all
observed events must have been background.  But the number of
events actually observed is significantly above the background expectation;
therefore, this point in the $(\mch, \tan\beta)$ plane can be excluded at a
high confidence level.

Both the CDF~\cite{bevensee98} and \dzero~\cite{klima98}
collaborations have carried out such an analysis.
The results
from CDF for the low $\tan\beta$ region are shown in
\figref{fg:cdfhiggs3}.  \dzero\ uses this method in both the
low and high $\tan\beta$ regions, as shown in
\figref{fg:d0higgs}.  Note that the \dzero\ results have not yet been
updated to take into account the three body Higgs decay and the
nonperturbative regions.

\begin{figure}
\epsfig{file=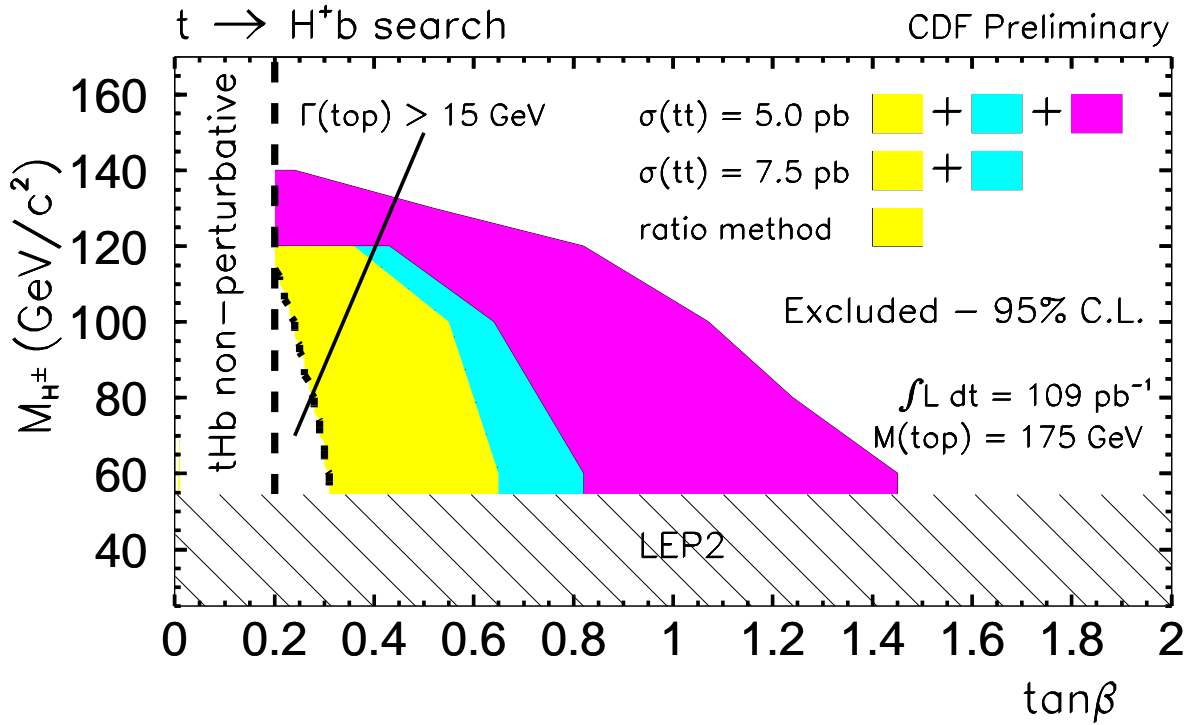,width=\hsize}
\caption{CDF 95\% exclusion limit contours in the $(\mch,\tan\beta)$
  plane for the indirect Higgs~boson
  search.  Results are shown for two
  different assumed top~quark cross sections.  The region labeled
  ``ratio method'' is excluded by using the lepton + jets channel to
  measure $\sigma_{\ttbar}$, rather than assuming a value.  The search
  does not apply in regions where $\tan\beta < 0.2$ or where the top
  quark width is predicted to be larger than $15\gevcc$.
  From Ref.~\protect\onlinecite{bevensee98}.
}
\label{fg:cdfhiggs3}
\end{figure}

\begin{figure}
\epsfig{file=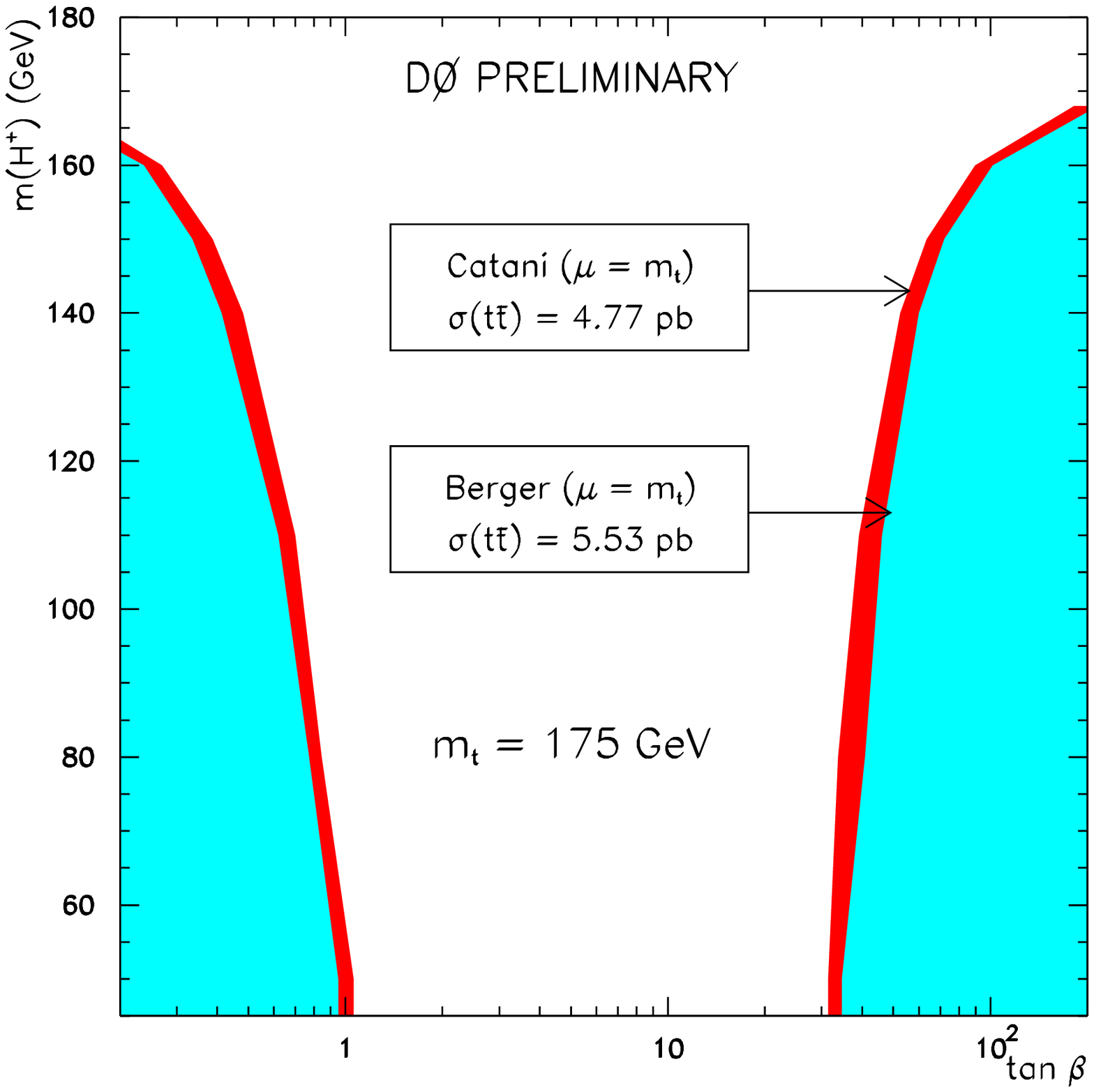,width=\hsize}
\caption{\dzero\ 95\% exclusion limit contours in the $(\mch,\tan\beta)$
  plane for the indirect Higgs~boson
  search.
  This result has not yet been updated to take into account the
  three-body Higgs boson decay or the nonperturbative regions; thus,
  the limits may not be reliable outside of $0.2 < \tan\beta < 170$
  or for $\mch > 135\gevcc$ in the low $\tan\beta$ region.
  From Ref.~\protect\onlinecite{klima98}.
}
\label{fg:d0higgs}
\end{figure}

CDF performs one other variation on this analysis, which avoids the
need to assume a $\ttbar$ cross section.  This is based on the
observation that a large $B(t \ra H^+ b \ra c\bar s b)$ will suppress
dilepton events more severely than lepton+jets events.  CDF first
measures $\sigma_{\ttbar}$ using lepton+jets events, then uses this
to find the expected number of dilepton events, taking into account
the selection efficiency as a function of $\mch$ and $\tan\beta$.
The results of this analysis are also shown in \figref{fg:cdfhiggs3}.

\subsection{$W$ Boson Helicity Fraction in Top Quark Decays}

The $W$~bosons from top quark decay can be either transversely
or longitudinally polarized.  The Standard Model prediction for the
fraction of longitudinally polarized $W$~bosons is
$F_0 = \mt^2 / (2M_W^2 + \mt^2) \approx 0.70\pm 0.01$
for $\mt = 173.8\pm 5.0\gevcc$.  The charged lepton from the decay of
a transversely polarized $W$~boson is preferentially antiparallel to
the direction of the boost from the top~quark rest frame to the $W$~boson
rest frame.  Longitudinally polarized $W$~bosons, on the other hand,
tend to emit the charged lepton perpendicular to the boost
direction.  The consequence is that the $\pt$ spectrum of leptons from
transverse $W$~bosons will be harder than that from longitudinal
$W$~bosons.  Thus, one can extract $F_0$ by fitting the observed
lepton $\pt$ spectrum to the sum of the contributions from
longitudinal $W$~bosons,
transverse $W$~bosons, and background.
(The sensitivity of this measurement could conceivably be improved
by using information from the mass fit.  However, that would
reduce the available statistics and introduce additional biases
which would need to be understood.)

This analysis has been carried out by the CDF
collaboration,\cite{bevensee98} which does a simultaneous fit for $F_0$
in both its $e\mu$ and lepton+jets data samples (of seven and 91
events, respectively).
The fit results are shown in \figref{fg:cdf-whelicity}; the extracted
$F_0$ is $\statsyst{0.55}{0.32}{0.12}$.

\begin{figure}
\epsfig{file=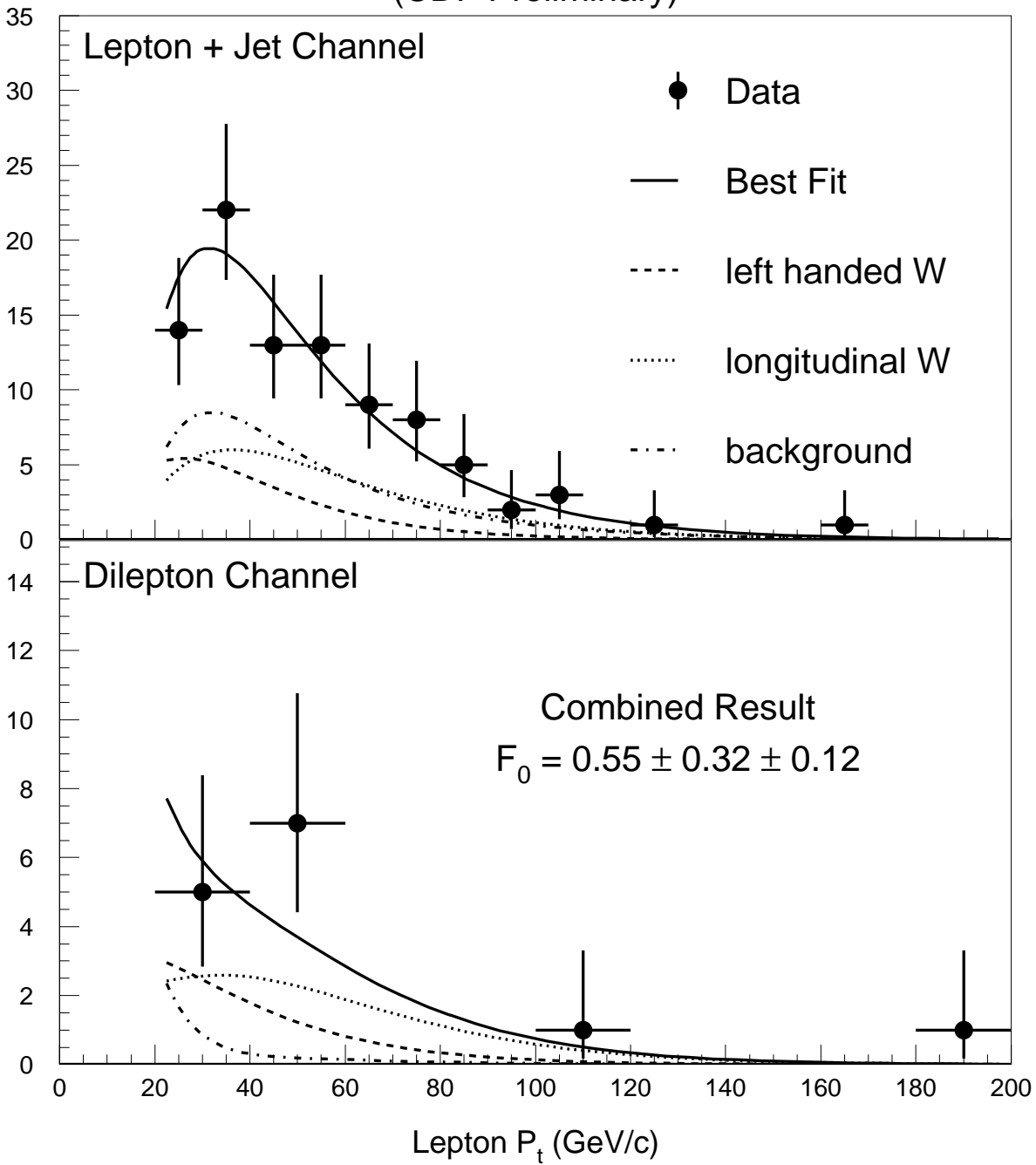}
\caption{The CDF lepton $\pt$ spectra for the lepton+jets and $e\mu$
         (dilepton) channels, fit to a sum of longitudinal $W$~boson
         decays, transverse (left-handed) $W$~boson decays, and
         background.
         From Ref.~\protect\onlinecite{bevensee98}.
}
\label{fg:cdf-whelicity}
\end{figure}

\subsection{Studies of $\ttbar$ Kinematics}

Armed with the results of the kinematic fit from the mass
analysis, one can study the
kinematics of $\ttbar$ events.  Typically, one picks the jet
configuration which gives the lowest $\chisq$ solution and then
calculates quantities based on those jet assignments.  For example,
the $\ttbar$ invariant mass from the CDF collaboration using its
tagged lepton+jets sample\cite{leone97} is shown
in \figref{fg:cdfmtt}.
It should be realized that the quantity
plotted is not what one would directly calculate for the invariant
mass of the two quarks, but is smeared by the effects of
QCD~radiation, jet misassignments, and detector effects.
\Figref{fg:d0topkin} shows a selection of kinematic results
from the \dzero\ collaboration\cite{d0ljtopmassprd} using its lepton+jets
samples.
In all cases, the data are in good agreement with the SM
expectations.

\begin{figure}
\epsfig{file=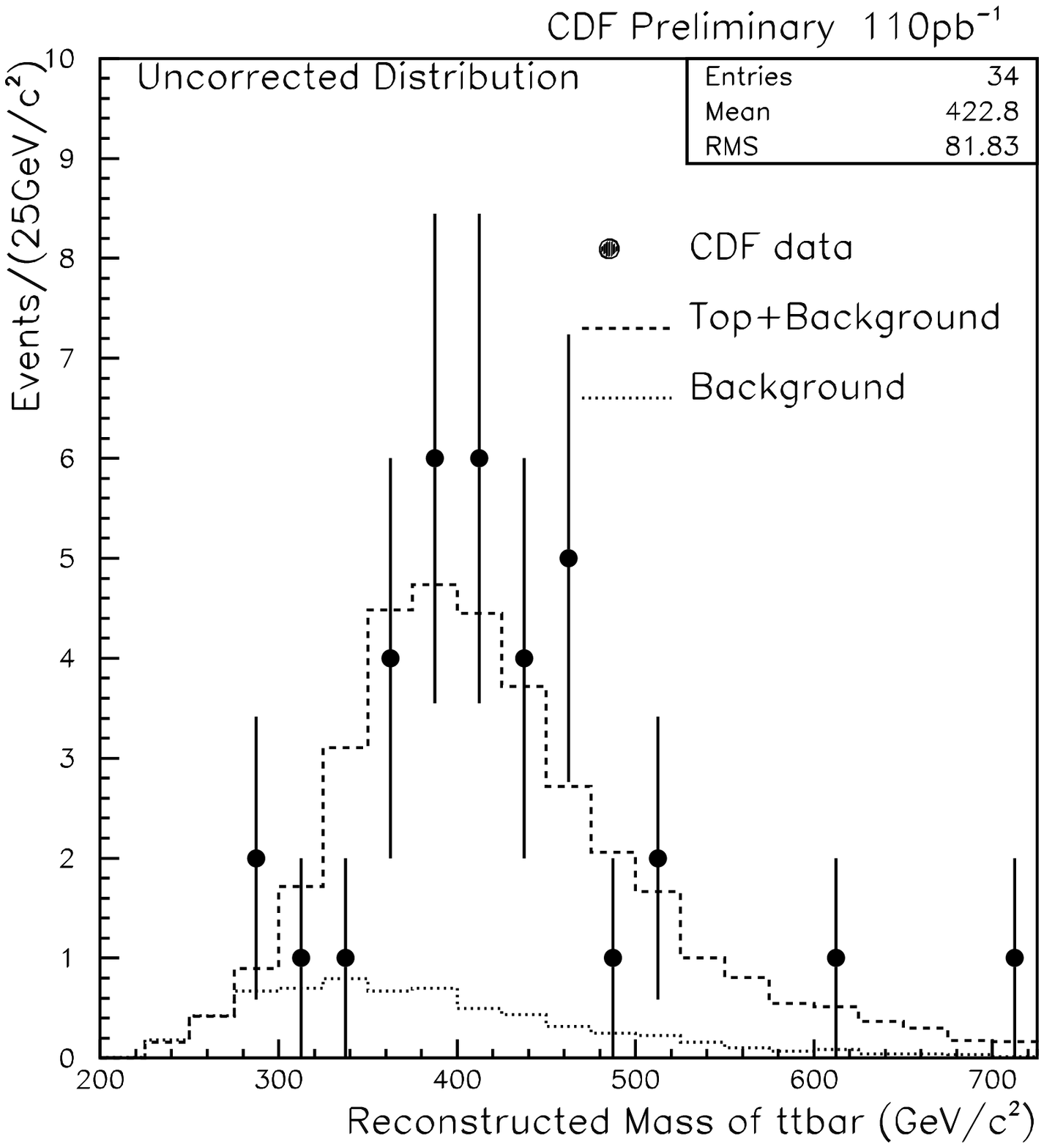,width=\hsize}
\caption{Invariant mass of the $\ttbar$ system in single lepton tagged
  events, from CDF.
  From Ref.~\protect\onlinecite{leone97}.
}
\label{fg:cdfmtt}
\end{figure}

\begin{figure}
\centering
\epsfig{file=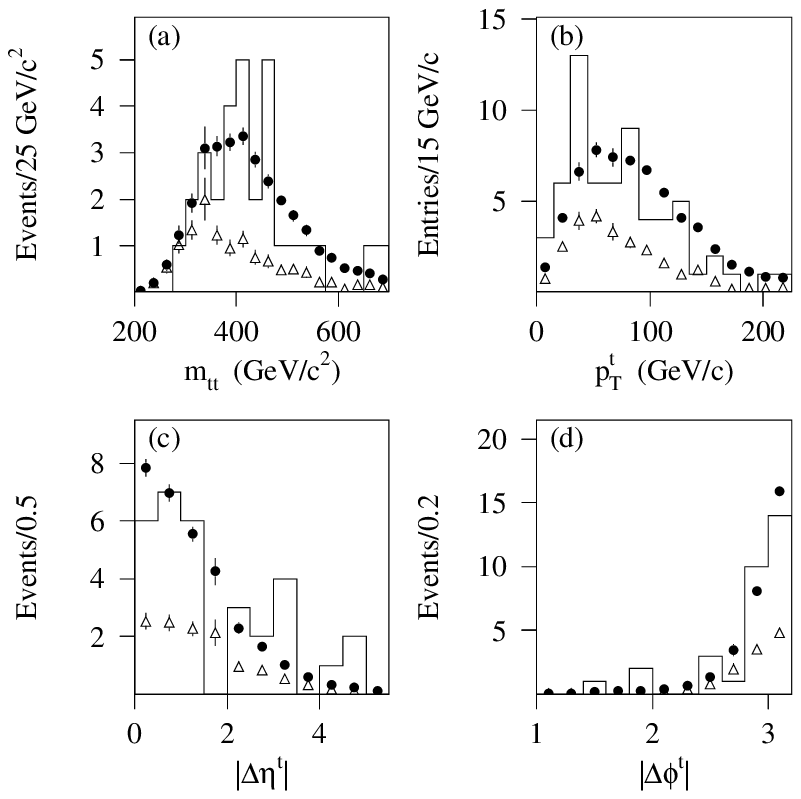}
\caption{Kinematic results from the \dzero\ lepton+jets samples.
  The histograms are data, open triangles are the expected background,
  and filled circles are the expected signal plus background
  ($\mt = 175\gevcc$).
  (a) Invariant mass of the $\ttbar$ pair.
  (b) The transverse momenta
  of the two top quarks (two entries per event).  
  (c) The difference in pseudorapidity $\Delta\eta$ between the two
  top quarks.
  (d) The difference in azimuthal angle $\Delta\phi$ between the two
  top quarks.
  From Ref.~\protect\onlinecite{d0ljtopmassprd}.
}
\label{fg:d0topkin}
\end{figure}

Recently, the CDF collaboration has also presented a measurement of the
mass of the hadronically-decaying $W$~boson in lepton+jets $\ttbar$
decays.\cite{cdfhadw97}
\Figref{fg:cdfhadw1} shows the distribution of the invariant
masses of all dijet pairs in lepton+jets events which
pass the cut $\et^\ell + \sum \et^{\text{jet}} > 310\gev$.
The contributions from non-$\ttbar$ background and incorrect
combinations are subtracted, and the result is then fit to a
Gaussian.  The result is
$\statsyst{77.1}{3.8}{3.6}\gevcc$, and the
significance of the excess is $2.8\sigma$.  They also have 11~events
with two $b$-tagged jets.  (When looking for the second tag,
the ``jet probability'' tag algorithm is used,\cite{alephzbb93}
in addition to the usual SVX and SLT tags.)  \Figref{fg:cdfhadw2}
shows the distribution of the invariant masses of the pair of untagged
jets in each event.  When fit to a Gaussian signal model plus
background, the resulting $W$~boson mass is
$\statsyst{78.1}{4.4}{2.9}\gevcc$.  The
significance of the peak is $2.9\sigma$.  Finally, these two analyses
are combined, giving a $W$~boson mass of
$\statsyst{77.2}{3.5}{2.9}\gevcc$ with a total
significance of $3.3\sigma$.

\begin{figure}
\epsfig{file=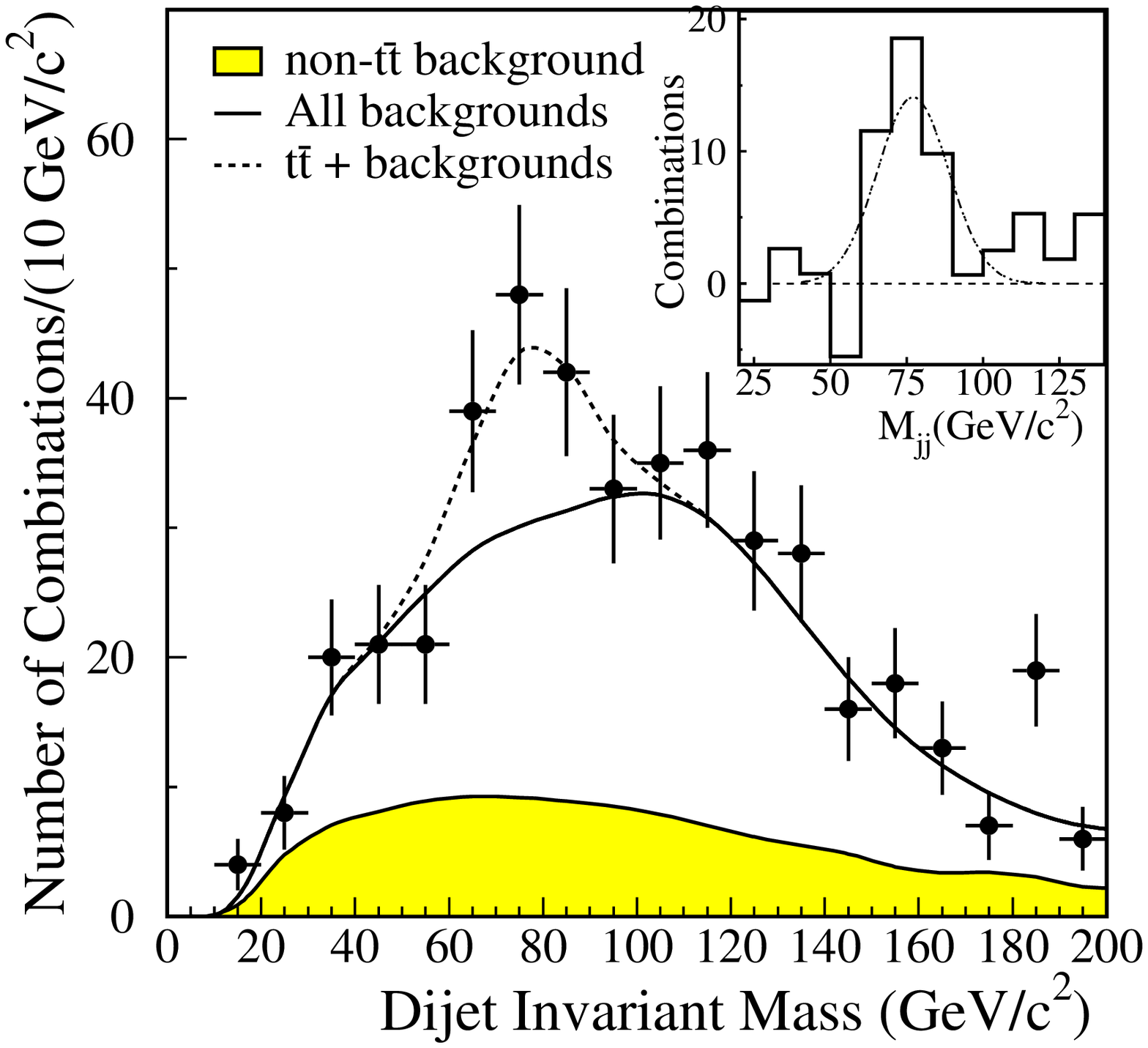,width=\hsize}
\caption{The dijet mass distribution from CDF data
  for $W + \ge 4~\jet$ events after
  the cut $\et^\ell + \sum \et^{\text{jet}} > 310\gev$.
  The points are the data, the shaded curve is the
  \progname{vecbos} $W+\jets$ background, and the solid curve is the
  sum of that and the $\ttbar$ combinatorial background.  The dotted
  curve is the result of a Gaussian fit to the $W$~boson mass peak.
  The inset plot shows the peak after background
  subtraction.
  From Ref.~\protect\onlinecite{cdfhadw97}.
}
\label{fg:cdfhadw1}
\end{figure}

\begin{figure}
\epsfig{file=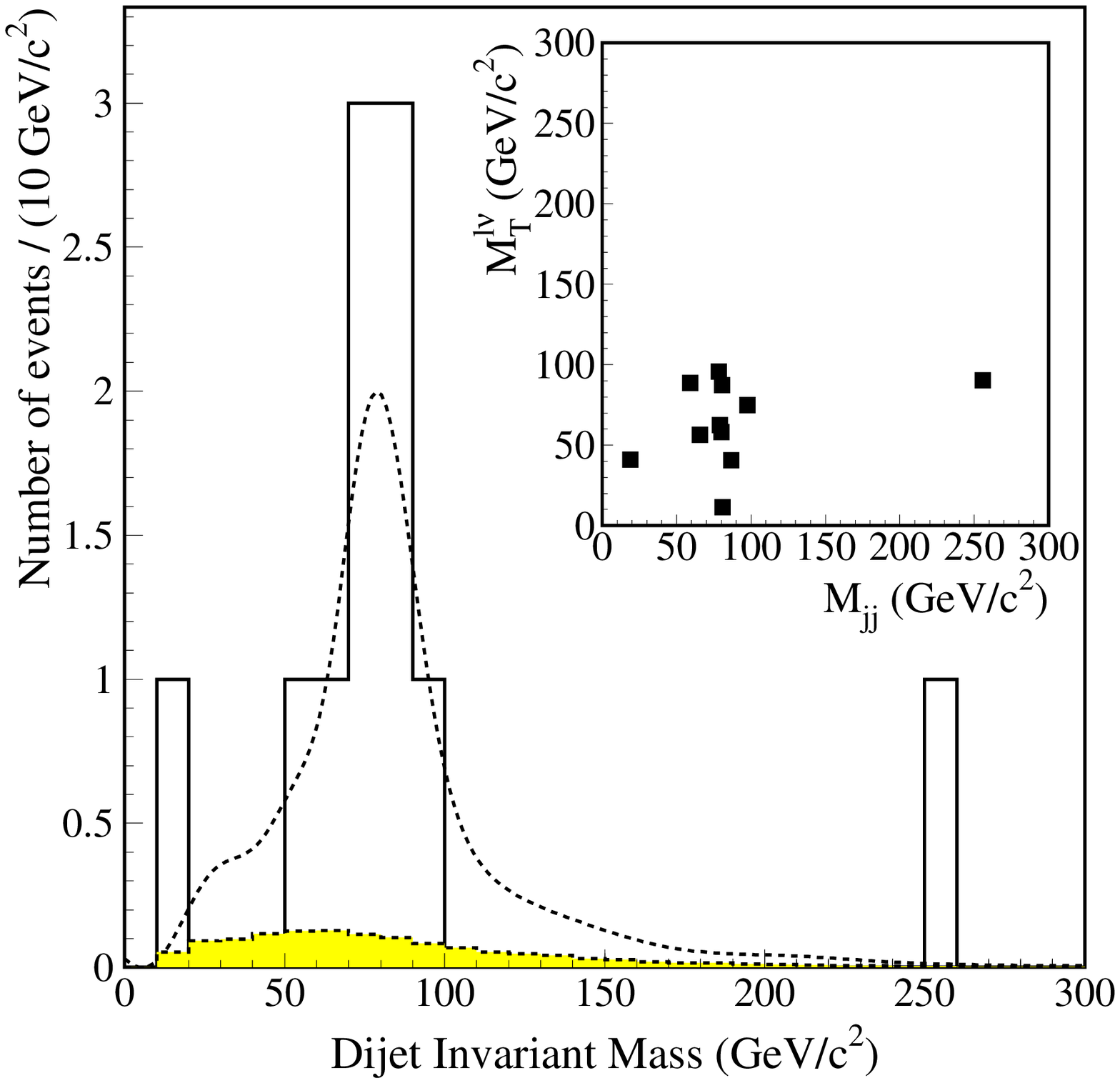,width=\hsize}
\caption{Dijet mass distribution of the two untagged jets in the
  CDF double $b$-tag sample.  The shaded curve is the expected
  background contribution, and the dashed curve is the expectation for
  $\ttbar$ signal plus background.  The inset compares the lepton-$\met$
  transverse mass to the dijet invariant mass for these
  eleven events.
  From Ref.~\protect\onlinecite{cdfhadw97}.
}
\label{fg:cdfhadw2}
\end{figure}

\section{Prospects for Run 2}
\label{prospects}

\subsection{Accelerator and Detector Upgrades}

The Tevatron collider Run~1 ended in February, 1996.  The accelerator
and both detectors are presently undergoing major upgrades for the
next run, due to start in the year~2000.

The major improvement to the accelerator complex will be the addition of
the Main Injector.  This $3319\unit{m}$ circumference machine
is composed of conventional magnets and has a maximum energy of
$150\gev$ and a minimum cycle time of $\sim 1.5\unit{s}$.
It will replace the Main Ring, and will serve as the injector
into the Tevatron.  It will also be used to produce antiprotons,
thus removing the detector deadtimes incurred by running beam
in the Main Ring during collider operations.
(It can also be used to support fixed target experiments
running concurrently with
the collider.)  The Main Injector is expected to increase
the peak luminosity of the Tevatron from 
the present value of
$\sim 2\times 10^{31}\, \text{cm}^{-2}\text{s}^{-1}$ to
$\sim 8\times 10^{31}\, \text{cm}^{-2}\text{s}^{-1}$.
A second machine, the Recycler, will  be built in the same tunnel as the Main
Injector.\cite{recyclertdr}
It will be constructed almost entirely
from permanent magnets, and will operate at a fixed energy of
$8\gev$.  Its principal function will be to serve as
a high-reliability repository
for antiprotons.  It will also be used to capture unused
antiprotons from the Tevatron at the end of a store, recool them,
and use them again in another collider store.  Use of the Recycler
is expected to increase the peak luminosity to
$\sim 2\times 10^{32}\, \text{cm}^{-2}\text{s}^{-1}$.
In addition, the superconducting magnets of the Tevatron will be
operated at a lower temperature; this should enable
an increase of the center-of-mass energy from $\sqrt{s} = 1.8\tev$
to $\sqrt{s} \approx 2.0\tev$.  This corresponds to about a $40\%$ increase
in the $\ttbar$ production cross section.  The goal for the integrated
luminosity in the next collider run is $2\ifb$ per experiment.  
Also envisioned
is a possible Run~3, with additional improvements to the proton and
antiproton sources, which could reach an integrated luminosity
of $\sim 30\ifb$.

Both collider detectors will be
upgraded\mcite{cdfupgrade,*cdfupgrade1,*d0upgrade}
for Run~2.
The \dzero\ detector
will acquire a central magnetic field, a silicon vertex detector,
and a scintillating fiber tracker,
while CDF will upgrade to an expanded vertex detector that will
provide coverage of the entire luminous region of the beam.  These
improvements will significantly increase the efficiency for tagging
$b$-jets in $\ttbar$ decays.  Many other improvements will also be
made to both detectors in outer tracking, calorimetry, and muon coverage.

\subsection{Cross Section and Mass Measurements}

CDF and \dzero\ have made impressively precise 
measurements of the top quark mass:
$m_t = 175.3\pm 6.4\gevcc$, and
$m_t = 172.1\pm 7.1\gevcc$, respectively. The corresponding cross sections
are $\sigtop=7.6^{+1.8}_{-1.5}\pb$ and  $5.6 \pm 1.8\pb$.
These results are based on samples that range in size from about five to
one hundred events. In the next run, 
sample sizes should increase by at least twenty-fold.

The potential for $\ttbar$ physics at the Tevatron during Run~2 and beyond
has been studied extensively.\mcite{tev2k,aspentop}  The expected
yields for each detector are on the order of 160~dilepton events,
1200 $\text{lepton} + \ge 4~\jets$ events,
and 500 double-tagged $\text{lepton} + \ge 4~\jets$
events.  For the cross section measurement, the statistical component
of the uncertainty
scales as $1/\sqrt{N}$.  The dominant contributions to the systematic
uncertainty are the uncertainties in the $\ttbar$ acceptance and background
estimates.  Both of these are amenable to study using control samples
from the data, and therefore can be also expected to scale as $1/\sqrt{N}$.
The limiting factor could be the error on the luminosity, which can be
measured to $\sim 5\%$ using the $W\rightarrow \ell\nu$ rate. With these
assumptions, the precision of the cross section measurement in Run~2
should be about $8$--$10\%$.

For the mass measurement, the statistical uncertainty should again scale
as $1/\sqrt{N}$.  However, the present measurements are already
limited by systematic uncertainties, so the challenge will be to reduce this
component.  One of the dominant uncertainties is that in the
jet energy scale.  However, given  large statistics, the
energy scale can be well characterized using $Z+\text{multijet}$ events.
With an adequate sample of double $b$-tagged events, it should be
possible to use the hadronically-decaying $W$~bosons in
$\ttbar \ra \ell +\jets$ channels
to perform an in situ calibration of the energy scale.
Therefore, the uncertainty on the jet scale should decrease as $1/\sqrt{N}$.
The other major uncertainty is that in modeling QCD~radiative effects
in $\ttbar$ decays.  Presently, these  are modeled using
parton-shower Monte Carlo programs such as \progname{herwig}.
However, some theoretical progress has recently been made in
understanding the phenomenology of gluon radiation in
$\ttbar$ events,\mcite{orr97,masuda96} and the situation should be greatly
helped by the availability of $\ttbar$ data  with sufficient
statistics to provide meaningful constraints on  models.
We therefore expect this uncertainty also to decrease.  The total achievable
uncertainty in the top quark mass measurement from Run~2 should be about
$3$--$4\gevcc$ for each experiment.

A precise measurement of the top quark mass, along with that of
the $W$~boson provides a crucial test of the electroweak theory 
as well as a strong constraint
on the mass of the Higgs boson.  \Figref{fig:mw_mt} shows the correlation
between the W boson mass and top quark mass for a range of Higgs boson masses. 
Also plotted are the current direct and indirect measurements of 
$m_t$ and $M_W$.   

\begin{figure}
\centering
\epsfig{file=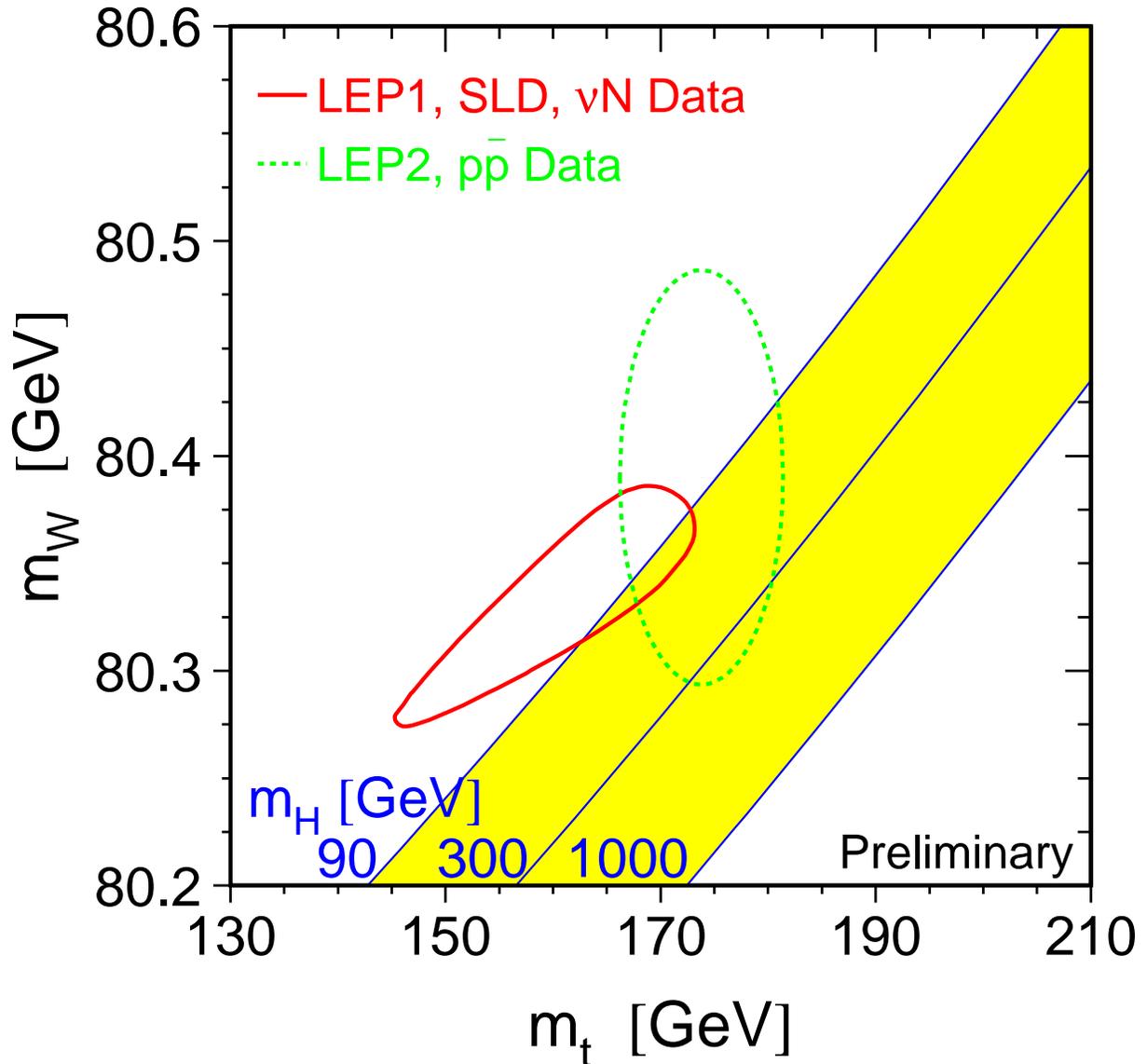,width=\hsize}
\caption{ The correlation between the $W$ boson mass and the top quark mass
as predicted by the electroweak theory of the Standard Model, for various
possible masses of the yet undiscovered Higgs boson mass.
From Ref.\protect\onlinecite{lepewg98}.
(Each line corresponds the mass value shown.)
Also shown 
are the measured $W$ boson mass and the direct and indirect measurements of
the top quark mass.  As the errors in the measurements of $M_W$ and $m_t$
decrease, the Higgs boson mass can be better constrained.
}
\label{fig:mw_mt}
\end{figure}

\subsection{Single Top Quark Production}
\label{singletop}

So far, this review has focussed on the QCD pair production of top~quarks.
As mentioned in \secref{production},
top~quarks can also be produced singly, through the electroweak ``$W^*$''
(\figref{fig:topprod-drellyan}) and $W$-gluon fusion
(\figref{fig:topprod-wgfusion}) processes.%
\mcite{heinson97b,heinson97a,tait97}  (There is also a
contribution from $\ppbar \ra tW$, but the rate for
that process is very small at
the Tevatron.)  The cross sections for these processes have been
calculated at NLO.  The results for $m_t = 175\gevcc$ and
$\sqrt{s} = 2\tev$ are 
$\sigma(p\pbar\rightarrow t\bbar (\tbar b) + X) = 0.9\pb$
for the $W^*$ process,\cite{smith96} and 
$\sigma(p\pbar\rightarrow tq (\tbar\qbar) + X) = 2.4\pb$
for the $W$-gluon fusion process.\cite{stelzer97}
These should be compared to the
resummed NLO $\ttbar$ cross section of $\sim 7\pb$.\cite{bonciani98}

The single top~quark production cross section is directly proportional to
$\Vtb^2$.  Therefore, a measurement of the cross section
gives a measurement of $\Vtb$, independent of any assumptions about
the unitarity of the CKM matrix.  Note that this strict
proportionality fails in the presence of certain forms of new physics,
such as anomalous couplings of the top~quark.  However, evidence for
such new physics should also be present in quantities which do not
depend on $\Vtb$, such as the ratio of the two subprocess cross
sections,\mcite{heinson97b,tait97} so the effects can, in principle,
be disentangled.

One would search for single top~quark events by selecting events
with exactly one high-$\pt$ lepton, large $\met$, and exactly two
jets, one of which has a $b$-tag.  This sort of selection would yield
a signal/background ratio of about $1:2$,\cite{tev2k} the dominant
background being $W\bbbar$ production, with smaller contributions from
$\ttbar$ pair production and other QCD processes.  The signal can be
further enhanced by plotting the distribution of $m(Wb)$ and looking
for an excess at around the known mass of the top~quark, as shown in
\figref{fig:singletop-mwb}.

\begin{figure}
\epsfig{figure=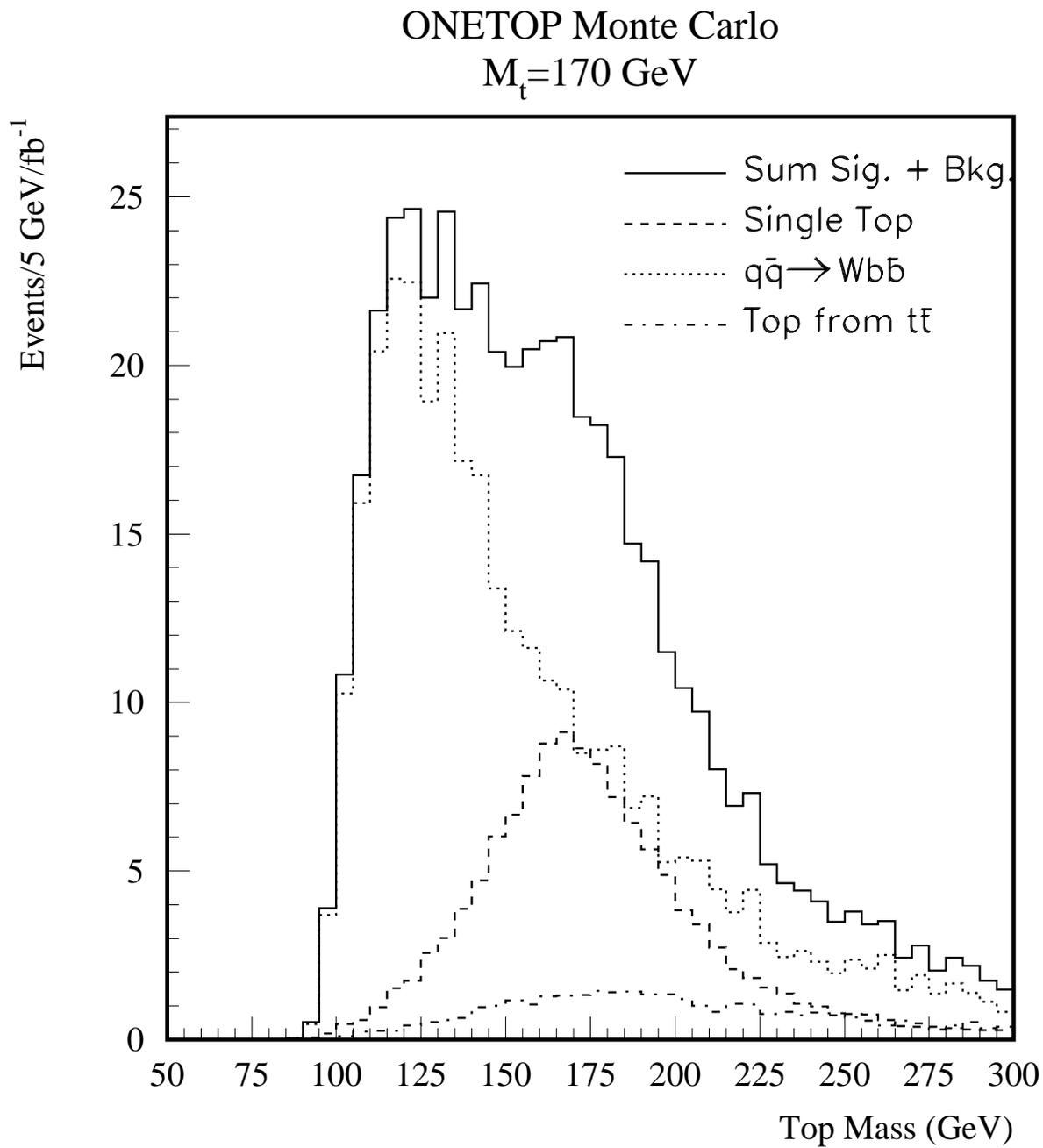,width=\hsize}
\caption{Reconstructed top~quark invariant mass for single top~quark
signal and background, after typical selection cuts.
For $\sqrt{s} = 2.0\tev$ and $m_t = 170\gevcc$.
The vertical scale is normalized to a total
integrated luminosity of $1\ifb$.  From Ref.\protect\onlinecite{tev2k}.}
\label{fig:singletop-mwb}
\end{figure}

In Run~2, with $2\ifb$, $\Vtb$ can probably be measured to a
precision
of $\sim 10\%$.\cite{tev2k}  (The limiting factor in this measurement
is expected to be the error on the total integrated luminosity, which
has been assumed to be $5\%$.)

\subsection{Top and New Physics}
\label{newphysics}

The large mass of the top quark is certainly
allowed in the Standard
Model,  but is not \emph{required}. In fact, the Standard Model offers no
explanation as to why the top quark is so heavy.  But because of its
large mass, the top quark may be a sensitive probe of new physics.
The heaviness of the top quark may, in fact, reflect the presence of
new physics at the electroweak scale. 
Furthermore, as we
discussed in \secref{introduction}, in supersymmetric theories, a
heavy top quark can induce the spontaneous breakdown of electroweak
symmetry.  If this were found to be true, it would represent a major
step forward in our understanding of the origin of mass.  We mention below
some popular theories of new physics, and outline the ways in
which its manifestations might be studied using the top quark.

A very promising class of theories is based on the concept of
supersymmetry.  (See Ref.~\onlinecite{dawson96} for a recent review.)
There are many valid and intriguing reasons for the prevailing
optimism concerning their veracity: 
supersymmetric theories permit the \emph{unification} of
the strong, weak, and electromagnetic coupling constants at a grand
unification (GUT) scale, consistent with the experimental bound
on the proton lifetime; they solve the problems of \emph{gauge
hierarchy} and \emph{naturalness}.  To add to their allure,
\emph{superstring theories}, which may lead to a theory of all
forces, predict supersymmetry!

Supersymmetry (SUSY) is a conjectured symmetry between fermions and
bosons that predicts the existence of a supersymmetric partner for
each particle of the Standard Model. (SUSY can also be construed, in a
more profound way, as an augmentation of spacetime with fermionic
dimensions.)  In some supersymmetric models, top quarks can be
produced in the decay of heavy supersymmetric particles; the top quark
itself may also decay into lighter superpartners.  The study and
precise measurement of top quark production rates can provide useful
information about such nonstandard processes.  It is expected that the
production of single top quarks, proceeding through electroweak
processes, will be particularly sensitive to new physics.
Supersymmetric theories usually impose conservation of ``$R$-parity,''
which is a quantum number defined such that it has a value
of $R = +1$ for normal particles and $R = -1$ for their SUSY partners.
However, $R$-parity conservation is not required, and
if it is not imposed, the
rate of single top quark production $\qqbar \ra b\,\tbar$ can be greatly
enhanced over that predicted by the Standard Model.  Such reactions
would be induced by the exchange of a supersymmetric quark (a
``squark'') in the $t$-channel, which violates baryon number, as well
as by the lepton-number violating exchange of a supersymmetric lepton
(``slepton'') in the $s$-channel.  The decay of a top quark into its
supersymmetric partner (called a top squark or a ``stop'') and the
lightest neutralino, a process predicted in $R$-parity violating SUSY
models, could be observed during the next run of the upgraded
Tevatron.

Another attractive, perhaps radical, approach to understanding
electroweak symmetry breaking and the origin of fermion masses and
mixings, is that of \emph{new strong dynamics}.  In these theories,
new strong interactions of fermions and bosons are invoked at the
scale $\Lambda \sim 1\tev$, and cause the \emph{dynamical} breaking of
electroweak symmetry.  Consequently, no elementary scalar bosons
(such as the Higgs) need exist.  The popular models of new strong
dynamics are those of \emph{technicolor}, \emph{topcolor}, and their
variants.  In the technicolor
model,\mcite{weinberg76,*susskind79,*farhi81,*king94} which is
analogous to QCD, the chiral symmetry is spontaneously broken through
a \emph{technifermion} condensation mechanism, thereby giving masses
to the gauge bosons.  The fermions acquire masses through new
interactions (\emph{extended technicolor}) with the technifermion
condensate.  To account for the heaviness of the top quark, one
postulates the existence of new \emph{topcolor}
interactions.\cite{topcolor} In these models, many signatures would
involve the top quark in the final state.  Particularly interesting
and unique are signatures of resonances that decay to~$\ttbar$.  In
technicolor models,\cite{eichten_lane} a spin-zero color-octet
resonance, the techni-eta, is produced in gluon-gluon collisions and
decays into $\ttbar$ or a gluon pair.  Such a resonance would be seen
as a distortion or broadening of the $\ttbar$ or two-jet invariant
mass distributions, but its effect on the $\bbbar$ mass distribution
would be negligible.  In topcolor models,\cite{hill_parke} a spin-one
\emph{coloron} (top-gluon), produced mainly by quark-antiquark
annihilations, would decay into $\ttbar$ and $\bbbar$ with roughly
equal probability, and would appear as a broad resonance in both
distributions.  If an enhancement were to be observed in the $\ttbar$
mass spectrum, it would be important to study it at the Large Hadron
Collider (LHC),\cite{LHC} due for completion in 2005, at CERN.
This machine will provide $pp$ collisions with a center-of-mass
energy of $\sqrt{s} = 14\tev$, with luminosities of up to
$10^{34}\ \lumunits$.  At the
LHC, 90\% of the top quark pairs are expected to be produced by
gluon-gluon fusion, in contrast with the Tevatron, where 90\% of the
top quark pairs come from quark-antiquark annihilations.  The
aforementioned models have very different predictions of rates and
consequences at the two colliders. The two colliders hold promise of
great discoveries and perhaps some surprises.

Finally, as Chris Quigg has noted recently,\cite{quigg97b} in spite of
its fleeting existence, the top quark may, in fact, have a profound
effect on the basic properties of the everyday world.  At the dawn of
the third millennium, with the improved Tevatron and upgraded CDF and
\dzero\ detectors, we expect to refine the measurements of the
properties of the top quark and to study its dynamics, not only for
its own sake, but also in the hope that we shall be afforded the good
fortune to unravel some of nature's mysteries.

\section*{Acknowledgments}

It is a pleasure to thank our collaborators at CDF and \dzero\ with whom 
we have shared the 
joy of discovering the top quark and making the many measurements reviewed
in this work.  Our special thanks go to Tom Ferbel, Herb Greenlee,
Hugh Montgomery, Chris Quigg, and Rob Roser
for carefully reading the manuscript
and for their many useful suggestions and comments.
We thank the Fermilab Accelerator, Computing, and Research Divisions, and
the support staffs at the collaborating institutions for their contributions
to the success of this work.
This work is supported in part by the U.S. Department of Energy
under contracts DE-AC02-76CHO3000 with Fermi National Accelerator Laboratory, 
DE-AC02-98CH10886 with Brookhaven National Laboratory, and 
DE-FG02-97ER41022 with Florida State University.

\makeatletter\def\mciftail#1#2#3{\mc@iftail{#1}{#2}{#3}}\makeatother
\begin{mcbibliography}{100}

\mcbibitem{cdfdiscovery}
{CDF} Collaboration, F. Abe {\it et~al.}, Phys. Rev. Lett. {\bf 74},  2626
  (1995)\relax
\relax
\mcbibitem{cdftopprd94}
\mciftail{cdftopprd94}{\textit{ibid}}{{CDF} Collaboration, F. Abe {\it
  et~al.}}, Phys. Rev.  {\bf D50},  2966  (1994)\relax
\relax
\mcbibitem{d0discovery}
{\dzero} Collaboration, S. Abachi {\it et~al.}, Phys. Rev. Lett. {\bf 74},
  2632  (1995)\relax
\relax
\mcbibitem{weinberg}
S. Weinberg, Phys. Rev. Lett. {\bf 19},  1264  (1967)\relax
\relax
\mcbibitem{salam}
A. Salam, {\em Elementary Particle Physics} (Almquvist and Wiksells, Stockholm,
  1968)\relax
\relax
\mcbibitem{glashow}
S. Glashow, J.Iliopoulos, and L. Maiani, Phys. Rev.  {\bf D2},  1285
  (1970)\relax
\relax
\mcbibitem{gross1}
D. Gross and F. Wilczek, Phys. Rev.  {\bf D8},  3633  (1973)\relax
\relax
\mcbibitem{gross2}
\mciftail{gross2}{\textit{ibid}}{D. Gross and F. Wilczek}, Phys. Rev. Lett.
  {\bf 30},  1343  (1973)\relax
\relax
\mcbibitem{politzer}
H.~D. Politzer, Phys. Rev. Lett. {\bf 30},  1346  (1973)\relax
\relax
\mcbibitem{lepewg1}
{LEP Electroweak Working Group},  Report No.~{CERN-PPE}/96-183 (1996)\relax
\relax
\mcbibitem{wimpenny96}
S.~J. Wimpenny and B.~L. Winer, Annu. Rev. Nucl. Part. Sci. {\bf 46},  149
  (1996)\relax
\relax
\mcbibitem{franklin97}
C. Campagnari and M. Franklin, Rev. Mod. Phys. {\bf 69},  137  (1997)\relax
\relax
\mcbibitem{pdg96}
R.~M. Barnett {\it et~al.}, Phys. Rev.  {\bf D54},  1  (1996), with updates
  from the 1997 WWW edition at \texttt{http://www.lbl.gov/}\relax
\relax
\mcbibitem{ellis97}
J. Ellis, Int. J. Mod. Phys.  {\bf A12},  5531  (1997)\relax
\relax
\mcbibitem{altarelli98}
G. Altarelli, R. Barbieri, and F. Caravaglios, Int. J. Mod. Phys.  {\bf A13},
  1031  (1998)\relax
\relax
\mcbibitem{wilczek98}
F. Wilczek, Int. J. Mod. Phys.  {\bf A13},  863  (1998)\relax
\relax
\mcbibitem{kane82}
G.~L. Kane and M.~E. Peskin, Nucl. Phys.  {\bf B195},  29  (1982)\relax
\relax
\mcbibitem{bean87}
{CLEO} Collaboration, A. Bean {\it et~al.}, Phys. Rev.  {\bf D35},  3533
  (1987)\relax
\relax
\mcbibitem{roy90}
D.~P. Roy and S.~U. Sankar, Phys. Lett.  {\bf B243},  296  (1990)\relax
\relax
\mcbibitem{albrecht87}
{ARGUS} Collaboration, H. Albrecht {\it et~al.}, Phys. Lett.  {\bf B192},  245
  (1987)\relax
\relax
\mcbibitem{albrecht94}
\mciftail{albrecht94}{\textit{ibid}}{{ARGUS} Collaboration, H. Albrecht {\it
  et~al.}}, Phys. Lett.  {\bf B324},  249  (1994)\relax
\relax
\mcbibitem{bartelt93}
{CLEO} Collaboration, J. Bartelt {\it et~al.}, Phys. Rev. Lett. {\bf 71},  1680
   (1993)\relax
\relax
\mcbibitem{quigg}
C. Quigg, {\em Gauge Theories of the Strong, Weak, and Electromagnetic
  Interactions} (Benjamin/Cummings, Reading, MA, 1990)\relax
\relax
\mcbibitem{bcharge1}
{PLUTO} Collaboration, {Ch.~Berger} {\it et~al.}, Phys. Lett.  {\bf B76},  243
  (1978)\relax
\relax
\mcbibitem{bcharge2}
J.~K. Bienlein {\it et~al.}, Phys. Lett.  {\bf B78},  360  (1978)\relax
\relax
\mcbibitem{bcharge3}
C.~W. Darden {\it et~al.}, Phys. Lett.  {\bf B76},  246  (1978)\relax
\relax
\mcbibitem{lepewg}
{LEP Collaborations}, {LEP Electroweak Group}, and {SLD Heavy Flavour Group},
  Report No.~{CERN-PPE/97-154} (1997)\relax
\relax
\mcbibitem{schaile}
D. Schaile and P.~M. Zerwas, Phys. Rev.  {\bf D45},  3262  (1992)\relax
\relax
\mcbibitem{chanowitz78}
M.~S. Chanowitz, M.~A. Furman, and I. Hinchliffe, Phys. Lett.  {\bf B78},  285
  (1978)\relax
\relax
\mcbibitem{chanowitz79}
\mciftail{chanowitz79}{\textit{ibid}}{M.~S. Chanowitz, M.~A. Furman, and I.
  Hinchliffe}, Nucl. Phys.  {\bf B153},  402  (1979)\relax
\relax
\mcbibitem{ehlq84}
E. Eichten, I. Hinchliffe, K. Lane, and C. Quigg, Rev. Mod. Phys. {\bf 56},
  579  (1984)\relax
\relax
\mcbibitem{martinez98}
M. Martinez, R. Miquel, L. Rolandi, and R. Tenchini,  Report No.~CERN-EP/98-027
  (1998)\relax
\relax
\mcbibitem{simmons}
E.~H. Simmons,  in {\em Beyond the Standard Model V, Balholm, Norway} (AIP,
  Woodbury, NY, 1997)\relax
\relax
\mcbibitem{ibanez}
L. {Ib\'a\~nez}, Nucl. Phys.  {\bf B218},  514  (1983)\relax
\relax
\mcbibitem{gaume}
L. {Alvarez-Gaum\'e}, J. Polchinski, and M.~B. Wise, Nucl. Phys.  {\bf B221},
  495  (1983)\relax
\relax
\mcbibitem{albajar90}
UA1 Collaboration, C. Albajar {\it et~al.}, Z. Phys.  {\bf C48},  1
  (1990)\relax
\relax
\mcbibitem{akesson90}
UA2 Collaboration, T. {\r{A}kesson} {\it et~al.}, Z. Phys.  {\bf C46},  179
  (1990)\relax
\relax
\mcbibitem{cacciari97}
M. Cacciari,  in {\em Symposium on Twenty Beautiful Years of Bottom Physics -
  {b20}} (AIP, Woodbury, NY, 1998)\relax
\relax
\mcbibitem{frixione97}
S. Frixione, M.~L. Mangano, P. Nason, and G. Ridolfi,  in {\em Heavy Flavors
  II}, edited by A.~J. Buras and M. Linder (World Scientific, River Edge, NJ,
  1997)\relax
\relax
\mcbibitem{collins86}
J.~C. Collins, D.~E. Soper, and G. Sterman, Nucl. Phys.  {\bf B263},  37
  (1986)\relax
\relax
\mcbibitem{georgi78}
H.~M. Georgi, S.~L. Glashow, M.~E. Machacek, and D.~V. Nanopoulos, Ann. Phys.
  {\bf 114},  273  (1978)\relax
\relax
\mcbibitem{jones78}
L.~M. Jones and H.~W. Wyld, Phys. Rev.  {\bf D17},  1782  (1978)\relax
\relax
\mcbibitem{gluck78}
M. {Gl\"uck}, J.~F. Owens, and E. Reya, Phys. Rev.  {\bf D17},  2324
  (1978)\relax
\relax
\mcbibitem{babcock78}
J. Babcock, D. Sivers, and S. Wolfram, Phys. Rev.  {\bf D18},  162
  (1978)\relax
\relax
\mcbibitem{hagiwara79}
K. Hagiwara and T. Yoshino, Phys. Lett.  {\bf B80},  282  (1979)\relax
\relax
\mcbibitem{combridge79}
B.~L. Combridge, Nucl. Phys.  {\bf B151},  429  (1979)\relax
\relax
\mcbibitem{nason88}
P. Nason, S. Dawson, and R.~K. Ellis, Nucl. Phys.  {\bf B303},  607
  (1988)\relax
\relax
\mcbibitem{nason89}
\mciftail{nason89}{\textit{ibid}}{P. Nason, S. Dawson, and R.~K. Ellis}, Nucl.
  Phys.  {\bf B327},  49  (1989)\relax
\relax
\mcbibitem{beenakker89}
W. Beenakker, H. Kuijf, W.~L. van Neerven, and J. Smith, Phys. Rev.  {\bf D40},
   54  (1989)\relax
\relax
\mcbibitem{beenakker91}
W. Beenakker {\it et~al.}, Nucl. Phys.  {\bf B351},  507  (1991)\relax
\relax
\mcbibitem{mangano92}
M.~L. Mangano, P. Nason, and G. Ridolfi, Nucl. Phys.  {\bf B373},  295
  (1992)\relax
\relax
\mcbibitem{altarelli88}
G. Altarelli, M. Diemoz, G. Martinelli, and P. Nason, Nucl. Phys.  {\bf B308},
  724  (1988)\relax
\relax
\mcbibitem{ellis91}
R.~K. Ellis, Phys. Lett.  {\bf B259},  492  (1991)\relax
\relax
\mcbibitem{laenen94}
E. Laenen, J. Smith, and W.~L. van Neerven, Phys. Lett.  {\bf B321},  254
  (1994)\relax
\relax
\mcbibitem{laenen92}
\mciftail{laenen92}{\textit{ibid}}{E. Laenen, J. Smith, and W.~L. van Neerven},
  Nucl. Phys.  {\bf B369},  543  (1992)\relax
\relax
\mcbibitem{kidonakis95}
N. Kidonakis and J. Smith, Phys. Rev.  {\bf D51},  6092  (1995)\relax
\relax
\mcbibitem{berger95}
E.~L. Berger and H. Contopanagos, Phys. Lett.  {\bf B361},  115  (1995)\relax
\relax
\mcbibitem{berger96}
\mciftail{berger96}{\textit{ibid}}{E.~L. Berger and H. Contopanagos}, Phys.
  Rev.  {\bf D54},  3085  (1996)\relax
\relax
\mcbibitem{berger97}
\mciftail{berger97}{\textit{ibid}}{E.~L. Berger and H. Contopanagos},  in {\em
  Deep Inelastic Scattering and {QCD}: 5th International Workshop, Chicago, Il}
  (AIP, Woodbury, NY, 1997)\relax
\relax
\mcbibitem{berger98}
\mciftail{berger98}{\textit{ibid}}{E.~L. Berger and H. Contopanagos}, Phys.
  Rev.  {\bf D57},  253  (1998)\relax
\relax
\mcbibitem{contopanagos93}
H. Contopanagos and G. Sterman, Nucl. Phys.  {\bf B400},  211  (1993)\relax
\relax
\mcbibitem{contopanagos94}
\mciftail{contopanagos94}{\textit{ibid}}{H. Contopanagos and G. Sterman}, Nucl.
  Phys.  {\bf B419},  77  (1994)\relax
\relax
\mcbibitem{catani96}
S. Catani, M.~L. Mangano, P. Nason, and L. Trentadue, Phys. Lett.  {\bf B378},
  329  (1996)\relax
\relax
\mcbibitem{bonciani98}
R. Bonciani, S. Catani, M.~L. Mangano, and P. Nason,  Report No.~CERN-TH/98-31
  (1998), hep-ph/9801375\relax
\relax
\mcbibitem{martin96}
A.~D. Martin, R.~G. Roberts, and W.~J. Stirling, Phys. Lett.  {\bf B387},  419
  (1996)\relax
\relax
\mcbibitem{martin93}
A.~D. Martin, W.~J. Stirling, and R.~G. Roberts, Phys. Lett.  {\bf B306},  145
  (1993)\relax
\relax
\mcbibitem{cteq95}
H.~L. Lai {\it et~al.}, Phys. Rev.  {\bf D51},  4763  (1995)\relax
\relax
\mcbibitem{heinson97b}
A.~P. Heinson, A.~S. Belyaev, and E.~E. Boos, Phys. Rev.  {\bf D56},  3114
  (1997)\relax
\relax
\mcbibitem{hollik97}
W. Hollik, W.~M. {M\"osle}, and D. Wackeroth, Nucl. Phys.  {\bf B516},  29
  (1998)\relax
\relax
\mcbibitem{lampe97}
B. Lampe, Phys. Lett.  {\bf B415},  63  (1997)\relax
\relax
\mcbibitem{moretti98}
S. Moretti, J. Phys.  {\bf G24},  525  (1998)\relax
\relax
\mcbibitem{orr91}
L.~H. Orr, Phys. Rev.  {\bf D44},  88  (1991)\relax
\relax
\mcbibitem{bigi86}
I. Bigi, Y. Dokshitzer, V. Khoze, and P. Zerwas, Phys. Lett.  {\bf B181},  157
  (1986)\relax
\relax
\mcbibitem{herwig}
G. Marchesini {\it et~al.}, Comp. Phys. Comm. {\bf 67},  465  (1992)\relax
\relax
\mcbibitem{altarelli77}
G. Altarelli and G. Parisi, Nucl. Phys.  {\bf B126},  298  (1977)\relax
\relax
\mcbibitem{frixione95}
S. Frixione, M.~L. Mangano, P. Nason, and G. Ridolfi, Phys. Lett.  {\bf B351},
  555  (1995)\relax
\relax
\mcbibitem{orr97}
L.~H. Orr, T. Stelzer, and W.~J. Stirling, Phys. Rev.  {\bf D56},  446
  (1997)\relax
\relax
\mcbibitem{mrenna97}
S. Mrenna and C.-P. Yuan, Phys. Rev.  {\bf D55},  120  (1997)\relax
\relax
\mcbibitem{pythia}
T. {Sj\"ostrand}, Comp. Phys. Comm. {\bf 82},  74  (1994)\relax
\relax
\mcbibitem{tevrev}
H.~T. Edwards, Annu. Rev. Nucl. Part. Sci. {\bf 35},  605  (1985)\relax
\relax
\mcbibitem{tev1rep}
Design Report {Tevatron} 1 Project, 1984, {Fermilab} internal note
  (unpublished)\relax
\relax
\mcbibitem{joeythesis}
W.~J. Thompson, Ph.D. thesis, State University of New York at Stony Brook,
  Stony Brook, New York, 1994\relax
\relax
\mcbibitem{apsource}
M.~P. Church and J.~P. Marriner, Annu. Rev. Nucl. Part. Sci. {\bf 43},  253
  (1993)\relax
\relax
\mcbibitem{stcool}
D. M{\"{o}}hl, G. Petrucci, L. Thorndahl, and S. van~der Meer, Phys. Rep. {\bf
  58},  73  (1980)\relax
\relax
\mcbibitem{cdfdetector}
{CDF} Collaboration, F. Abe {\it et~al.}, Nucl. Instrum. Methods Phys. Res.,
  Sect.  {\bf A271},  387  (1988)\relax
\relax
\mcbibitem{cdfsvx}
D. Amidei {\it et~al.}, Nucl. Instrum. Methods Phys. Res., Sect.  {\bf A350},
  73  (1994)\relax
\relax
\mcbibitem{cdfsvx2}
J. Antos {\it et~al.}, Nucl. Instrum. Methods Phys. Res., Sect.  {\bf A360},
  118  (1995)\relax
\relax
\mcbibitem{d0detector}
{\dzero} Collaboration, S. Abachi {\it et~al.}, Nucl. Instrum. Methods Phys.
  Res., Sect.  {\bf A338},  185  (1994)\relax
\relax
\mcbibitem{d0topprd}
\mciftail{d0topprd}{\textit{ibid}}{{\dzero} Collaboration, S. Abachi {\it
  et~al.}}, Phys. Rev.  {\bf D52},  4877  (1995)\relax
\relax
\mcbibitem{d0ljtopmassprd}
{\dzero} Collaboration, B. Abbott {\it et~al.}, Phys. Rev.  {\bf D58},  052001
  (1998)\relax
\relax
\mcbibitem{snowmass}
J.~E. Huth {\it et~al.},  in {\em Research directions for the decade:
  {Snowmass} '90: proceedings}, edited by E.~L. Berger (World Scientific, River
  Edge, NJ, 1990), pp.\ 134--136\relax
\relax
\mcbibitem{cdf4jet93}
{CDF} Collaboration, F. Abe {\it et~al.}, Phys. Rev.  {\bf D47},  4857
  (1993)\relax
\relax
\mcbibitem{cafix96}
{\dzero} Collaboration, R. Kehoe,  in {\em Frascati 1996, Calorimetetry in High
  Energy Physics} (World Scientific, River Edge, NJ, 1996), pp.\ 349--358\relax
\relax
\mcbibitem{cafix98}
{\dzero} Collaboration, B. Abbott {\it et~al.},  Report
  No.~{Fermilab}-Pub-97/330-E (1998), hep-ex/9805009, submitted to Nucl.\
  Instrum.\ Methods\relax
\relax
\mcbibitem{d0wmassprd1b}
\mciftail{d0wmassprd1b}{\textit{ibid}}{{\dzero} Collaboration, B. Abbott {\it
  et~al.}},  Report No.~{Fermilab}-Pub-97/422-E (1997), hep-ex/9712029,
  submitted to Phys.\ Rev.\ D\relax
\relax
\mcbibitem{tb90l1a}
{\dzero} Collaboration, S. Abachi {\it et~al.}, Nucl. Instrum. Methods Phys.
  Res., Sect.  {\bf A324},  53  (1993)\relax
\relax
\mcbibitem{tb90l1b}
{\dzero} Collaboration, H. Aihara {\it et~al.}, Nucl. Instrum. Methods Phys.
  Res., Sect.  {\bf A325},  393  (1993)\relax
\relax
\mcbibitem{cdftoptotau97}
{CDF} Collaboration, F. Abe {\it et~al.}, Phys. Rev. Lett. {\bf 79},  3585
  (1997)\relax
\relax
\mcbibitem{cdfxs98}
\mciftail{cdfxs98}{\textit{ibid}}{{CDF} Collaboration, F. Abe {\it et~al.}},
  Phys. Rev. Lett. {\bf 80},  2773  (1998)\relax
\relax
\mcbibitem{cdfblifetime}
\mciftail{cdfblifetime}{\textit{ibid}}{{CDF} Collaboration, F. Abe {\it
  et~al.}}, Phys. Rev.  {\bf D57},  5382  (1998)\relax
\relax
\mcbibitem{d0xsecprl97}
{\dzero} Collaboration, S. Abachi {\it et~al.}, Phys. Rev. Lett. {\bf 79},
  1203  (1997)\relax
\relax
\mcbibitem{barger93}
V. Barger, J. Ohnemus, and R.~J.~N. Phillips, Phys. Rev.  {\bf D48},  3953
  (1993)\relax
\relax
\mcbibitem{baer89}
H. Baer, V. Barger, and R.~J.~N. Phillips, Phys. Rev.  {\bf D39},  3310
  (1989)\relax
\relax
\mcbibitem{likhoded90}
A.~K. Likhoded and S.~R. Slabospitskii, Yad. Fiz. {\bf 52},  1106  (1990)\relax
\relax
\mcbibitem{herb77}
S.~W. Herb {\it et~al.}, Phys. Rev. Lett. {\bf 39},  252  (1977)\relax
\relax
\mcbibitem{behrend84}
CELLO Collaboration, H.-J. Behrend {\it et~al.}, Phys. Lett.  {\bf B144},  297
  (1984)\relax
\relax
\mcbibitem{bartel79a}
JADE Collaboration, W. Bartel {\it et~al.}, Phys. Lett.  {\bf B88},  171
  (1979)\relax
\relax
\mcbibitem{bartel79b}
\mciftail{bartel79b}{\textit{ibid}}{JADE Collaboration, W. Bartel {\it
  et~al.}}, Phys. Lett.  {\bf B89},  136  (1979)\relax
\relax
\mcbibitem{bartel81}
\mciftail{bartel81}{\textit{ibid}}{JADE Collaboration, W. Bartel {\it et~al.}},
  Phys. Lett.  {\bf B99},  277  (1981)\relax
\relax
\mcbibitem{barber79}
MARK-J Collaboration, D.~P. Barber {\it et~al.}, Phys. Lett.  {\bf B85},  463
  (1979)\relax
\relax
\mcbibitem{barber80}
\mciftail{barber80}{\textit{ibid}}{MARK-J Collaboration, D.~P. Barber {\it
  et~al.}}, Phys. Rev. Lett. {\bf 44},  1722  (1980)\relax
\relax
\mcbibitem{adeva83a}
MARK-J Collaboration, B. Adeva {\it et~al.}, Phys. Rev. Lett. {\bf 50},  799
  (1983)\relax
\relax
\mcbibitem{adeva83b}
\mciftail{adeva83b}{\textit{ibid}}{MARK-J Collaboration, B. Adeva {\it
  et~al.}}, Phys. Rev. Lett. {\bf 51},  443  (1983)\relax
\relax
\mcbibitem{adeva85}
\mciftail{adeva85}{\textit{ibid}}{MARK-J Collaboration, B. Adeva {\it et~al.}},
  Phys. Lett.  {\bf B152},  439  (1985)\relax
\relax
\mcbibitem{adeva86}
\mciftail{adeva86}{\textit{ibid}}{MARK-J Collaboration, B. Adeva {\it et~al.}},
  Phys. Rev.  {\bf D34},  681  (1986)\relax
\relax
\mcbibitem{berger79}
PLUTO Collaboration, C. Berger {\it et~al.}, Phys. Lett.  {\bf B86},  413
  (1979)\relax
\relax
\mcbibitem{brandelik82}
TASSO Collaboration, R. Brandelik {\it et~al.}, Phys. Lett.  {\bf B113},  499
  (1982)\relax
\relax
\mcbibitem{althoff84a}
TASSO Collaboration, M. Althoff {\it et~al.}, Z. Phys.  {\bf C22},  307
  (1984)\relax
\relax
\mcbibitem{althoff84b}
\mciftail{althoff84b}{\textit{ibid}}{TASSO Collaboration, M. Althoff {\it
  et~al.}}, Phys. Lett.  {\bf B138},  441  (1984)\relax
\relax
\mcbibitem{sagawa88}
AMY Collaboration, H. Sagawa {\it et~al.}, Phys. Rev. Lett. {\bf 60},  93
  (1988)\relax
\relax
\mcbibitem{igarashi88}
AMY Collaboration, S. Igarashi {\it et~al.}, Phys. Rev. Lett. {\bf 60},  2359
  (1988)\relax
\relax
\mcbibitem{adachi88}
TOPAZ Collaboration, I. Adachi {\it et~al.}, Phys. Rev. Lett. {\bf 60},  97
  (1988)\relax
\relax
\mcbibitem{yoshida87}
VENUS Collaboration, H. Yoshida {\it et~al.}, Phys. Lett.  {\bf B198},  570
  (1987)\relax
\relax
\mcbibitem{abe90}
VENUS Collaboration, K. Abe {\it et~al.}, Phys. Lett.  {\bf B234},  382
  (1990)\relax
\relax
\mcbibitem{abrams89}
MARK II Collaboration, G.~S. Abrams {\it et~al.}, Phys. Rev. Lett. {\bf 63},
  2447  (1989)\relax
\relax
\mcbibitem{decamp90}
ALEPH Collaboration, D. Decamp {\it et~al.}, Phys. Lett.  {\bf B236},  511
  (1990)\relax
\relax
\mcbibitem{abreu90}
DELPHI Collaboration, P. Abreu {\it et~al.}, Phys. Lett.  {\bf B242},  536
  (1990)\relax
\relax
\mcbibitem{akrawy90}
OPAL Collaboration, M.~Z. Akrawy {\it et~al.}, Phys. Lett.  {\bf B236},  364
  (1990)\relax
\relax
\mcbibitem{isrppbar}
{CERN ISR Division},  Report No.~{CERN/ISR-DI/82-02} (1982)\relax
\relax
\mcbibitem{arnison84}
UA1 Collaboration, G. Arnison {\it et~al.}, Phys. Lett.  {\bf B147},  493
  (1984)\relax
\relax
\mcbibitem{albajar88}
UA1 Collaboration, C. Albajar {\it et~al.}, Z. Phys.  {\bf C37},  505
  (1988)\relax
\relax
\mcbibitem{cdfejets90}
{CDF} Collaboration, F. Abe {\it et~al.}, Phys. Rev. Lett. {\bf 64},  142
  (1990)\relax
\relax
\mcbibitem{cdfemu90}
\mciftail{cdfemu90}{\textit{ibid}}{{CDF} Collaboration, F. Abe {\it et~al.}},
  Phys. Rev. Lett. {\bf 64},  147  (1990)\relax
\relax
\mcbibitem{cdfejets91}
\mciftail{cdfejets91}{\textit{ibid}}{{CDF} Collaboration, F. Abe {\it et~al.}},
  Phys. Rev.  {\bf D43},  664  (1991)\relax
\relax
\mcbibitem{cdftopprl92}
\mciftail{cdftopprl92}{\textit{ibid}}{{CDF} Collaboration, F. Abe {\it
  et~al.}}, Phys. Rev. Lett. {\bf 68},  447  (1992)\relax
\relax
\mcbibitem{cdftopprd92}
\mciftail{cdftopprd92}{\textit{ibid}}{{CDF} Collaboration, F. Abe {\it
  et~al.}}, Phys. Rev.  {\bf D45},  3921  (1992)\relax
\relax
\mcbibitem{d0xsecprl94}
{\dzero} Collaboration, S. Abachi {\it et~al.}, Phys. Rev. Lett. {\bf 72},
  2138  (1994)\relax
\relax
\mcbibitem{cdftopprl94}
{CDF} Collaboration, F. Abe {\it et~al.}, Phys. Rev. Lett. {\bf 73},  225
  (1994)\relax
\relax
\mcbibitem{d0xsecprl95}
{\dzero} Collaboration, S. Abachi {\it et~al.}, Phys. Rev. Lett. {\bf 74},
  2422  (1995)\relax
\relax
\mcbibitem{kim95}
{CDF} Collaboration, S.-B. Kim,  in {\em Physics in Collision 15: Proceedings}
  (World Scientific, River Edge, NJ, 1996), pp.\ 1--26\relax
\relax
\mcbibitem{pushpadpf}
{\dzero} Collaboration, P.~C. Bhat,  in {\em Proceedings of the 8th Meeting of
  the Division of Particles and Fields of the American Physical Society,
  Albuquerque, New Mexico, USA} (World Scientific, River Edge, NJ, 1994), p.\
  705\relax
\relax
\mcbibitem{isajet}
F. Paige and S. Protopopescu, {BNL} Report {38034}, {Brookhaven, 1986}
  (unpublished)\relax
\relax
\mcbibitem{vecbos}
F.~A. Berends, H. Kuijf, B. Tausk, and W.~T. Giele, Nucl. Phys.  {\bf B357},
  32  (1991)\relax
\relax
\mcbibitem{cdfdilep98}
{CDF} Collaboration, F. Abe {\it et~al.}, Phys. Rev. Lett. {\bf 80},  2779
  (1998)\relax
\relax
\mcbibitem{cdfalljets}
\mciftail{cdfalljets}{\textit{ibid}}{{CDF} Collaboration, F. Abe {\it et~al.}},
  Phys. Rev. Lett. {\bf 79},  1992  (1997)\relax
\relax
\mcbibitem{tauola}
R. Decker {\it et~al.}, Comp. Phys. Comm. {\bf 76},  361  (1993)\relax
\relax
\mcbibitem{gerdes97}
{CDF} Collaboration, D.~W. Gerdes,  in {\em Proceedings of the 32nd Rencontres
  de Moriond: Electroweak Interactions and Unified Theories, Les Arcs, France}
  (Editions Frontieres, Gif-sur-Yvette, France, 1997), pp.\ 69--76\relax
\relax
\mcbibitem{pushpapbarp}
{\dzero} Collaboration, P.~C. Bhat,  in {\em 10th Topical Workshop on
  Proton-Antiproton Collider Physics: Batavia, Il} (AIP, Woodbury, NY, 1995),
  p.\ 308\relax
\relax
\mcbibitem{rgs}
{\dzero} Collaboration, H. Prosper {\it et~al.},  in {\em Proceedings of the
  International Conference on Computing in High Energy Physics '95, Rio de
  Janeiro, Brazil} (World Scientific, River Edge, NJ, 1996)\relax
\relax
\mcbibitem{amos98}
{\dzero} Collaboration, N. Amos,  in {\em Proceedings of the 33rd Rencontres de
  Moriond, QCD and High Energy Hadronic Interactions} (Editions Frontieres,
  Gif-sur-Yvette, France, 1998)\relax
\relax
\mcbibitem{klima98}
{\dzero} Collaboration, B. Klima,  in {\em Results and Perspectives in Particle
  Physics, La Thuile, Italy} (Editions Frontieres, Gif-sur-Yvette, France,
  1998), hep-ex/9804017, Fermilab-Conf-98/137-E\relax
\relax
\mcbibitem{d0alljetsprd}
{\dzero} Collaboration, B. Abbott {\it et~al.},  Report
  No.~{Fermilab}-Pub-98/130-E (1998), hep-ex/9808034, submitted to Phys.\ Rev.\
  D\relax
\relax
\mcbibitem{tkachov95}
F. Tkachov, Int. J. Mod. Phys.  {\bf A12},  5411  (1997)\relax
\relax
\mcbibitem{giele97}
W.~T. Giele and E.~W.~N. Glover,  Report No.~{Fermilab-Pub-97/413-T} (1997),
  submitted to Phys. Rev. Lett\relax
\relax
\mcbibitem{MaxLikelihood}
F. James, Comp. Phys. Comm. {\bf 20},  29  (1980)\relax
\relax
\mcbibitem{prosper95}
H.~B. Prosper, Phys. Lett.  {\bf B335},  515  (1994)\relax
\relax
\mcbibitem{roser}
R. Roser, private communication\relax
\relax
\mcbibitem{yao98}
{CDF} Collaboration, W. Yao,  in {\em Proceedings of the {XXIX} International
  Conference on High Energy Physics, Vancouver, Canada} (World Scientific,
  River Edge, NJ, 1998)\relax
\relax
\mcbibitem{cdftopmass98}
{CDF} Collaboration, F. Abe {\it et~al.}, Phys. Rev. Lett. {\bf 80},  2767
  (1998)\relax
\relax
\mcbibitem{d0jetshape}
{\dzero} Collaboration, S. Abachi {\it et~al.}, Phys. Lett.  {\bf B357},  500
  (1995)\relax
\relax
\mcbibitem{FredJames}
F. James, Rep. Prog. Phys. {\bf 43},  1145  (1980)\relax
\relax
\mcbibitem{kondo88}
K. Kondo, J. Phys. Soc. Japan {\bf 57},  4126  (1988)\relax
\relax
\mcbibitem{kondo91}
\mciftail{kondo91}{\textit{ibid}}{K. Kondo}, J. Phys. Soc. Japan {\bf 60},  836
   (1991)\relax
\relax
\mcbibitem{kondo93}
\mciftail{kondo93}{\textit{ibid}}{K. Kondo}, J. Phys. Soc. Japan {\bf 62},
  1177  (1993)\relax
\relax
\mcbibitem{dalitz92a}
R.~H. Dalitz and G.~R. Goldstein, Phys. Rev.  {\bf D45},  1531  (1992)\relax
\relax
\mcbibitem{dalitz92b}
\mciftail{dalitz92b}{\textit{ibid}}{R.~H. Dalitz and G.~R. Goldstein}, Phys.
  Lett.  {\bf B287},  225  (1992)\relax
\relax
\mcbibitem{d0llmass}
{\dzero} Collaboration, B. Abbott {\it et~al.}, Phys. Rev. Lett. {\bf 80},
  2063  (1998)\relax
\relax
\mcbibitem{d0llmassprd}
\mciftail{d0llmassprd}{\textit{ibid}}{{\dzero} Collaboration, B. Abbott {\it
  et~al.}},  Report No.~{Fermilab}-Pub-98/261-E (1998), hep-ex/9808029,
  submitted to Phys.\ Rev.\ D\relax
\relax
\mcbibitem{bayes}
G. D'Agostini,  Report No.~{DESY-95-242} (1995)\relax
\relax
\mcbibitem{holmstroem95}
L. {Holmstr\"om}, S.~R. Sain, and H.~E. Miettinen, Comp. Phys. Comm. {\bf 88},
  195  (1995)\relax
\relax
\mcbibitem{d0ljmassprl}
{\dzero} Collaboration, S. Abachi {\it et~al.}, Phys. Rev. Lett. {\bf 79},
  1197  (1997)\relax
\relax
\mcbibitem{strovink95}
{\dzero} Collaboration, M. Strovink,  in {\em Proceedings of the {XI} Symposium
  on Hadron Collider Physics, Padua, Italy} (World Scientific, River Edge, NJ,
  1996), {Fermilab}-Conf-96/336-E\relax
\relax
\mcbibitem{bayesplb}
P.~C. Bhat, H.~B. Prosper, and S. Snyder, Phys. Lett.  {\bf B407},  73
  (1997)\relax
\relax
\mcbibitem{boaz}
B. Klima, private communication\relax
\relax
\mcbibitem{cabibbo63}
N. Cabibbo, Phys. Rev. Lett. {\bf 10},  531  (1963)\relax
\relax
\mcbibitem{km73}
M. Kobayashi and T. Maskawa, Prog. Theor. Phys. {\bf 49},  652  (1973)\relax
\relax
\mcbibitem{tartarelli97}
{CDF} Collaboration, G.~F. Tartarelli,  in {\em Proceedings of International
  Europhysics Conference on High-Energy Physics (HEP 97), Jerusalem, Israel}
  (World Scientific, River Edge, NJ, 1997), {Fermilab}-Conf-97/401-E\relax
\relax
\mcbibitem{heinson97a}
A.~P. Heinson,  in {\em Honolulu 1997, {$B$} Physics and {CP} Violation} (World
  Scientific, River Edge, NJ, 1997), pp.\ 369--376, {Fermilab}-Conf-97/238-E,
  hep-ex/9707026\relax
\relax
\mcbibitem{tev2k}
The TeV-2000 Group Report, 1996, {Fermilab}-Pub-96/082\relax
\relax
\mcbibitem{swain97}
J. Swain and L. Taylor,  Report No.~hep-ph/9712420 (1997)\relax
\relax
\mcbibitem{gim}
S.~L. Glashow, J. Iliopoulos, and L. Maiani, Phys. Rev.  {\bf D2},  1285
  (1970)\relax
\relax
\mcbibitem{eilam91}
G. Eilam, J.~L. Hewett, and A. Soni, Phys. Rev.  {\bf D44},  1473  (1991)\relax
\relax
\mcbibitem{mele98}
B. Mele, S. Petrarca, and A. Soddu, Phys. Lett.  {\bf B435},  401  (1998)\relax
\relax
\mcbibitem{fritzsch89}
H. Fritzsch, Phys. Lett.  {\bf B224},  423  (1989)\relax
\relax
\mcbibitem{atwood95}
D. Atwood, L. Reina, and A. Soni, Phys. Rev.  {\bf D53},  1199  (1996)\relax
\relax
\mcbibitem{luke93}
M. Luke and M.~J. Savage, Phys. Lett.  {\bf B307},  387  (1993)\relax
\relax
\mcbibitem{diaz93}
J.~L. Diaz-Cruz and G.~L. Castro, Phys. Lett.  {\bf B301},  405  (1993)\relax
\relax
\mcbibitem{grzadkowski91}
B. Grzadkowski, J.~F. Gunion, and P. Krawczyk, Phys. Lett.  {\bf B268},  106
  (1991)\relax
\relax
\mcbibitem{agashe95}
K. Agashe and M. Graesser, Phys. Rev.  {\bf D54},  4445  (1995)\relax
\relax
\mcbibitem{yang94}
J.~M. Yang and C.~S. Li, Phys. Rev.  {\bf D49},  3412  (1994)\relax
\relax
\mcbibitem{li94}
C.~S. Li, R.~J. Oakes, and J.~M. Yang, Phys. Rev.  {\bf D49},  293
  (1994)\relax
\relax
\mcbibitem{yang97}
A. Datta, J.~M. Yang, B.-L. Young, and X. Zhang, Phys. Rev.  {\bf D56},  3107
  (1997)\relax
\relax
\mcbibitem{buchmuller89}
W. {Buchm\"uller} and M. Gronau, Phys. Lett.  {\bf B220},  641  (1989)\relax
\relax
\mcbibitem{han96}
T. Han, K. Whisnant, B.-L. Young, and X. Zhang, Phys. Rev.  {\bf D55},  7241
  (1997)\relax
\relax
\mcbibitem{han95}
T. Han, R. Peccei, and X. Zhang, Nucl. Phys.  {\bf B454},  527  (1995)\relax
\relax
\mcbibitem{cdffcnc97}
{CDF} Collaboration, F. Abe {\it et~al.}, Phys. Rev. Lett. {\bf 80},  2525
  (1998)\relax
\relax
\mcbibitem{hhguide}
J.~F. Gunion, H.~E. Haber, G. Kane, and S. Dawson, {\em The {Higgs} Hunter's
  Guide} (Addison-Wesley, New York, 1990)\relax
\relax
\mcbibitem{ma98}
E. Ma, D.~P. Roy, and J. Wudka, Phys. Rev. Lett. {\bf 80},  1162  (1998)\relax
\relax
\mcbibitem{bevensee98}
{CDF} Collaboration, B. Bevensee,  in {\em Proceedings of the 33rd Rencontres
  de Moriond, QCD and High Energy Hadronic Interactions} (Editions Frontieres,
  Gif-sur-Yvette, France, 1998), {Fermilab}-Conf-98/155-E\relax
\relax
\mcbibitem{barger90}
V. Barger, J.~L. Hewett, and R.~J.~N. Phillips, Phys. Rev.  {\bf D41},  3421
  (1990)\relax
\relax
\mcbibitem{delphihiggs97}
{DELPHI} Collaboration, P. Abreu {\it et~al.}, Phys. Lett.  {\bf B420},  140
  (1998)\relax
\relax
\mcbibitem{opalhiggs97}
{OPAL} Collaboration, K. Ackerstaff {\it et~al.}, Phys. Lett.  {\bf B426},  180
   (1998)\relax
\relax
\mcbibitem{alephhiggs97}
{ALEPH} Collaboration, R. Barate {\it et~al.}, Phys. Lett.  {\bf B418},  419
  (1998)\relax
\relax
\mcbibitem{opalhiggs96}
{OPAL} Collaboration, G. Alexander {\it et~al.}, Phys. Lett.  {\bf B370},  174
  (1996)\relax
\relax
\mcbibitem{delphihiggs94}
{DELPHI} Collaboration, P. Abreu {\it et~al.}, Z. Phys.  {\bf C64},  183
  (1994)\relax
\relax
\mcbibitem{l3higgs92}
{L3} Collaboration, O. Adriani {\it et~al.}, Phys. Lett.  {\bf B294},  457
  (1992)\relax
\relax
\mcbibitem{alephsearches92}
{ALEPH} Collaboration, D. Decamp {\it et~al.}, Phys. Rep. {\bf 216},  253
  (1992)\relax
\relax
\mcbibitem{dejong98}
S. de~Jong, Higgs searches at {LEP}, talk at 1998 Rencontres de Moriond\relax
\relax
\mcbibitem{ua1higgs91}
{UA1} Collaboration, C. Albajar {\it et~al.}, Phys. Lett.  {\bf B257},  459
  (1991)\relax
\relax
\mcbibitem{ua2higgs92}
{UA2} Collaboration, J. Alitti {\it et~al.}, Phys. Lett.  {\bf B280},  137
  (1992)\relax
\relax
\mcbibitem{cdfhiggs96}
{CDF} Collaboration, F. Abe {\it et~al.}, Phys. Rev.  {\bf D54},  735
  (1996)\relax
\relax
\mcbibitem{cdfhiggs94}
\mciftail{cdfhiggs94}{\textit{ibid}}{{CDF} Collaboration, F. Abe {\it et~al.}},
  Phys. Rev. Lett. {\bf 73},  2667  (1994)\relax
\relax
\mcbibitem{guchait97}
M. Guchait and D.~P. Roy, Phys. Rev.  {\bf D55},  7263  (1997)\relax
\relax
\mcbibitem{stahl97}
A. Stahl and H. Voss, Z. Phys.  {\bf C74},  73  (1997)\relax
\relax
\mcbibitem{cleopenguin95}
{CLEO} Collaboration, M.~S. Alam {\it et~al.}, Phys. Rev. Lett. {\bf 74},  2885
   (1995)\relax
\relax
\mcbibitem{goto96}
T. Goto and Y. Okada, Prog. Theor. Phys. Suppl. {\bf 123},  213  (1996)\relax
\relax
\mcbibitem{cdfhiggs97}
{CDF} Collaboration, F. Abe {\it et~al.}, Phys. Rev. Lett. {\bf 79},  357
  (1997)\relax
\relax
\mcbibitem{leone97}
{CDF} Collaboration, S. Leone, Nucl. Phys. Proc. Suppl. {\bf 64},  406
  (1998)\relax
\relax
\mcbibitem{cdfhadw97}
{CDF} Collaboration, F. Abe {\it et~al.}, Phys. Rev. Lett. {\bf 80},  5720
  (1998)\relax
\relax
\mcbibitem{alephzbb93}
{ALEPH} Collaboration, D. Buskulic {\it et~al.}, Phys. Lett.  {\bf B313},  535
  (1993)\relax
\relax
\mcbibitem{recyclertdr}
G. Jackson {\it et~al.}, {Fermilab-TM}-1991, {Fermilab, 1996}
  (unpublished)\relax
\relax
\mcbibitem{cdfupgrade}
{CDF} Collaboration, C. Newman-Holmes,  Report No.~{Fermilab-Conf-96/218-E}
  (1996)\relax
\relax
\mcbibitem{cdfupgrade1}
{CDF} Collaboration, R. Blair {\it et~al.},  Report
  No.~{Fermilab-Conf-96/390-E} (1996)\relax
\relax
\mcbibitem{d0upgrade}
{\dzero} Collaboration, S. Abachi {\it et~al.},  Report
  No.~{Fermilab-Pub-96/357-E} (1996)\relax
\relax
\mcbibitem{aspentop}
R. Frey {\it et~al.},  Report No.~{Fermilab}-Conf-97/085 (1997),
  hep-ph/9704243\relax
\relax
\mcbibitem{masuda96}
B. Masuda, L.~H. Orr, and W.~J. Stirling, Phys. Rev.  {\bf D54},  4453
  (1996)\relax
\relax
\mcbibitem{lepewg98}
{LEP Collaborations}, {LEP Electroweak Group}, and {SLD Heavy Flavour Group},
  1998, http://www.cern.ch/LEPEWWG/plots/summer98\relax
\relax
\mcbibitem{tait97}
T. Tait and C.-P. Yuan,  Report No.~{MSUHEP}-71015, hep-ph/9710372 (1997)\relax
\relax
\mcbibitem{smith96}
M.~C. Smith and S. Willenbrock, Phys. Rev.  {\bf D54},  6696  (1996)\relax
\relax
\mcbibitem{stelzer97}
T. Stelzer, Z. Sullivan, and S. Willenbrock, Phys. Rev.  {\bf D56},  5919
  (1997)\relax
\relax
\mcbibitem{dawson96}
S. Dawson,  in {\em Techniques and Concepts of High-Energy Physics {IX}},
  edited by T. Ferbel (Plenum Press, New York, 1997), hep-ph/9612229\relax
\relax
\mcbibitem{weinberg76}
S. Weinberg, Phys. Rev.  {\bf D13},  974  (1976)\relax
\relax
\mcbibitem{susskind79}
L. Susskind, Phys. Rev.  {\bf D20},  2619  (1979)\relax
\relax
\mcbibitem{farhi81}
E. Farhi and L. Susskind, Phys. Rep. {\bf 74},  277  (1981)\relax
\relax
\mcbibitem{king94}
S.~F. King, Rep. Prog. Phys. {\bf 58},  263  (1995)\relax
\relax
\mcbibitem{topcolor}
C. Hill, Phys. Lett.  {\bf B345},  483  (1995)\relax
\relax
\mcbibitem{eichten_lane}
E. Eichten and K. Lane, Phys. Lett.  {\bf B327},  129  (1994)\relax
\relax
\mcbibitem{hill_parke}
C. Hill and S. Parke, Phys. Rev.  {\bf D49},  4454  (1994)\relax
\relax
\mcbibitem{LHC}
{See for example J. P. Gourber},  Report No.~{CERN-LHC-Project-Report-167}
  (1998)\relax
\relax
\mcbibitem{quigg97b}
C. Quigg, Physics Today {\bf 50},  20  (1997), (No. 5)\relax
\relax
\end{mcbibliography}

\end{document}